\documentclass[onecolumn,amsmath,amssymb,prd,reprint,longbibliography,floatfix,8pt]{revtex4-1}

\usepackage{amsmath}
\usepackage{tikz}
\usepackage{cancel}

\usetikzlibrary{arrows.meta, positioning, arrows, decorations.pathmorphing, backgrounds, fit, petri}
\usepackage{braket}
\usepackage{enumitem}
\usepackage[utf8x]{inputenc}
\usepackage{graphicx}
\usepackage{dcolumn}
\usepackage{bm}
\usepackage[switch]{lineno}
\usepackage{float} 
\usepackage{tikz}
\usepackage{multirow}
\usepackage{booktabs}
\usepackage{siunitx}
\begin{document}

\title{A machine-learning framework for accelerating spin-lattice relaxation simulations}
%\title{Accelerating ab initio simulations of spin-lattice relaxation with machine learning force fields}

\author{Valerio Briganti}
\author{Alessandro Lunghi}
\email{lunghia@tcd.ie}
\affiliation{School of Physics, AMBER and CRANN Institute, Trinity College, Dublin 2, Ireland}

\begin{abstract}
{\bf Molecular and lattice vibrations are able to couple to the spin of electrons and lead to their relaxation and decoherence. Ab initio simulations have played a fundamental role in shaping our understanding of this process but further progress is hindered by their high computational cost. Here we present an accelerated computational framework based on machine-learning models for the prediction of molecular vibrations and spin-phonon coupling coefficients. We apply this method to three open-shell coordination compounds exhibiting long relaxation times and show that this approach achieves semi-to-full quantitative agreement with ab initio methods reducing the computational cost by about 80\%. Moreover, we show that this framework naturally extends to molecular dynamics simulations, paving the way to the study of spin relaxation in condensed matter beyond simple equilibrium harmonic thermal baths.}

%The simulation of spin phonon relaxation is a crucial tool to understand the interplay of geometry, magnetism and electronic structure at the molecular scale.
%State of the art calculations in this field are based on open quantum systems theory and require highly expensive full ab initio calculations of phonons spectra, posing severe limitations on the number of systems that can be studied. 

%Here we present a framework to circumvent this cost by fitting machine learning force fields for fast and accurate prediction of phonons requiring minimal user intervention and only a few \textit{ab initio} calculations. Its robustness and accuracy is tested on three different open shell compounds, two with Co$^{2+}$ and one with Dy$^{3+}$ as a metal center, which are potential candidates for quantum technologies application. Furthermore, we deploy machine learning to predict the correlation functions of tensorial quantities related to spin phonon relaxation, such as the zero field splitting tensor and tesseral operators, over machine learning generated molecular dynamics trajectories of several ns. We show that this framework guarantees a substantial computational speed up with a negligible accuracy loss with respect to a full ab initio approach, paving the way for the possibility to rapidly predict spin phonon relaxation and speed up the discovery of highly performing single molecule magnets.}

\end{abstract}

\maketitle

\section{Introduction}

The study of spin dynamics of coordination compounds is central to the most recent advancements in different fields such as information storage, sensing, information processing, spintronics and paramagnetic resonance for biosystems \cite{coronado_nature_mat_review_spintronics,molecular_approach_to_quantum_sensing,quantum_internet_laorenza}. Understanding how spin relaxes towards equilibrium in molecules represents a long-standing puzzle whose solution would lead to a better interpretation of experiments as well as powering computational design of new magnetic materials \cite{Chapter_Ale_Rajaraman_book}. 

Ab initio spin dynamics simulations based on open quantum systems theory, density functional theory (DFT) and multiconfigurational quantum chemistry techniques are nowadays considered a gold standard tool to perform spin relaxation simulations \cite{Chapter_Ale_Rajaraman_book}. This methodology has now been successfully applied to a large number of magnetic systems, including spin-1/2\cite{garlatti2023critical,mariano2024role,non_adiabatic_relaxation_qubits_shushkov_2024}, single-molecule magnets\cite{Unravelling_JACS, reta_2021_ab_initio,briganti_2021_complete,science_advances_Ale_2022}, solid-state defects\cite{sourav_npj_2023,cambria2023temperature}, and surface-adsorbed atoms\cite{garai2023microscopic}. However, these simulations require the numerical calculation of phonons for hundreds of atoms and derivatives with multi-reference ab initio methods, making predictions extremely expensive\cite{Chapter_Ale_Rajaraman_book}. Moreover, whilst full ab initio calculation of a few systems at the time is possible through modern-day high-performance computing resources, the discovery of new materials with long relaxation times will inevitably require the characterization of many compounds\cite{high_throughput_cobalt,Ale_nature_rev_chem_comp_design}, making a brute force ab initio approach unsuitable.

Machine learning (ML) has gained increasing popularity in science over the last years as a tool to overcome known computational bottlenecks of ab initio methods or to extract hidden patterns in big data.  
Relevant to this work is its application in the generation of force fields (FF) from a dataset of ab initio data, referred to as a training set. A MLFF is specified by a parametric function that links atomic coordinates and the total energy of a system. The MLFF's parameters are determined by minimization of the differences between the model's predictions and the full quantum mechanical calculations. Once the parameters of the model are determined, i.e. the model is trained, the MLFF can rapidly output the energy and atomic forces for any molecular configuration, thus avoiding the expensive step of solving the Schr\"odinger equation. 
This field has extended considerably in the last years and a plethora of different MLFFs are now available \cite{NN_Behler_Parrinello,CHMIELA201938, DeepMD,ANI_Isayev,GAP,GPR_Ceriotti,GPR_Rasmussen, Physnet, Schnet, FCHL19, moment_tensor_potential, Cormorant,Nequip_Kozinsky, Schutt2017_Message_passage_neur_netw} and capable of reaching chemical accuracy in their predictions of several material properties  \cite{Review_Unke,Review_Tkatch,Review_keith_Tkatchenko,Review_Musil_Ceriotti}. Importantly, MLFFs have been used to predict phonon spectra of a large variety of compounds such as molecular crystals \cite{review_results_global_machine_learning_FF_for_molecular_crystals}, organic molecules \cite{organic_chains_dual_cutoff} and inorganic solid compounds \cite{universal_harmonic_potential}, and new schemes are currently under study to achieve high transferability \cite{phonons_transferability,accelerating_high_throughput_phon_wolverton}.

While the accuracy that the state-of-the-art ML models can achieve has been largely shown \cite{beyond_RMSE_Kovacs}, their applicability in real-world scenarios has been explored to a lesser extent. Only recently, studies have been conducted to investigate the possibility of performing long-scale MD simulation with state-of-the-art ML models and to find key metrics beyond accuracy on benchmark sets \cite{stocker_robust_MD_2022,forces_are_not_enough}. Arguably, the full potential of MLFFs in addressing practical chemistry or physics problems has yet to be fully unleashed and their successful integration to tackle them is still far from trivial.

We aim to fill this gap in the field of spin relaxation by presenting a general ML workflow for the efficient and robust geometrical optimization and prediction of harmonic molecular vibrations for gas-phase molecular systems, a necessary first step towards a future application in condensed matter. Our work builds on previous efforts to use equivariant ML models to predict spin Hamiltonian parameters\cite{zaverkin2021thermally,annie_tensors} and molecular potential energy surfaces\cite{Lunghi2019-ug}. The molecular potential energy surface is approximated as linear in the bispectrum components of the atomic environments and the training is performed with an on-the-fly active learning (AL) strategy requiring minimal user intervention\cite{active_learning_Valerio}. The method is applied to three compounds of interest for the molecular magnetism community, chosen for their challenging electronic structure and diversified complexity of the potential energy surface. We show that this scheme only requires up to $20 \%$ of the total number of calculations needed by a full ab initio simulation of spin relaxation. Excellent results on predictions of spin-phonon relaxation times are achieved by using these phonon spectra and only a minimal accuracy loss is observed when employing ML for both molecular vibrations and spin-phonon coupling. Finally, we show that the same framework applies to molecular dynamics (MD) simulations, thus allowing us to open a window on spin relaxation beyond the conventional equilibrium harmonic lattice.

\section{Methods}

\textbf{Spin-phonon relaxation simulations.} Magnetic properties of single-ion coordination compounds can be described by mapping their electronic structure onto an effective Hamiltonian. For compounds with quenched orbital angular momentum, as most transition metal complexes, this effective Hamiltonian often takes the form
\begin{equation}
    \hat{H}_{\mathrm{S}}=\vec{\textbf{S}}\cdot \textbf{D} \cdot \vec{\textbf{S}},
    \label{TM_Spin_Hamiltonian}
\end{equation}
where $\vec{\textbf{S}}$ is the spin and \textbf{D} is the zero-field splitting tensor.
In the case of compounds with un-quenched angular momentum, as for the Lanthanide compounds, the form of the effective Hamiltonian is instead
\begin{equation}
    \hat{H}_{\mathrm{CF}}=\sum_{l=2,4,6} \sum_{m=-l}^l B_m^l \hat{O}_m^l,
    \label{Lanthanide_Spin_Hamiltonian}
\end{equation}
where $\hat{O}_m^l$ are tesseral functions of the total angular momentum \textbf{J}. From now on, both Hamiltonians are indicated with $\hat{H}_0$.

To describe spin-phonon relaxation, we need to take into account the Orbach and Raman mechanisms, involving respectively one and two contemporary phonon exchanges between the spin and lattice. The relaxation rate $\hat{W}_{b a}$ between the $\hat{H}_0$ eigenstates $\ket{a}$ and $\ket{b}$ due to the Orbach process is given by \cite{Unravelling_JACS}

\begin{equation}
 \hat{W}_{b a}^{1-\mathrm{ph}}=\frac{2 \pi}{\hbar^2} \sum_\alpha\left|\left\langle b\left|\left(\frac{\partial \hat{H}_0}{\partial Q_\alpha}\right)\right| a\right\rangle\right|^2 G^{1-\mathrm{ph}}\left(\omega_{b a}, \omega_\alpha\right),
 \label{Orbach_equation}
\end{equation}

where $\hbar \omega_{ba}=(E_{b} -E_{a})$,  $E_{a}$ and $E_{b}$ being the eigenvalues of $\hat{H}_0$ , $Q_{\alpha}$ and $\omega_{\alpha}$ are the phonon modes and frequencies respectively.
$G^{1-ph}$ is the Fourier transform of the correlation function of the phonon bath
\begin{equation}
    G^{1-ph}(\omega,\omega_{\alpha})=\delta(\omega-\omega_{\alpha})\overline{n}_{\alpha}+\delta(\omega+\omega_{\alpha})(\overline{n}_{\alpha}+1),
\end{equation}
where $\overline{n}_{\alpha}$ is the Bose-Einstein distribution population of the phonon with frequency $\omega_{\alpha}$ and $\delta$ is the Dirac-delta function. For computational purposes, the Dirac-delta is approximated as a normalized Gaussian function where the smearing, i.e. the standard deviation, can be changed by the user. In this work, the smearing value was selected based on the convergence of the relaxation time as advised in previous works\cite{lunghi2020multiple}. 

For the Raman relaxation process, the transition rate reads\cite{science_advances_Ale_2022}
\begin{equation}
\hat{W}_{ba}^{2-\mathrm{ph}}=\frac{2 \pi}{\hbar^2} \sum_{\alpha \beta}\left|T_{b a}^{\alpha \beta,+}+T_{b a}^{\beta \alpha,-}\right|^2 G^{2-\mathrm{ph}}\left(\omega_{b a}, \omega_\alpha, \omega_\beta\right),
\label{Raman_equation}
\end{equation}
where
\begin{equation}
T_{b a}^{\alpha \beta, \pm}=\sum_c \frac{\left\langle b\left|\left(\partial \hat{H}_{\mathrm{0}} / \partial Q_\alpha\right)\right| c\right\rangle\left\langle c\left|\left(\partial \hat{H}_{\mathrm{0}} / \partial Q_\beta\right)\right| a\right\rangle}{E_{\mathrm{c}}-E_{\mathrm{a}} \pm \hbar \omega_\beta} \:.
\label{Raman_relaxation_equation}
\end{equation}
The function $G^{2-ph}$ takes into account all the possible processes involving two phonons: double absorption, double emission and contemporary absorption and emission. For example, for the contemporary absorption of phonon of frequency $\omega_{\alpha}$ and emission of a phonon $\omega_{\beta}$, $G^{2-ph}$ takes the following form:

\begin{equation}
    G^{2\text{-ph}}(\omega_{ba}, \omega_{\alpha}, \omega_{\beta}) = \delta(\omega_{ba} - \omega_{\alpha} + \omega_{\beta}) \bar{n}_{\alpha} (\bar{n}_{\beta} + 1)
\end{equation}

It is important to note that in deriving Eq. \ref{Raman_equation}, it is assumed that the spectrum of $\widehat{H}_0$ is non-degenerate. To fulfil this condition, simulations of Raman relaxation are performed in the presence of a small magnetic field (0.1 T) along the easy axis.
Spin-phonon couplings, i.e. $\partial \hat{H}_{\mathrm{0}} / \partial Q_\alpha$, are calculated as
\begin{equation}
   \left(\frac{\partial \hat{H}_{\mathrm{0}}}{\partial Q_\alpha}\right)=\sum_i^{3 N} \sqrt{\frac{\hbar}{2 \omega_\alpha m_i}} L_{\alpha i}\left(\frac{\partial \hat{H}_{\mathrm{0}}}{\partial X_i}\right),
\label{spin_phonon_couplings}
\end{equation}
where $i$ runs over all the Cartesian degrees of freedom, $m_i$ are the masses of atoms, $L_{\alpha i}$ are the eigenvectors of the Hessian matrix and $X_i$ are the Cartesian degrees of freedom. The quantities $\partial \hat{H}_{\mathrm{0}} / \partial X_i$ are calculated by symmetric derivative with a displacement of 0.01 \AA. Simulations of spin-phonon relaxation are performed with the software  MolForge, freely available at github.com/LunghiGroup/MolForge \cite{science_advances_Ale_2022}. A comprehensive derivation and discussion of the equations reported above is available in literature \cite{Chapter_Ale_Rajaraman_book}. 

\textbf{Electronic structure calculations.} To calculate the magnetic properties of the compounds we use the software ORCA 4.2.1 \cite{orca_software}. Firstly, a multireference wavefunction is obtained at the level of Complete Active Space Self Consistent Field (CASSCF) theory. Spin-orbit coupling is included at the level of quasi-degenerate perturbation theory to calculate the zero-field splitting tensor \textbf{D} and the coefficients $B^{l}_{m}$. For Dy$^{3+}$ ions, we employ an active space of 7 4f orbitals with 9 electrons using all the solutions with multiplicity 6. For Co$^{2+}$ we consider 7 electrons in 5 3d orbitals (7,5) using 10 solutions with multiplicity 4 and 40 solutions with multiplicity 2. Relativistic effects of the CASSCF calculations are treated by the Douglas–Kroll–Hess method \cite{DKH_Hamiltonian}. The RIJCOSX approximation for coulomb and exchange integrals is used for both ions. DKH-def2-TZVPP basis set is used for all elements with an "AutoAux" generated auxiliary set, except for Dy for which SARC-DKH-TZVPP is used as a basis set.

To evaluate the expressions in Eqs. \ref{Orbach_equation}-\ref{spin_phonon_couplings}, it is necessary to calculate the phonon spectrum. For this purpose, firstly we use ORCA to optimize the geometrical configuration at hand at the level of density functional theory (DFT). Tight optimization settings, exchange-correlation functional BP86 with D3BJ \cite{D3BJ_grimme} corrections along with def2-TZVPP and def2/J as basis and auxilary set are used. Large integration grid settings are deployed (Grid 7, GridPruning 0), with increased precision for cobalt (SpecialGridIntAcc 10) and very tight SCF convergence criteria. For the dysprosium compounds, DFT calculations are carried out replacing Dy with Y.
Once the optimized structure is obtained, $6N$ DFT single point calculations with the same settings used for optimization are performed for configurations where atoms are individually displaced along the Cartesian axes by 0.001 \AA. Atomic forces are extracted from these calculations and provided as input to MolForge to determine the molecular vibrational spectra.

\textbf{Machine learning force fields.} MLFFs are deployed to substitute the expensive ab initio geometrical optimization and calculation of molecular vibrations. In this work, we use the linear ML interatomic potential named   Spectral Neighbour Analysis Potential (SNAP) \cite{SNAP} . According to this model, the total energy of a system is written as the sum of single atomic energy contributions, further expanded in a linear combination of atomic descriptors called bispectrum components:
\begin{equation}
    E=\sum^{N_i}_{i} E_i=\sum^{N_i}_{i}\sum^{N_k}_{k}c_{k}(\alpha_i)B_{k}(i) \:,
\label{SNAP_equation}    
\end{equation}
where  $B_{k}(i)$ is the $k$-th bispectrum component of atom $i$, while $N_k$ and $N_i$ are the number of bispectrum components in the expansion and the number of atoms in the system, respectively. The coefficients $c_{k}(\alpha_i)$ depend on the atom species identified by the index $\alpha_i$ and are determined by Ridge regression. Atomic forces are rapidly evaluated by analytical derivation of the total energy.
To fully avail of the efficiency of MLFFs, we implement an AL scheme for the construction of training sets for linear models\cite{active_learning_Valerio}. The ultimate goal of AL is to build a training set for a ML model according to two different criteria: one to include a structure in the training set (1) and one to declare the AL over (2). In the context of this paper, including a structure in the training set is a synonym of triggering an ab initio calculation to add the energy/forces data to the learning dataset.
For (1), we choose the comparison between the uncertainty $s$ on the predicted quantity and a threshold defined by the user through $\delta$
\begin{equation}
 s > \delta \cdot s_z,
 \label{criterion_stop}
\end{equation}
where $s_z$ is the error on the training set and $\delta$ is set by the user. $\delta$ is strictly larger or equal to 1 and the smaller the more frequently it will individuate new structures. The criterium (2) instead is dependent on the particular task at hand and it will be specified later.

We provide a complete overview of a full ML approach to predict spin-phonon relaxation by predicting also \textbf{D} and $B^l_m$ with an equivariant extension of SNAP \cite{annie_tensors}. We briefly present here the equation underpinning this model.
Any physical property can be written as a combination of spherical tensor components $T_m^l$ of order $l$ and $2l+1$ components. According to this principle, \textbf{D} and $B^l_m$ are decomposed in a combination of spherical tensors $T_m^l$, furtherly written in terms of single atomic contributions $T_m^l(a)$
\begin{equation}
    T_m^l = \sum_{a=1}^{N_a} T_m^l(a) = \sum_{i}^{N_i} \sum_{k}^{N_k} c_k(\alpha_i) B_k(i) \overline{Y}_m^l(i),
    \label{equivariant_SNAP_equation}
\end{equation}
where $\overline{Y}_m^l(i)$ is the spherical harmonic of the $i$ th atomic environment and is defined as 
$\overline{Y}_m^l(i) = \sum_{j}^{N_j} Y^{l}_{m}(\hat{r}_{ij})$, 
where $j$ runs over $N_j$ neighbouring atoms within $R_{cut}$ from atom $a$. 
$Y^{l}_{m}$ is the standard definition of a complex spherical harmonic and $\hat{r}_{ij}$ are the coordinates of the atom $j$ rescaled by the coordinates of the atom $i$.
The predictions of the spin Hamiltonian tensors are then used to evaluate $\partial \hat{H}_{\mathrm{0}} / \partial X_i$ for the calculation of $\partial \hat{H}_{\mathrm{0}} / \partial Q_\alpha$ according to Eq.\ref{spin_phonon_couplings}. By doing so, we avoid performing 6N ab initio calculations for the calculation of the symmetric derivatives.
In order to build the training sets for tensorial prediction, we deploy an AL scheme where uncertainty is evaluated analogously to the case of forces in \cite{active_learning_Valerio}. For a tensor of order $l$, we simultaneously predict $2l+1$ components and evaluate a $(2l+1) \times (2l+1)$ matrix, whose diagonal elements represent the variances, i.e. uncertainties, of the predictions on the single tensorial components. The value of $s$ to use in Eq. \ref{criterion_stop} is the square root of the largest value of its diagonal elements. The interested reader can find the specific expressions to evaluate the uncertainty on the predictions in \cite{active_learning_Valerio}. 

Spin-phonon couplings are calculated through Eq. \ref{spin_phonon_couplings} assuming that molecular vibrations can be described harmonically. It is possible to go beyond the harmonic approximation by extracting the spin-phonon couplings directly from MD simulations as shown in \cite{Lunghi2020_molecular_tumbling}. For example, we expand at the linear order the variation of the zero field splitting tensor elements $D_{ij}$ with respect to their mean values, i.e. $
\delta D_{ij} = D_{ij} - \langle D_{ij} \rangle$, due to normal mode vibrations
\begin{equation}
    \delta D_{ij}(t) = \sum_{\alpha} \left( \frac{\partial D_{ij}}{\partial Q_{\alpha}} \right) \hat{Q}_{\alpha}(t),
    \label{linear_approximation_deltag}
\end{equation}
where $i$ and $j$ label the Cartesian axis for tensor \textbf{D}.
It is then possible to establish a proportionality between the half Fourier transform of the correlation functions of $\delta D_{ij}$ and the spin-phonon couplings

\begin{equation}
    \int_0^{\infty} \langle \delta D_{ij}(0) \delta D_{ij}(t) \rangle e^{i\omega t} \, dt \propto \sum_{\alpha} \left( \frac{\partial D_{ij}}{\partial Q_{\alpha}} \right)^2 \delta(\omega - \omega_{\alpha}).
\label{correlation_function_coupling}
\end{equation}

Analogous expressions can be written for lanthanide complexes.
Achieving an accurate evaluation of the LHS of Eq. \ref{correlation_function_coupling} requires long MD simulations together with the evaluation of the tensor of interest. To address this resource-intensive process, we make use again of AL generated ML FFs for the MD simulations and the equivariant extension of SNAP augmented with AL for fast prediction of the tensors over the MD trajectories.  

All the training and AL for the scalar version of SNAP are performed targeting both energies and forces. With reference to the notation in \cite{active_learning_Valerio}, all the hyperparameters are kept fixed across the different compounds ($2J_{max}=12$,\,$\lambda=0.1$,\,$R_{cut}=4.0$ \AA ). Only the relative weight for the training of energy and forces depends on the specific compound and is set to $3N$, where $N$ is the number of atoms in the system in order to equally weight energies and forces during the training. Optimizations performed with MLFFs employ the Adam algorithm \cite{kingma2014adam}. Using the notation of \cite{kingma2014adam}, $\alpha=10^{-4}$, $\beta_1=0.9$, $\beta_2=0.999$ and $\epsilon=10^{-7}$. \\
For equivariant SNAP, the hyperparameters are set to $2J_{max}=8$, \,$\lambda=0.1$,\ and $R_{cut}=4.0$ \AA \, for all compounds.

MD trajectories for the study the correlation functions are produced with LAMMPS \cite{LAMMPS}, setting the timestep to 1 fs and geometrical configurations are dumped every 5 fs for study of the correlation functions.

\section{Results}

\begin{figure}[H]
      \centering
      \includegraphics[scale=0.21]{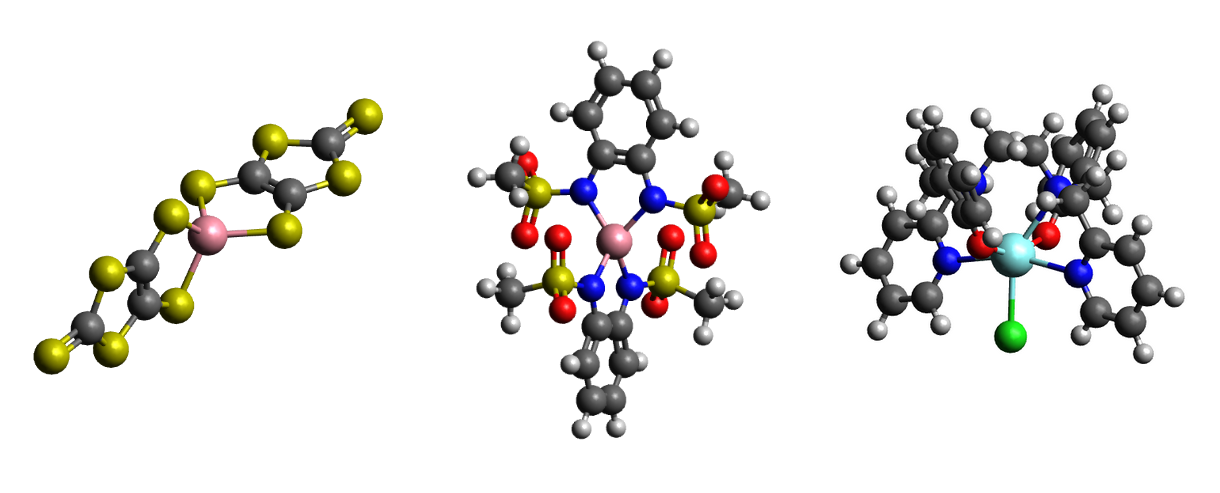}
      \caption{\textbf{Molecular structures of the three investigated compounds }  From left to right, the two Co$^{2+}$ compounds \textbf{1,2} and the Dy$^{3+}$ compound \textbf{3}. Colour code: cobalt in pink, dysprosium in cyan, oxygen in red, carbon in dark grey, sulphur in yellow, nitrogen in blue, hydrogen in white and chlorine in green.}
      \label{molecule_to_study}
\end{figure}

Here we present the results related to the implementation of ML models for 1) the prediction of molecular vibrations and spin Hamiltonian tensors, and the profiles of spin-phonon relaxation time as a function of temperature, and 2) the calculation of correlation functions of the spin Hamiltonian tensors over MD trajectories. Simulations are presented for three molecules whose spin relaxation profiles have been fully characterized both experimentally and with full quantum mechanical simulations \cite{Unravelling_JACS}:  [Co(C$_3$S$_5$)$_2$](Ph$_4$P)$_2$ (\textbf{1}) \cite{Cobalto_1},[CoL$_2$][(HNEt$_3$)$_2$] \cite{cobalto_2}, where H$_2$L = 1,2-bis(methane-sulfonamido) (\textbf{2}) and [Dy(bbpen)Cl] (\textbf{3}) \cite{dysprosium_1}. The two Co$^{2+}$ complexes exhibit a ground state with spin $S=3/2$ described through Eq. \ref{TM_Spin_Hamiltonian}, while the Dy$^{3+}$ complex possesses a ground state with total angular momentum $J=15/2$, described by Eq. \ref{Lanthanide_Spin_Hamiltonian}. The chemical structures of \textbf{1}-\textbf{3} are reported in Fig. \ref{molecule_to_study}.

\textbf{Machine learning of molecular vibrations.} The AL strategy presented in the Methods section is here applied to the optimization of molecular structures and calculation of molecular vibrations. The goal of the protocol is to train a MLFF able to accurately predict the forces for the configurations used to calculate the molecular vibrations and frequencies. For this purpose, we designed the following workflow  %
\begin{itemize}
    \item Geometrical optimization: the initial training set is the sole molecular crystallographic structure and its energy and atomic forces are computed with electronic structure methods. A geometrical optimization process is started with on-the-fly AL;
    \item MD simulation at 50 K: once the optimization is complete, the training set is reset. The only structure in the new training set is the one obtained from the final geometry of the ML-aided optimization. Also here an on-the-fly AL scheme is again employed during MD; 
    \item Molecular vibration prediction: using the ML force field trained through the above steps, vibrational spectra are predicted.
\end{itemize} 
In particular, we note the importance of the second step, where an MD at 50 K is performed, to achieve good accuracy in the prediction of frequencies. This step in fact guarantees a sufficient sampling of the harmonic portion of the potential, which is not automatically performed during optimization.

The optimization routine is declared over when the following conditions are contemporarily fulfilled: (1) the RMS of the gradient vector is smaller then 0.01 kcal/mol/\AA, (2) the maximum value of the elements in the gradient is smaller than 0.1  kcal/mol/\AA \ and (3) the variation in energy between two consecutive minimization steps is smaller than 0.0001 kcal/mol. These conditions are chosen in order to closely resemble the "TightOpt" option as implemented in ORCA, therefore making it possible to measure the performance of the ML workflow compared to the full ab initio approach. Geometry optimization is performed by testing different values of $\delta$ to prove the robustness of the protocol.
With the ML models trained at the end of optimization, we have calculated the molecules' vibrational modes and frequencies. To check whether the results for RMSE values on frequencies and hessian elements could be improved, we augment the training set by sampling more structures with AL during MD simulation at 50 K, thus guaranteeing a more effective sampling of structures around the energy minimum.
Since all of the ML models trained after optimization lead to similar results, 
%with the RMSE on vibrational frequencies being between 30 and 57 cm$^{-1}$ for \textbf{1}, 23 and 38 cm$^{-1}$ for \textbf{2} and 51 and 59 cm$^{-1}$
(see Tab. \ref{SI:optimization_phonons_3_compounds}),
only the optimized structure obtained for $\delta=30$ is considered for further exploration during MD at 50 K, because of the smallest final training set size.
Parity plots for the comparison of the predictions between ML and the ab initio approach are shown in Figs. S1-S12.
Different values of $\delta$ are also tested during MD at 50 K. The AL-augmented MD stops when 100 ps are simulated without finding new configurations to include in the training set. %The parity plots in Figs. S13-S24 and the RMSE values in Tab. \ref{SI:Phonons_after_MD_3_compounds} show the excellent agreement between the ML predicted frequencies after MD at 50K and the ones calculated with a full quantum mechanical approach.
As an example, we present here only the results for $\delta=2.5$ during MD in Tab. \ref{optimization_MD2.5_phonons}. The RMSE on vibrational frequencies and Hessian elements is always lower after MD at 50 K and equivalent results are obtained for all the values of $\delta$ (see Tab. \ref{SI:Phonons_after_MD_3_compounds} and Figs. S13-S24). 
%ML results on molecular vibrations and hessian elements at the end of optimization are reported in Tab. \ref{SI:optimization_phonons_3_compounds} and Figs. S1-S12. ML results on molecular vibrations and hessian elements after MD at 50 K are reported in Tab. \ref{SI:Phonons_after_MD_3_compounds} and Figs. S13-S24.
%and Fig. \ref{phonon_parity_plots_MD_2.5}.
Beyond accuracy, a key metric to judge the ML performance is the efficiency factor (EF), defined as $[1-(N_{ML}/N_{DFT})]\cdot 100$, where $N_{ML}$ and $N_{DFT}$ are the number of ab initio calculations performed with the aid of ML and with a traditional full quantum mechanical method, respectively. $N_{ML}$ is given by the sum of the final training set sizes for optimization and MD.
%and in Tab.\ref{SI:Phonons_after_MD_3_compounds} for all values of $\delta$.
For the three molecules, the efficiency factor is always around 80 \% (see Tab. \ref{optimization_MD2.5_phonons}) and higher efficiency factors are obtained for other values of $\delta$ (see Tab. \ref{SI:Phonons_after_MD_3_compounds}), clearly pointing out that ML, while preserving near-to-full ab initio accuracy, also leads to a consistent computational saving.

%\begin{table*} [t]
%\caption{\textbf{Optimization and phonon results for the selected compounds} The RMSE for phonons and hessians are reported in cm$^{-1} $ and kcal/mol/\AA$^{2} $.
%The training set size (TSS) selected by the active learning algorithm is also reported for different values of the threshold parameter $\delta$. For comparison with the machine learning approach, in parentheses next to TSS for the smallest value of $\delta$, the number of \textit{ab initio} calculations required in a full \textit{ab initio} optimization. Phonon results after molecular dynamics at 50 K. The RMSE for phonons and hessians are reported in cm$^{-1} $ and kcal/mol/\AA$^{2} $.}
%\centering
%\begin{tabular}{c |c c c | c c c}
%\toprule
%\textbf{Compound}  \hspace{0.1cm}   & \hspace{0.1cm} \textbf{TSS} \hspace{0.1cm} & \hspace{0.1cm} \textbf{RMSE Phonons } \hspace{0.1cm} & \hspace{0.1cm} \textbf{RMSE Hessian} \hspace{0.1cm} & \hspace{0.1cm} \textbf{TSS} \hspace{0.1cm} & \hspace{0.1cm} \textbf{RMSE Phonons } \hspace{0.1cm} & \hspace{0.1cm} \textbf{RMSE Hessian} \hspace{0.1cm}\\\hline
%\midrule
% \textbf{1}& 3 (24) & 42.17 & 0.019 & 22 & 2.47 & 0.004 \\\hline 
%\midrule
%\textbf{2}& 9 (19) & 38.3 & 0.009 & 54 & 14.8 & 0.005 \\\hline
%\midrule
%\textbf{3}& 15 (15) & 51.2 & 0.011 & 53 & 23.29 &0.006 \\ %\hline
%\bottomrule 
%\end{tabular}
%\label{optimization_phonons_3_compounds}
%\end{table*}

\begin{table} [t]
\caption{\textbf{Optimization and molecular vibrations for the selected molecules} The RMSE for molecular vibrations and hessians are reported in cm$^{-1} $ and kcal/mol/\AA$^{2} $. The two rows for each molecule (Mol.), correspond to optimization ($\delta=30$) and MD at 50K ($\delta=2.5$), respectively. Training set sizes (TSS) are reported for both optimization and MD. The efficiency factor (E.F.) is reported as a percentage.
%The training set sizes (TSS) selected by the active learning algorithm are reported for $\delta=30$ for optimization and $\delta=2.5$ for MD.
}
\centering
\begin{tabular}{c | c c c c}
\toprule
\textbf{Mol}  \hspace{0.1cm} & \hspace{0.1cm} \textbf{E.F.} \hspace{0.1cm}  & \hspace{0.1cm} \textbf{TSS} \hspace{0.1cm} & \hspace{0.1cm} \textbf{RMSE Phon} \hspace{0.1cm} & \hspace{0.1cm} \textbf{RMSE Hess} \hspace{0.1cm} \\\hline
\midrule
 \multirow{2}*{\textbf{1}}& \multirow{2}*{80\%} & 3 & 42.17 & 0.019 \\
 & & 22 & 2.47 & 0.004\\\hline
\midrule
\multirow{2}*{\textbf{2}}& \multirow{2}*{81\%} & 9 & 38.3 & 0.009 \\
& & 54 & 14.8 & 0.005 \\\hline
\midrule
\multirow{2}*{\textbf{3}}& \multirow{2}*{83\%} & 15 &51.2 & 0.011 \\
& & 53 & 23.29 & 0.006 \\ \hline
\bottomrule 
\end{tabular}
\label{optimization_MD2.5_phonons}
\end{table}

% \begin{figure}[H]
%     \centering
%     \includegraphics[scale=0.65]{Images/phonons/freq_phonons_Mol1.png}\\
%     \includegraphics[scale=0.65]{Images/phonons/freq_phonons_Mol6.png}\\
%     \includegraphics[scale=0.65]{Images/phonons/freq_phonons_Mol9.png}
%     \caption{\textbf{Phonon parity plots}. Comparison between molecular vibration frequencies calculated with ML and full ab initio approach. Results are presented for \textbf{1},\textbf{2} and \textbf{3} in order from top to bottom ($\delta=30$ for optimization, $\delta=2.5$ for MD).}
%     \label{phonon_parity_plots_MD_2.5}
% \end{figure}

\begin{figure}
    \centering
    \includegraphics[scale=0.65]{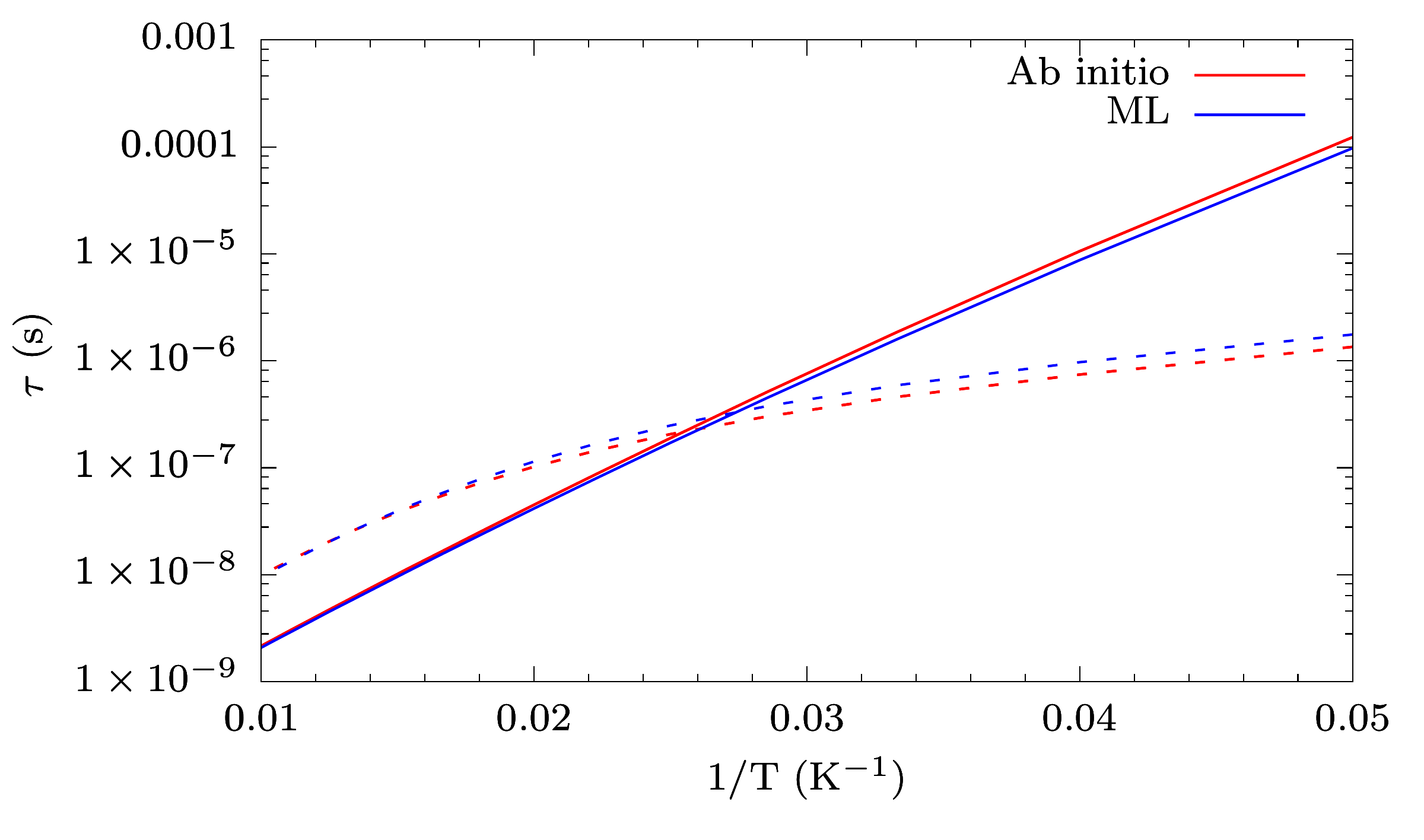}\\
    \includegraphics[scale=0.65]{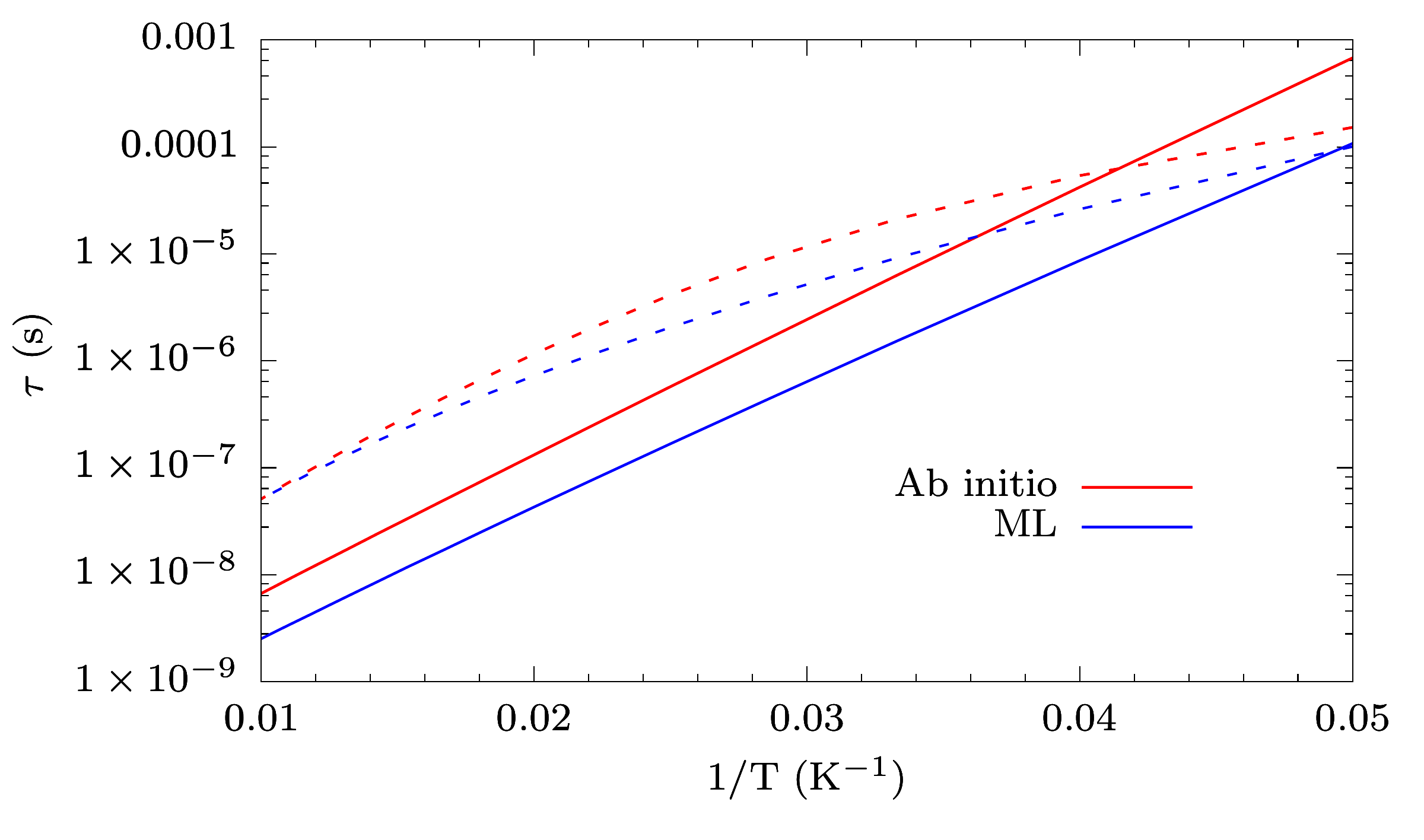}\\
    \includegraphics[scale=0.65]{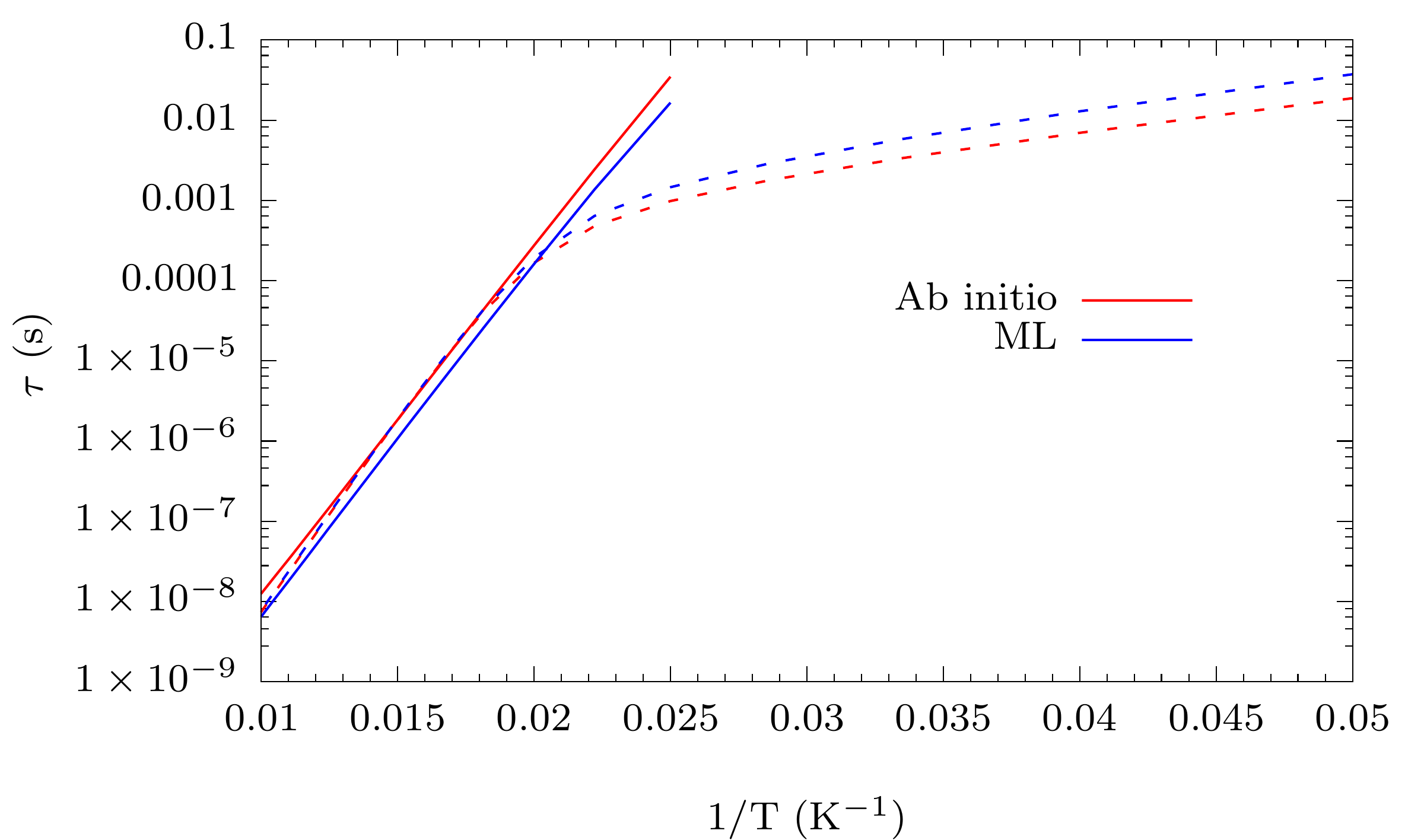}
    \caption{\textbf{Spin relaxation profiles} Comparison between spin-phonon relaxation profiles as a function of temperature calculated with phonons from ML (blue) and  with full ab initio approach (red). Results are presented for \textbf{1},\textbf{2} and \textbf{3} in order from top to bottom.}
    \label{spin_relaxation_profiles}
\end{figure}

Given a robust and accurate protocol for the prediction of vibrational spectra with ML, we proceed to test its integration within the framework of spin-phonon relaxation simulations to speed them up. Spin-phonon relaxation profiles show two different curves related to the Orbach and Raman mechanisms.
In the Orbach mechanism, an initially fully polarized spin state $M_S$ is excited to an intermediate spin state via a series of resonant phonon absorptions. Subsequently, a series of resonant phonon emissions leads the system towards the spin state $-M_S$. This mechanism leads to a relaxation time that exponentially increases in temperature as the phonon population grows according to the Bose-Einstein distribution. For the Raman mechanism, let us consider the case of contemporary absorption and emission of two phonons.  In Raman relaxation, a fully polarized spin state $M_S$ relaxes to the spin state $-M_S$ mediated by virtual excited states. Low-energy phonons are usually found to drive this mechanism\cite{Unravelling_JACS}. In general, therefore, the spin relaxation profiles show a regime at high temperatures where the Orbach process leads to relaxation and a Raman-dominated regime at low temperatures. Combining the goal of achieving a consistent acceleration of these simulations with the need for near-to-chemical accuracy for the quantities in Eqs. \ref{Orbach_equation} and \ref{Raman_equation} represents a serious test-bed for any ML model. Results for $\delta=2.5$ during MD are reported in Fig. \ref{spin_relaxation_profiles}, where the discrepancy between the full quantum mechanical and ML approach is always within a factor of 10 for all the compounds, i.e. within the upper bound of the typical error that affects these simulations \cite{Unravelling_JACS}.
%(see Tabs.  \ref{SI:optimization_phonons_3_compounds}, \ref{SI:Phonons_after_MD_3_compounds} and Figs. S1-S12)

\begin{figure}
    \centering
    \includegraphics[scale=0.65]{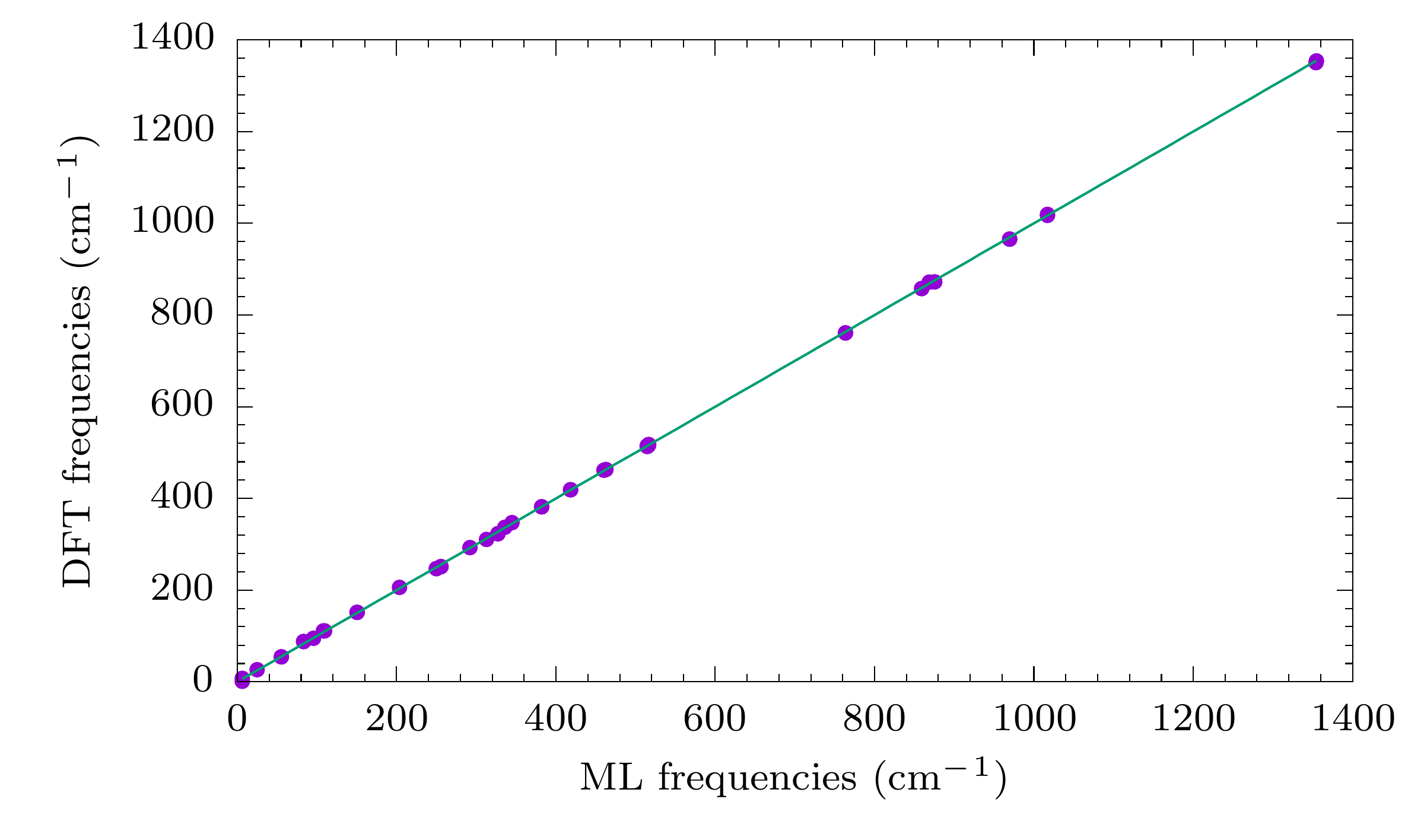}\\
    \includegraphics[scale=0.65]{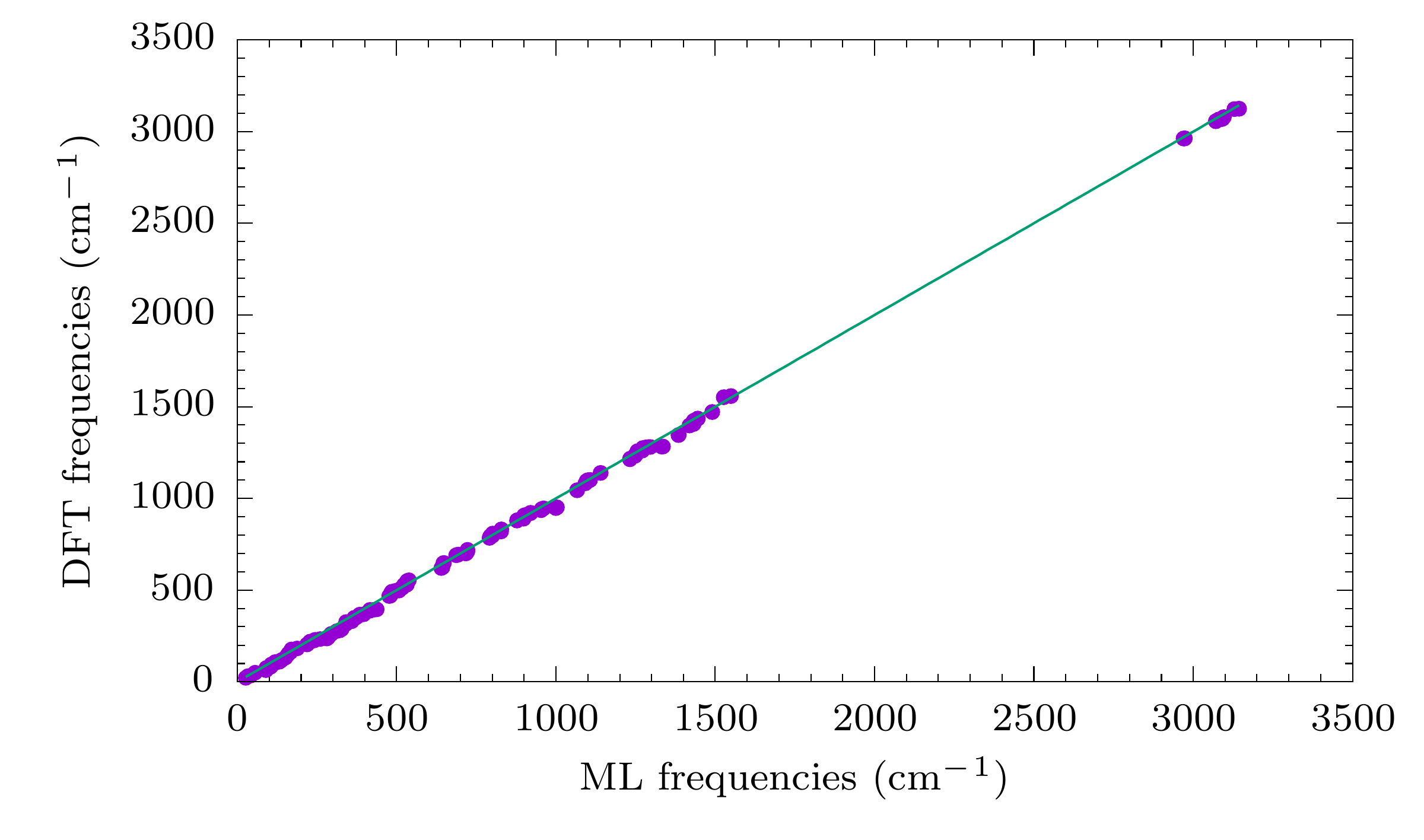}\\
    \includegraphics[scale=0.65]{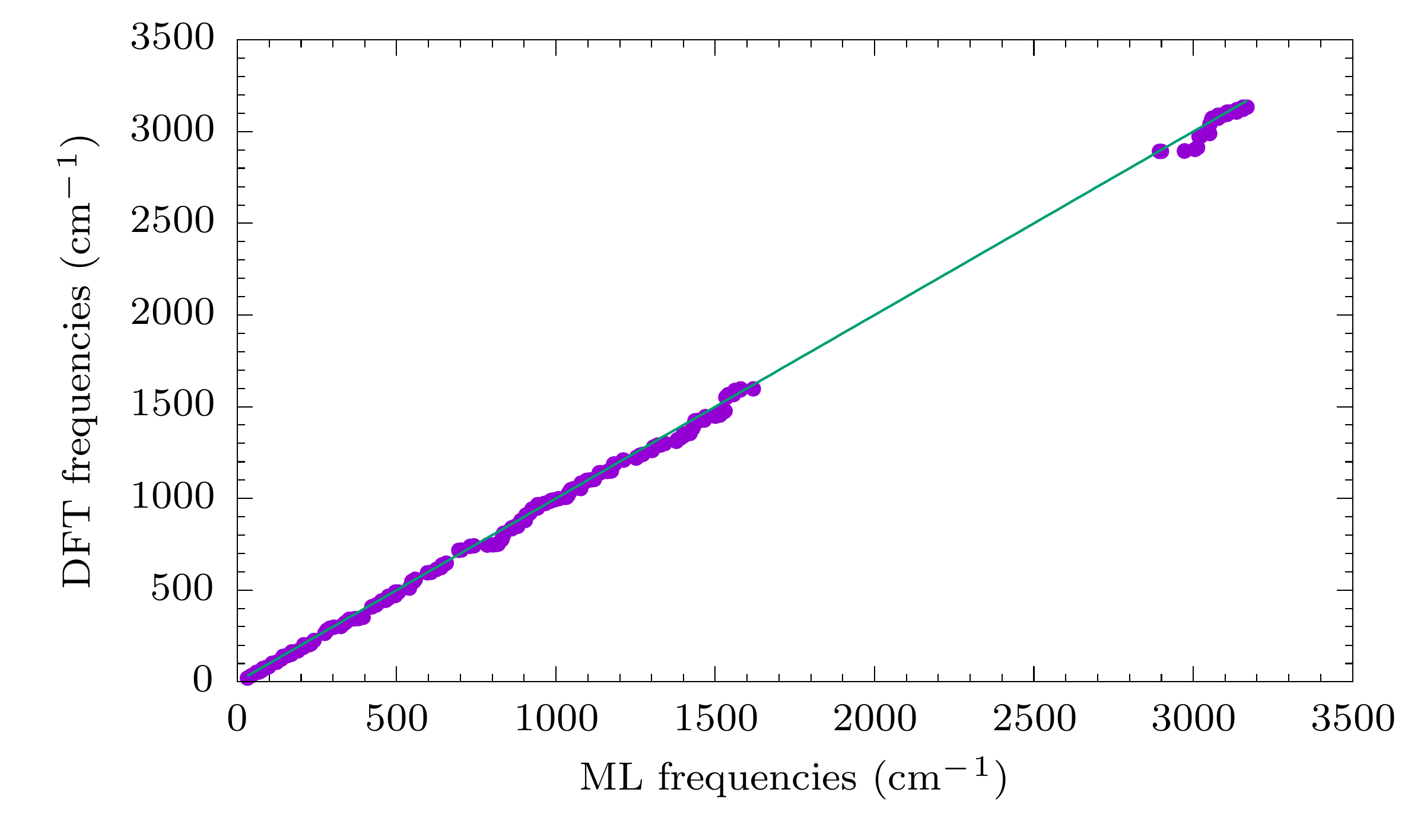}
    \caption{\textbf{Phonon parity plots}. Comparison between molecular vibration frequencies calculated with ML and full ab initio approach. Results are presented for \textbf{1},\textbf{2} and \textbf{3} in order from top to bottom ($\delta=30$ for optimization, $\delta=2.5$ for MD).}
    \label{phonon_parity_plots_MD_2.5}
\end{figure}

\begin{figure}
    \centering
    \includegraphics[scale=0.65]{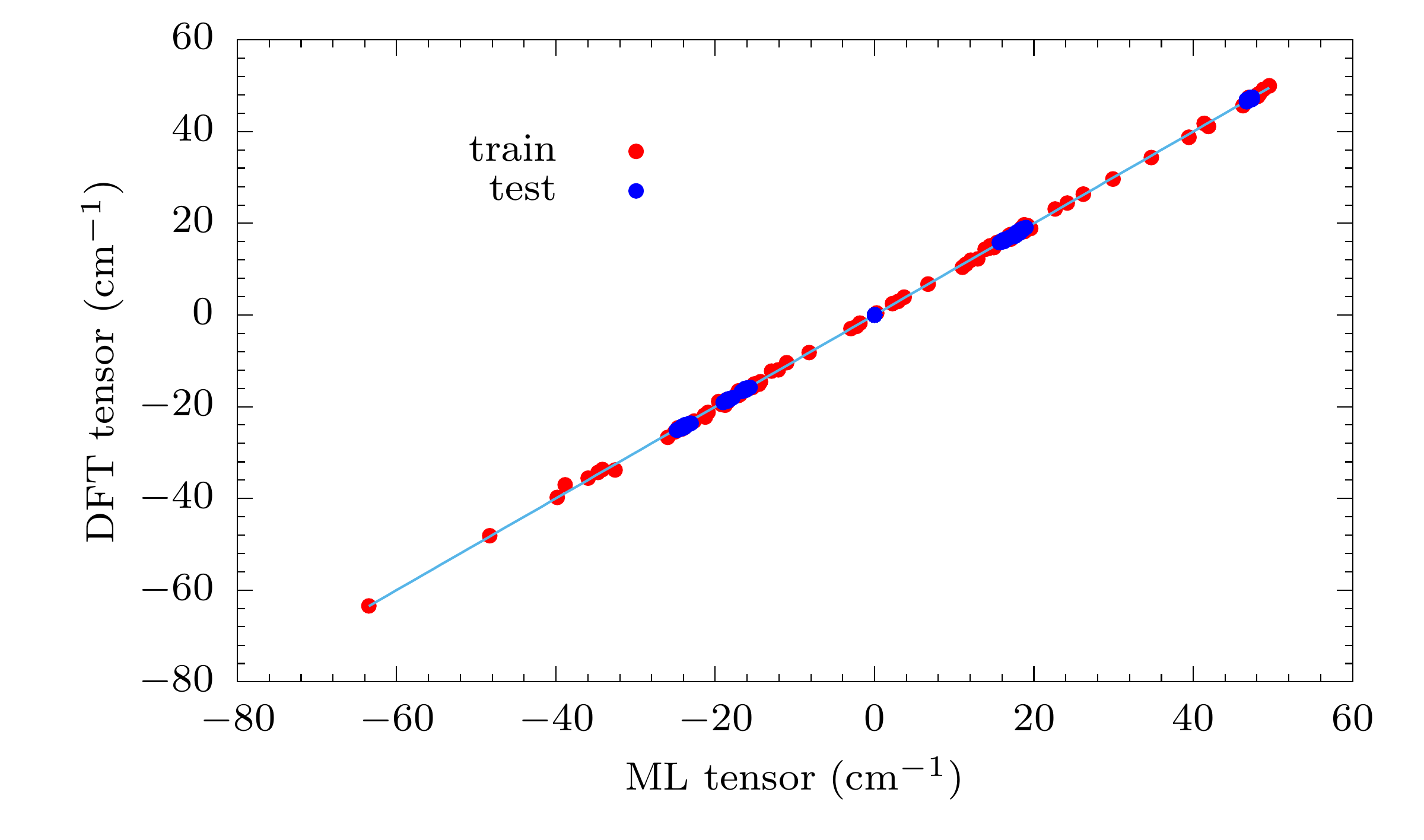}\\
    \includegraphics[scale=0.65]{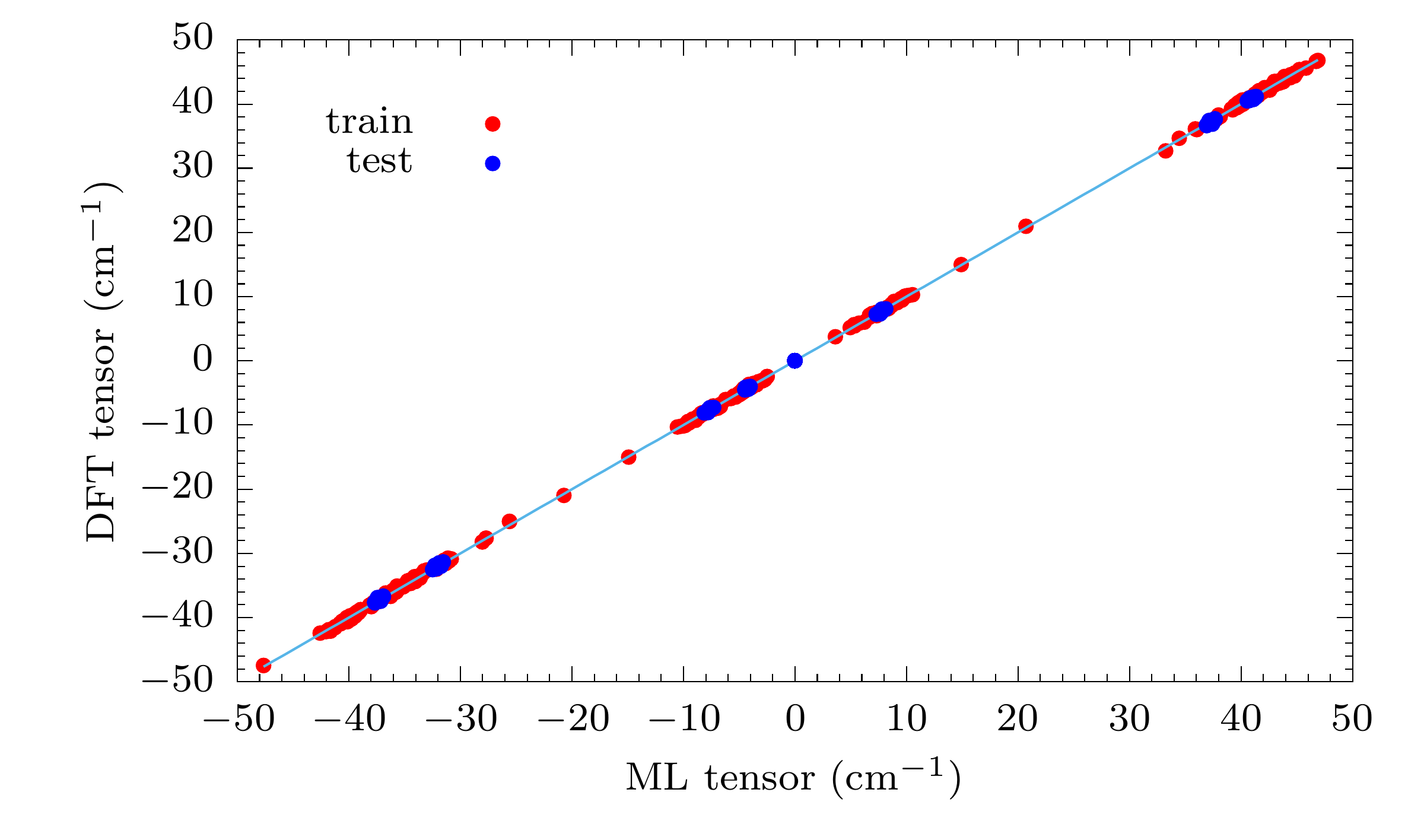}\\
    \includegraphics[scale=0.65]{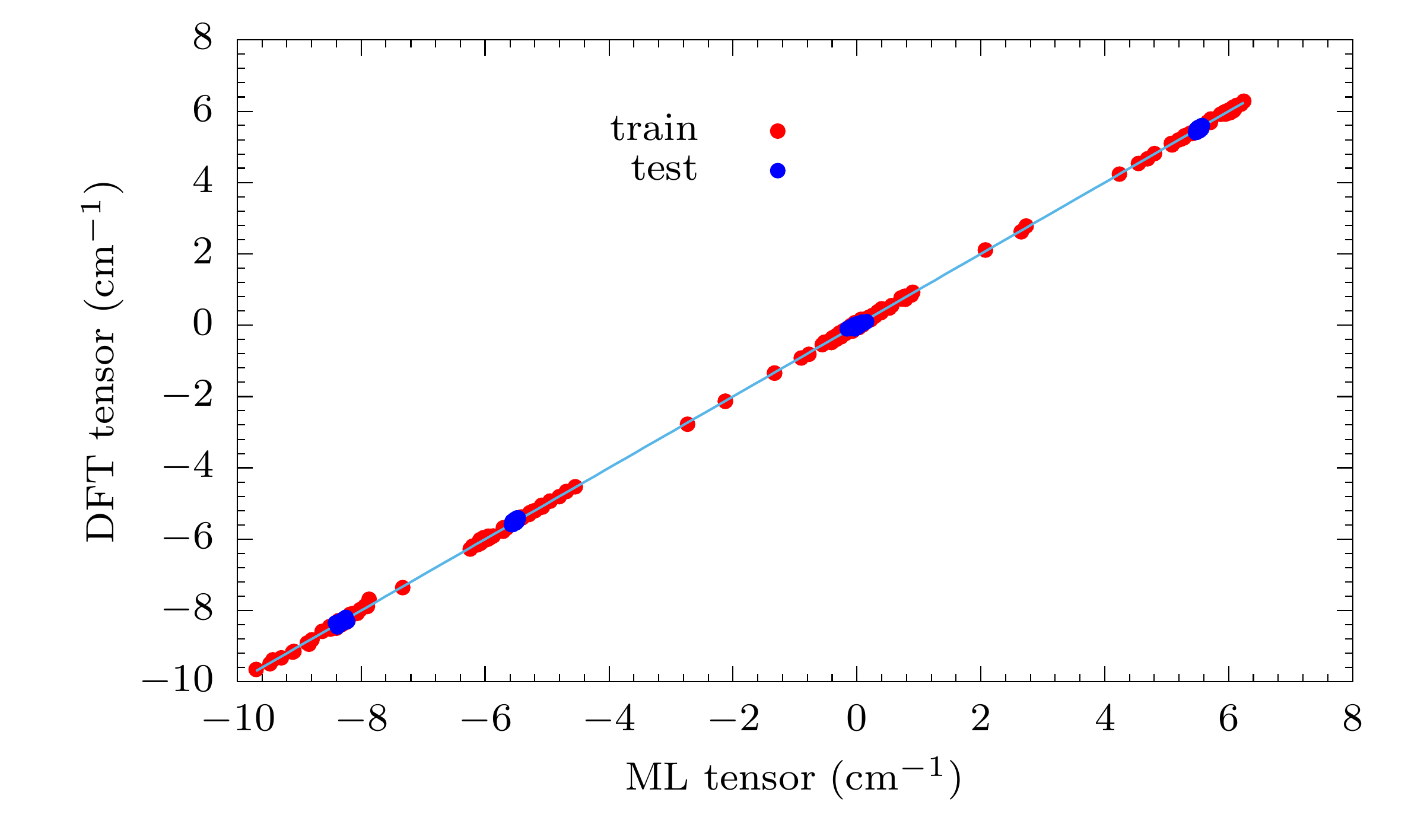}
    \caption{\textbf{Fit of spin Hamiltonian tensors.} Training and test error for the fit of the spherical tensor components in which the spin Hamiltonian tensors are decomposed for \textbf{1},\textbf{2} and \textbf{3}, in order. For \textbf{3}, only the plot for the tesseral functions for $l=2$ is reported. The plots for $l=4,6$ are reported in SI. }
    \label{fit_tensor_spin_relaxation}
\end{figure}

\textbf{Machine learning of spin-phonon coupling.} 
The optimal results shown on spin relaxation predictions have motivated us to further predict the spin Hamiltonian tensors in order to realize a complete ML scheme. For this purpose, we use the training sets built during MD with $\delta=2.5$ as learning datasets to predict the spin Hamiltonian tensors with equivariant SNAP and evaluate the spin-phonon couplings. In Fig. \ref{fit_tensor_spin_relaxation} we report the training and test sets predictions for all compounds (see RMSE values in Tab. \ref{SI:tensor_training_test_prediction}) , where the test set is made of the 6N configurations used to evaluate the derivatives $\partial \hat{H}_{\mathrm{0}} / \partial X_i$ appearing in Eq. \ref{spin_phonon_couplings}. 
The plots relative to the fit of higher-order tensors of \textbf{3} are reported in Fig. \ref{SI:fit_tensor_spin_relaxation}. 
%and Tab. \ref{SI:tensor_training_test_prediction}. 
%0.36 (0.19) cm$^{-1}$ for \textbf{1}, 0.18 (0.10) cm$^{-1}$ for \textbf{6}, $3.5 \cdot10^{-2}$($2.0 \cdot10^{-2}$) cm$^{-1}$, $1.2 \cdot10^{-3}$($9.5 \cdot10^{-4}$) cm$^{-1}$ and $2.0\cdot10^{-5}$($1.0\cdot10^{-5}$) cm$^{-1}$ for \textbf{3} with $l=2,4,6$ respectively .

\begin{figure}
    \centering
    \includegraphics[scale=0.65]{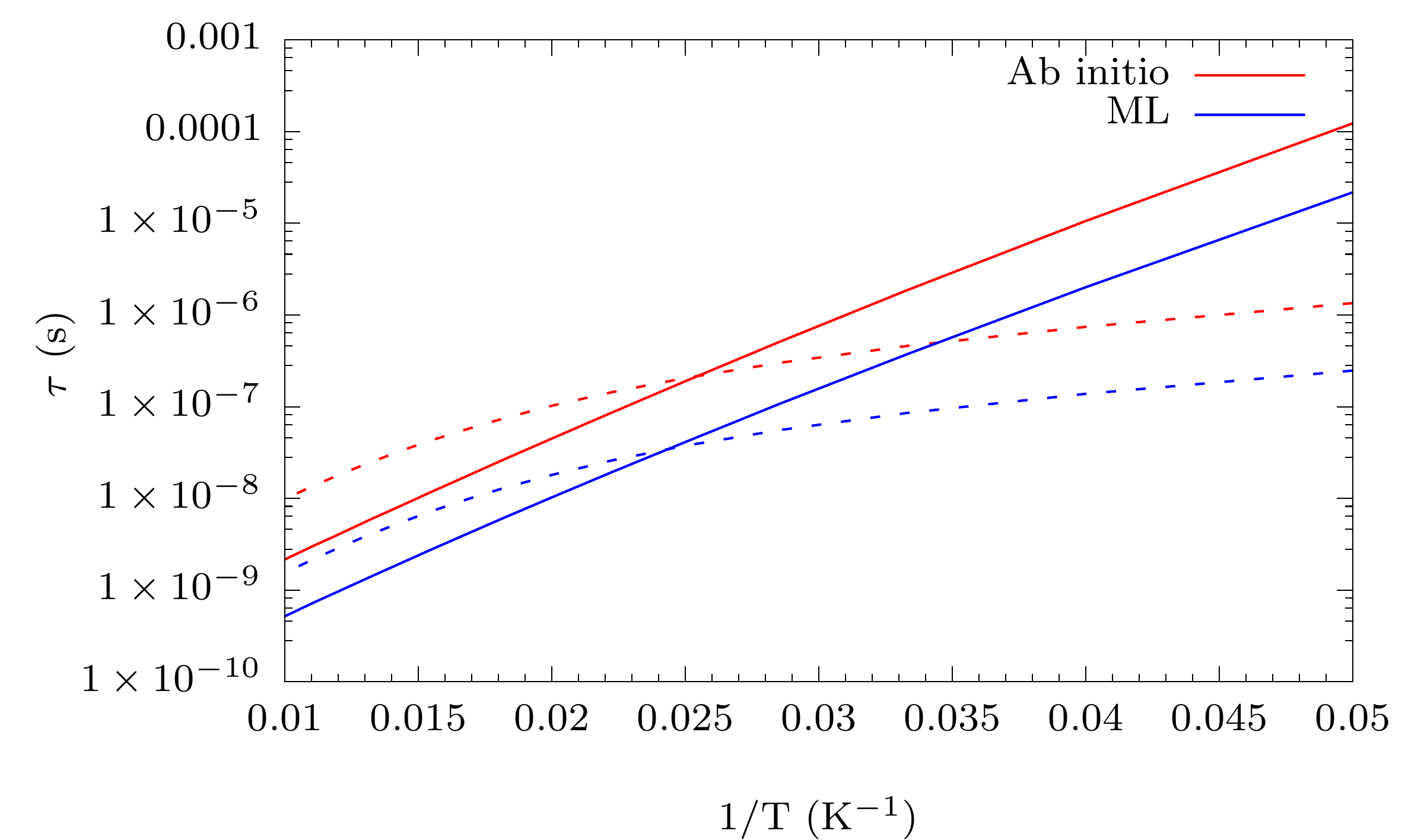}\\
    \includegraphics[scale=0.65]{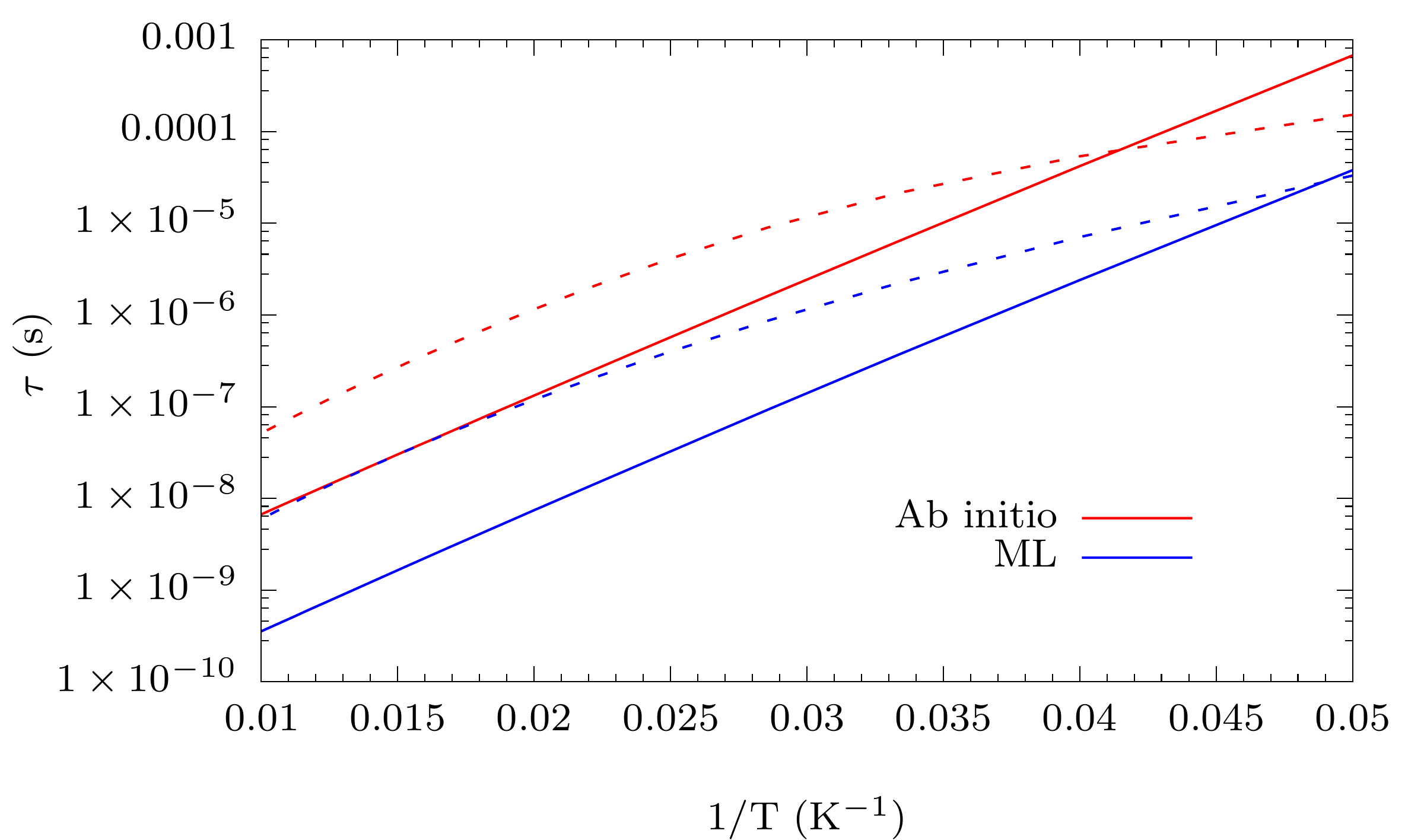}\\
    \includegraphics[scale=0.65]{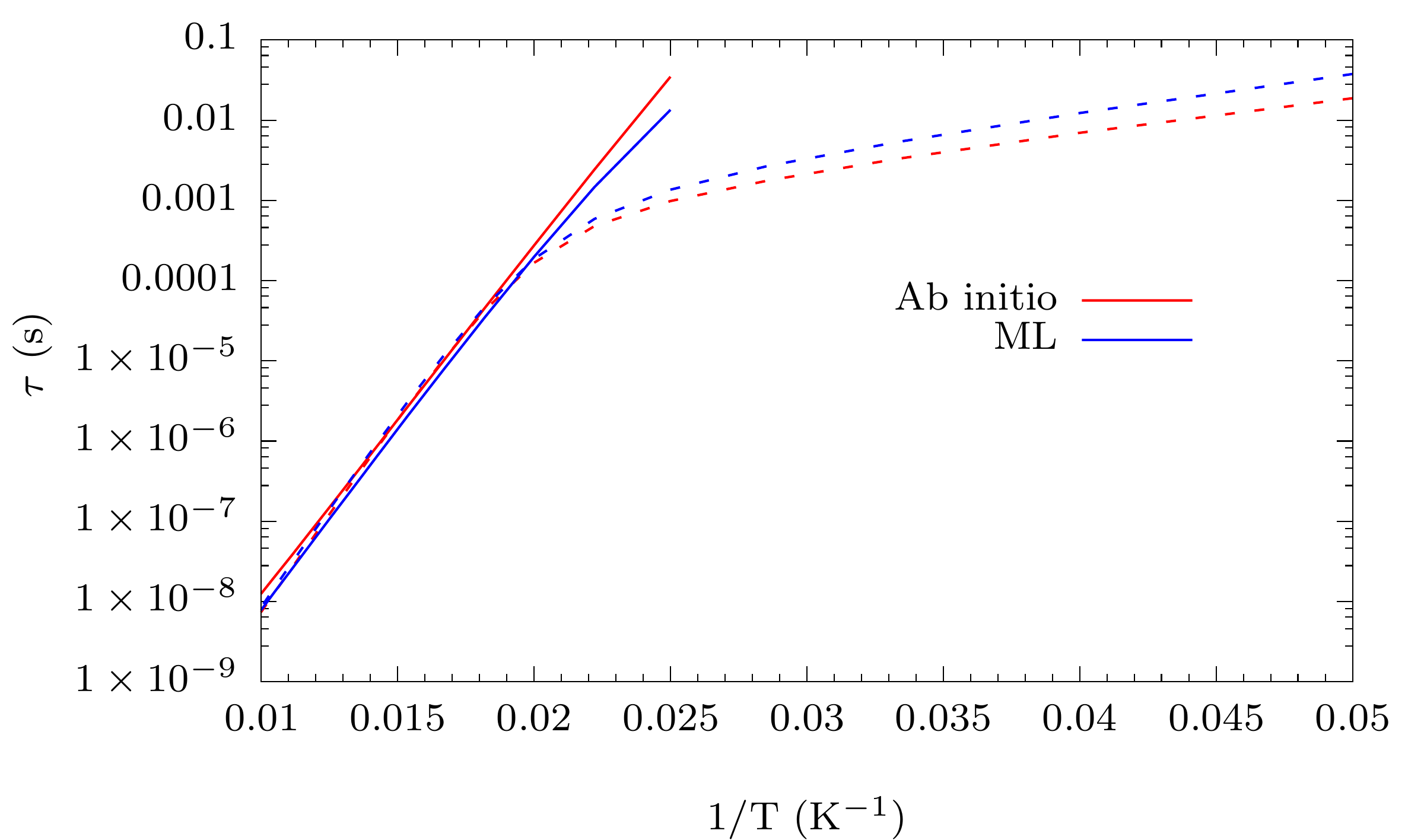}
    \caption{\textbf{Spin relaxation profiles} Comparison between spin-phonon relaxation profiles as a function of temperature calculated with a full ML (blue) and  with full ab initio approach (red). Results are presented for \textbf{1},\textbf{2} and \textbf{3} in order from top to bottom.}
    \label{spin_relaxation_profiles_full_ML}
\end{figure}

 The spin-phonon couplings appearing in Eq. \ref{spin_phonon_couplings} are evaluated using the predictions of equivariant SNAP and subsequently used to evaluate the Orbach and Raman relaxation times as a function of temperature. 
 %To check the accuracy of the spin-phonon couplings, in Fig. \ref{SI:spin_phonon_DOS_ML_CAS},  we report a comparison of the spin-phonon density, defined as  $\sum_{\alpha} \left( \partial D_{ij} / \partial Q_{\alpha} \right)^2 \delta(\omega - \omega_{\alpha})$ averaged over the Cartesian degrees of freedom, $i$ and $j$, for the Co$^{2+}$ complexes and $\sum_{\alpha} \left( \partial B^{l}_{m} / \partial Q_{\alpha} \right)^2 \delta(\omega - \omega_{\alpha})$ averaged over the allowed values of $m$ for $l=2,4,6$ for \textbf{3}, calculated with ML predicted tensors and with full quantum mechanical approach.
 A brief comparison of the spin relaxation profiles in Fig. \ref{spin_relaxation_profiles_full_ML} obtained within the full ML approach and Fig. \ref{spin_relaxation_profiles} shows that every compound is differently affected by using ML to obtain the spin-phonon couplings. 
 While accuracy is maintained for \textbf{3}, the discrepancy between ML and ab initio approach gets approximately 10 times worse for \textbf{1} and \textbf{2}. The larger discrepancy is due to the limited accuracy of the ML model in predicting $\partial \hat{H}_{\mathrm{0}} / \partial X_i$  with $R^2=0.69$ for \textbf{1}, $R^2=0.52$ for \textbf{2} and $R^2=0.32,0.03,\num{-7.5e-3}$ for \textbf{3}, order $l=2,4,6$ respectively (see Fig. \ref{SI:cartesian_gradients_ML_predictions}) and the spin-phonon density, defined as  $\sum_{\alpha} \left( \partial D_{ij} / \partial Q_{\alpha} \right)^2 \delta(\omega - \omega_{\alpha})$ averaged over the Cartesian degrees of freedom, $i$ and $j$, for the Co$^{2+}$ complexes and $\sum_{\alpha} \left( \partial B^{l}_{m} / \partial Q_{\alpha} \right)^2 \delta(\omega - \omega_{\alpha})$ averaged over the allowed values of $m$ for $l=2,4,6$ for \textbf{3} (see Fig. \ref{SI:spin_phonon_DOS_ML_CAS}). 
 Considering the Co$^{2+}$ complexes for example, error analysis shows that the relative error on the terms $\partial \hat{H}_{\mathrm{0}} / \partial X_i$ is proportional to $\sigma_{D}/\Delta D$, where $\sigma_{D}$ is the error in the prediction of \textbf{D} and $\Delta D$ the variation of the tensor component at the displaced geometries when evaluating the symmetric derivatives. Going to larger values of the derivative step in order to increase $\Delta D$ may improve the fit, but it could also invalidate the linear approximation underlying the derivative calculation. Manual tuning of the model's hyperparameters has not led to significant improvements.

\textbf{Machine learning the full potential energy surface and its spin Hamiltonian.} 
%As briefly mentioned in the Introduction, it is possible to extract the spin-phonon couplings according to other schemes then Eq. \ref{spin_phonon_couplings} to account also for anharmonic effects. This approach requires extended MD simulations (scale of  few tens of ns) for which we calculate the Spin Hamiltonian tensors by  performing millions of CASSCF calculations. In \cite{Lunghi2020_molecular_tumbling}, only the use of ML techniques has made possible the aforementioned calculations for an organic radical in solution. Here we make a step forward with respect to \cite{Lunghi2020_molecular_tumbling} by challenging the capabilities of the ML model on the dynamics of more complex systems and by deploying an AL scheme for the construction of the training dataset to generate MD simulations. Furthermore, equivariant SNAP is combined with the AL scheme in \cite{active_learning_Valerio} to achieve a rational construction of the training set also for the prediction of \textbf{D} and $B$.\\ 
Spin-phonon couplings can be extracted through analysis of MD simulations according to Eq. \ref{correlation_function_coupling}. However, evaluation of the LHS of Eq. \ref{correlation_function_coupling} requires long MD simulations in order to accurately calculate the averages, e.g. $\langle \delta D_{ij}(0) \delta D_{ij}(t) \rangle$. To tackle this challenge, we use the following workflow, which combines ML models for force fields and spin Hamiltonian tensors together with AL

\begin{itemize}
    \item to speed up the construction of the learning set, we decide to generate the ML model with the set  built for molecular vibrations with $\delta=2.5$. This is not a strict requirement and starting from scratch is possible;
    \item new configurations are added to the training set during an AL-augmented MD performed at 300K (for Co$^{2+}$ complexes) and 100K (for Dy$^{3+}$ complex) with $\delta$ set to 1.75;
    \item using the ML force field trained through the above steps,  MD trajectories approximately 40-60 ns long at 25, 50, 75 and 100 K are generated. Correlation functions for the spin Hamiltonian tensors are evaluated on these MD simulations.
\end{itemize}

Analysis of the correlation functions is done only up to 100 K because experimental spin relaxation data is commonly available only at low temperatures. AL-augmented MD is performed at a lower temperature for the Dy$^{3+}$ compound as the FF becomes substantially less accurate in its predictions if trained at 300 K, possibly due to the limited stability of this molecule at high-temperature and gas phase. The AL is over when no new structures are included after 0.5 ns of MD.
To assess the accuracy of the ML models, we benchmark them on test sets drawn from MD simulations at all 4 different temperatures. The test set is made of 50 structures sampled evenly during 2 ns of dynamics.  
%RMSE values for both the training and test sets are reported in Tabs. \ref{SI:training_set_RMSE_correlations} and \ref{SI:test_set_RMSE_correlation}.
The generated FFs achieve chemical accuracy on test sets for \textbf{1} and \textbf{3} with an upper bound on the RMSE for energies (forces) of 0.30 (0.99) and 0.57 (2.24) respectively (energies in kcal/mol and forces in kcal/mol/\AA). For \textbf{2}, the RMSE upper bounds on energies (forces) of 2.38 (1.68) kcal/mol (kcal/mol/\AA) are instead slightly above chemical accuracy. For results at all temperatures for all compounds and on the training set, see Tabs. \ref{SI:training_set_RMSE_correlations} and \ref{SI:test_set_RMSE_correlation} and Figs. S28-S39. %A slightly larger error is observed in theenergy for compound \textbf{2}, due to the combination of the high temperature for the sampling of structures and the complexity of the compound compared to \textbf{1}.
%As we certify the near-to-ab initio accuracy of the MD trajectories, we proceed to the prediction of tensors 

To evaluate the correlation functions of the spin Hamiltonian tensors over the MD trajectories, we use the following protocol:
\begin{itemize}
    \item with no \textit{a priori} training dataset we perform an AL ($\delta=1.5$) targeting \textbf{D} and $B^l_m$ over 2 ns of trajectory;
    \item with the ML models trained through the above step, we predict the spin Hamiltonian tensors over approximately 10M structures for every compound at each temperature, hence their correlation functions along the MD trajectories.
\end{itemize}

%Here it is worth mentioning a technical detail that has a high impact on the rapidity of this task. With respect to the AL performed during MD where the trajectory is generated along with the training set, here AL performs a scan of the existing configurations. 
%Since $s=c(\overrightarrow{\textbf{x}})\cdot s_z$, where $\overrightarrow{\textbf{x}}$ is the vector of Cartesian coordinates for a single structure, the term $s_z$ does not appear anymore no CASSCF calculation needs to be actually performed during AL since the criterium in \ref{criterion_stop} does not require any knowledge of the target quantity, but only the chemical desciptors of equivariant SNAP. This permits to easily perform in a single shot all the CASSCF calculations on the training set in parallel. 
Also in this case, we validate the accuracy of the ML model by calculating the RMSE of its predictions for the same configurations in the test set used for MD.
RMSE values on test sets for spin Hamiltonian tensors for \textbf{1} range between 1.49 and 2.90 cm$^{-1}$, and 0.97 and 1.90 cm$^{-1}$ for \textbf{2}. For \textbf{3}, RMSE test set values depend on the order $l$: it ranges between 0.31 and 0.74 cm$^{-1}$ ($l=2$), \num{2.7e-3} and \num{5.9e-3} cm$^{-1}$ ($l=4$) and \num{5e-5} and \num{1.2e-4} cm$^{-1}$ ($l=6$). For more details on RMSE values on the training sets and for specific RMSE values on the test set, see Tabs. \ref{SI:training_set_RMSE_correlations} and \ref{SI:test_set_RMSE_correlation}, respectively. Parity plots related to the results on tensorial predictions are shown in Figs. S28-S39. 

\begin{figure}
    \centering
    \includegraphics[scale=0.65]{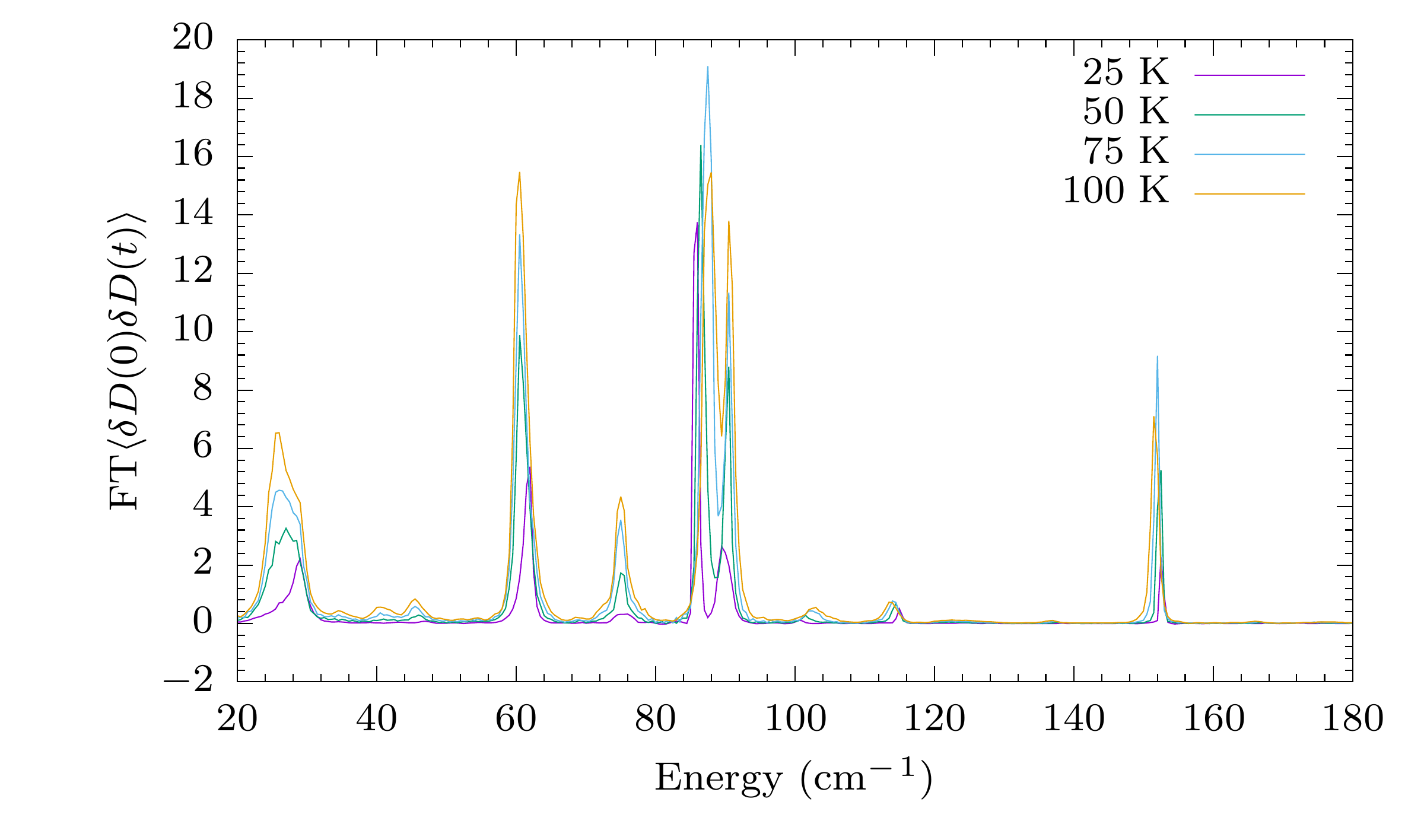}\\
    \includegraphics[scale=0.65]{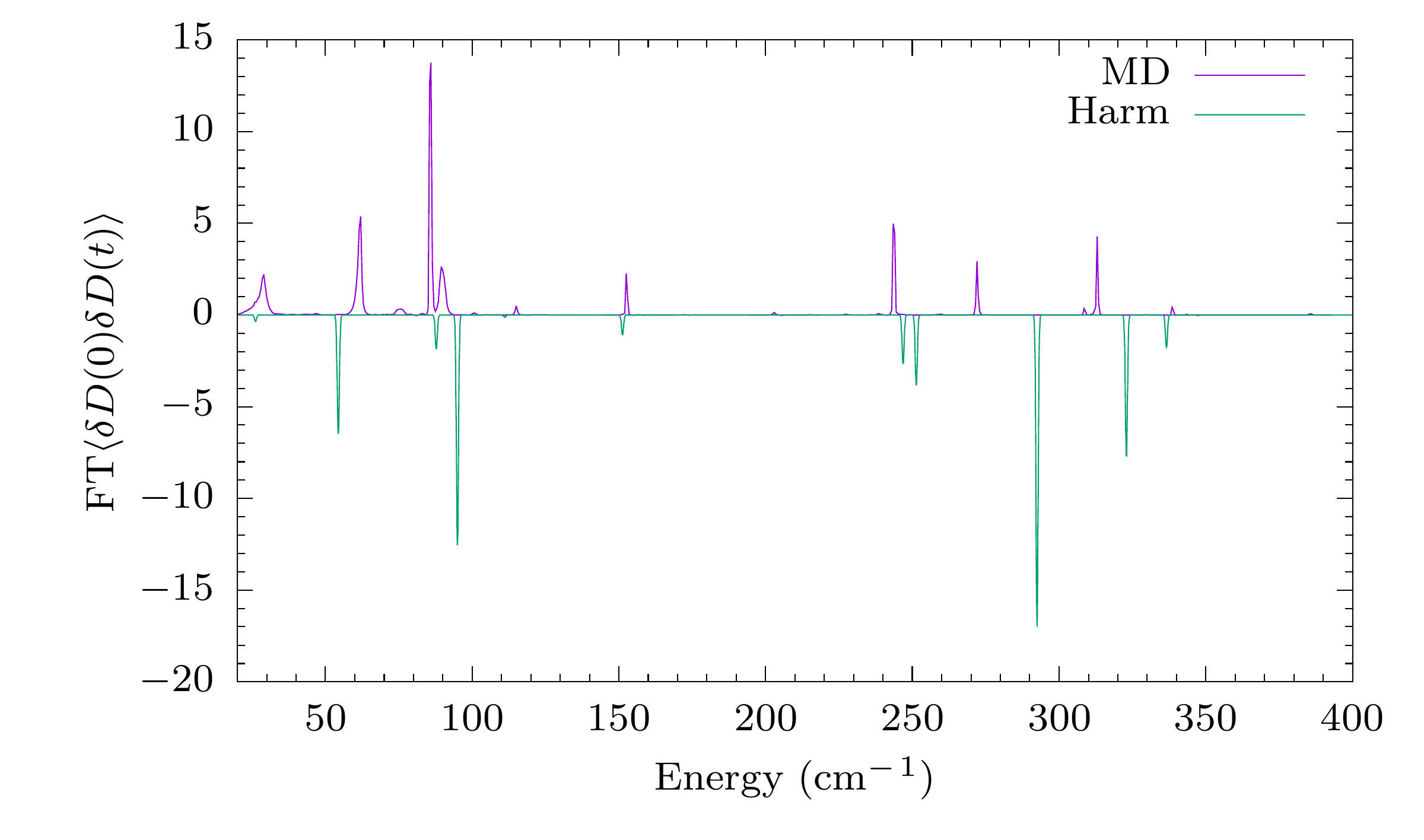}
    \caption{\textbf{Fourier transform of correlation functions for compound 1}. 
    Top: Average of the Fourier transforms of the correlation functions of $\delta D_{ij}$ at four different temperatures.
    Bottom: Average of the Fourier transforms of the correlation functions of $\delta D_{ij}$ calculated as in LHS of Eq.\ref{correlation_function_coupling} at $T=25$ K.}
    \label{correlation_profiles_vs_T}
\end{figure}

%RMSE results are reported in  Tabs. \ref{SI:training_set_RMSE_correlations} and \ref{SI:test_set_RMSE_correlation}.\\
% We are now in the position to calculate spin Hamiltonian tesnors over approximately 10M structures for every compound at each temperature and their correlation functions along the MD trajectories.
%From the Fourier transform of the correlation functions, spin-phonon couplings can be extracted 
%All the relevant plots for the correlation functions and their Fourier transforms compared to the spin-phonon densities from first-principles calculations in harmonic approximation are reported in the SI.
Fig. \ref{correlation_profiles_vs_T}, reports the plot for element $D_{11}$ of compound \textbf{1} at four different temperatures. As expected, the heights of the peaks directly correlate with the temperature since the proportionality factor omitted in Eq.\ref{correlation_function_coupling} is the thermal population of phonons. As can be seen from these results, the simulation of spin-phonon coupling from MD automatically accounts for the presence and the thermal evolution of the phonons' frequencies and linewidths. Fig. \ref{correlation_profiles_vs_T} and Figs. S40-S43 show that such features cannot be reproduced with a simple equilibrium harmonic approximation of the lattice dynamics. 

The correlation functions could in principle be used to estimate the relaxation time. However, a proper molecular dynamics spin relaxation theory is not available yet, and its derivation is beyond the scope of this work. Indeed, whilst MD is able to include non-equilibrium effects and phonons' dissipation, it is based on a classical notion of nuclei, which is expected to be of limited accuracy at such low temperatures. Moreover, MD has only been adapted to second-order perturbation theory so far, whilst Raman relaxation will require the development of a fourth-order theory.

\section{Discussion and conclusions}

In this study, we have presented a ML workflow based on AL that leads to computational savings of at least 80 \% in the prediction of molecular vibrations and spin-phonon couplings keeping near-to-full ab initio accuracy. The precision achieved on vibrational spectra has been proven to be sufficient to reach excellent results also in the prediction of spin-phonon relaxation. Due to the non-trivial dependence of the spin-phonon relaxation rate on molecular vibrations, this has to be considered a remarkable result.

We have also tested an equivariant extension of SNAP to predict the Spin Hamiltonian tensors proving its successful integration within the spin relaxation framework. This aspect of the work has room for refinement and we believe that different and complete chemical descriptors, improvable \textit{ad libitum} \cite{moment_tensor_potential,Dratzl_ACE}, represent a viable solution to this problem. However, the higher number of parameters associated with these descriptors is usually linked to larger training sets, sometimes by orders of magnitude, therefore possibly reducing the computational savings offered by the presented workflow. A detailed benchmark of different descriptors and their trade-off between efficiency and accuracy is a necessary next step. 

We note that vibrational spectra could have also been obtained through analytical differentiation. However, in general, the memory requirements of the analytical approach scale unfavourably with the system size and are not necessarily more efficient than the numerical differentiation\cite{bykov2015efficient}. In view of the applications of this method to large molecules and crystals, the favourable scaling of energy and gradients is key. Moreover, analytical frequencies are not necessarily always available for any level of theory, e.g. CASSCF or Coupled Cluster, and in any software. The present workflow on the other hand can be adapted to any electronic structure method used to produce the ab initio reference data. Similarly, the analytical calculation of spin-phonon coupling coefficients has been recently proposed\cite{staab2022analytic}, but it lacks the inclusion of spin-orbit coupling derivatives, and a numerical differentiation remains advisable for the time being.

Aiming to go beyond the equilibrium harmonic approximation for molecular vibrations, we have shown how SNAP and its equivariant version can be deployed to study the correlation functions of the spin Hamiltonian tensors, a task demanding tens of millions of tensorial predictions and generation of MD simulations on the scale of 50 ns. While this is far beyond the reach of any ab initio approach, regardless of the availability of analytical Hessians and gradients, it has required the execution of only hundreds of first-principles calculations.
In \cite{Lunghi2020_molecular_tumbling}, only the use of ML techniques has made possible the aforementioned calculations for an organic radical in solution. Here we have made a step forward with respect to \cite{Lunghi2020_molecular_tumbling} by challenging the capabilities of the ML model on the dynamics of more complex systems and by deploying an AL scheme for the construction of the training dataset to generate MD simulations. These kinds of simulations represent a challenging testbed for the AL framework since even a single inaccurate prediction during MD can lead to catastrophic results. Furthermore, equivariant SNAP is combined with the AL scheme in \cite{active_learning_Valerio} to achieve a rational construction of the training set also for the prediction of \textbf{D} and $B^l_m$.

In conclusion, we have presented here a ML method for the prediction of phonon spectra and spin-phonon couplings for molecules in the gas phase with computational savings of at least 80 \% compared to the full ab initio approach. We anticipate that this work, along with its future extensions to periodic systems, will significantly expand the scope of numerical simulations for the study of spin relaxation in condensed matter and non-equilibrium settings.

\vspace{0.2cm}
\noindent
\textbf{Authors Contribution}\\
The project was conceived by A.L. V.B. performed all simulations and their analysis under the supervision of A.L. All authors contributed to the discussion of results and to writing the manuscript. 

\vspace{0.2cm}
\noindent
\textbf{Acknowledgements and Funding}\\
This project has received funding from the European Research Council (ERC) under the European Union’s Horizon 2020 research and innovation programme (grant agreement No. [948493]). Computational resources were provided by the Trinity College Research IT and the Irish Centre for High-End Computing (ICHEC).

\vspace{0.2cm}
\noindent
\textbf{Conflict of interests}\\
The authors declare that they have no competing interests.

%\bibliographystyle{naturemag}
%\bibliography{biblio}

\clearpage

\setcounter{section}{0}
\setcounter{figure}{0}
\setcounter{table}{0}
\setcounter{page}{1}
% \documentclass[onecolumn,amsmath,amssymb,prd,reprint,longbibliography,floatfix,8pt]{revtex4-1}

% \usepackage{amsmath}
% \usepackage{tikz}
% \usepackage{cancel}
% \usetikzlibrary{arrows.meta, positioning, arrows, decorations.pathmorphing, backgrounds, fit, petri}
% \usepackage{braket}
% \usepackage{enumitem}
% \usepackage[utf8x]{inputenc}
% \usepackage{graphicx}
% \usepackage{dcolumn}
% \usepackage{bm}
% \usepackage[switch]{lineno}
% \usepackage{float} 
% \usepackage{multirow}
% \usepackage{booktabs}
% \usepackage{caption}
% \usepackage{subfig}
% \usepackage{siunitx}
% %\usepackage{multicol}
% \usepackage{hyperref}
% \usepackage{titlesec}

\renewcommand{\thefigure}{S\arabic{figure}}
\renewcommand{\thetable}{S\arabic{table}}

%\titlespacing{\subsection}{0pt}{*0}{0pt}
\sisetup{
  inter-unit-product=\ensuremath{{\cdot}},
  tight-spacing=true,
}
\sisetup{exponent-product = \cdot, output-product = \cdot}

%\begin{document}

\title{Supplementary Information}

\author{Valerio Briganti}
\author{Alessandro Lunghi}
\email{lunghia@tcd.ie}
\affiliation{School of Physics, AMBER and CRANN Institute, Trinity College, Dublin 2, Ireland}
\maketitle

\section{Machine learning of molecular vibrations - Optimization}
\subsection{Compound 1}

\begin{figure} [H]
    \centering
    \includegraphics[scale=0.65]{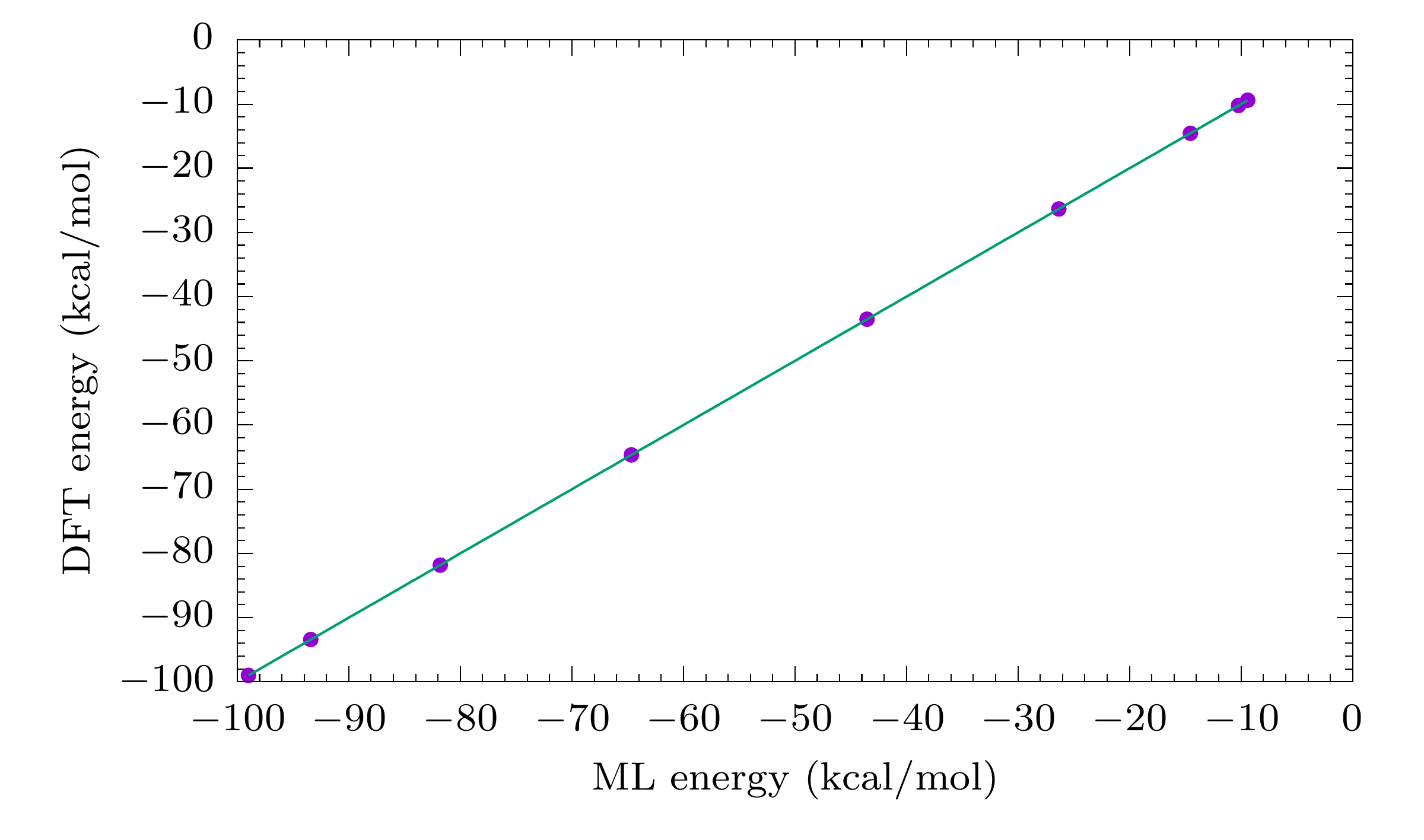}\\
    \includegraphics[scale=0.65]{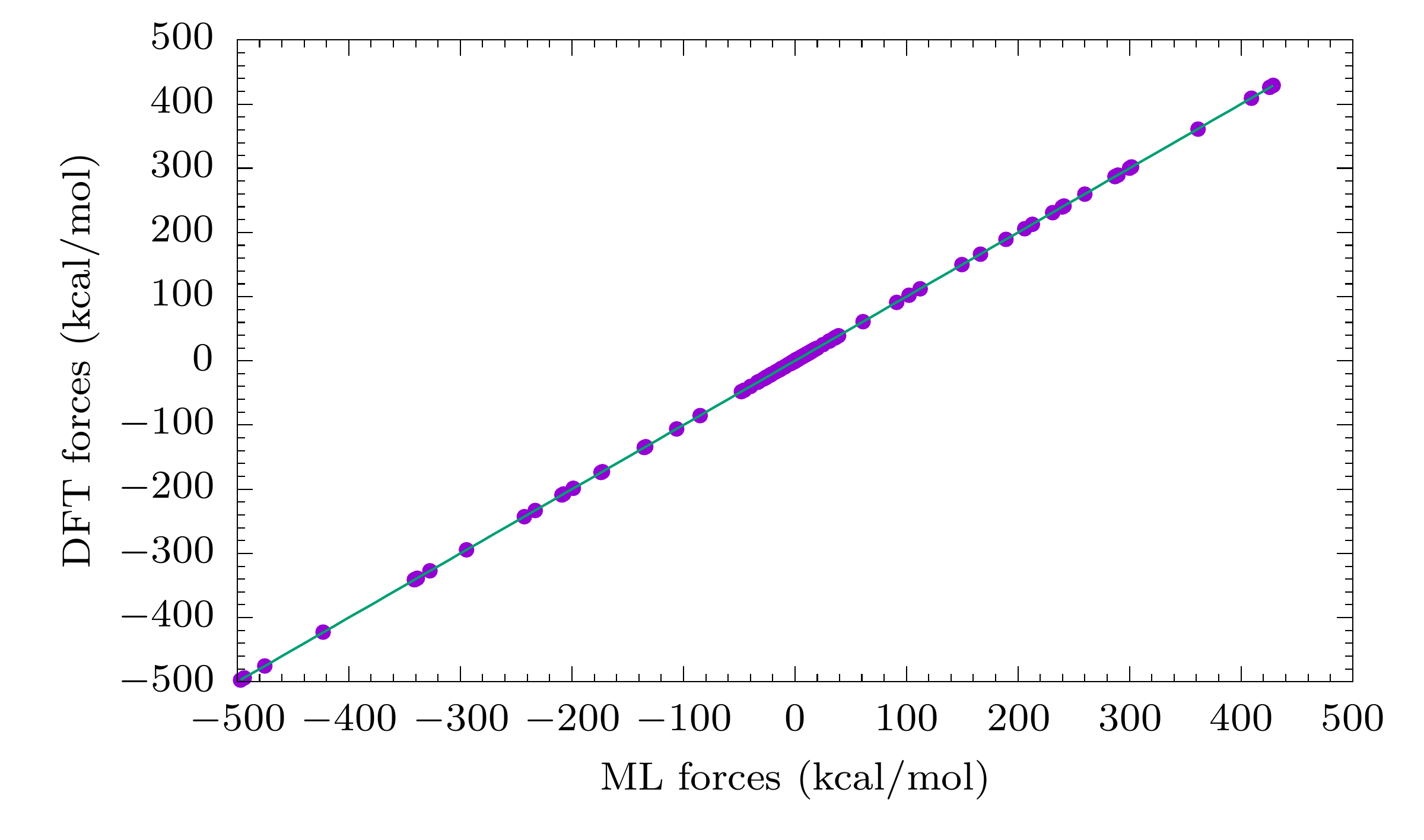}\\
    \includegraphics[scale=0.65]{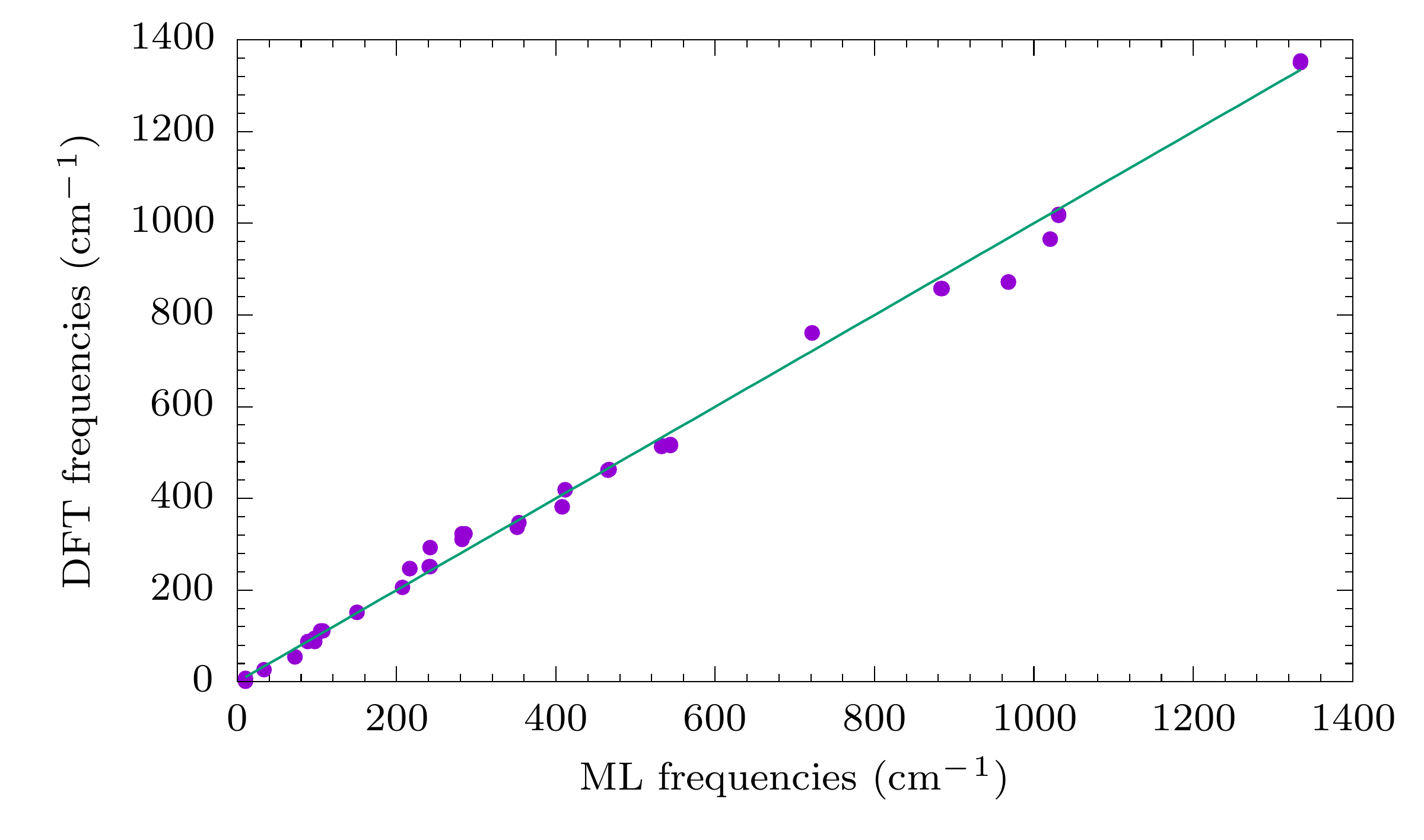}
    \includegraphics[scale=0.65]{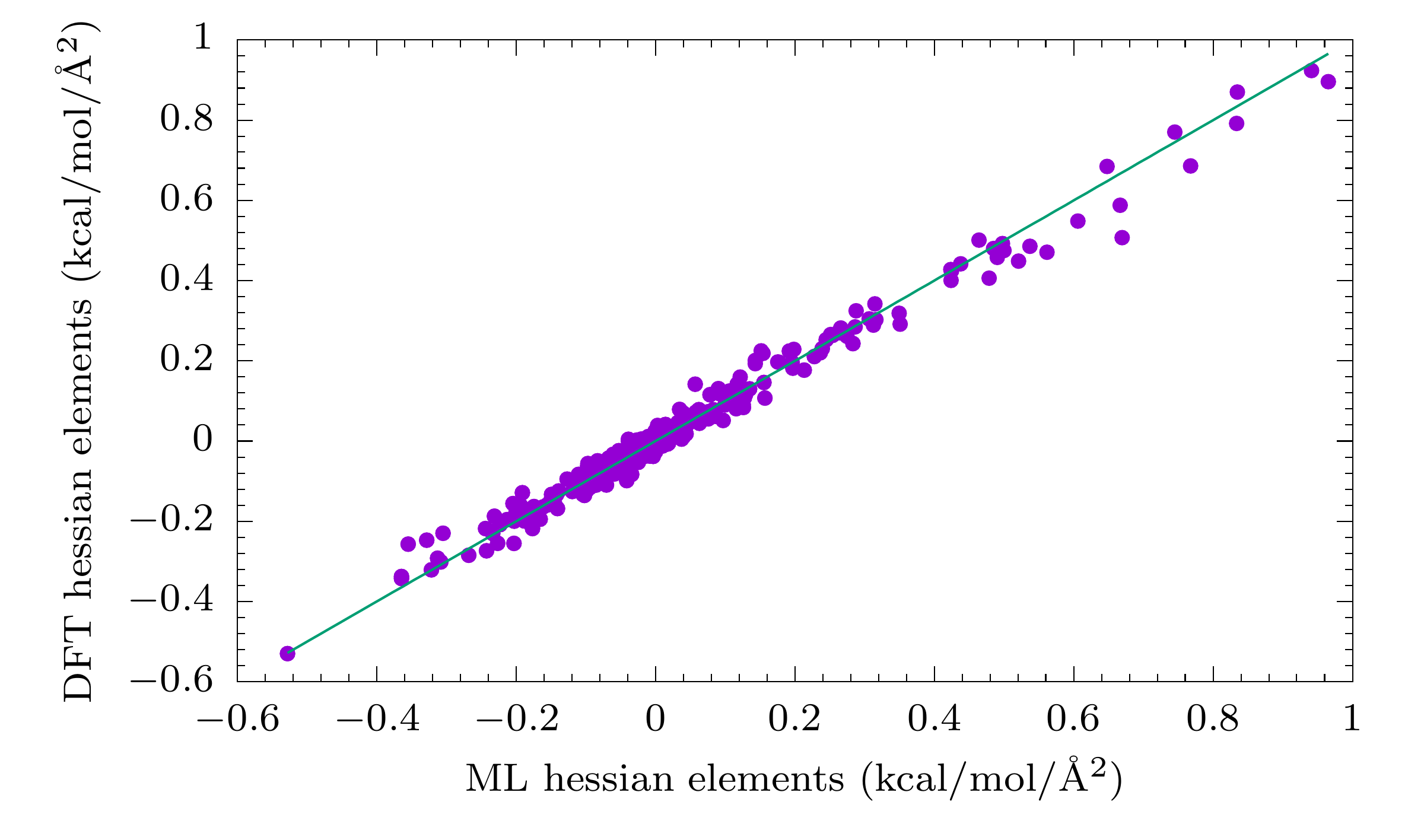}
    \caption{\textbf{Energy, forces, phonons and hessian.} $\delta =5$}
\end{figure}

\begin{figure} [H]
    \centering
    \includegraphics[scale=0.65]{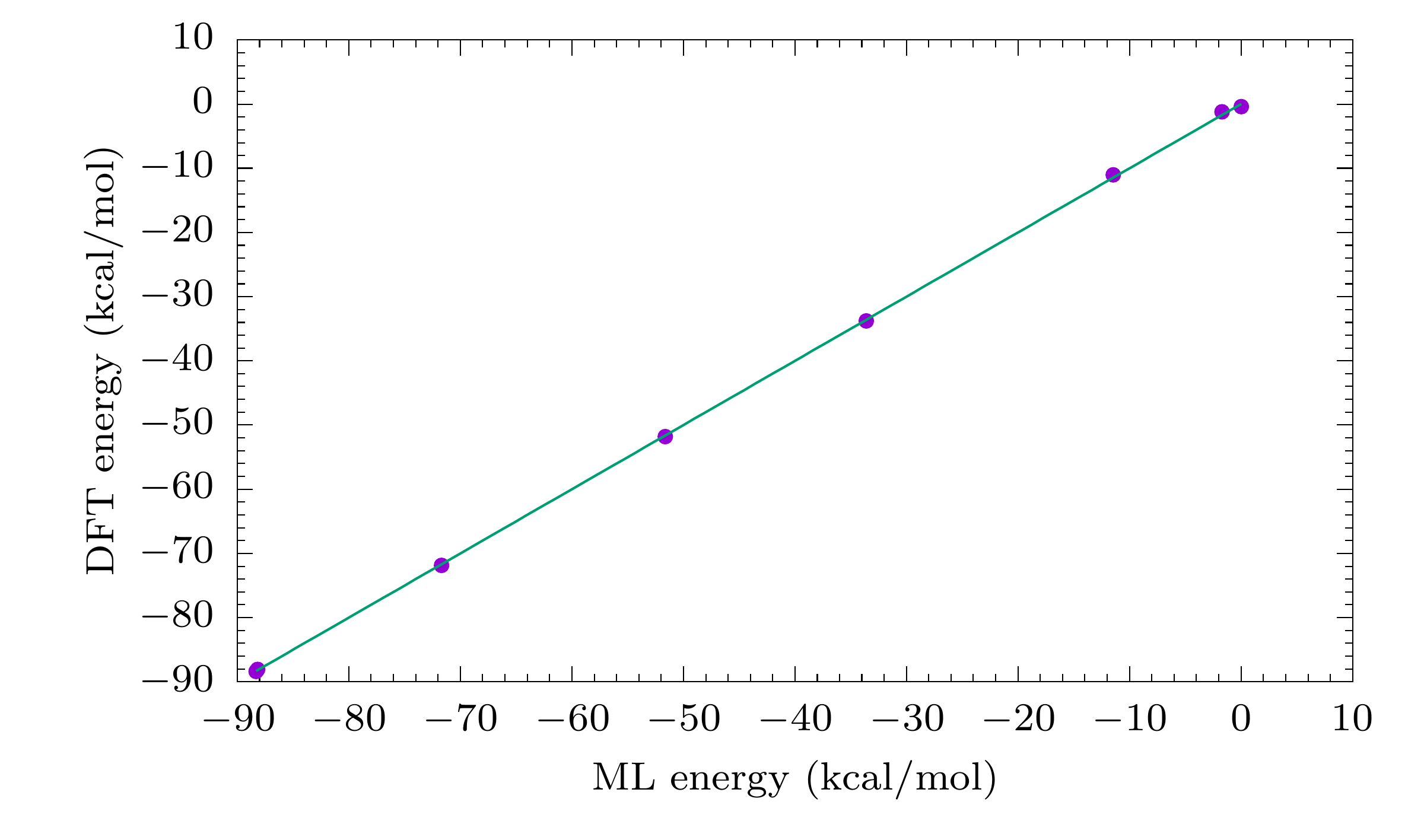}\\
    \includegraphics[scale=0.65]{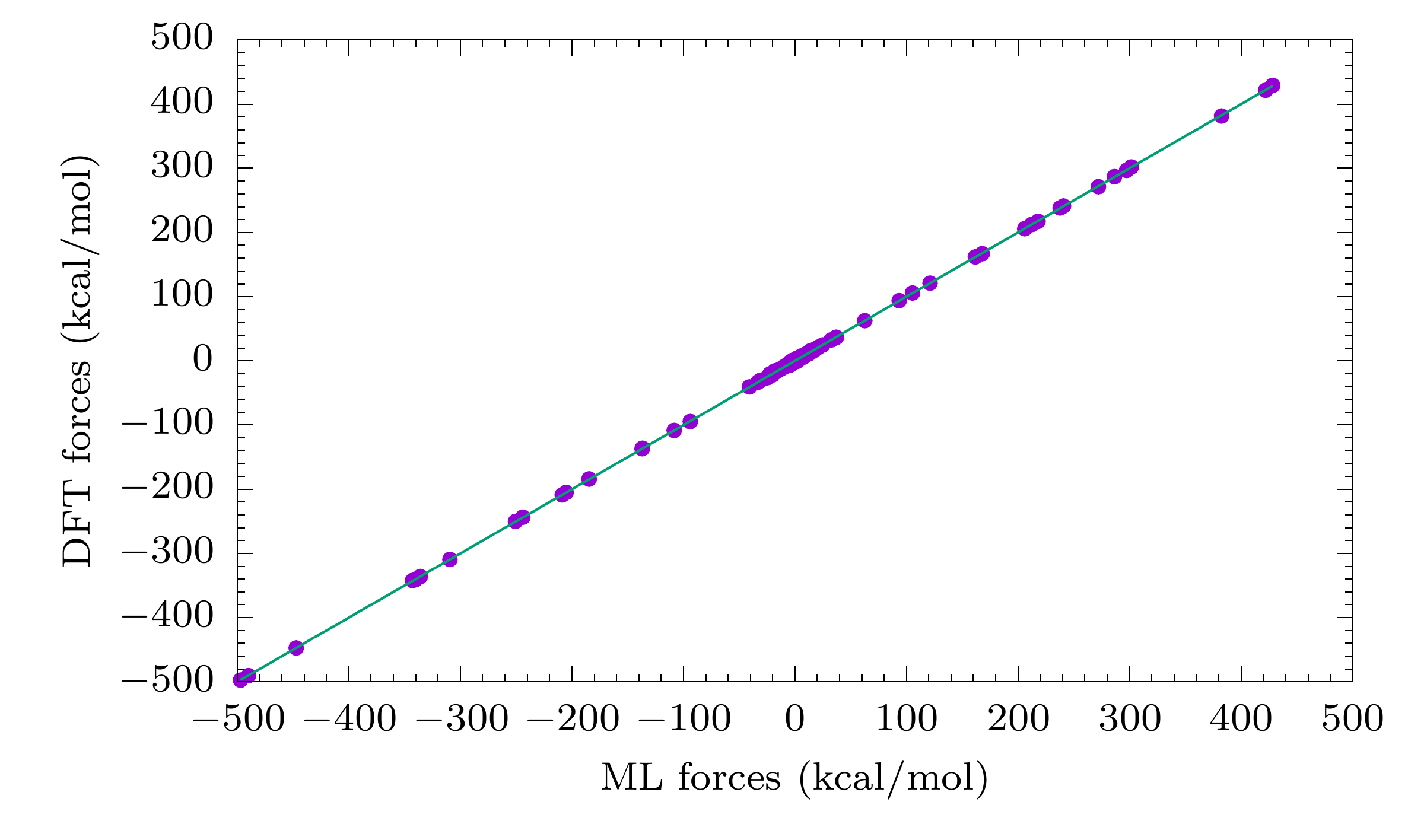}\\
    \includegraphics[scale=0.65]{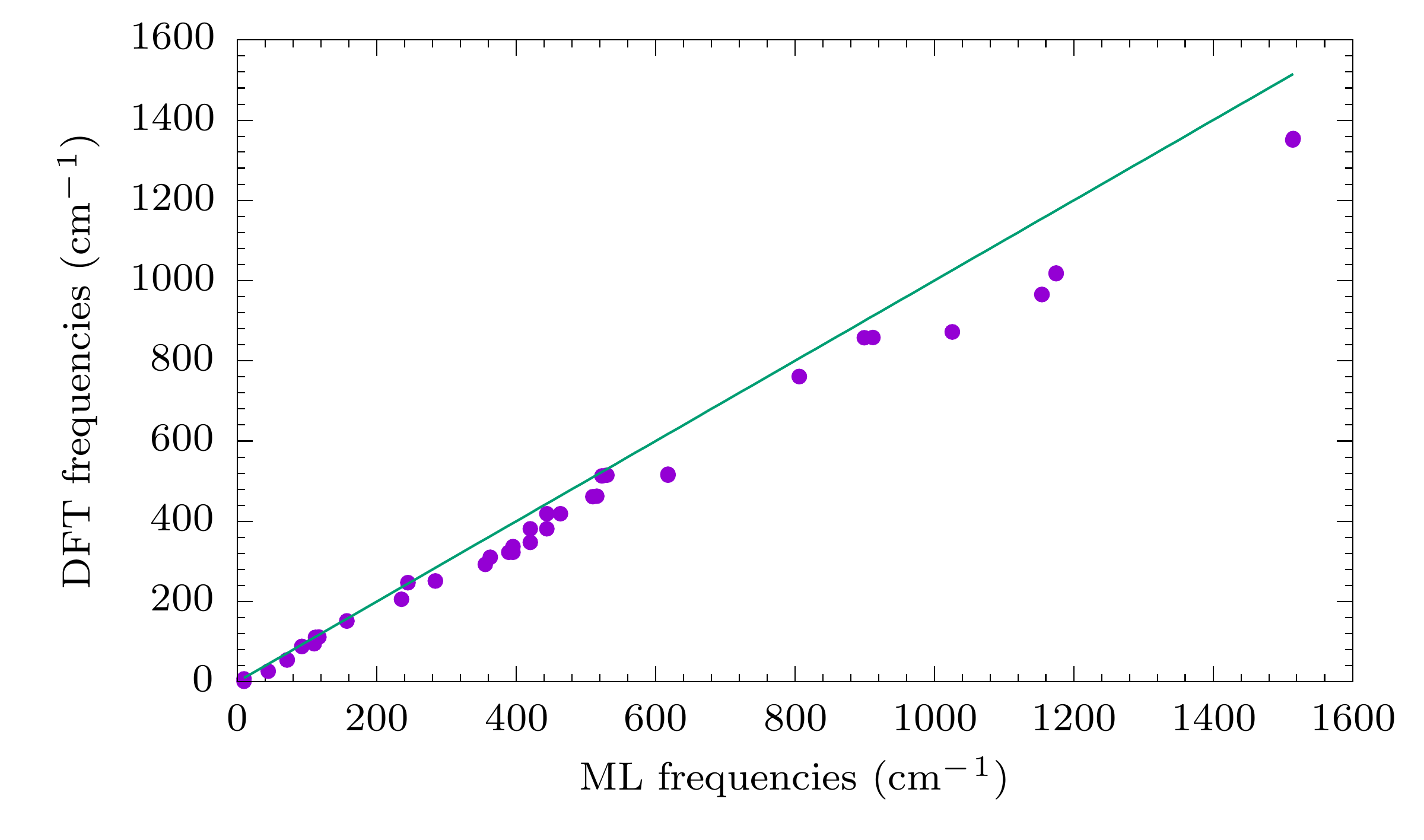}
    \includegraphics[scale=0.65]{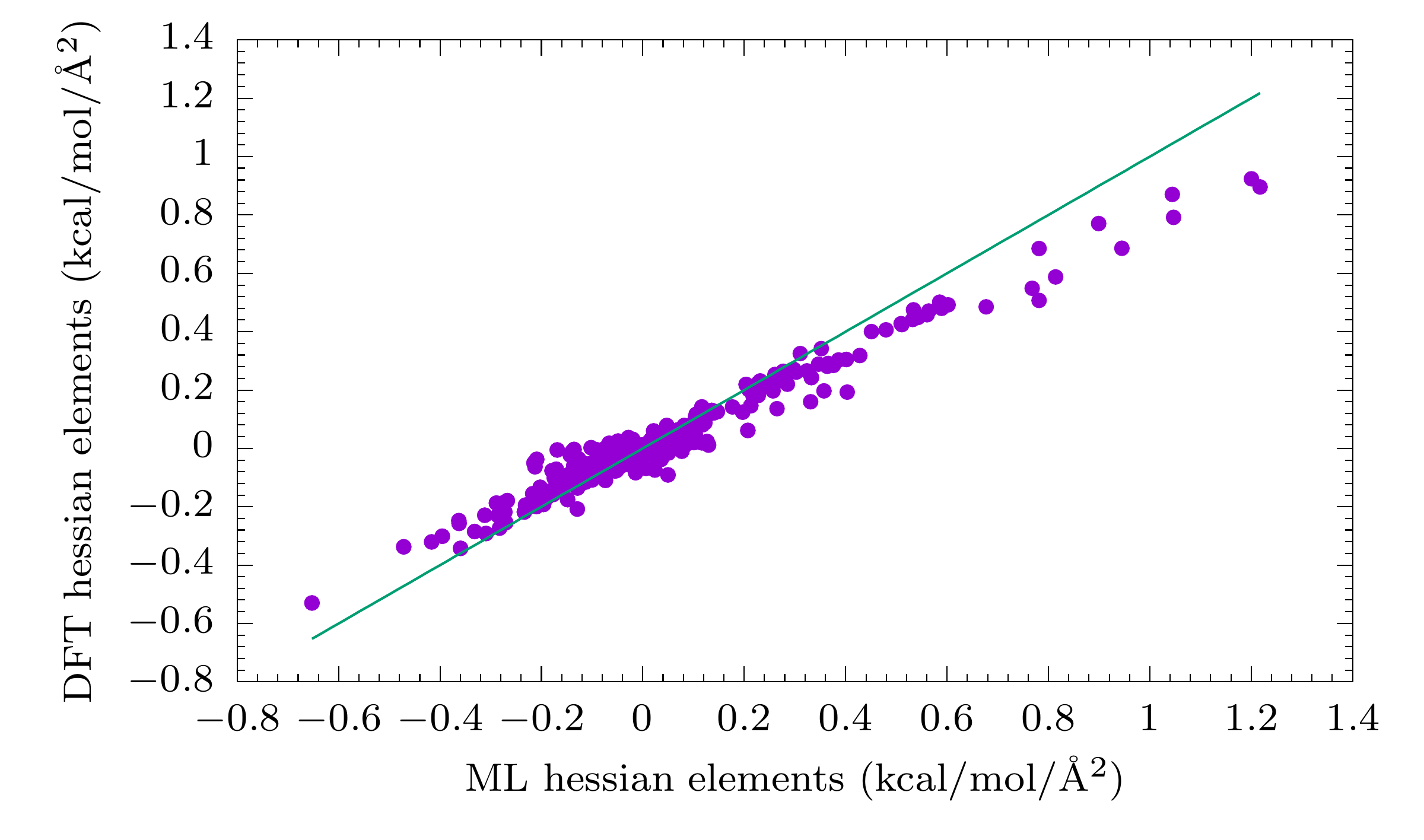}
    \caption{\textbf{Energy, forces, phonons and hessian.} $\delta =10$}
\end{figure}

\begin{figure} [H]
    \centering
    \includegraphics[scale=0.65]{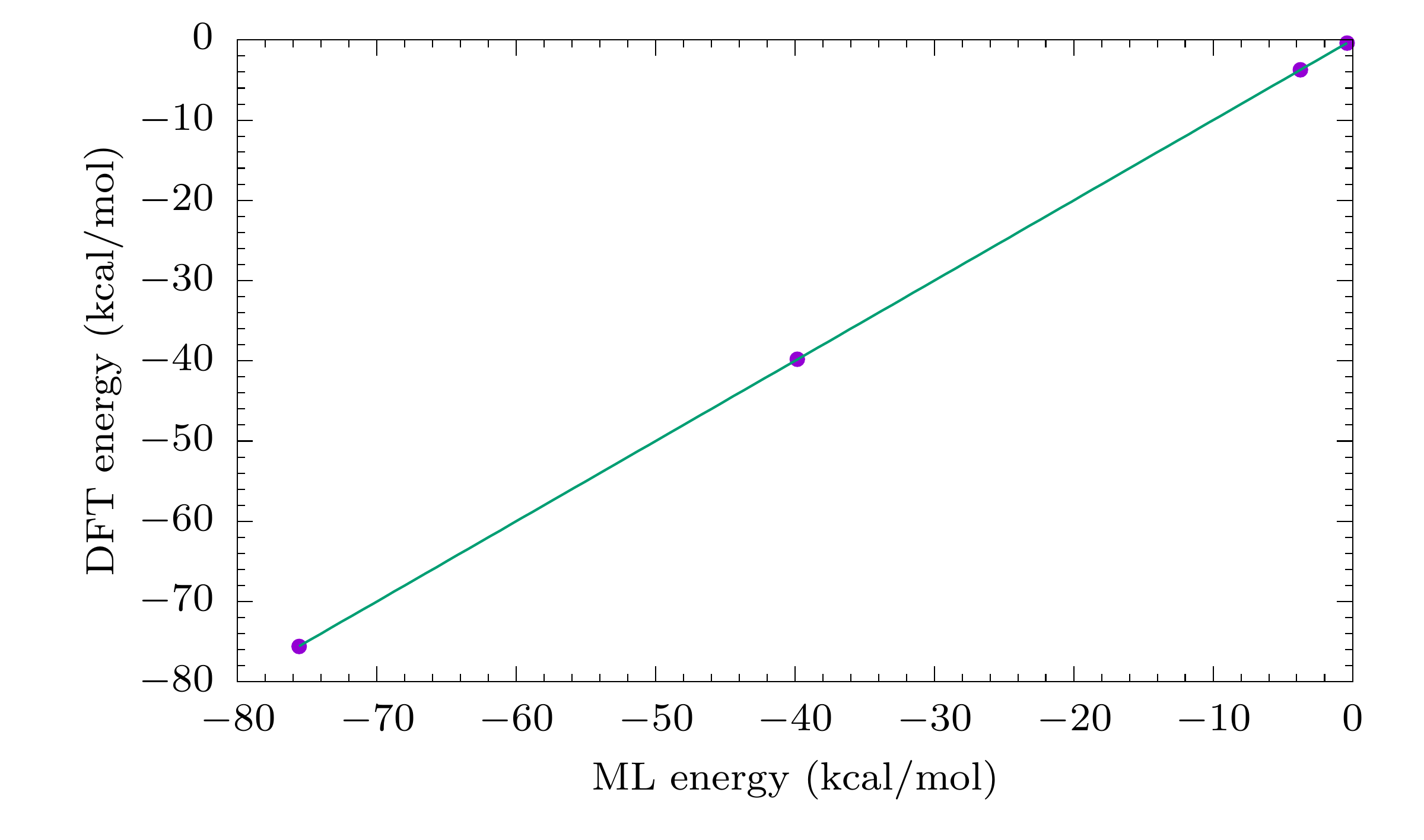}\\
    \includegraphics[scale=0.65]{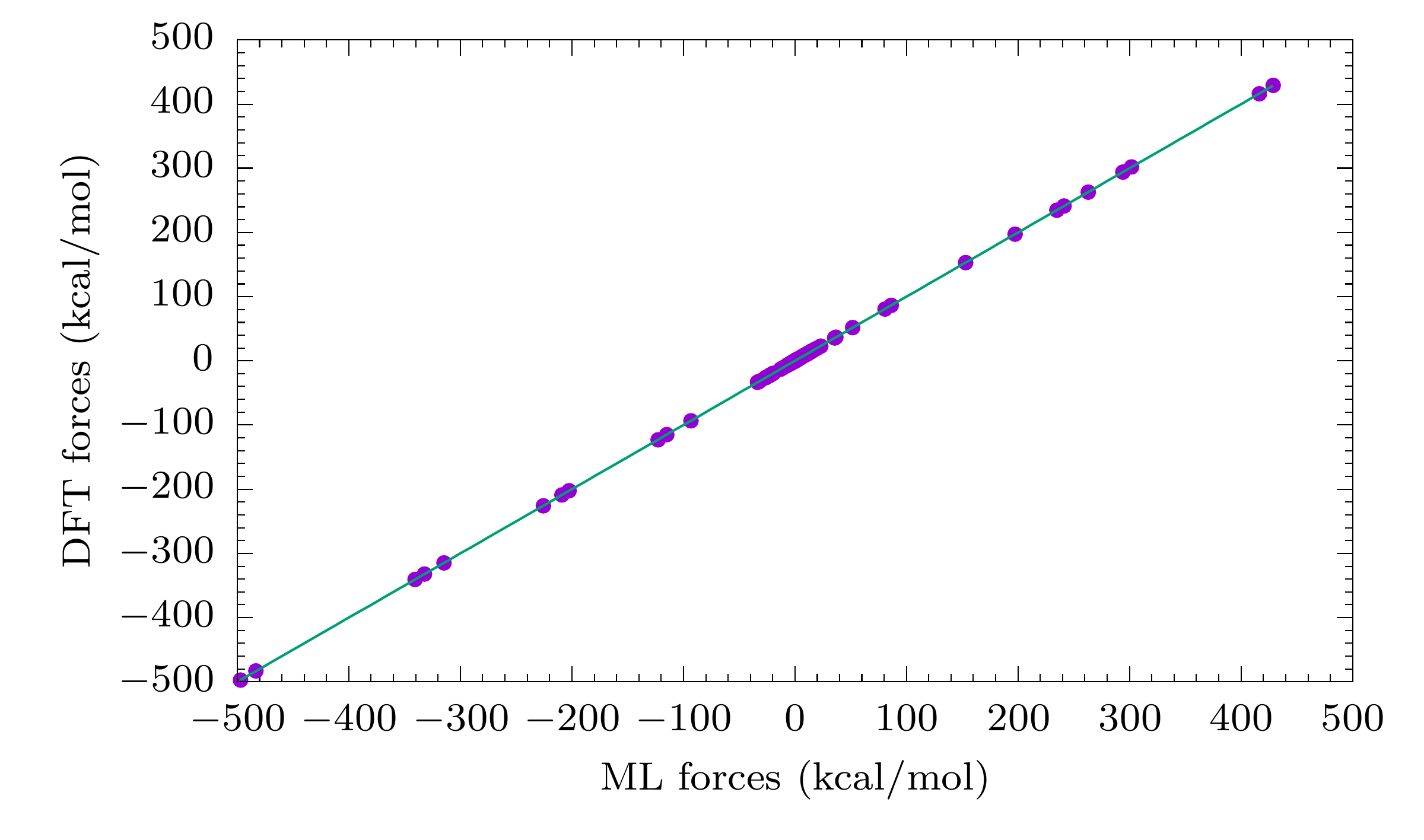}\\
    \includegraphics[scale=0.65]{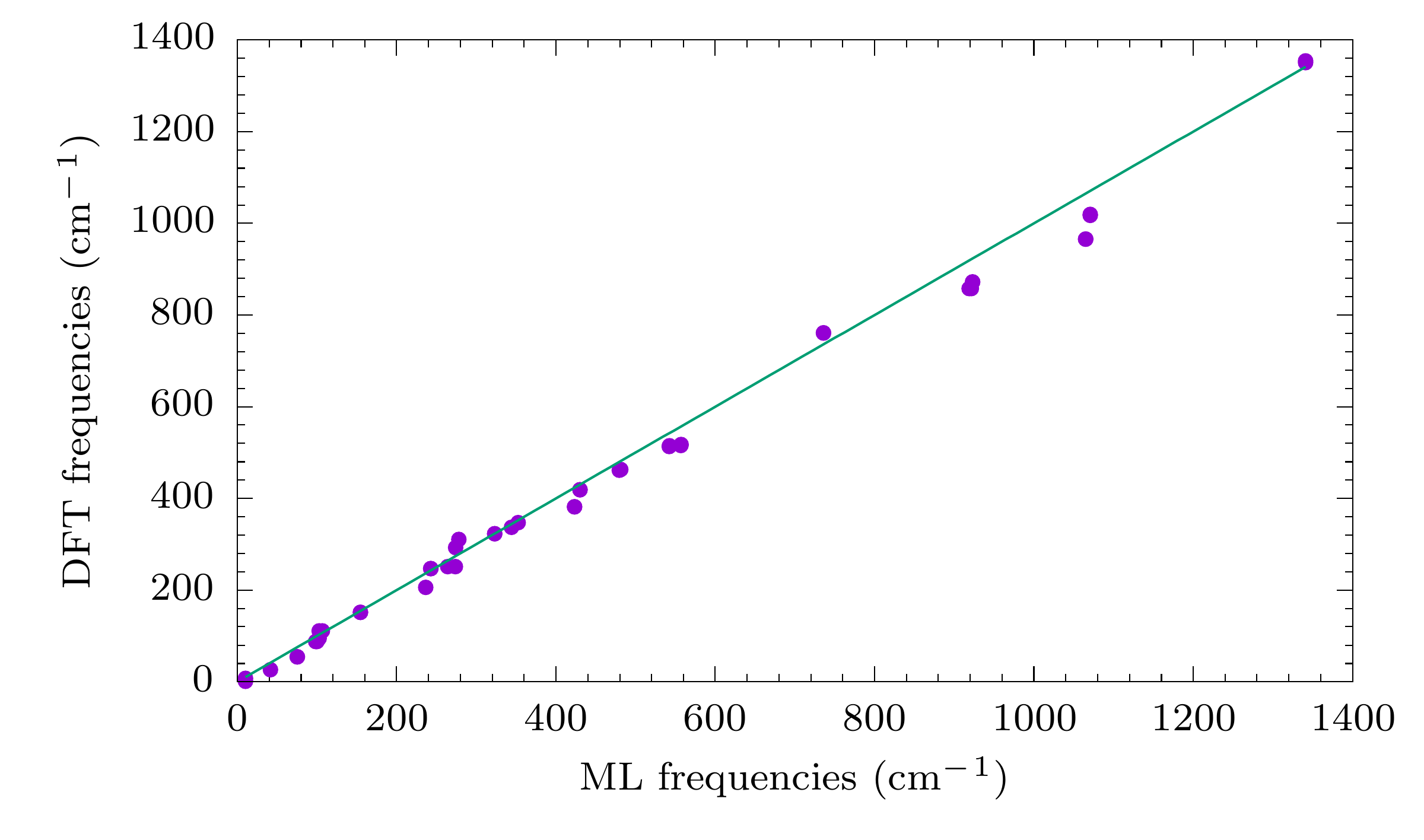}
    \includegraphics[scale=0.65]{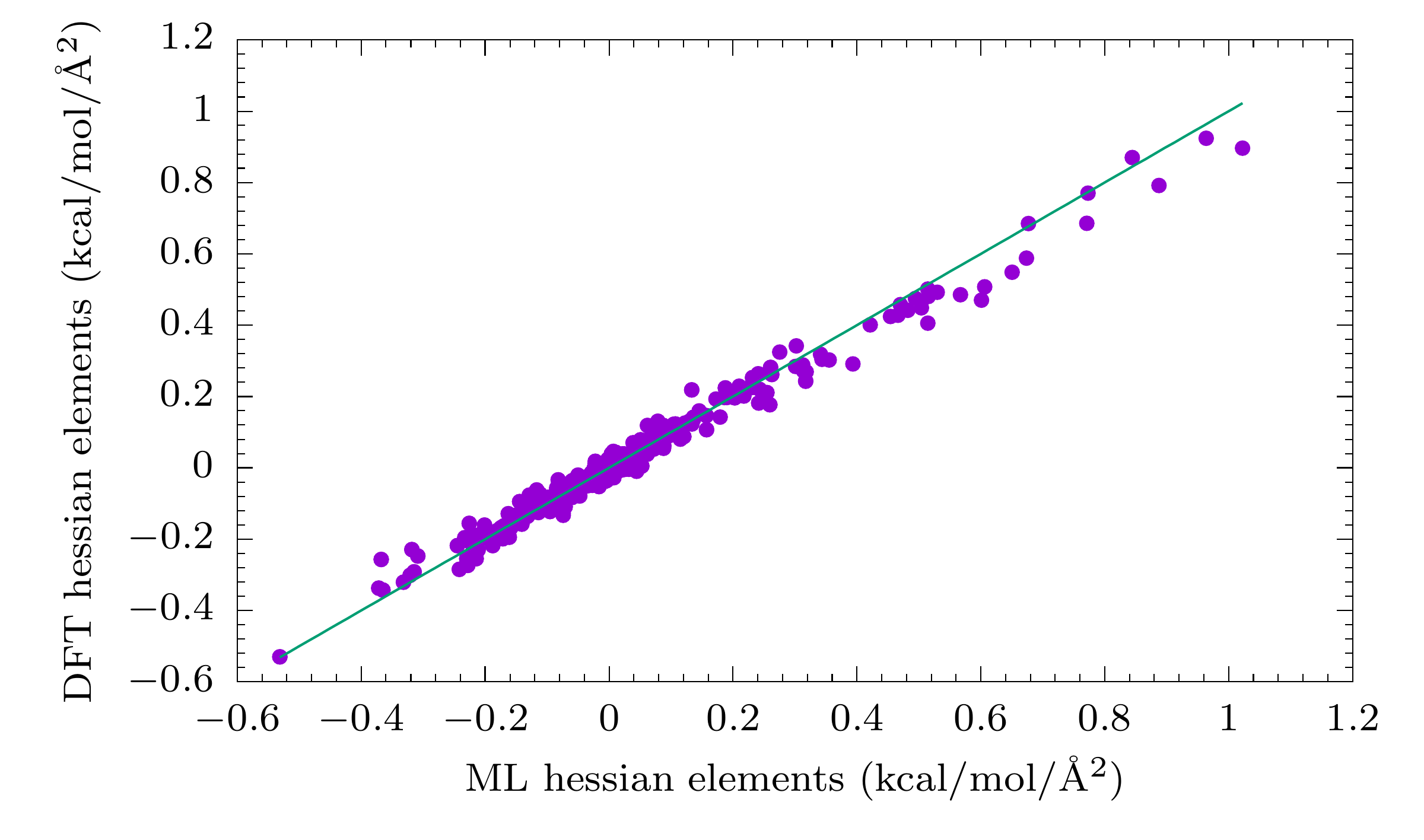}
    \caption{\textbf{Energy, forces, phonons and hessian.} $\delta =20$}
\end{figure}

\begin{figure} [H]
    \centering
    \includegraphics[scale=0.65]{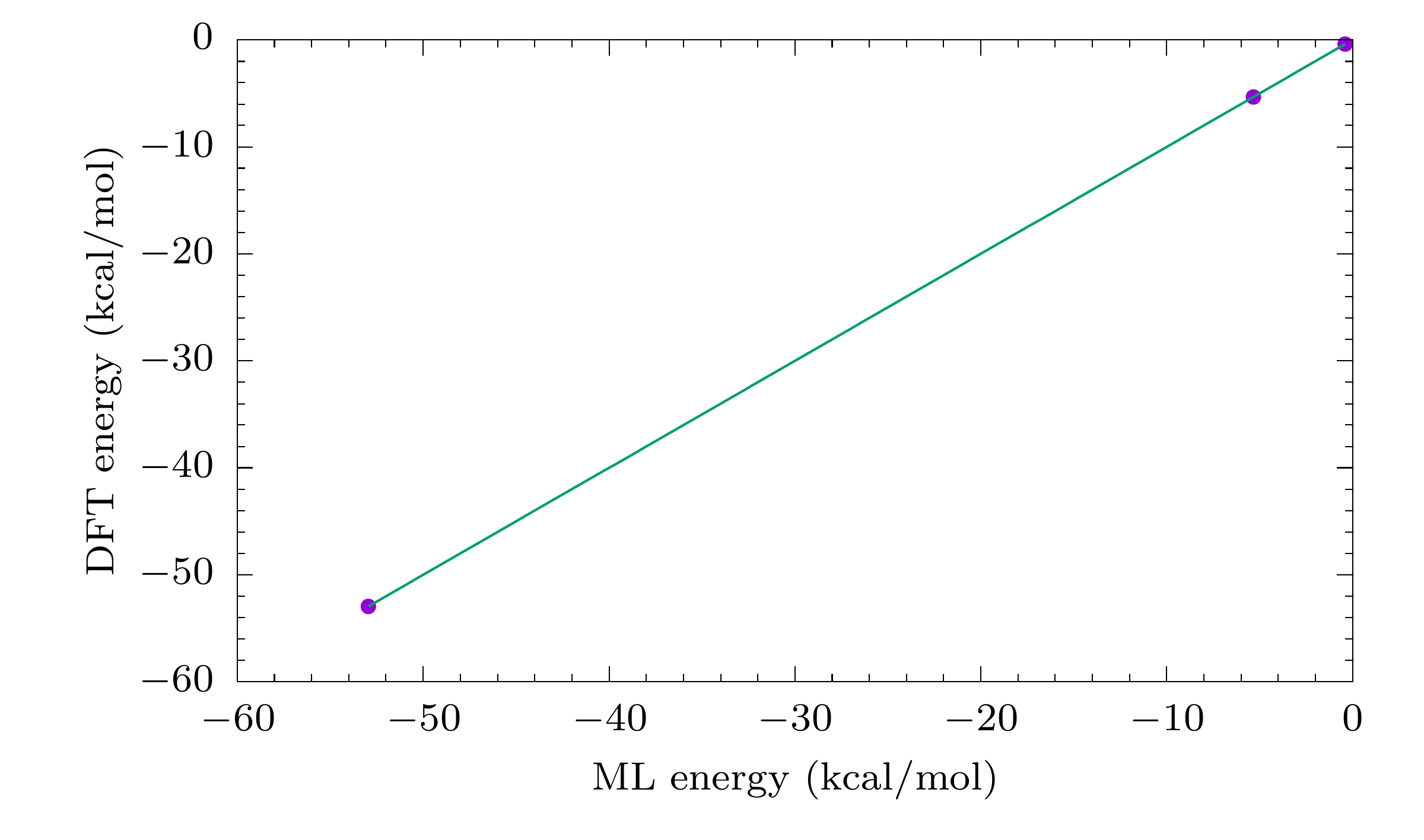}\\
    \includegraphics[scale=0.65]{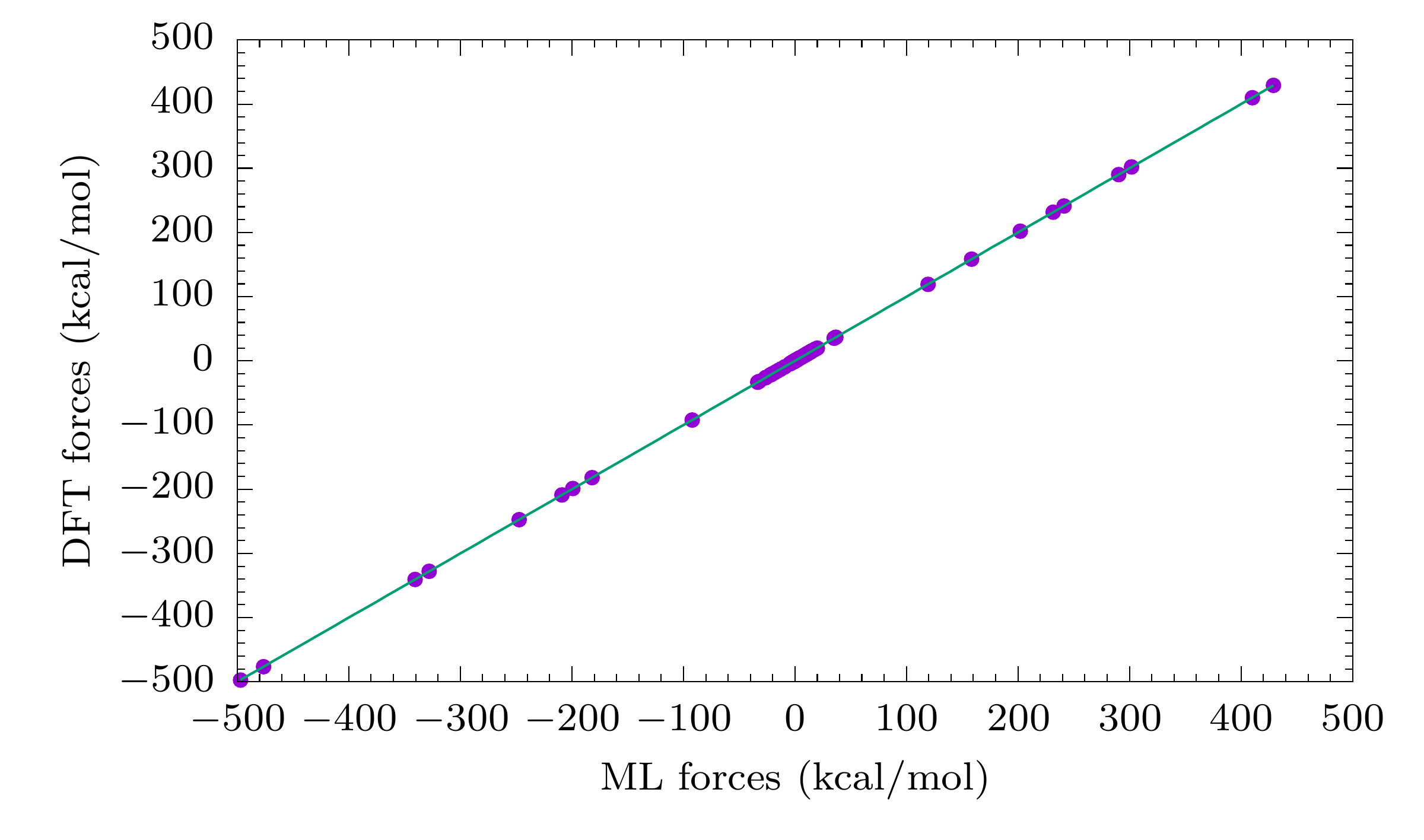}\\
    \includegraphics[scale=0.65]{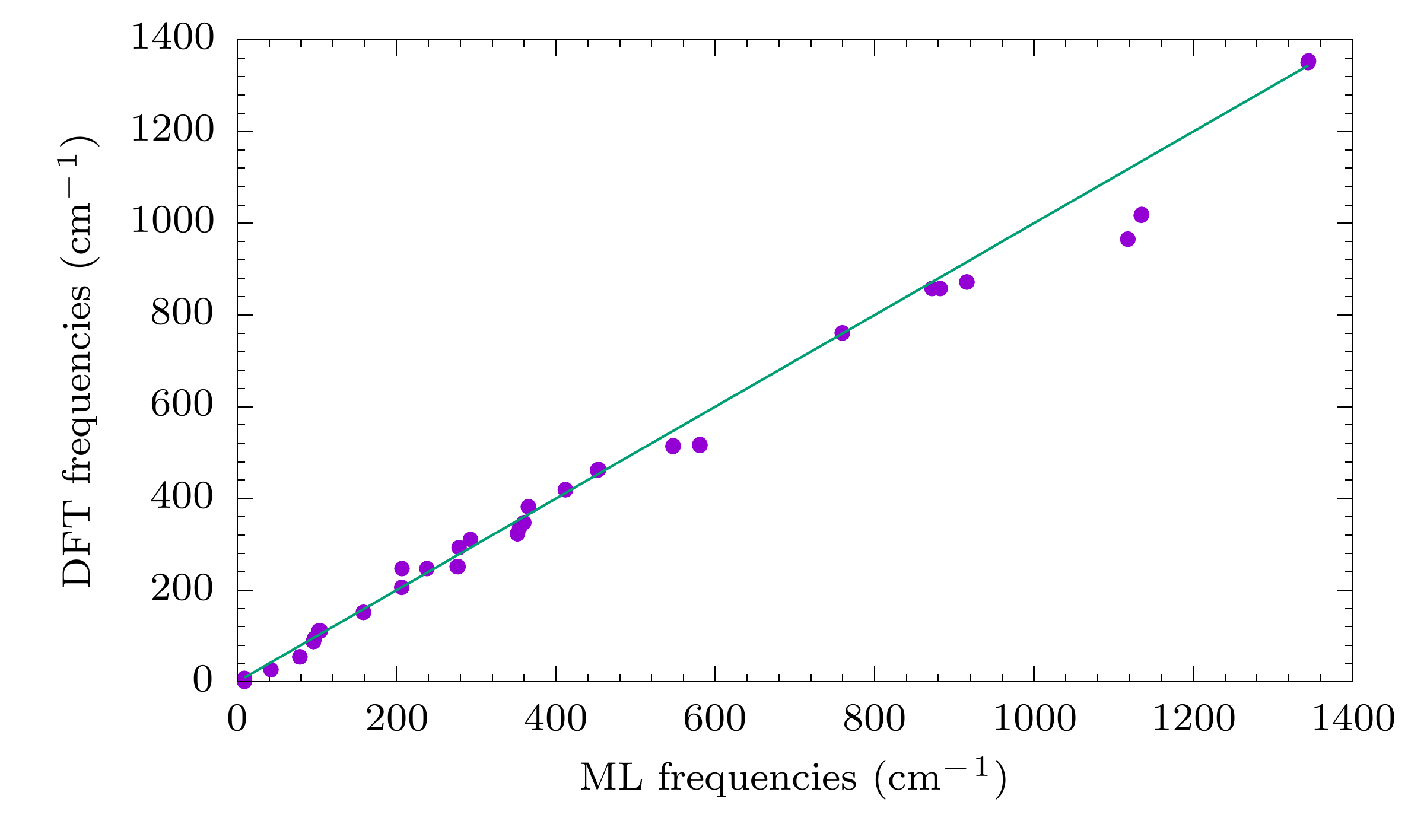}
    \includegraphics[scale=0.65]{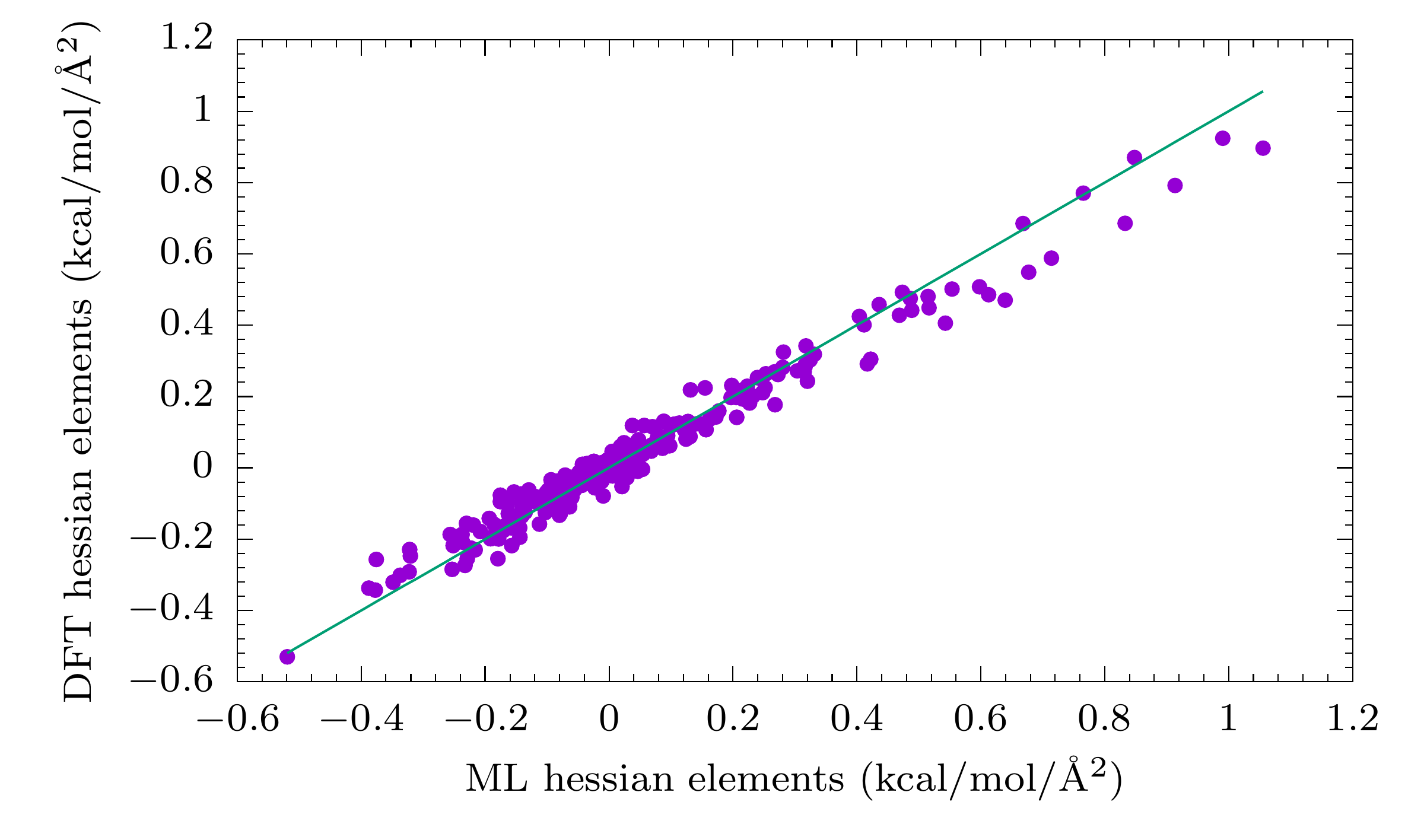}
    \caption{\textbf{Energy, forces, phonons and hessian.} $\delta =30$}
\end{figure}
\subsection{Compound 2}

\begin{figure} [H]
    \centering
    \includegraphics[scale=0.65]{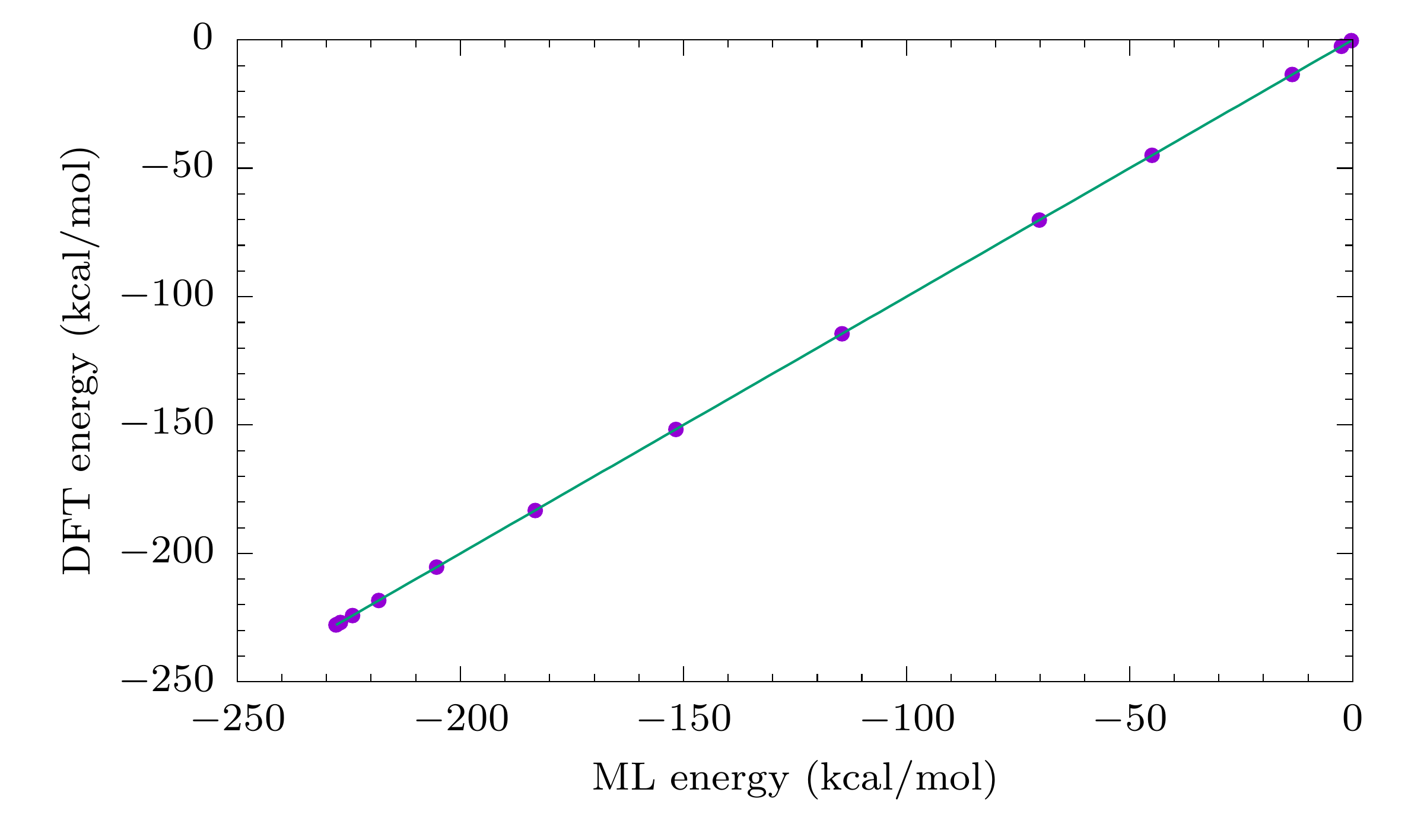}\\
    \includegraphics[scale=0.65]{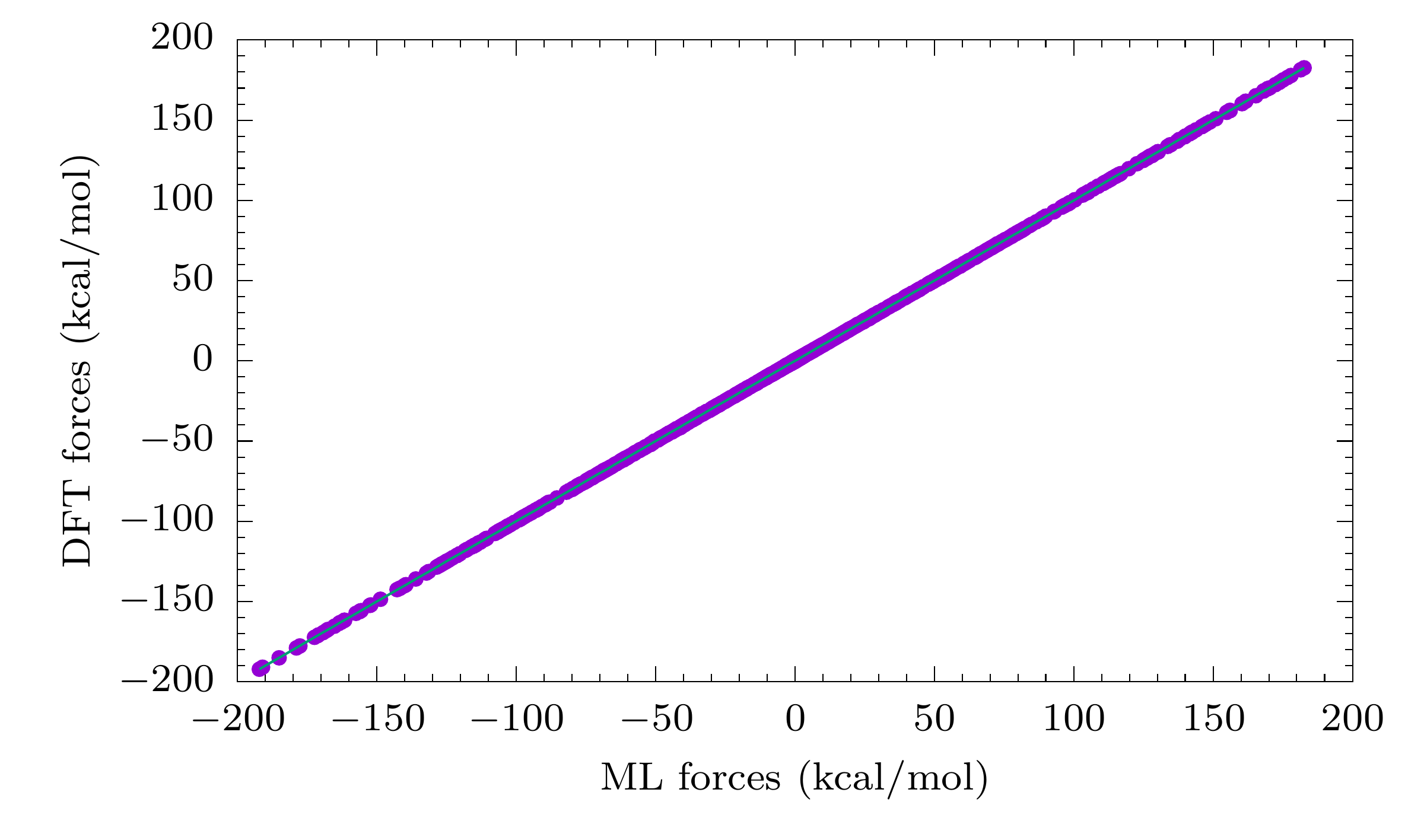}\\
    \includegraphics[scale=0.65]{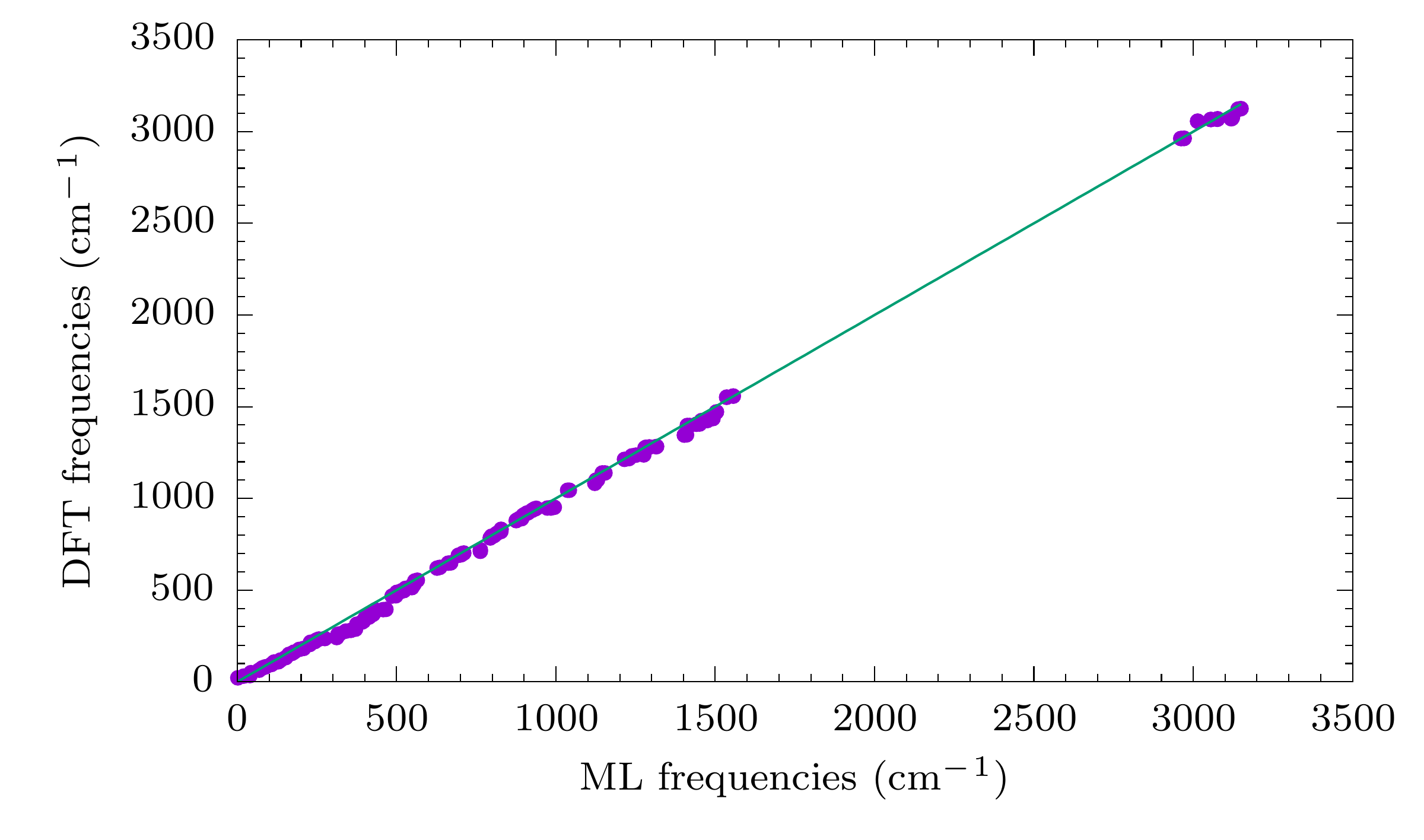}
    \includegraphics[scale=0.65]{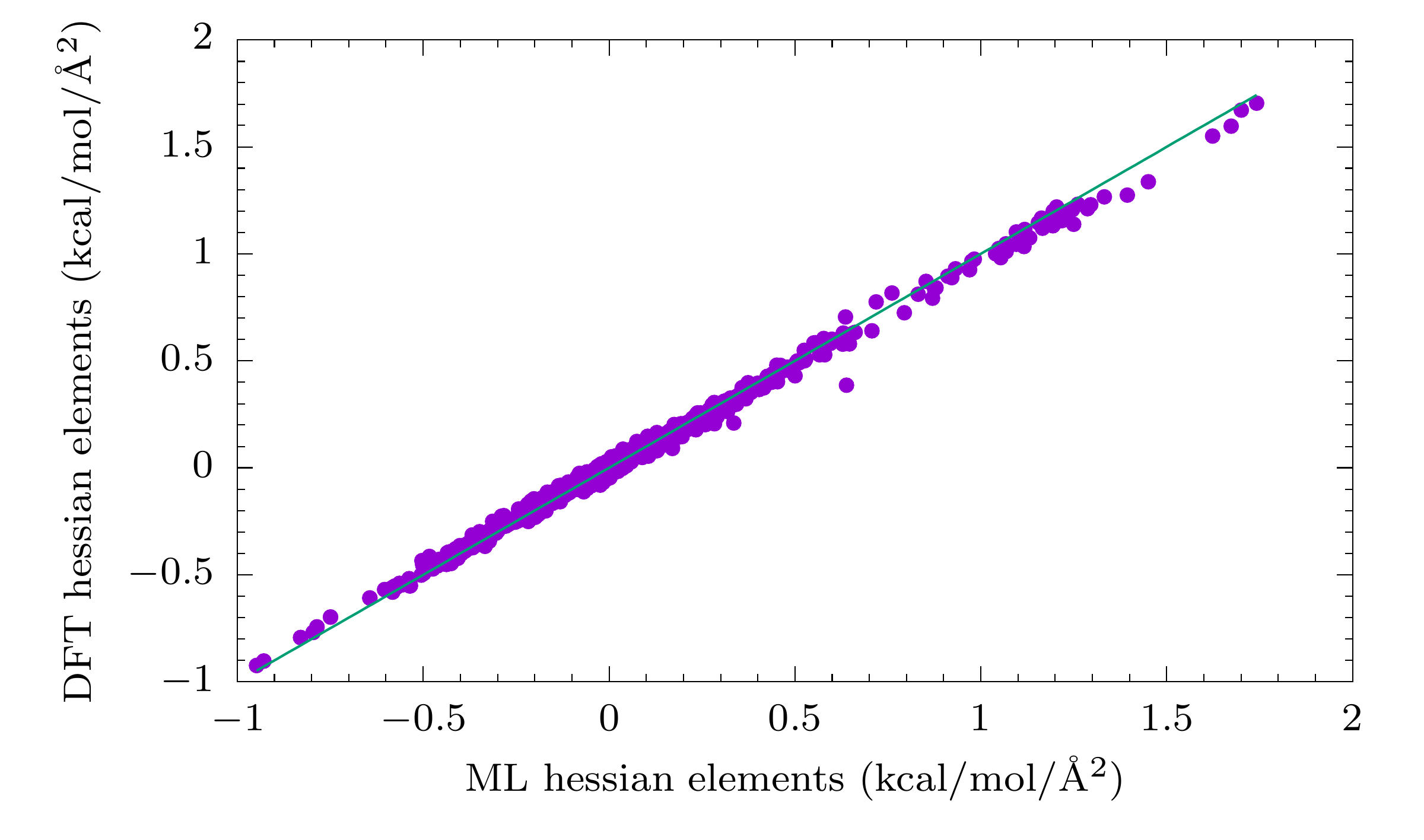}
    \caption{\textbf{Energy, forces, phonons and hessian.} $\delta =5$}
\end{figure}

\begin{figure} [H]
    \centering
    \includegraphics[scale=0.65]{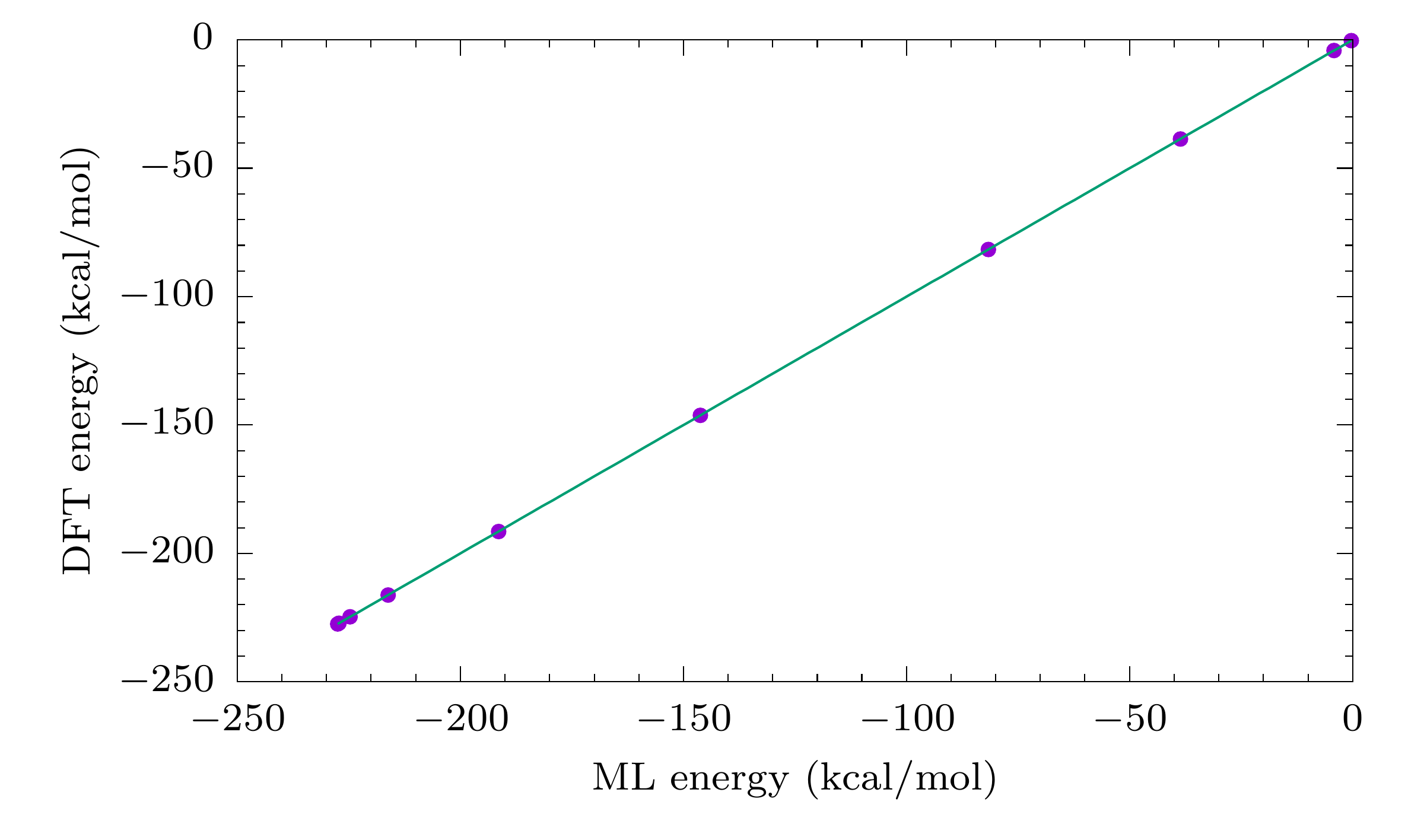}\\
    \includegraphics[scale=0.65]{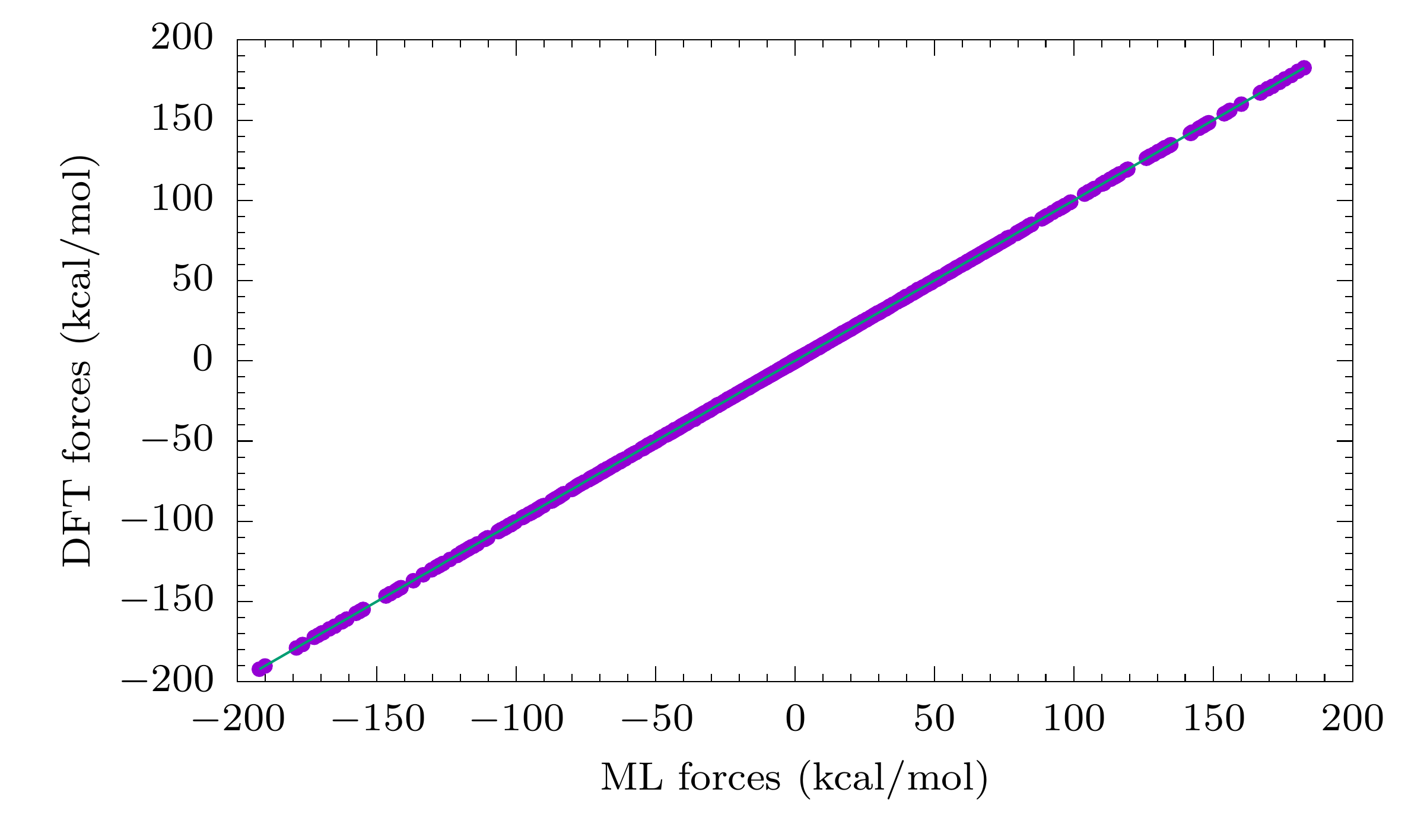}\\
    \includegraphics[scale=0.65]{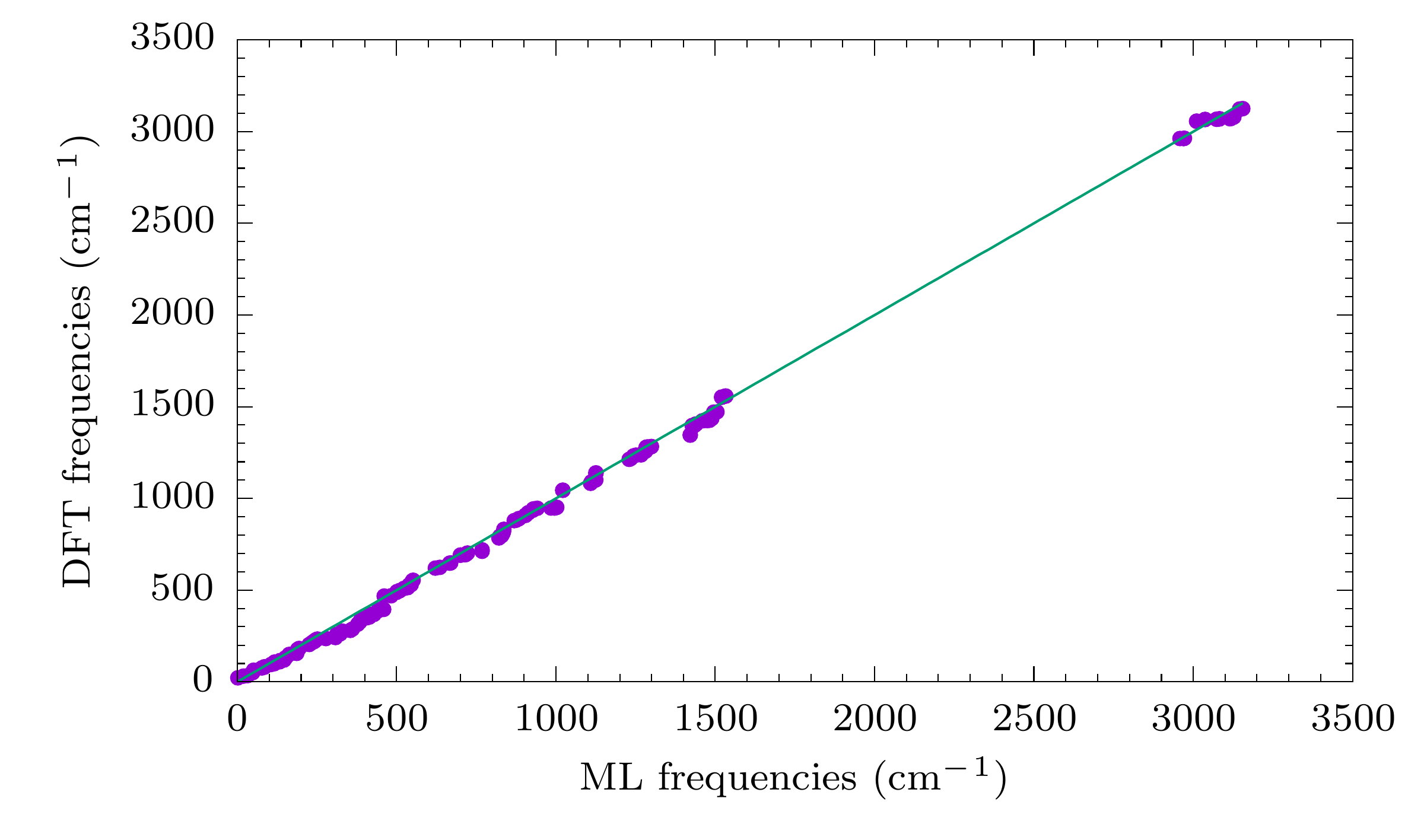}
    \includegraphics[scale=0.65]{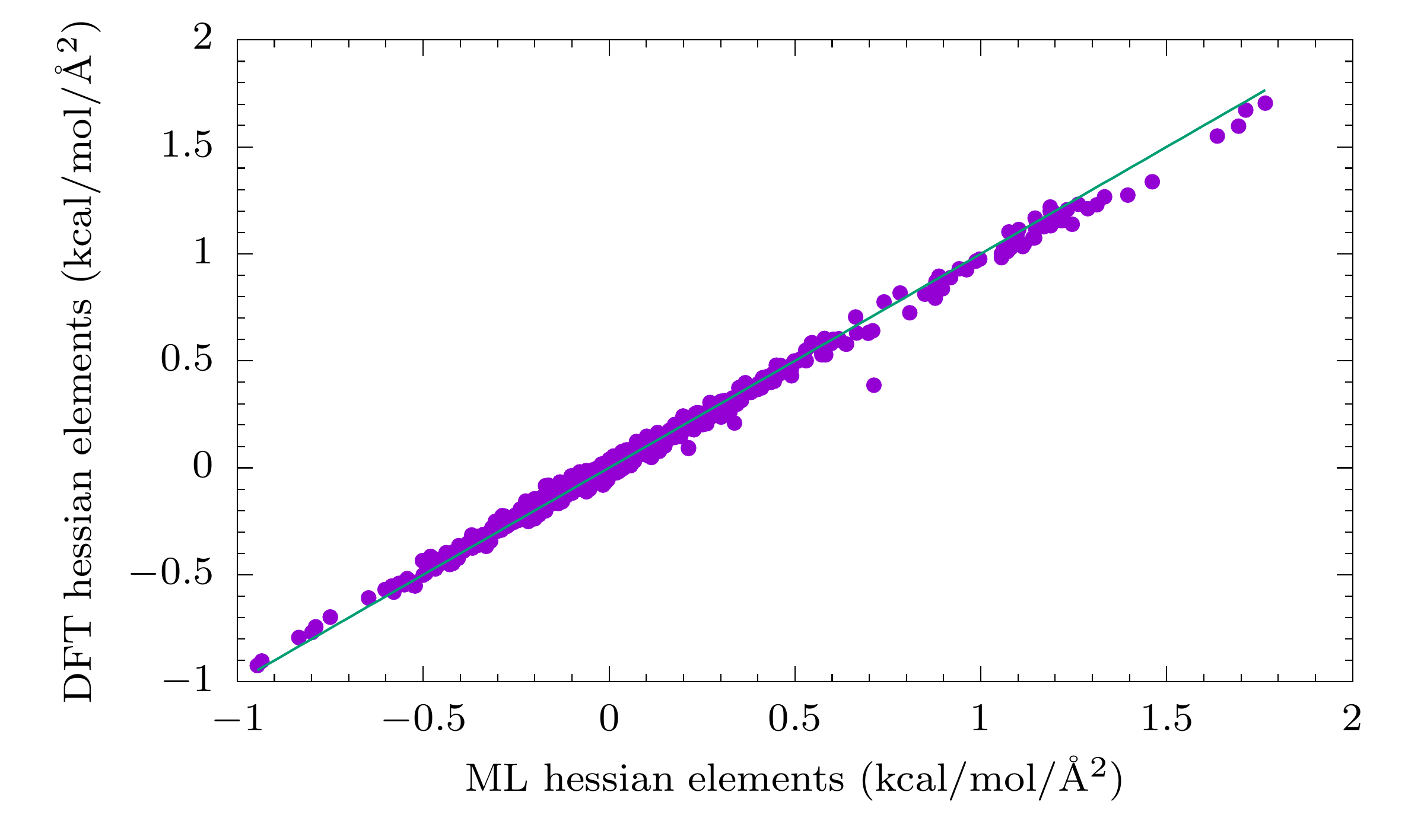}
    \caption{\textbf{Energy, forces, phonons and hessian.} $\delta =10$}
\end{figure}

\begin{figure}[H]
    \centering
    \includegraphics[scale=0.65]{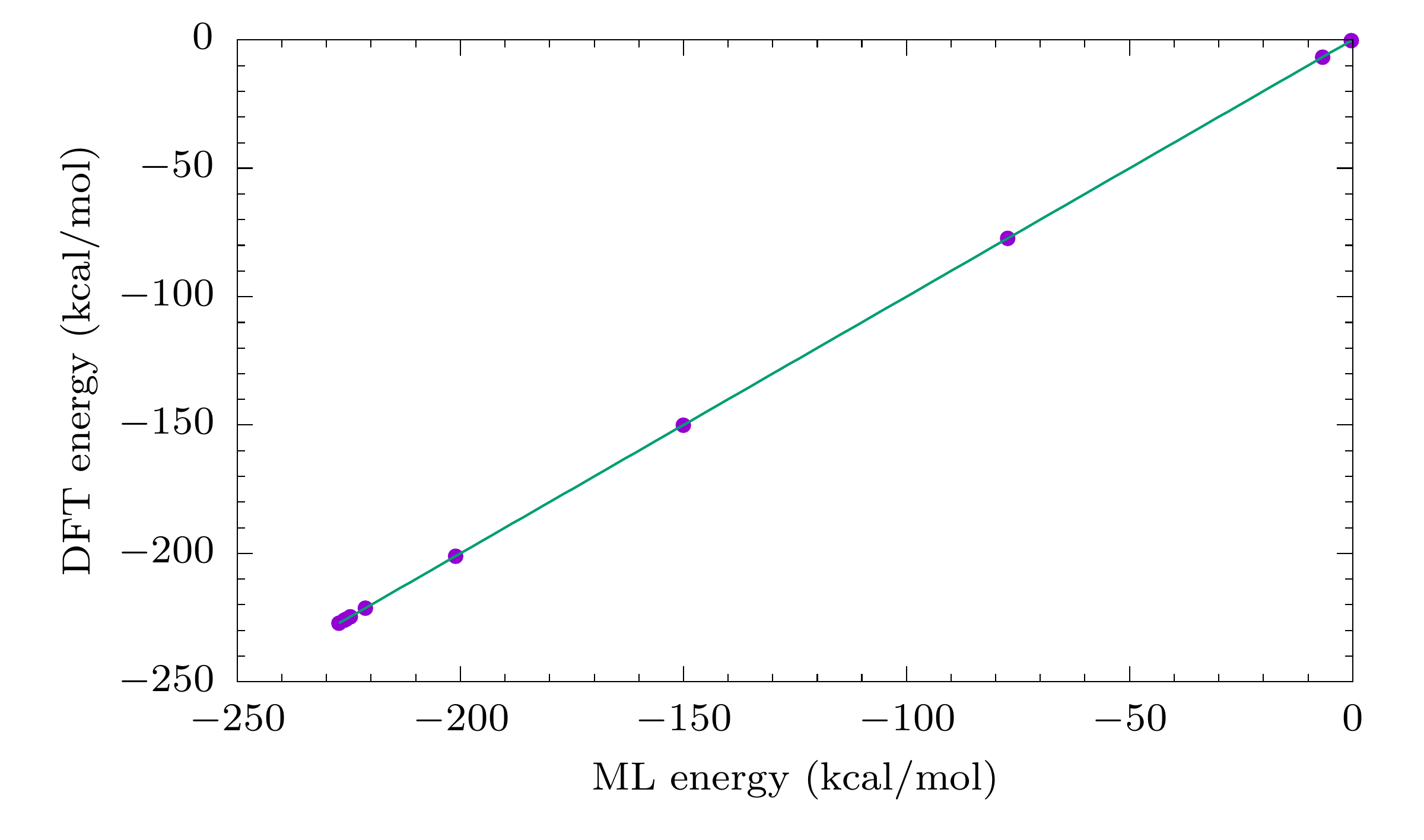}\\
    \includegraphics[scale=0.65]{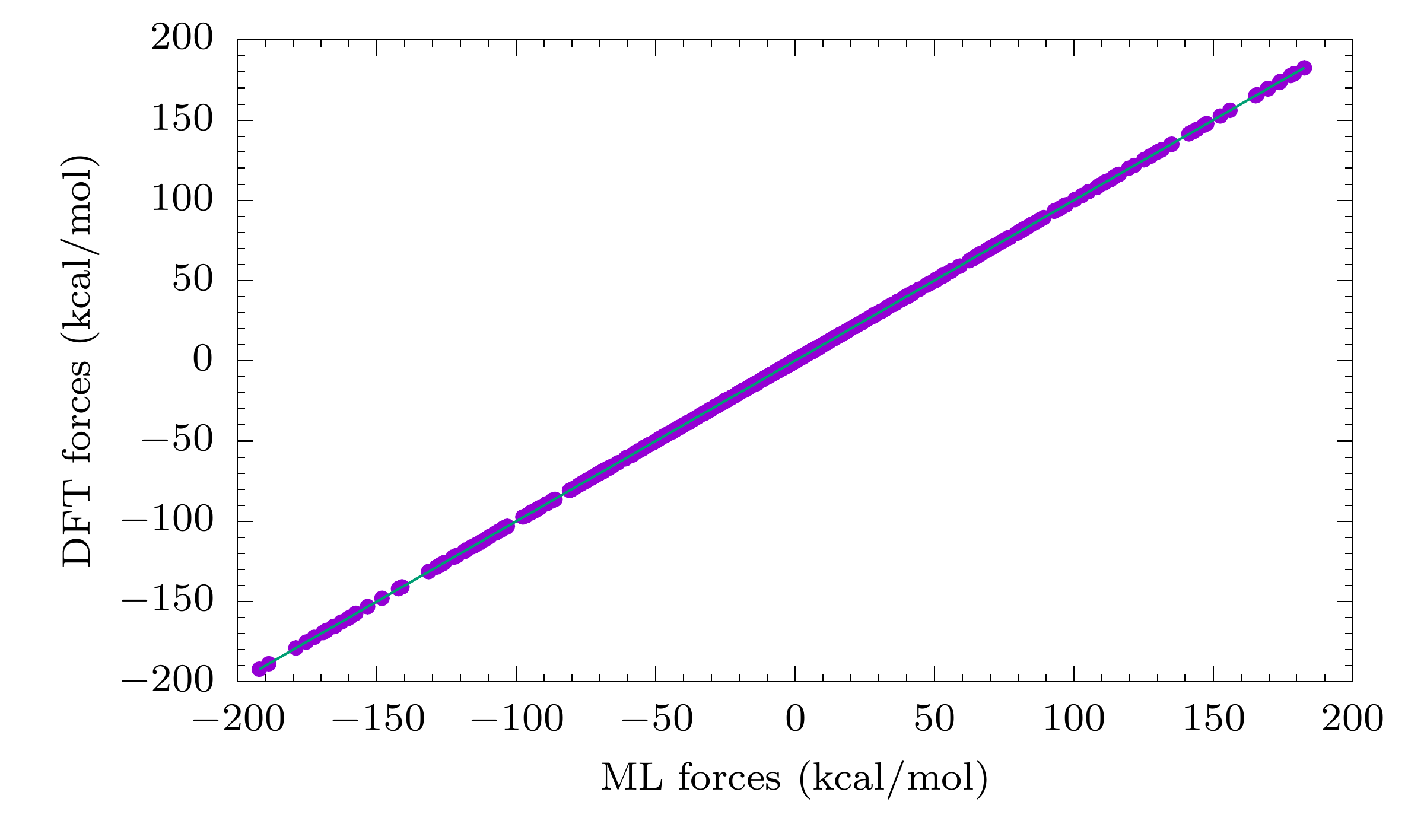}\\
    \includegraphics[scale=0.65]{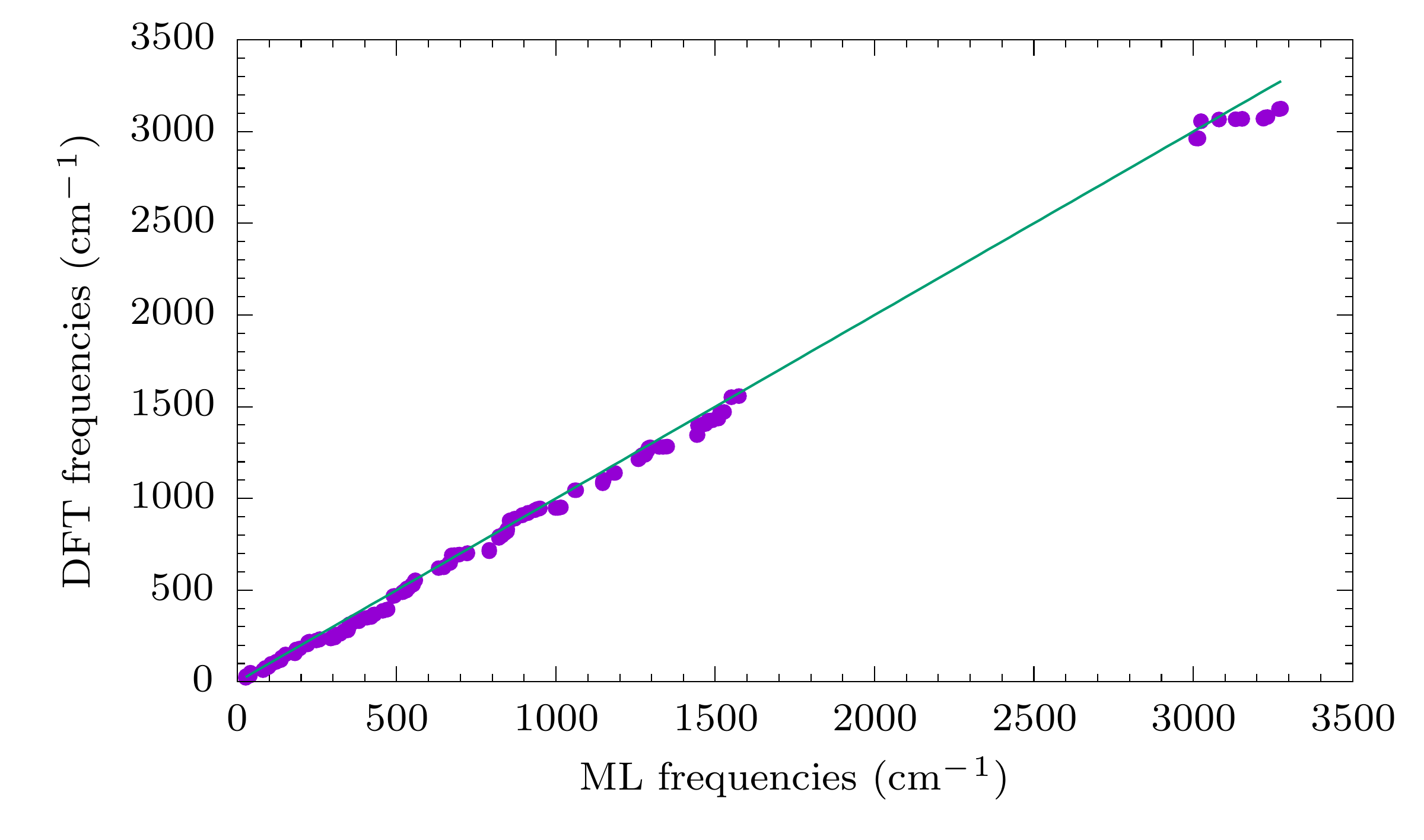}
    \includegraphics[scale=0.65]{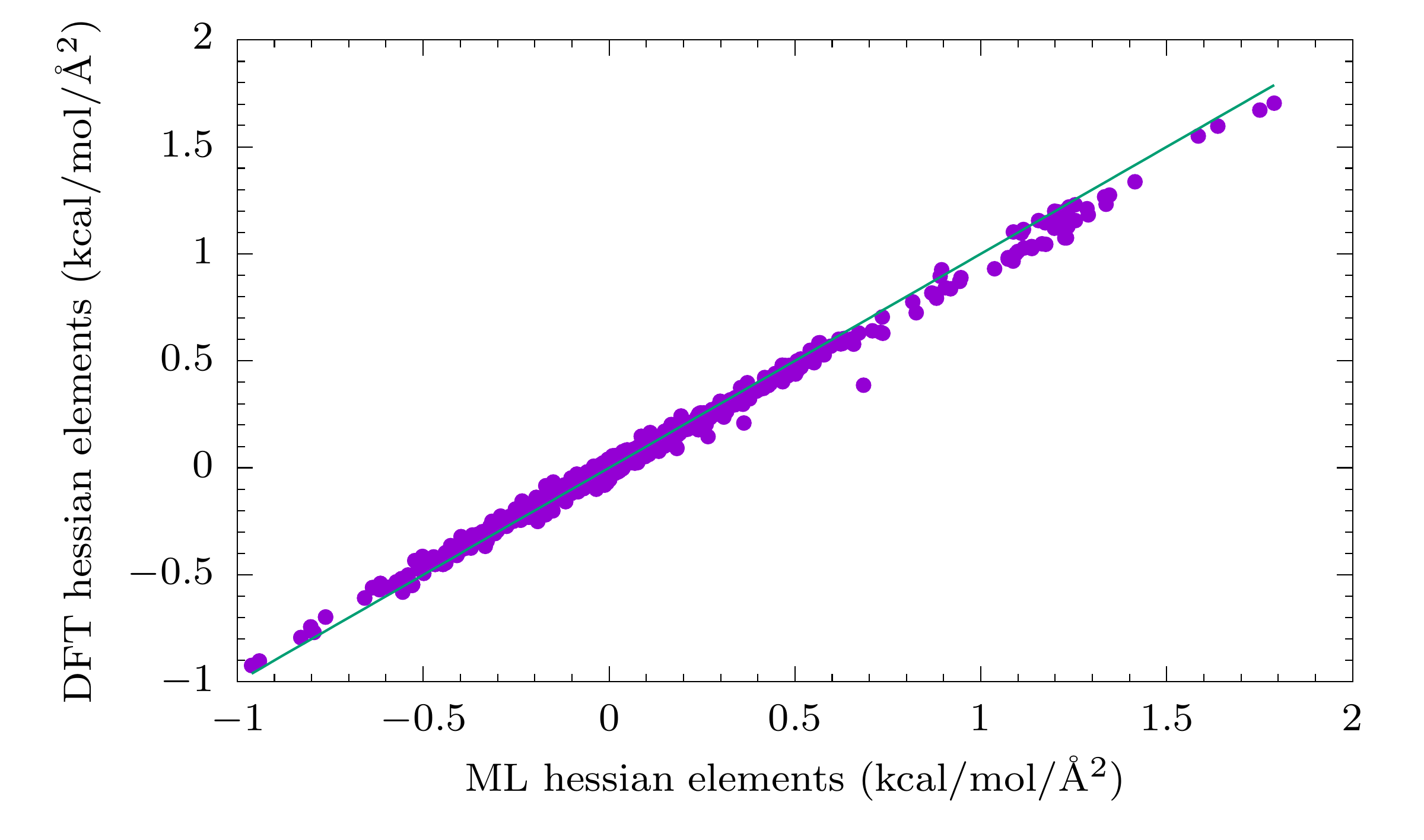}
    \caption{\textbf{Energy, forces, phonons and hessian.} $\delta =20$}
\end{figure}

\begin{figure}[H]
    \centering
    \includegraphics[scale=0.65]{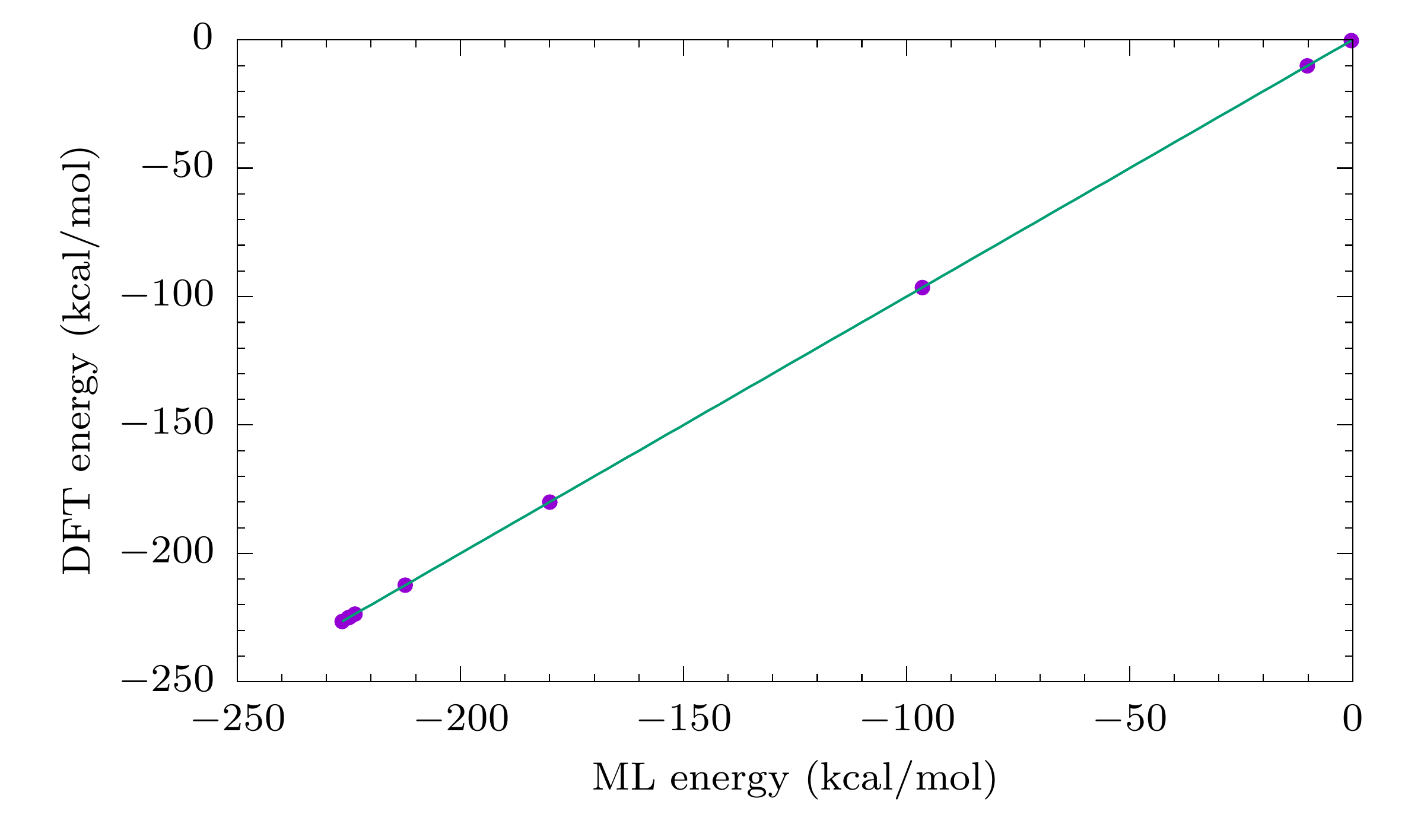}\\
    \includegraphics[scale=0.65]{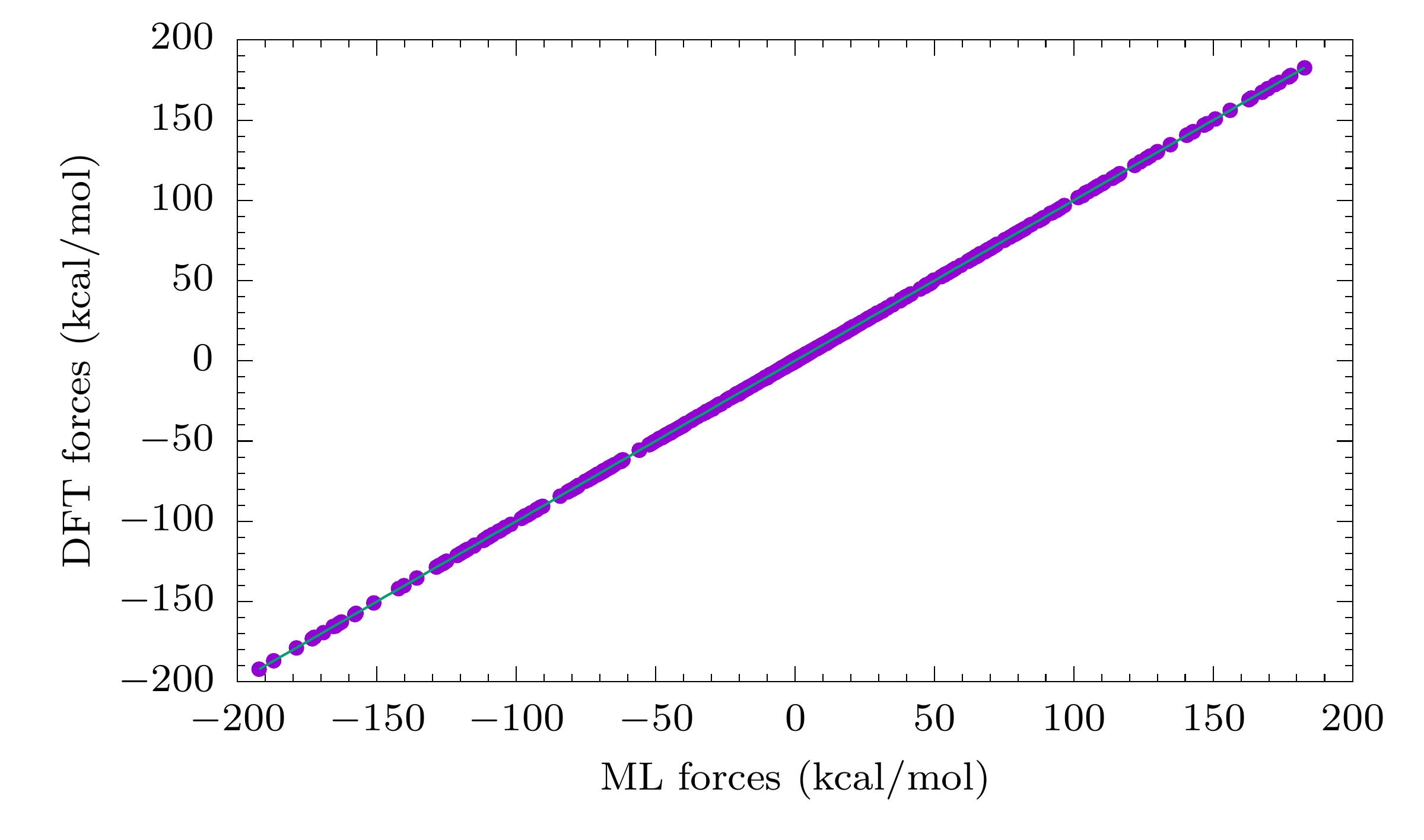}\\
    \includegraphics[scale=0.65]{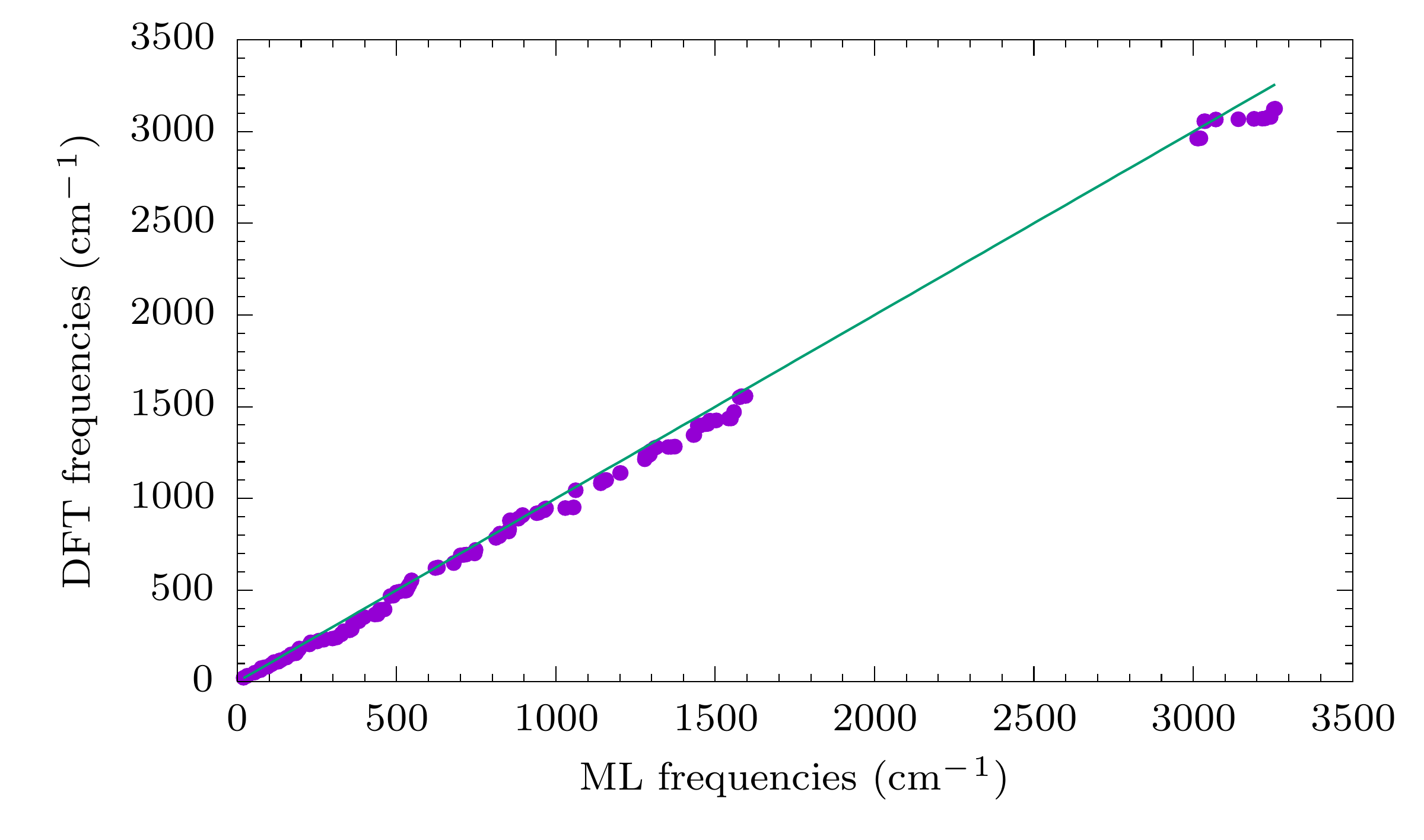}
    \includegraphics[scale=0.65]{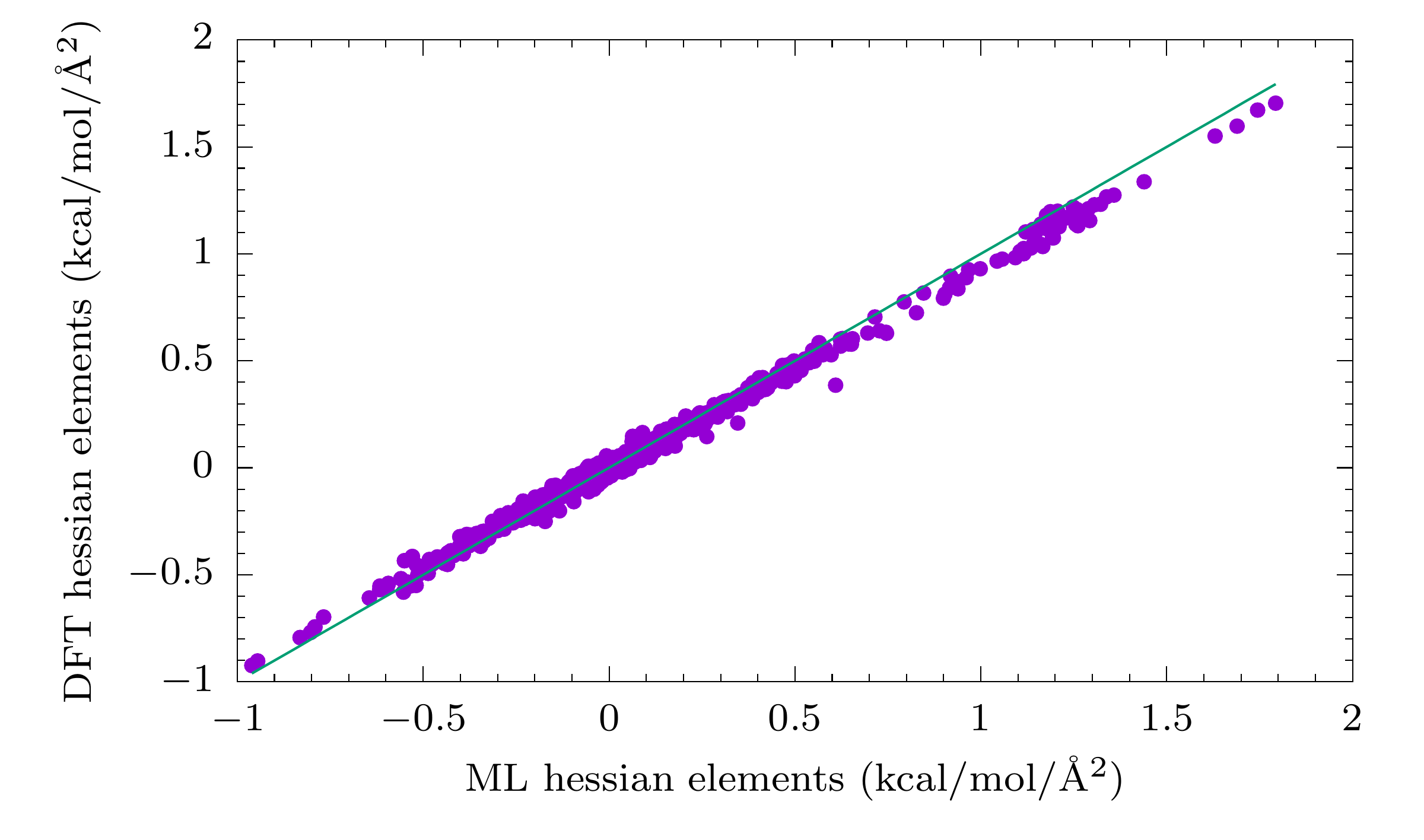}
    \caption{\textbf{Energy, forces, phonons and hessian.} $\delta =30$}
\end{figure}

\subsection{Compound 3}

\begin{figure}[H]
    \centering
    \includegraphics[scale=0.65]{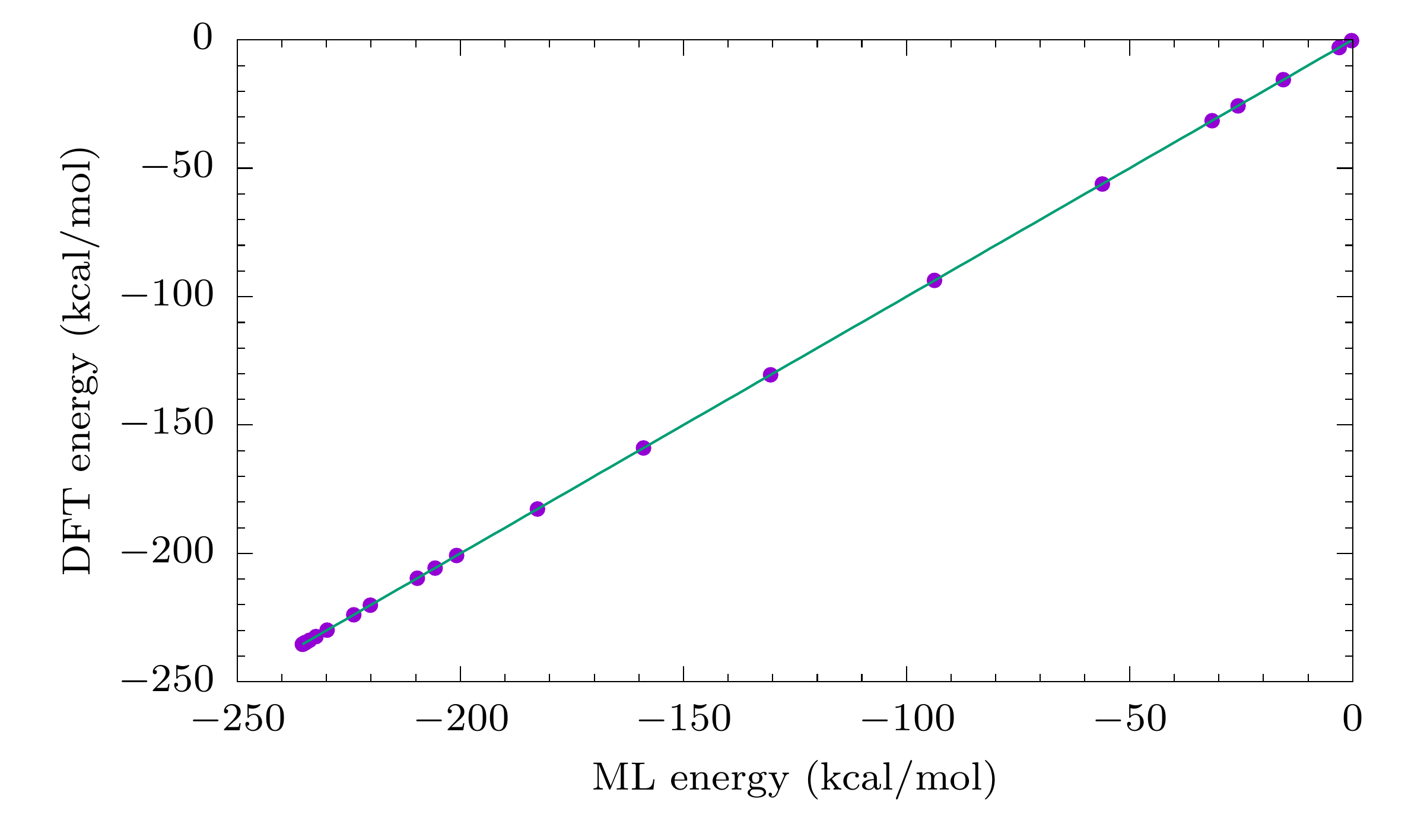}\\
    \includegraphics[scale=0.65]{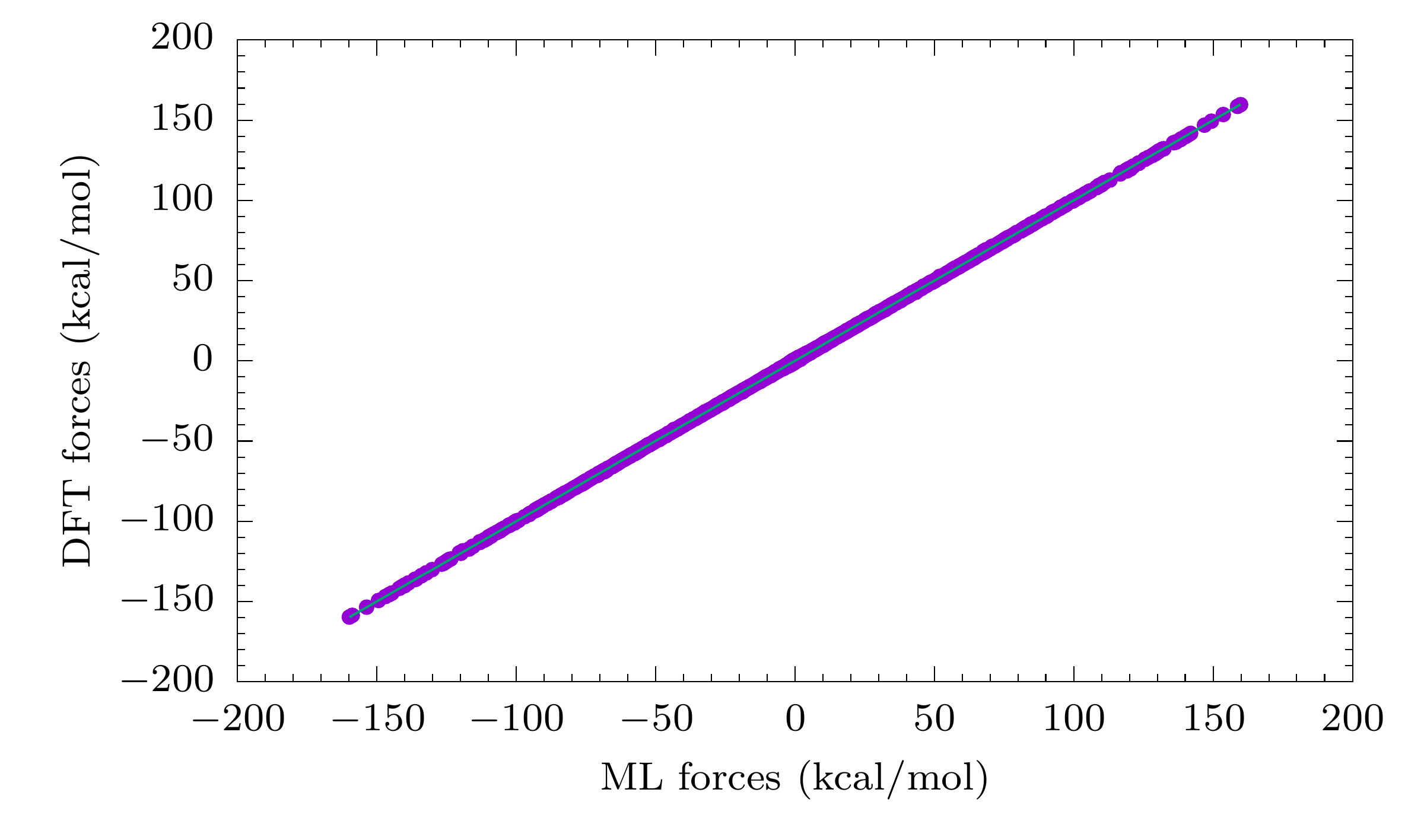}\\
    \includegraphics[scale=0.65]{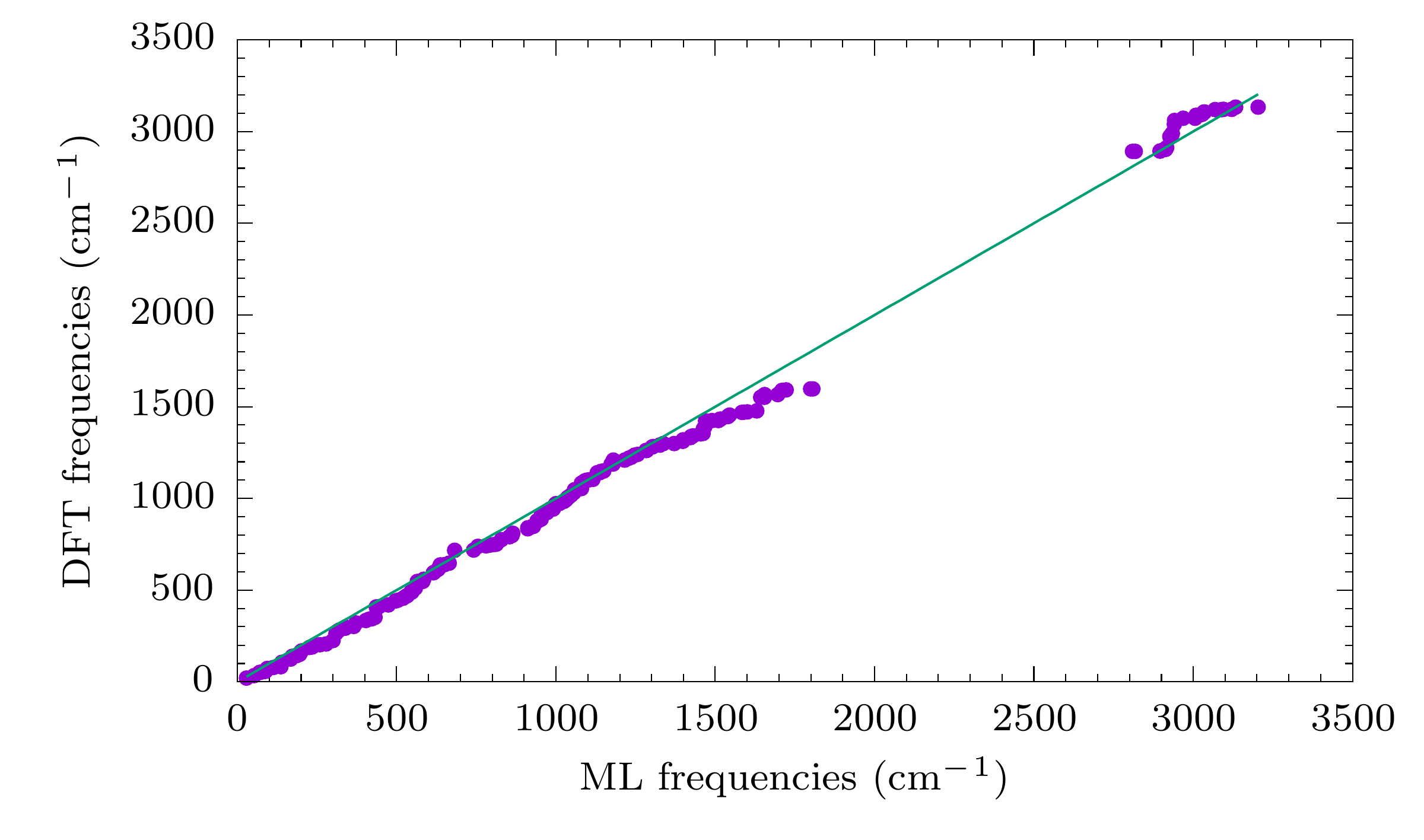}
    \includegraphics[scale=0.65]{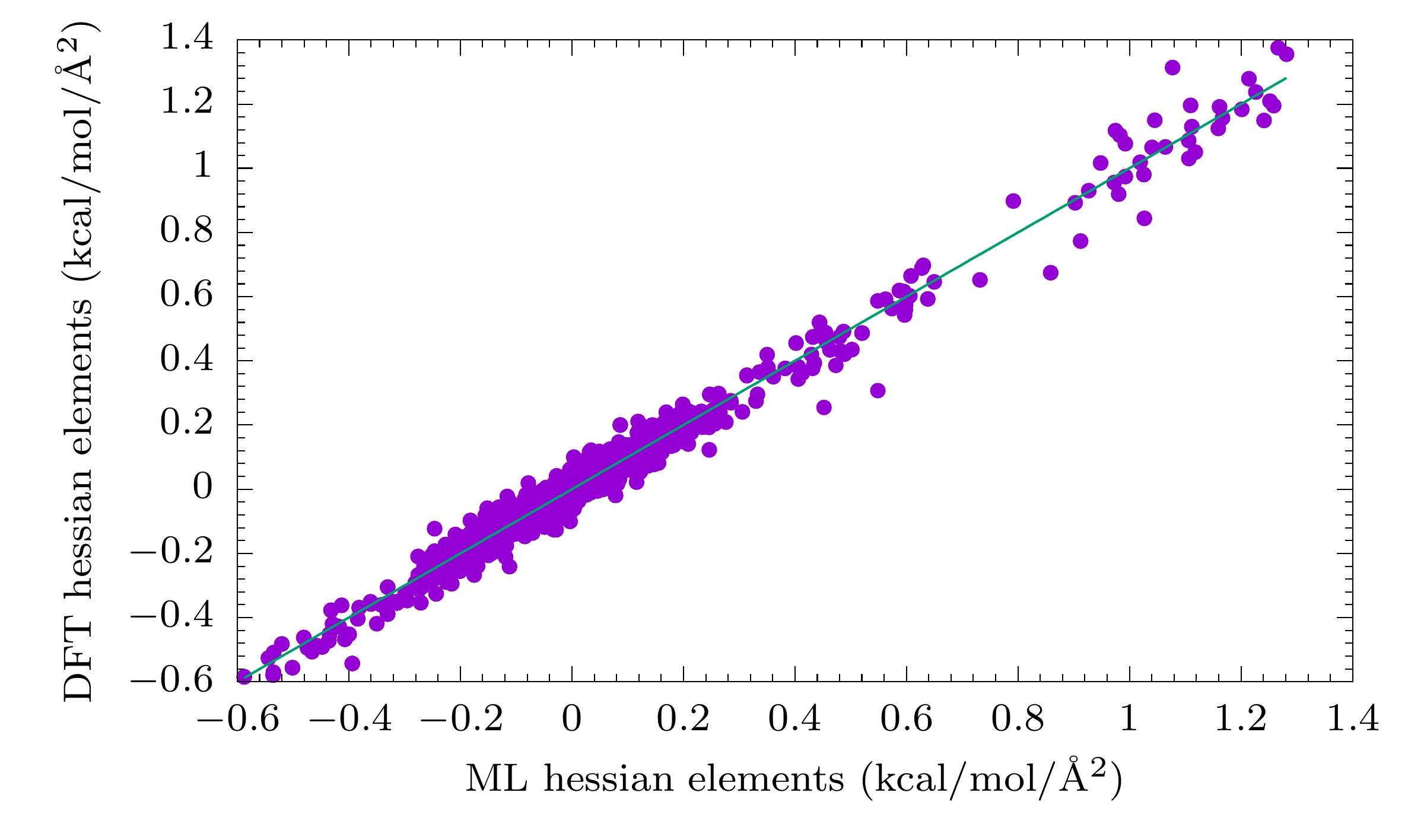}
    \caption{\textbf{Energy, forces, phonons and hessian.} $\delta =5$}
\end{figure}

\begin{figure}[H]
    \centering
    \includegraphics[scale=0.65]{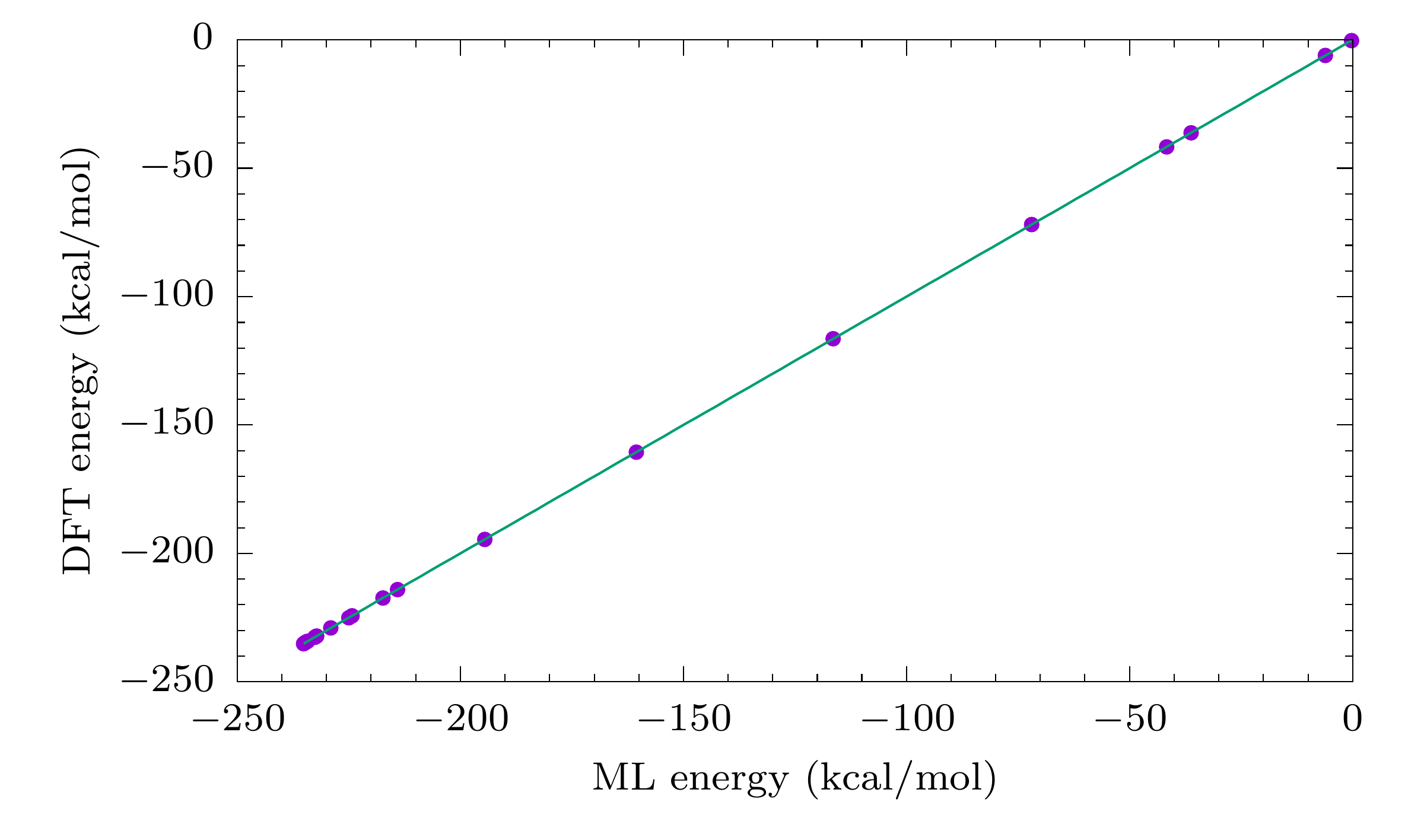}\\
    \includegraphics[scale=0.65]{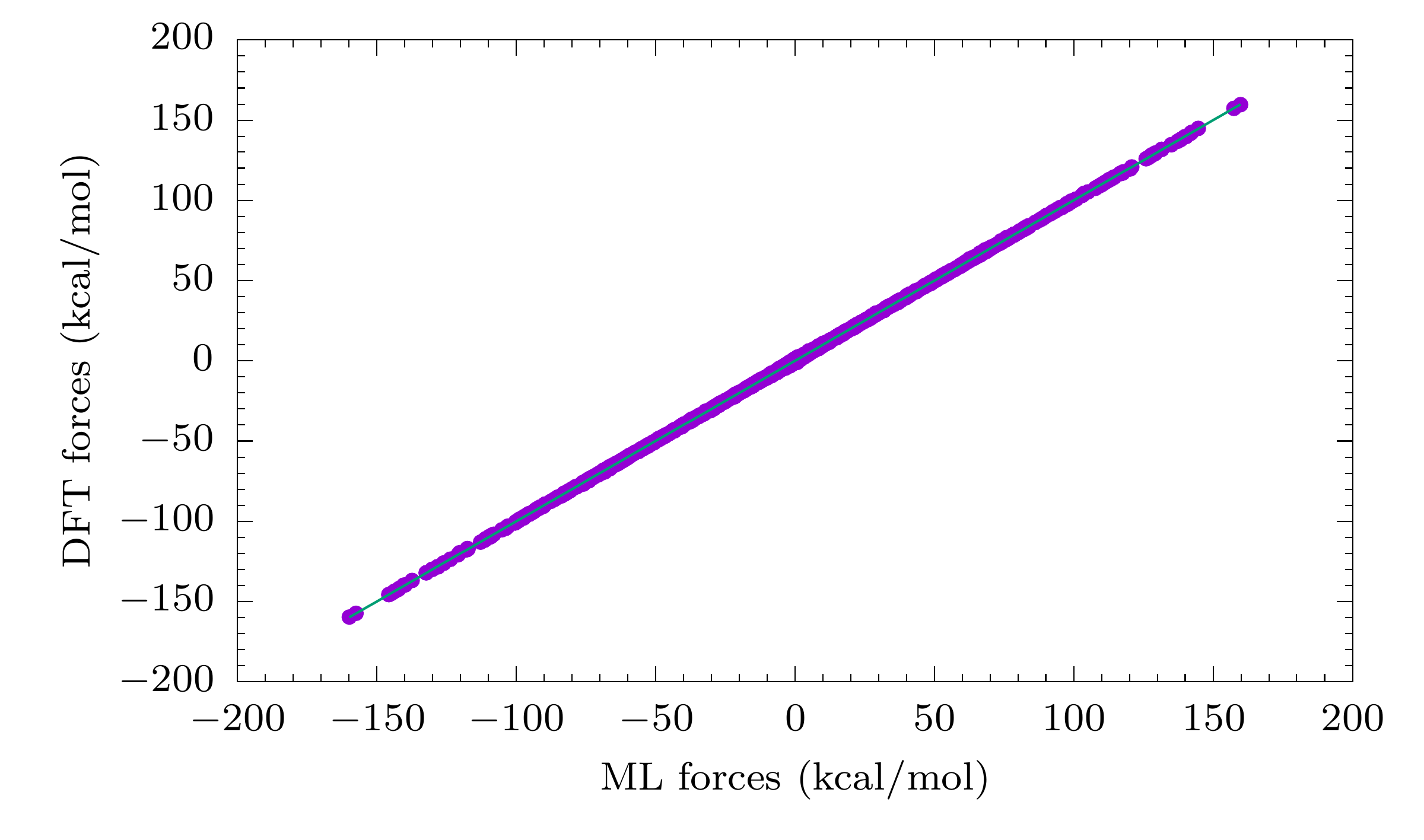}\\
    \includegraphics[scale=0.65]{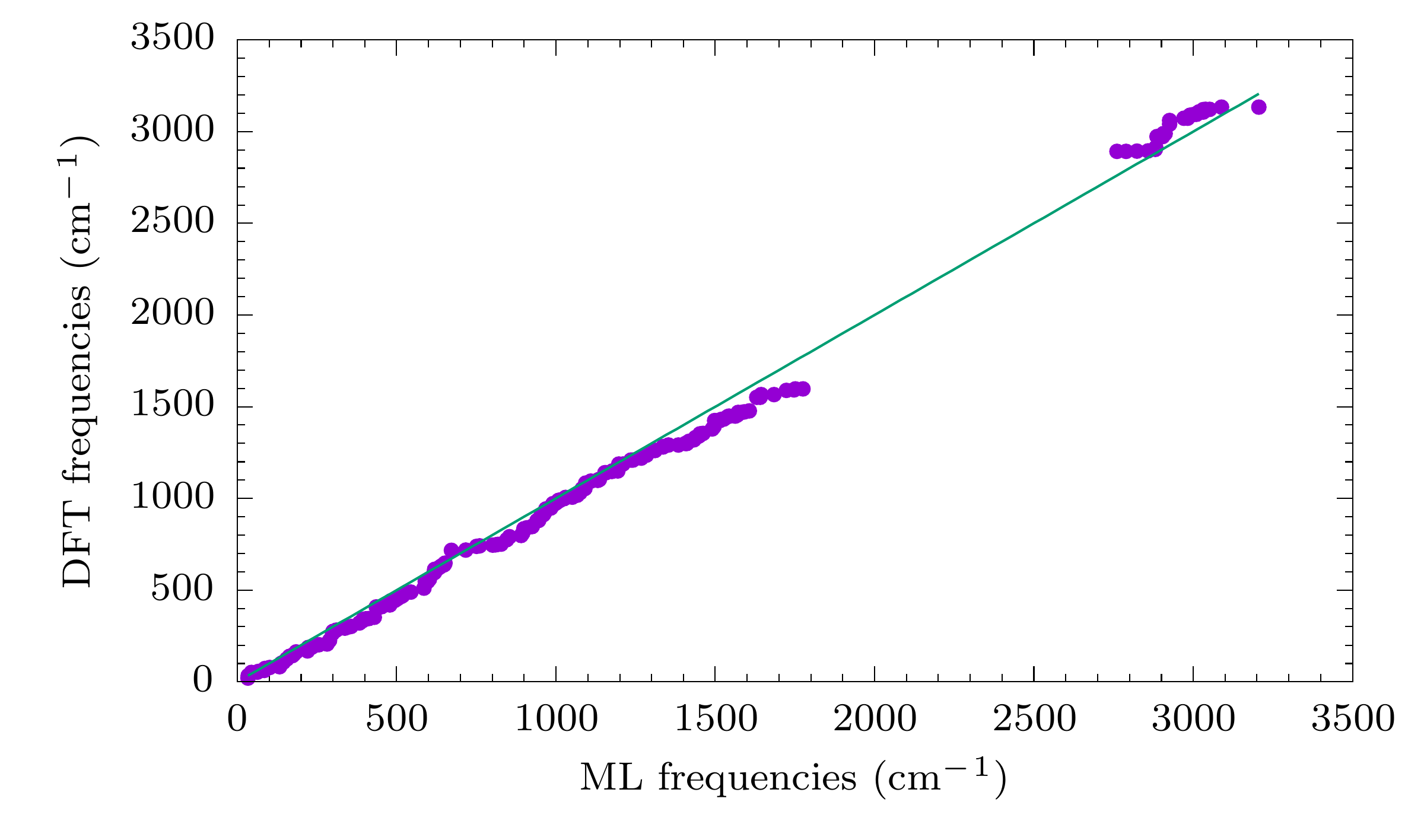}
    \includegraphics[scale=0.65]{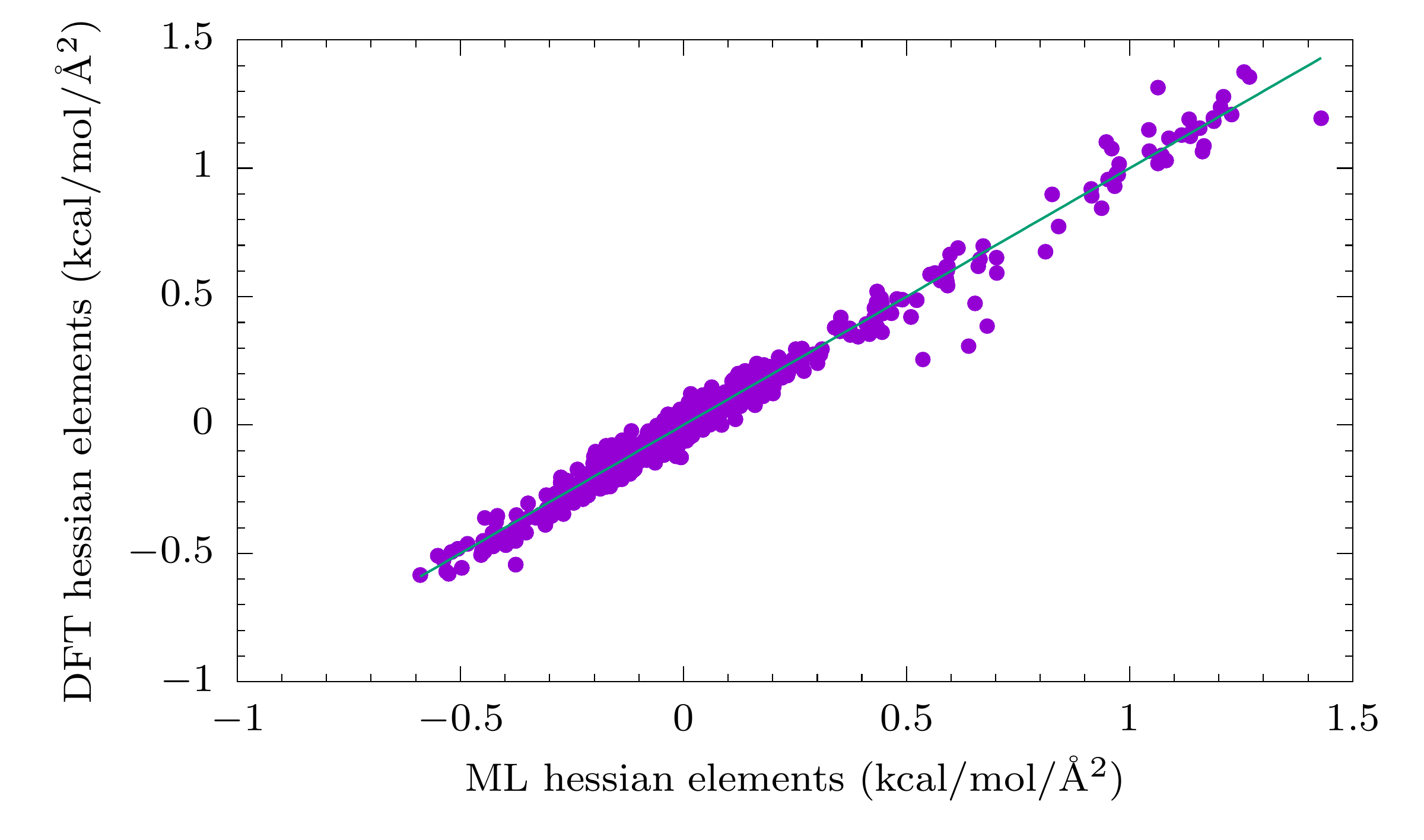}
    \caption{\textbf{Energy, forces, phonons and hessian.} $\delta =10$}
\end{figure}

\begin{figure}[H]
    \centering
    \includegraphics[scale=0.65]{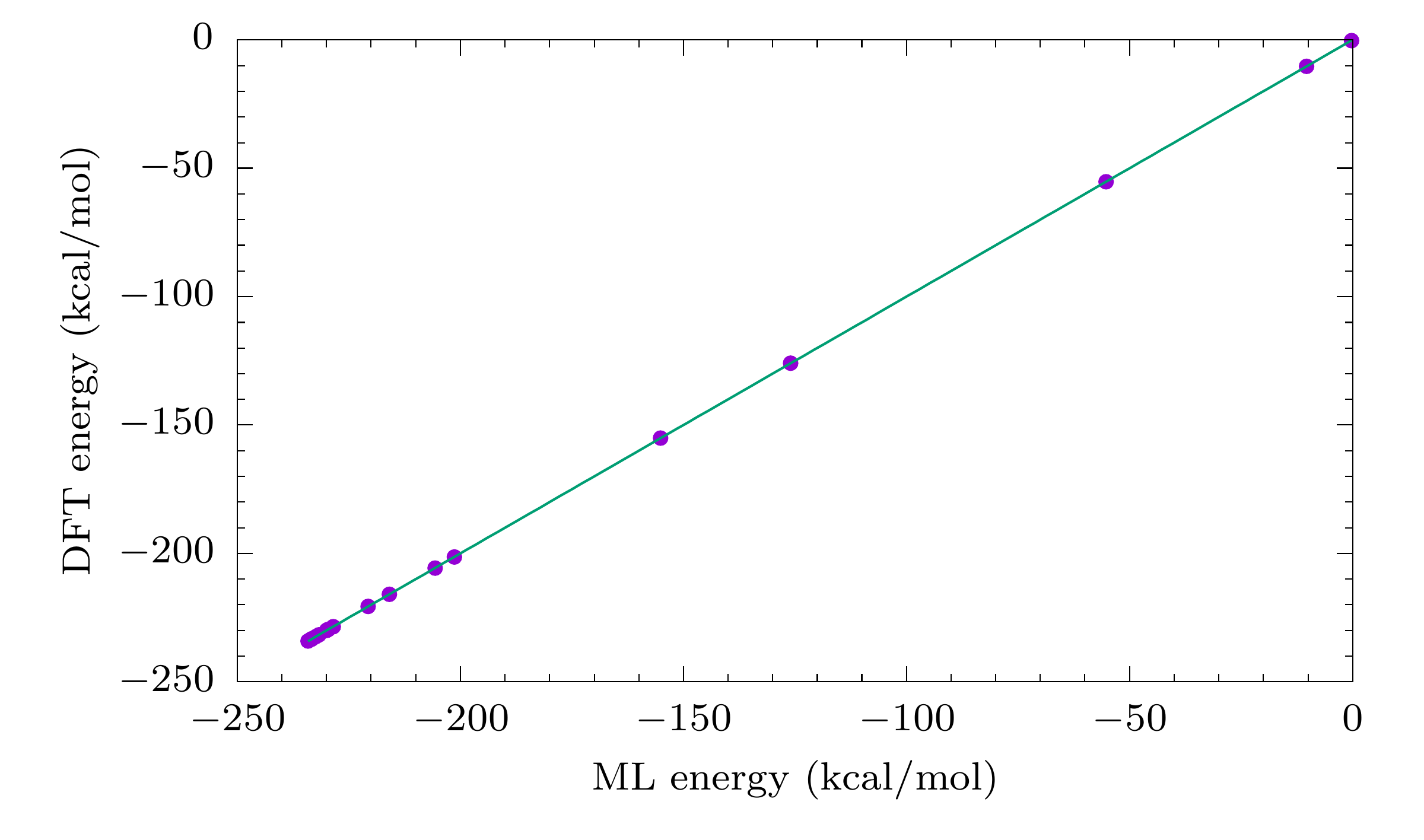}\\
    \includegraphics[scale=0.65]{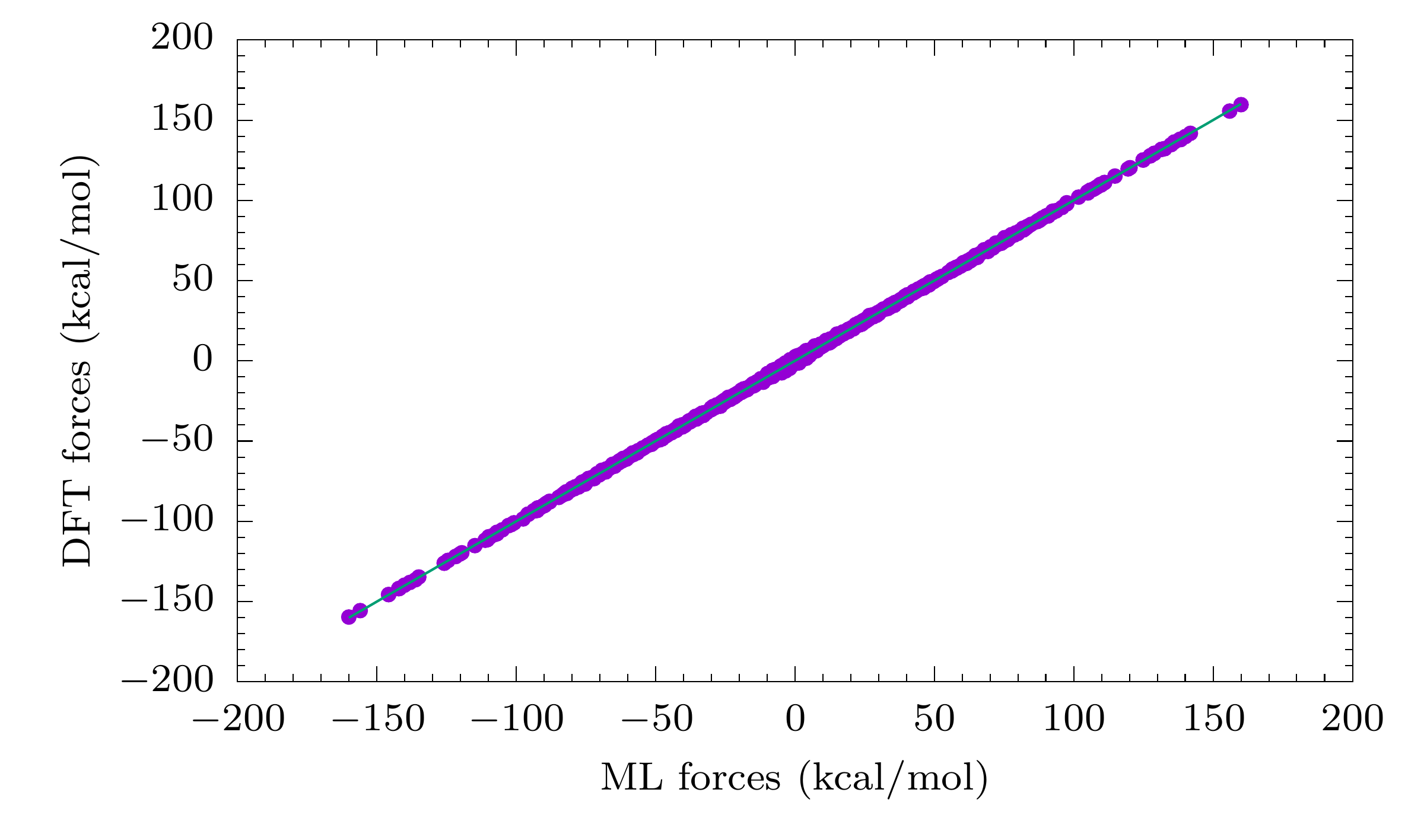}\\
    \includegraphics[scale=0.65]{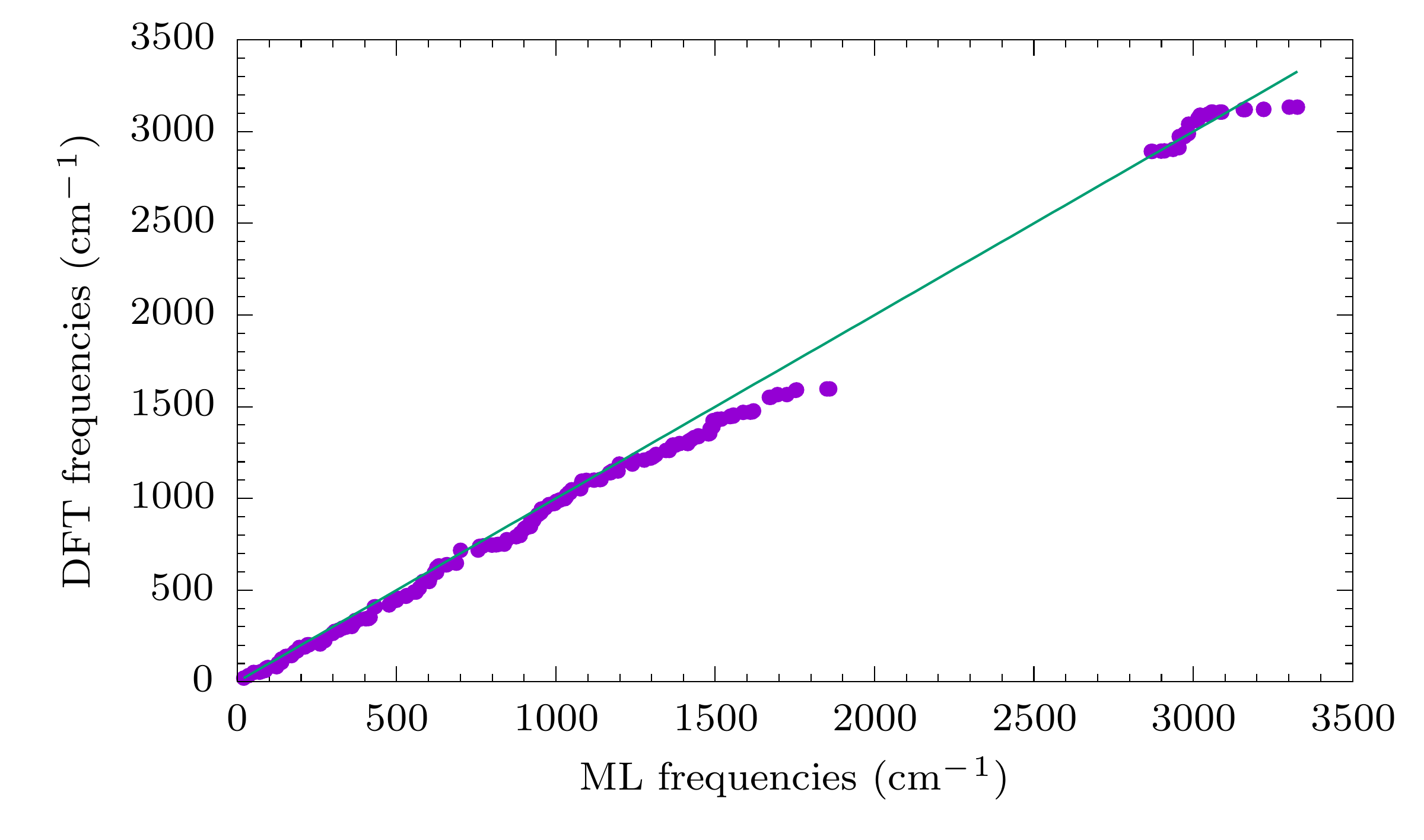}
    \includegraphics[scale=0.65]{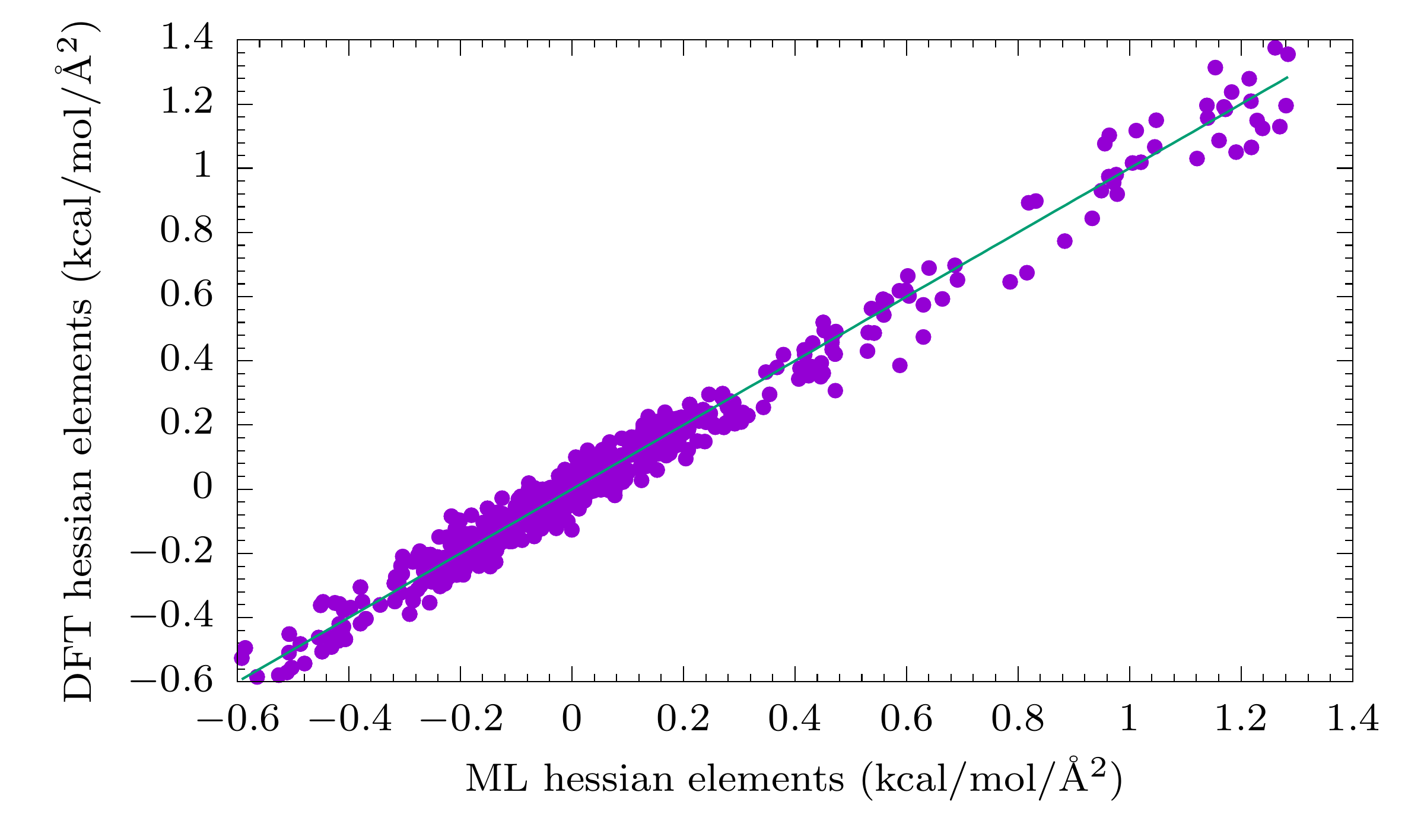}
    \caption{\textbf{Energy, forces, phonons and hessian.} $\delta =20$}
\end{figure}

\begin{figure}[H]
    \centering
    \includegraphics[scale=0.65]{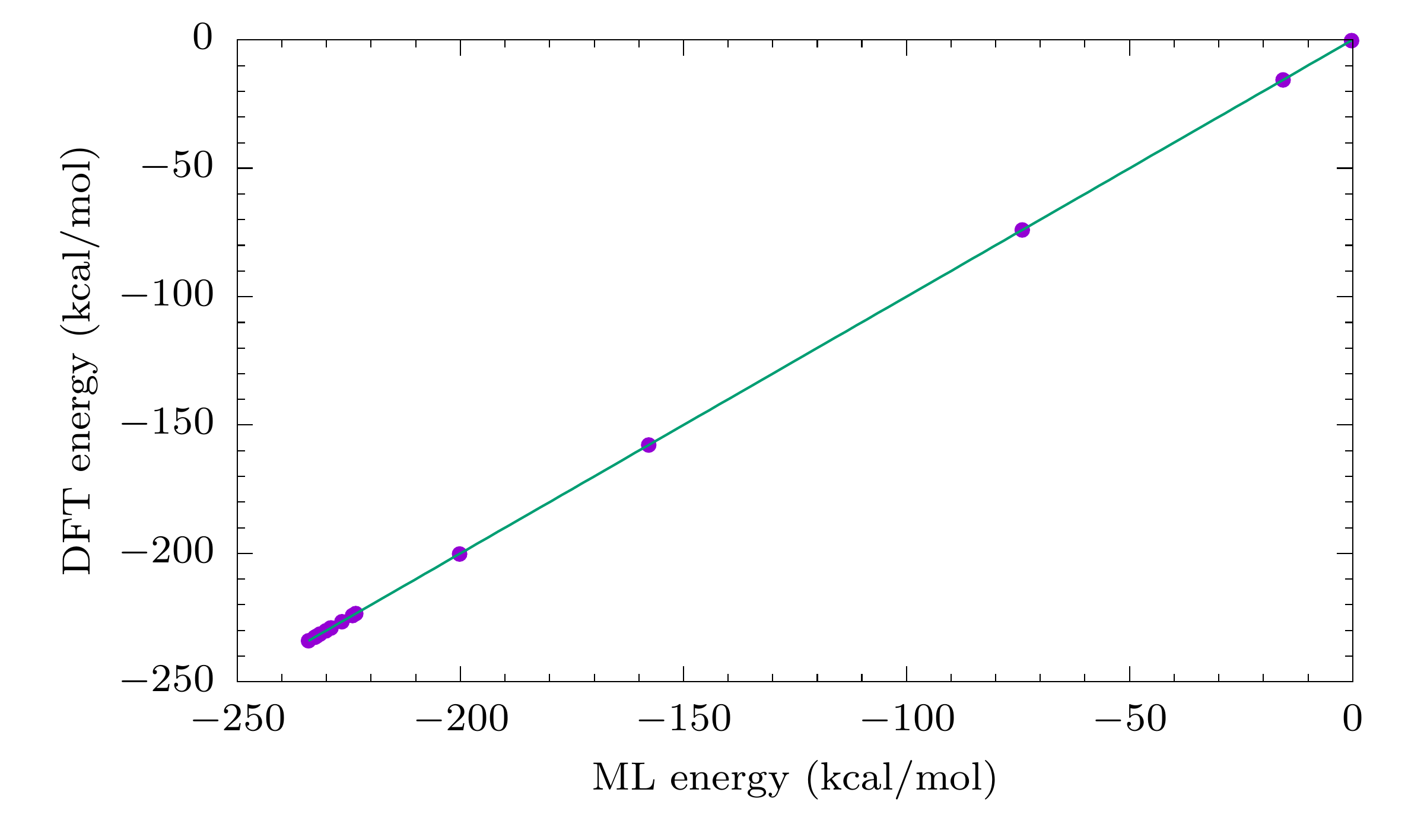}\\
    \includegraphics[scale=0.65]{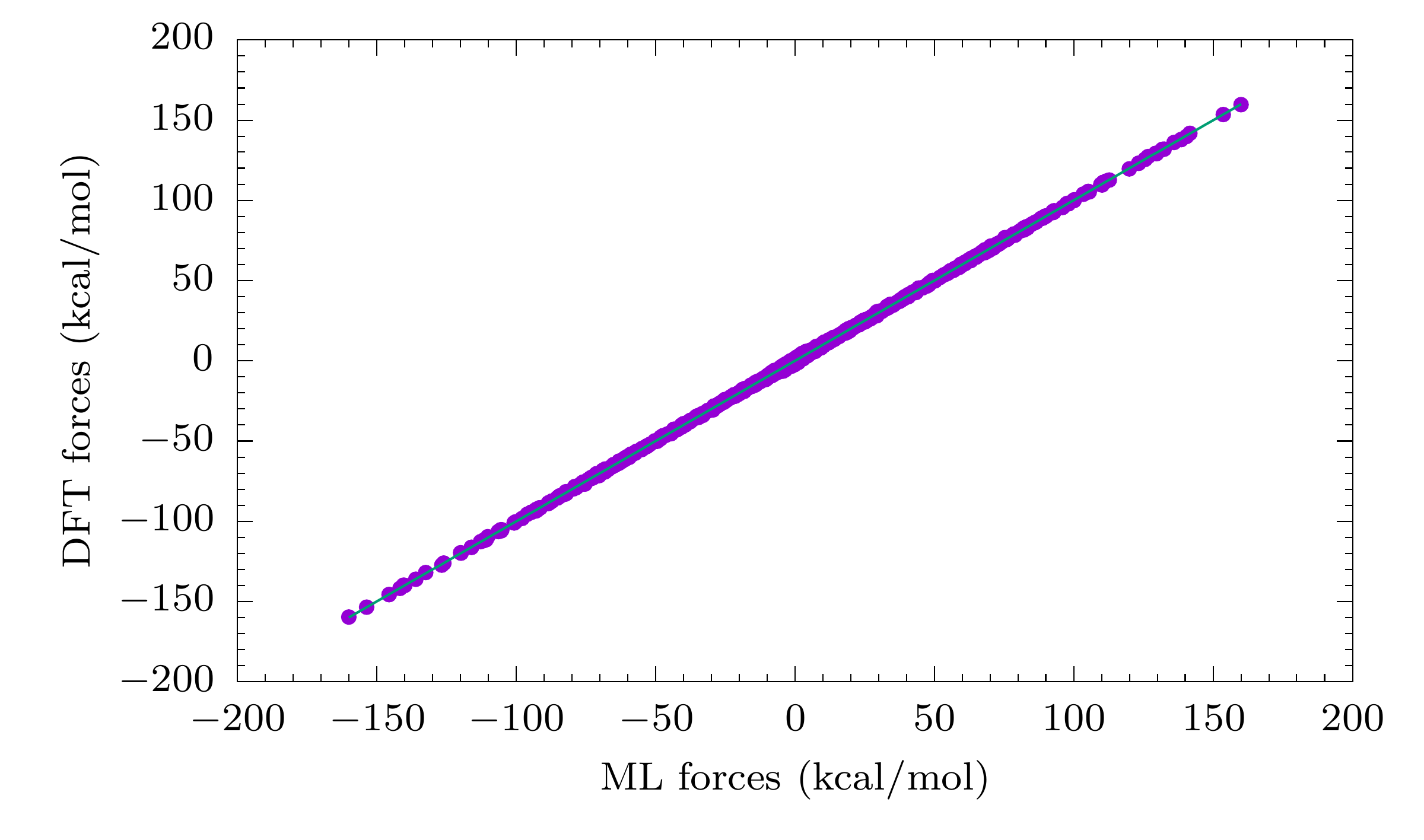}\\
    \includegraphics[scale=0.65]{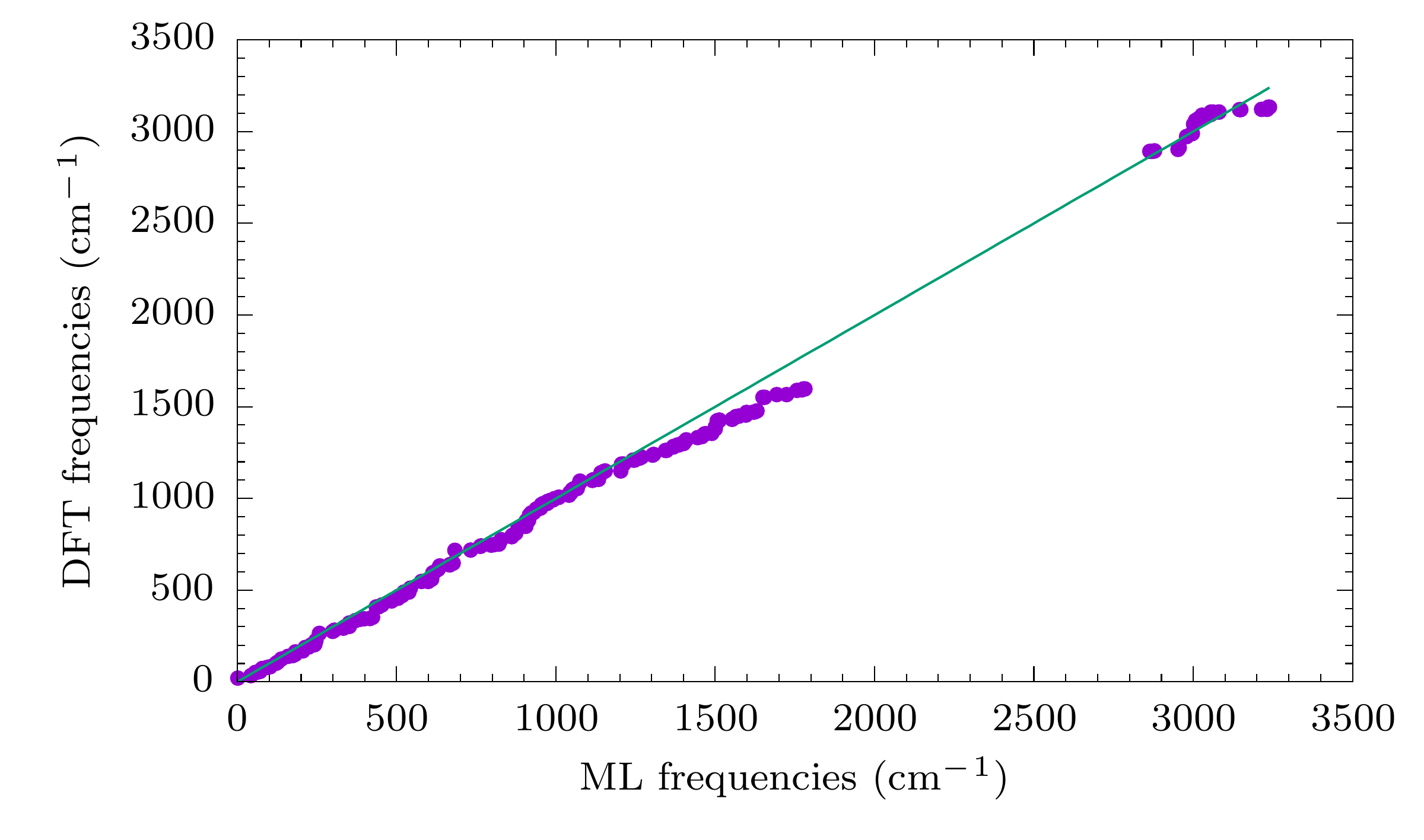}
    \includegraphics[scale=0.65]{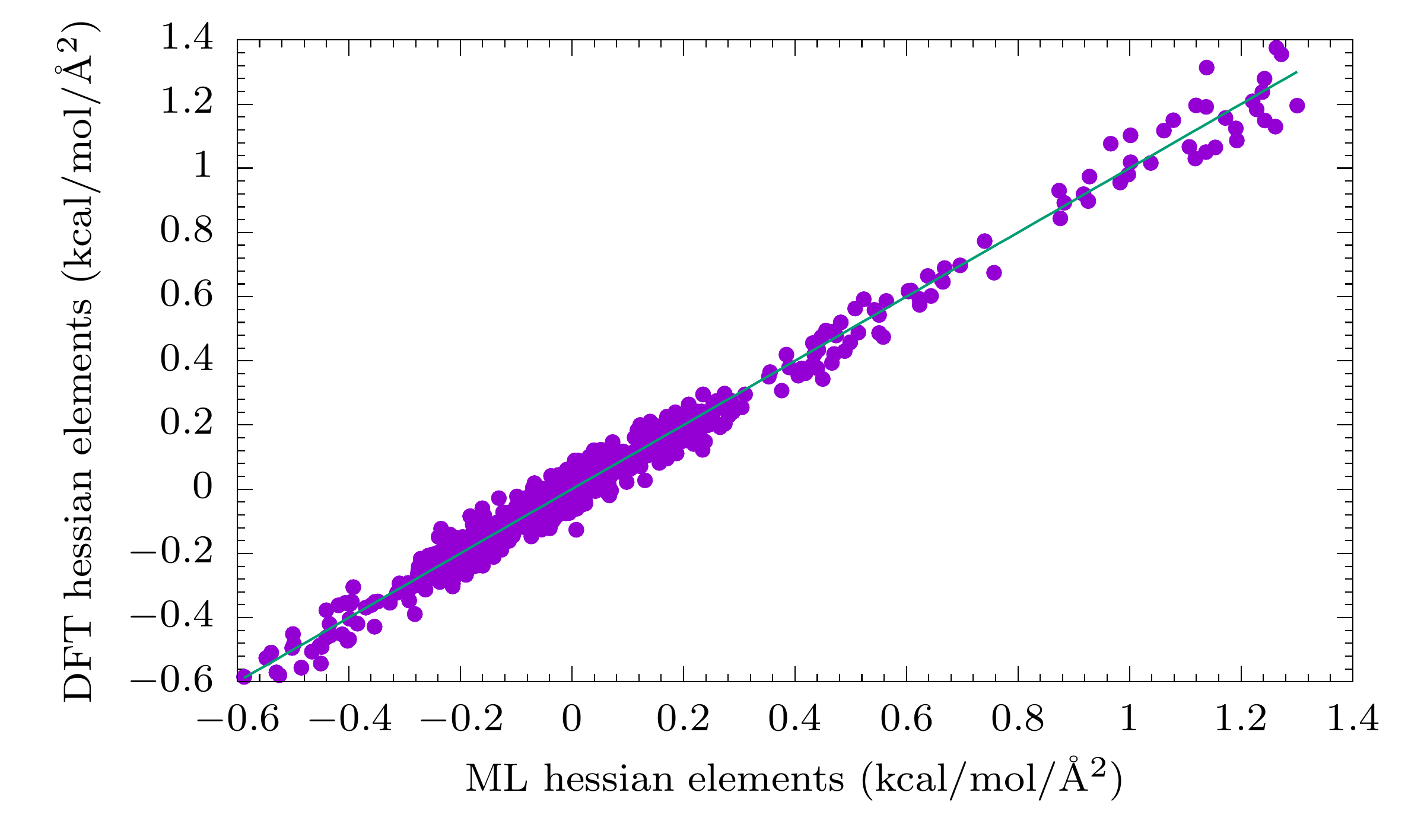}
    \caption{\textbf{Energy, forces, phonons and hessian.} $\delta =30$}
\end{figure}

\section{Machine learning of molecular vibrations - MD @50 K}

\subsection{Compound 1}

\begin{figure}[H]
    \centering
    \includegraphics[scale=0.65]{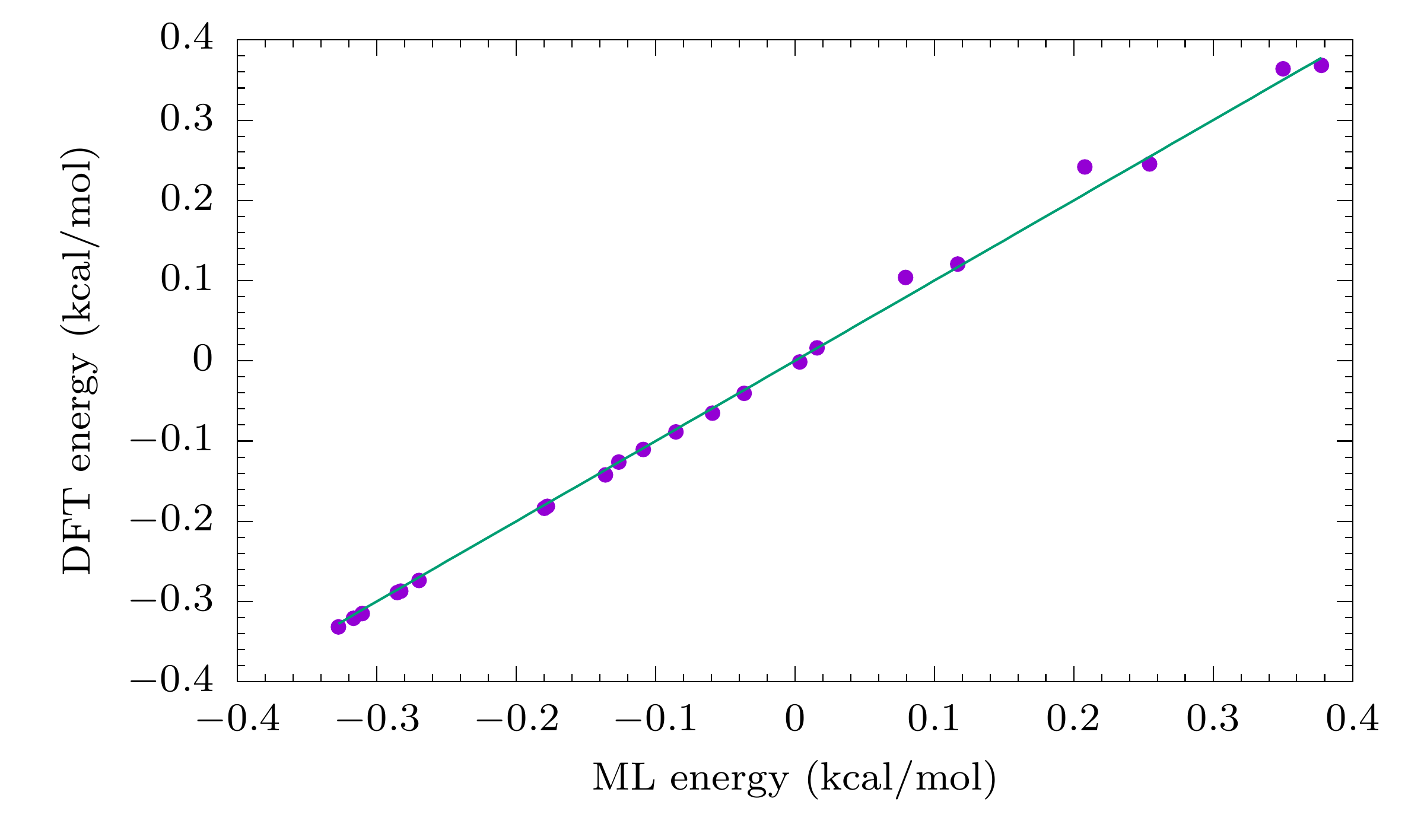}\\
    \includegraphics[scale=0.65]{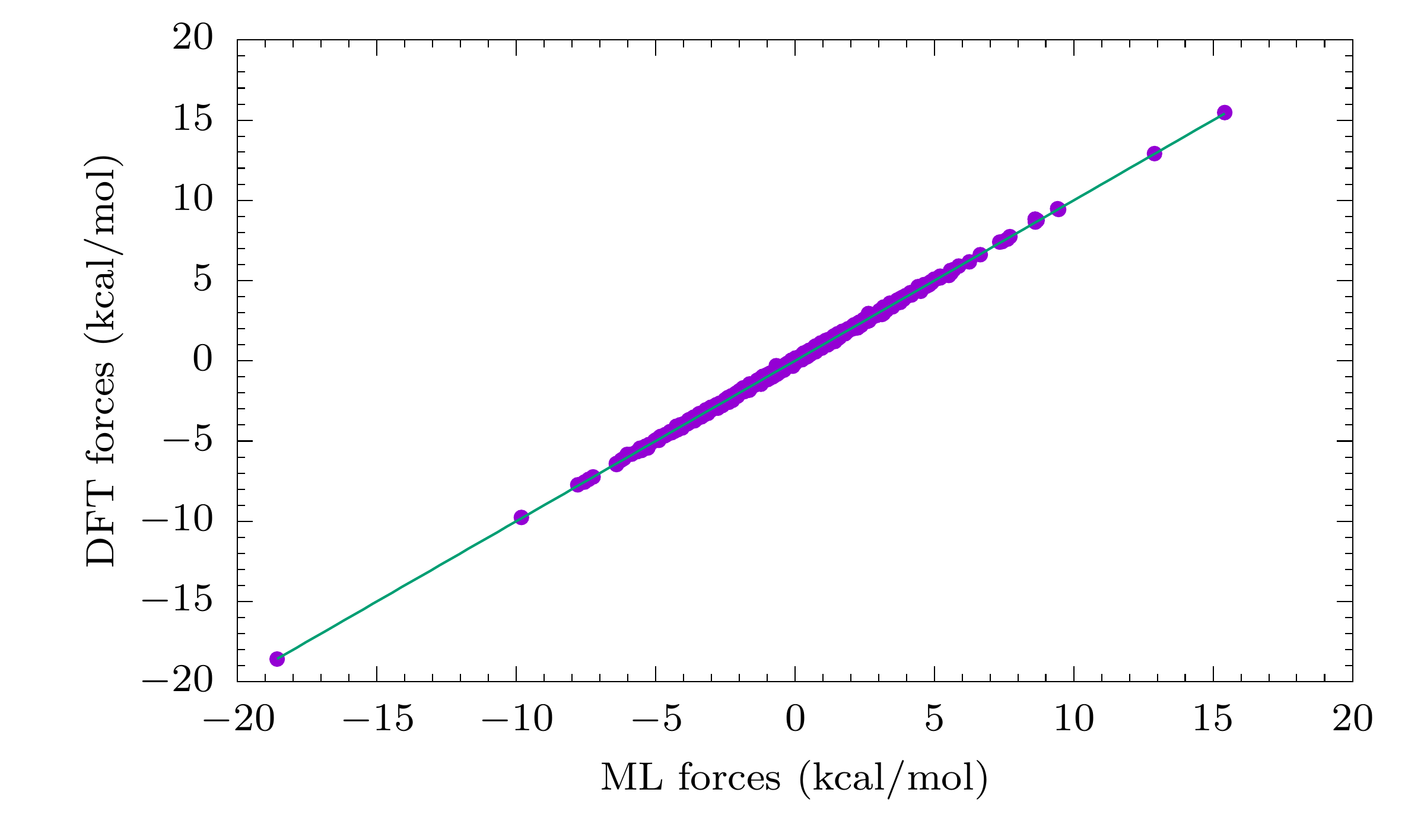}\\
    \includegraphics[scale=0.65]{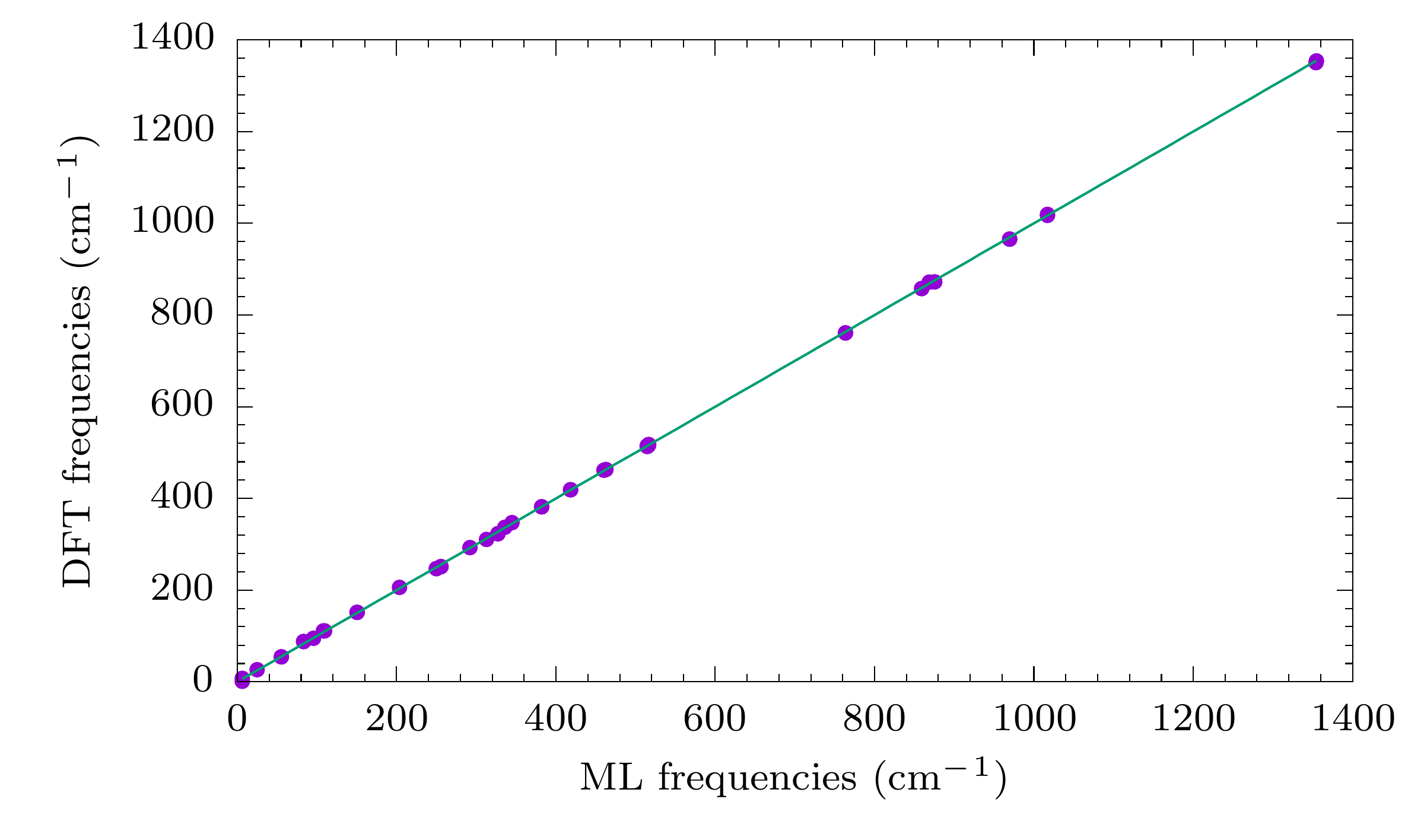}
    \includegraphics[scale=0.65]{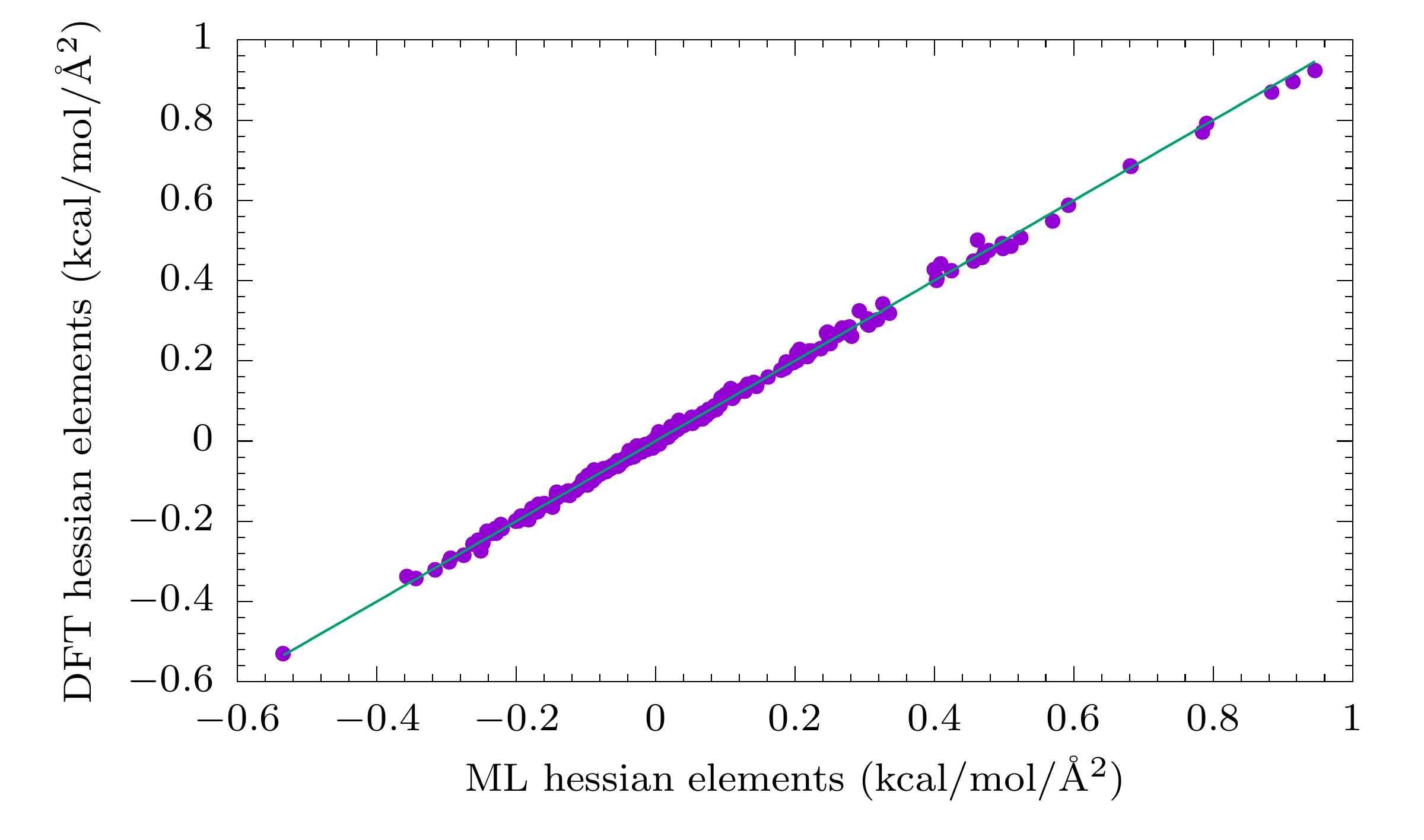}
    \caption{\textbf{Energy, forces, phonons and hessian.} $\delta =2.5$}
\end{figure}

\begin{figure}[H]
    \centering
    \includegraphics[scale=0.65]{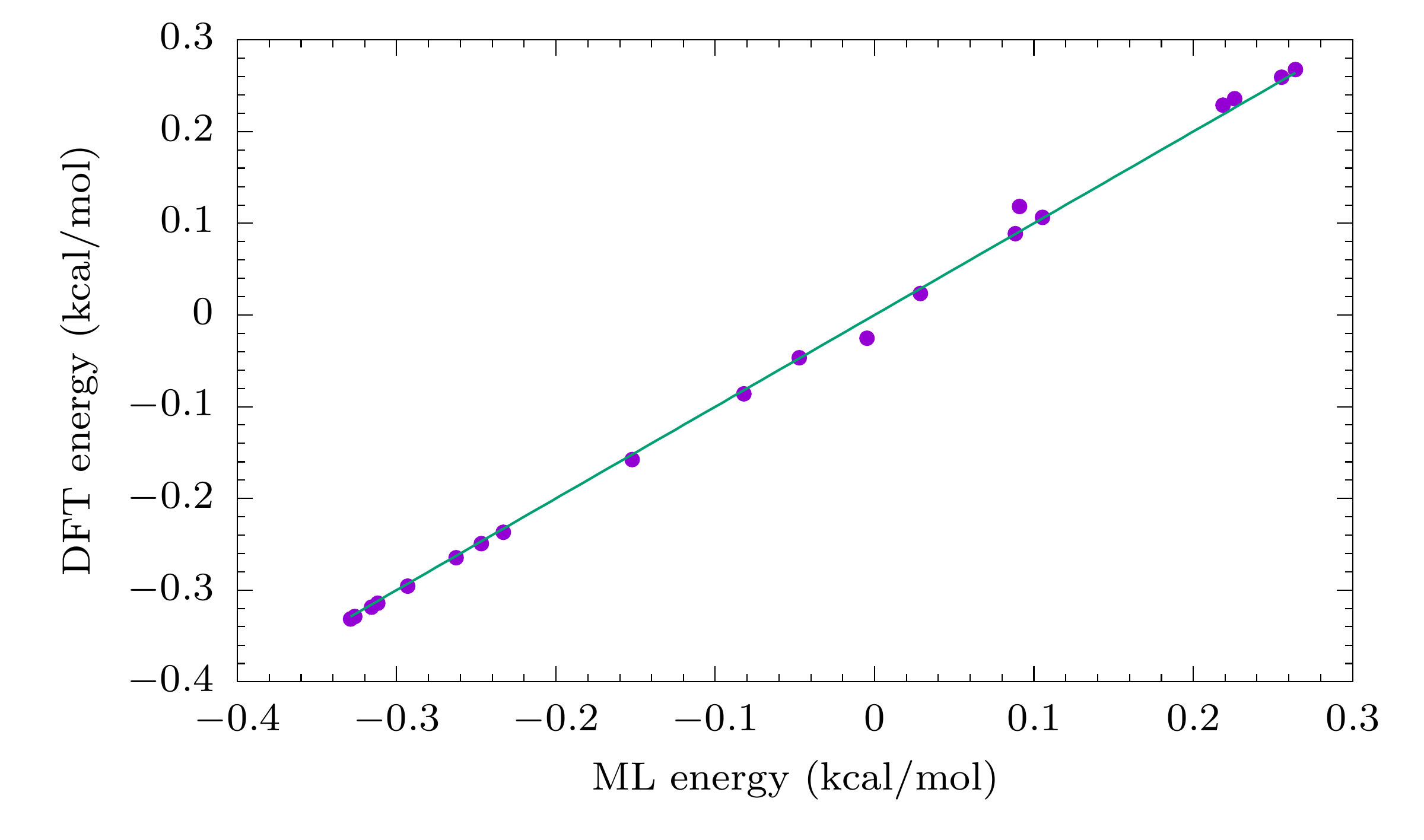}\\
    \includegraphics[scale=0.65]{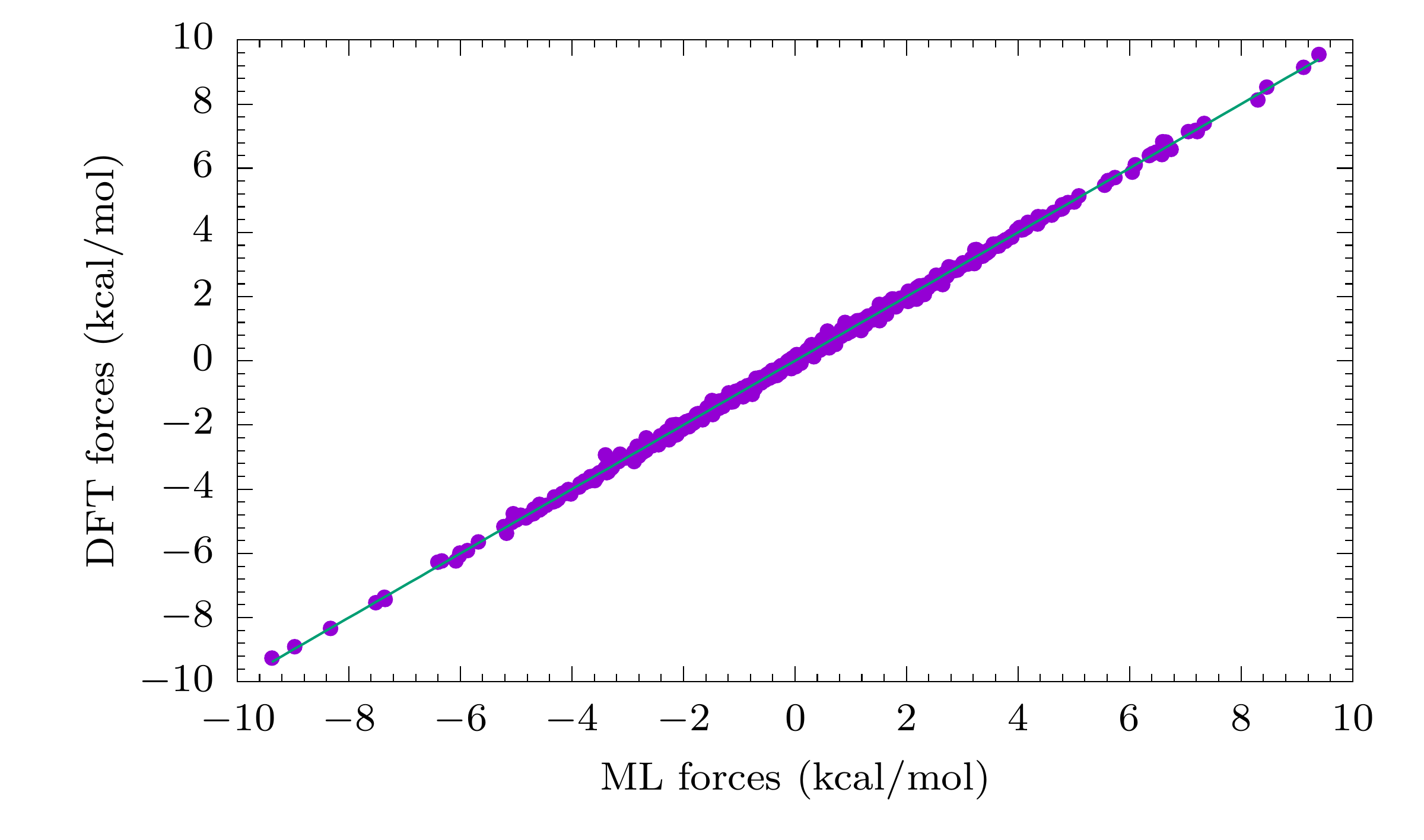}\\
    \includegraphics[scale=0.65]{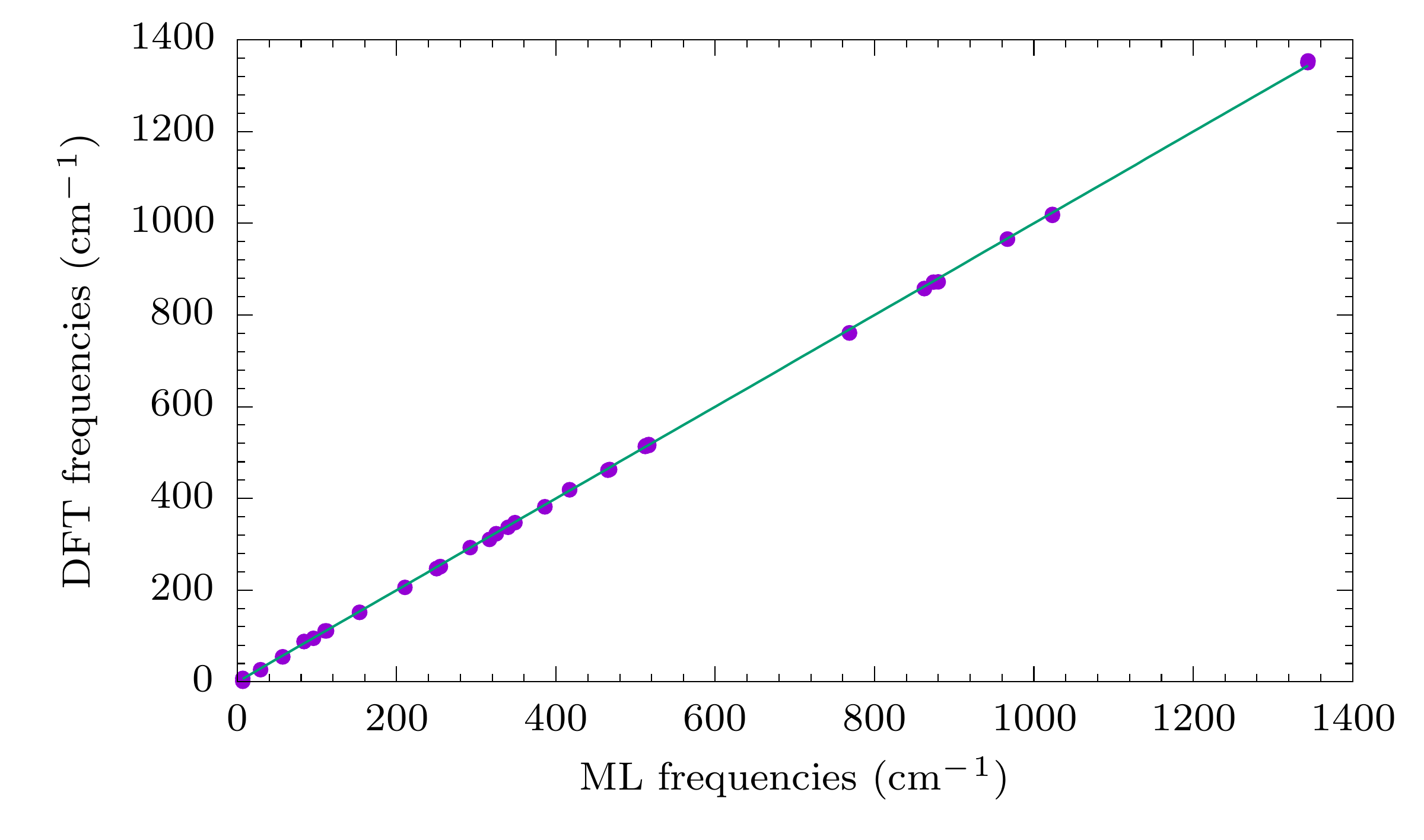}
    \includegraphics[scale=0.65]{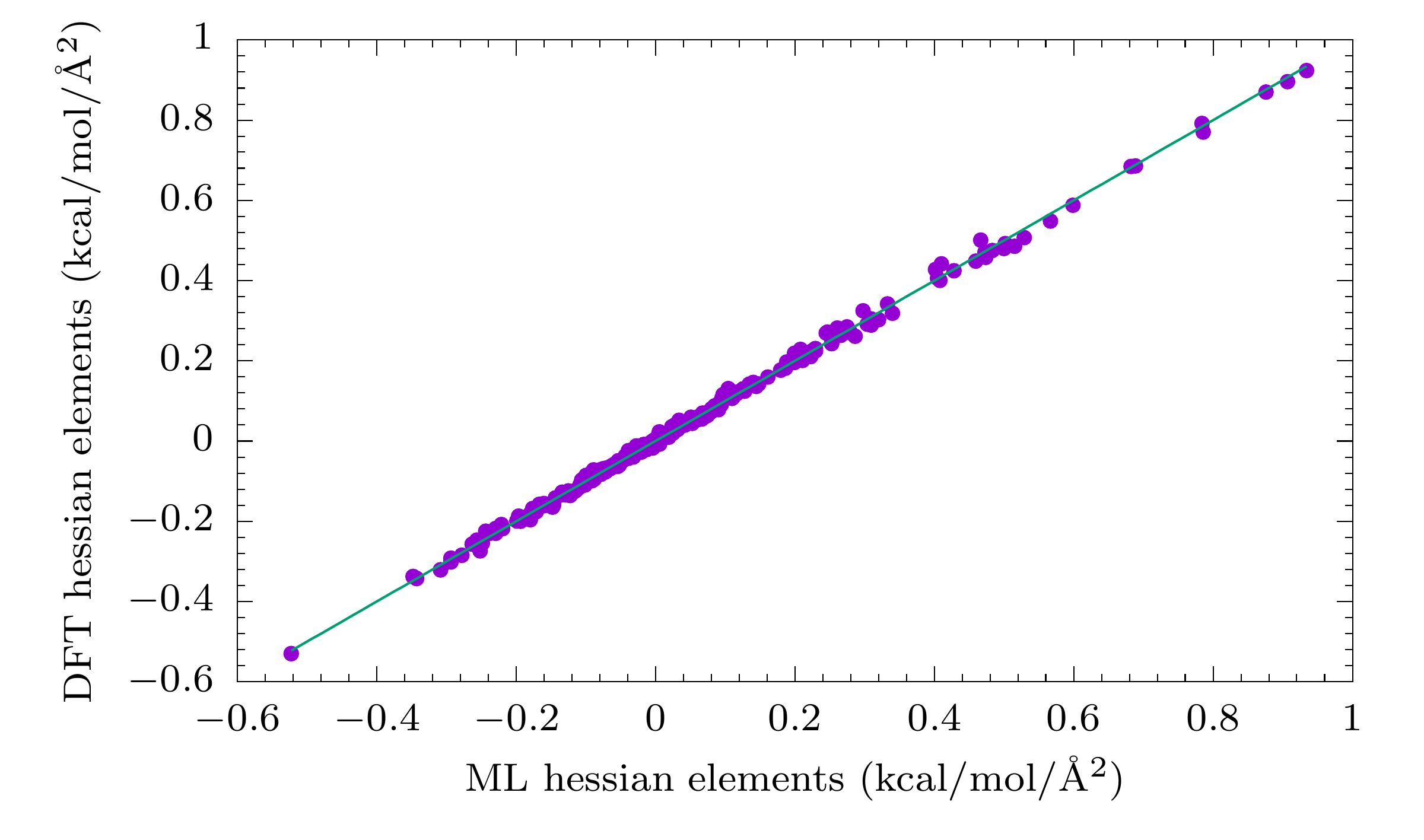}
    \caption{\textbf{Energy, forces, phonons and hessian.} $\delta =2.75$}
\end{figure}

\begin{figure}[H]
    \centering
    \includegraphics[scale=0.65]{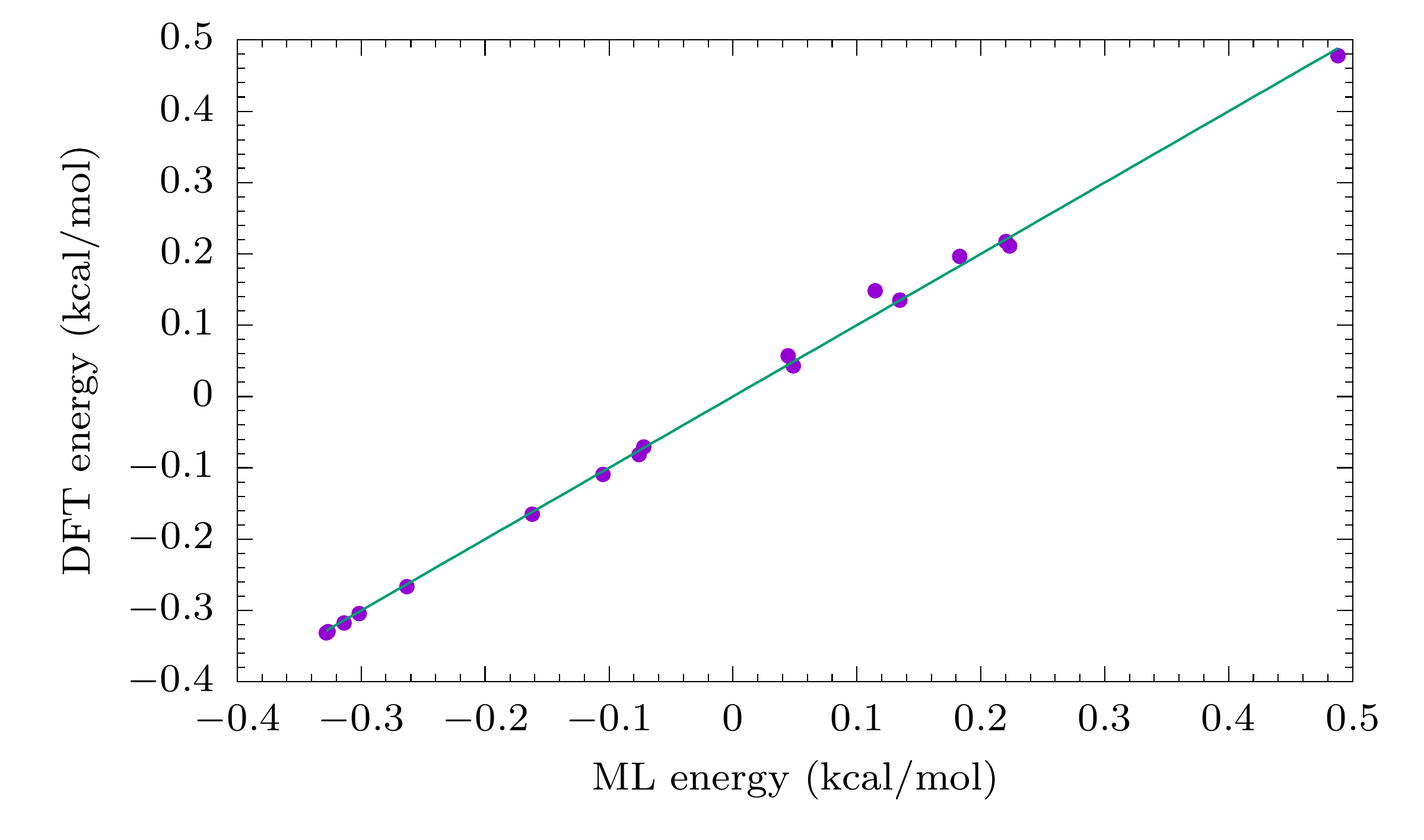}\\
    \includegraphics[scale=0.65]{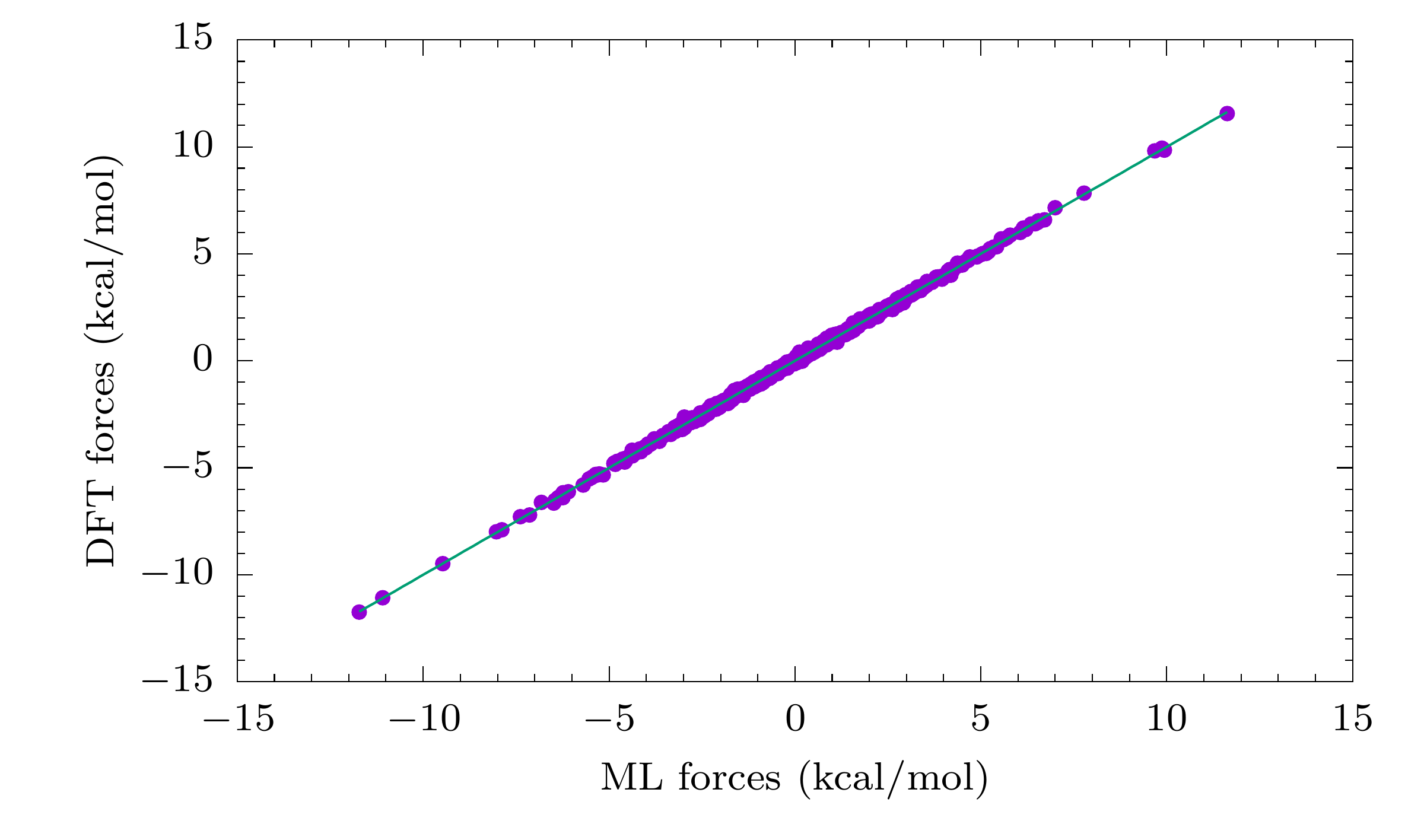}\\
    \includegraphics[scale=0.65]{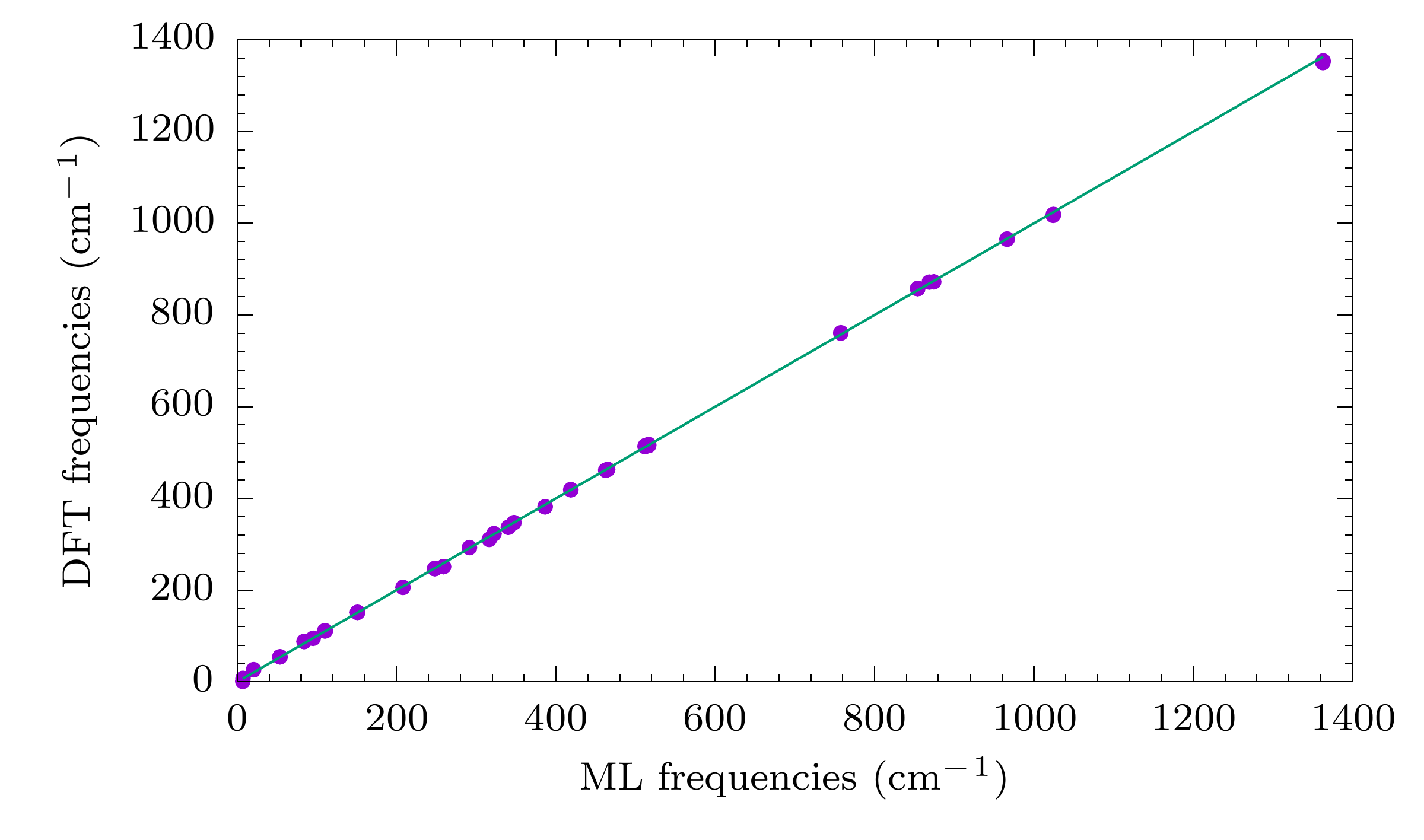}
    \includegraphics[scale=0.65]{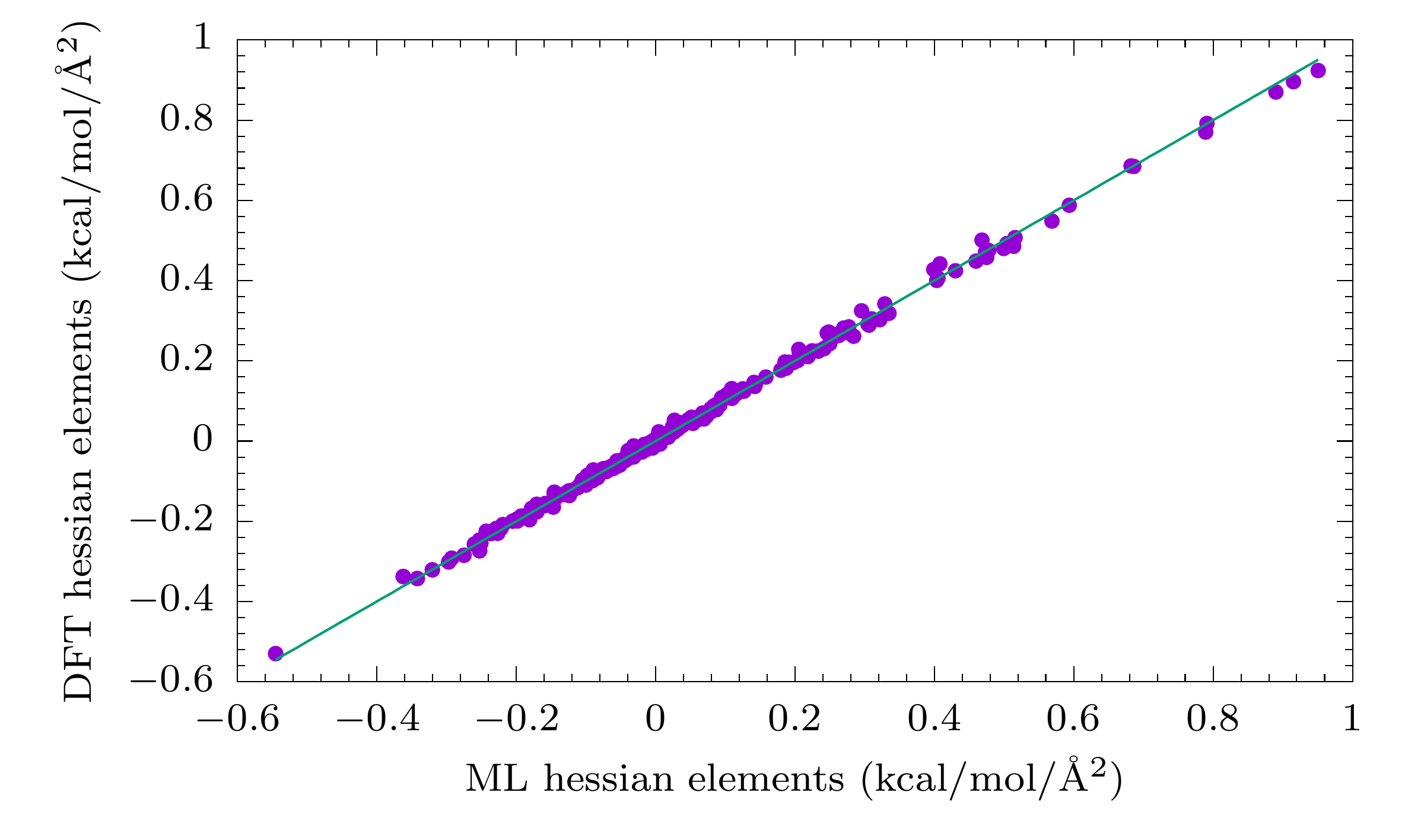}
    \caption{\textbf{Energy, forces, phonons and hessian.} $\delta =3$}
\end{figure}

\begin{figure}[H]
    \centering
    \includegraphics[scale=0.65]{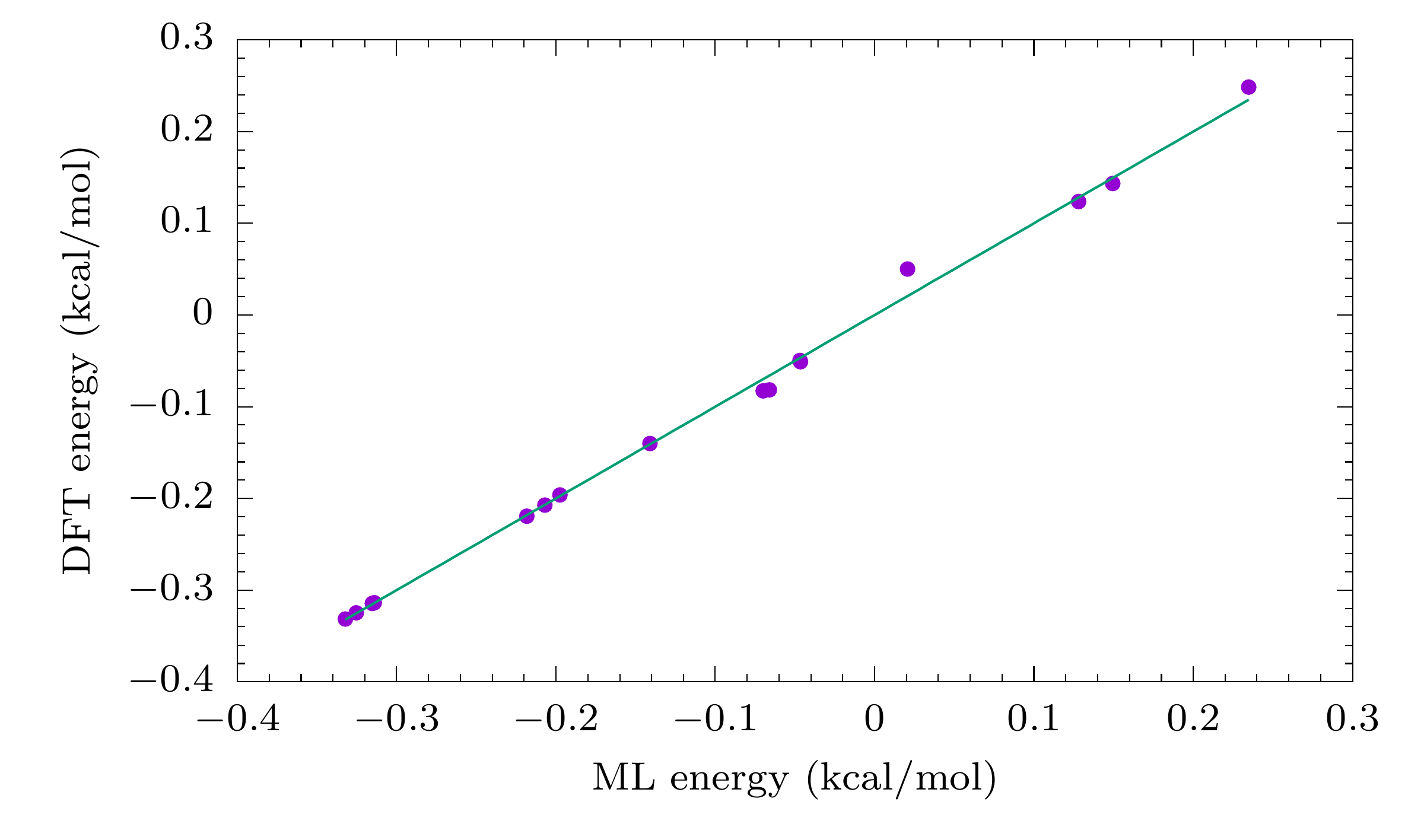}\\
    \includegraphics[scale=0.65]{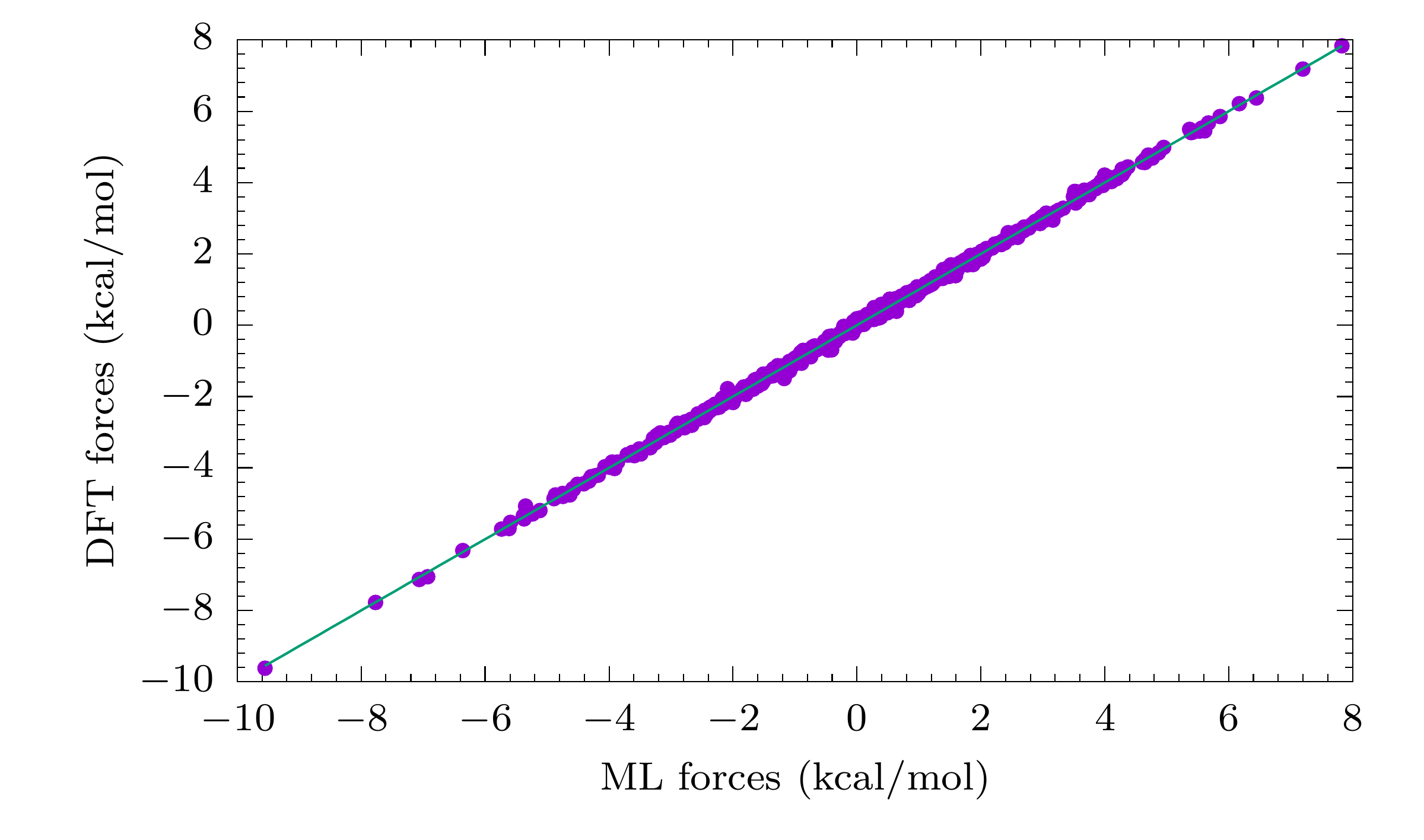}\\
    \includegraphics[scale=0.65]{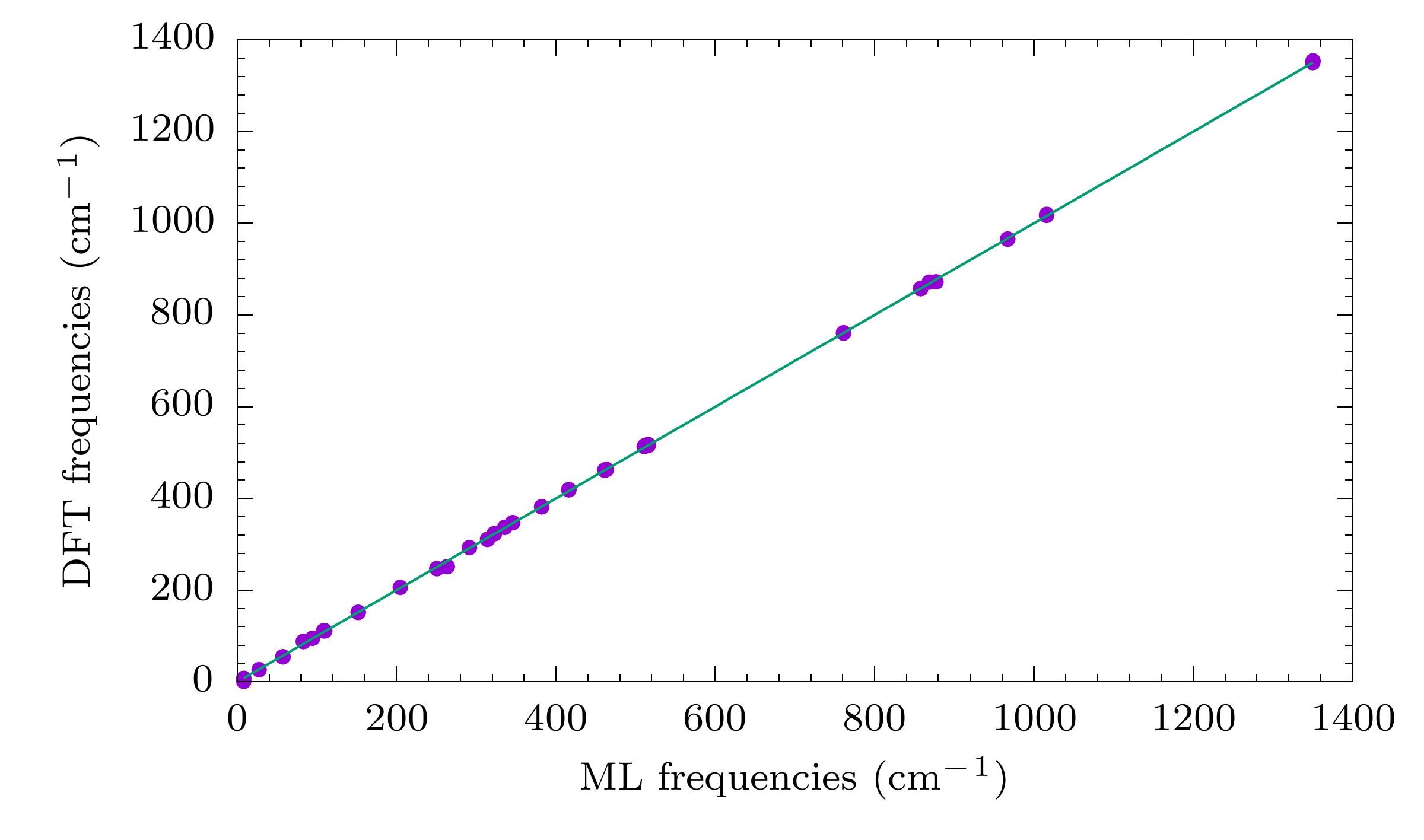}
    \includegraphics[scale=0.65]{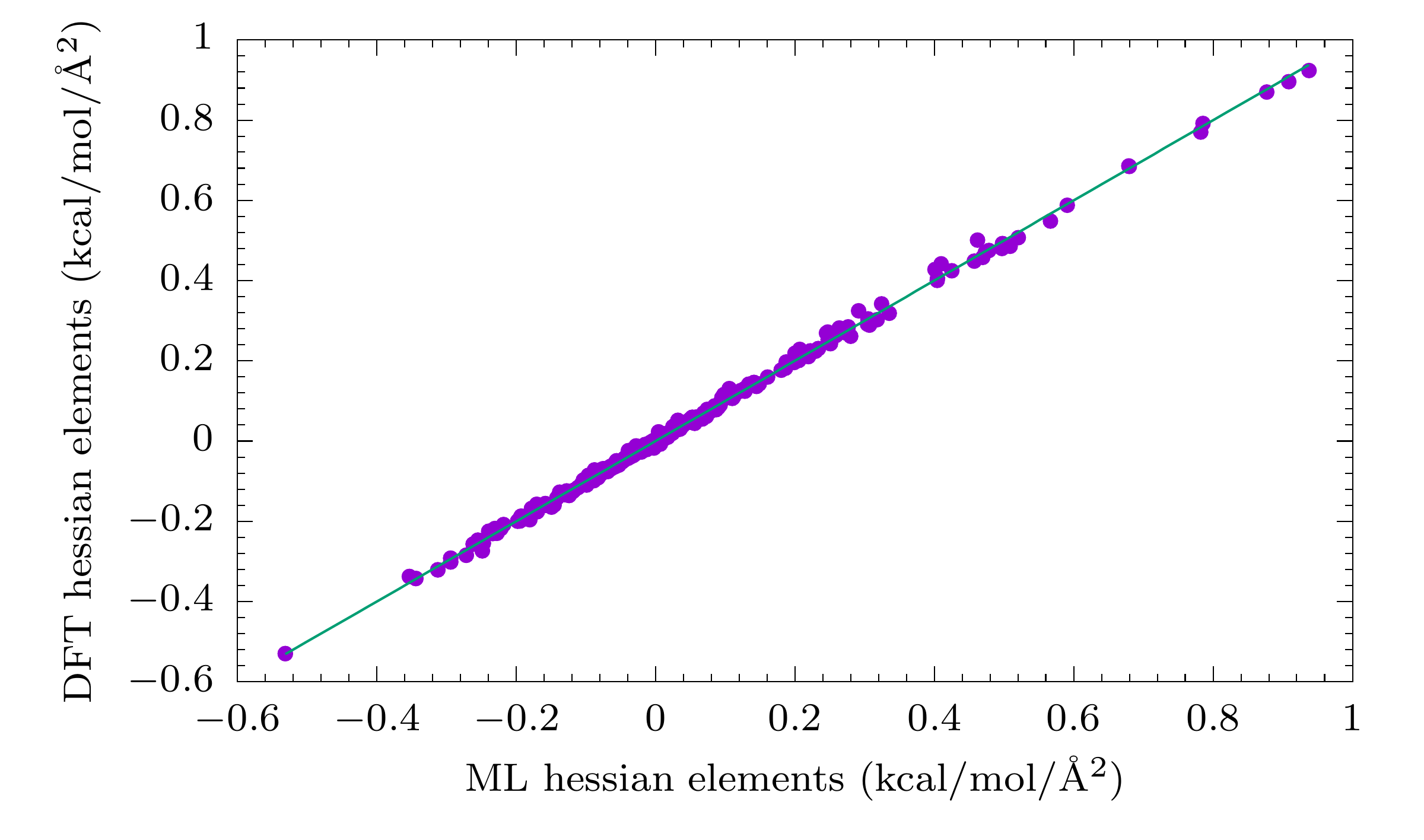}
    \caption{\textbf{Energy, forces, phonons and hessian.} $\delta =3.25$}
\end{figure}

\subsection{Compound 2}

\begin{figure}[H]
    \centering
    \includegraphics[scale=0.65]{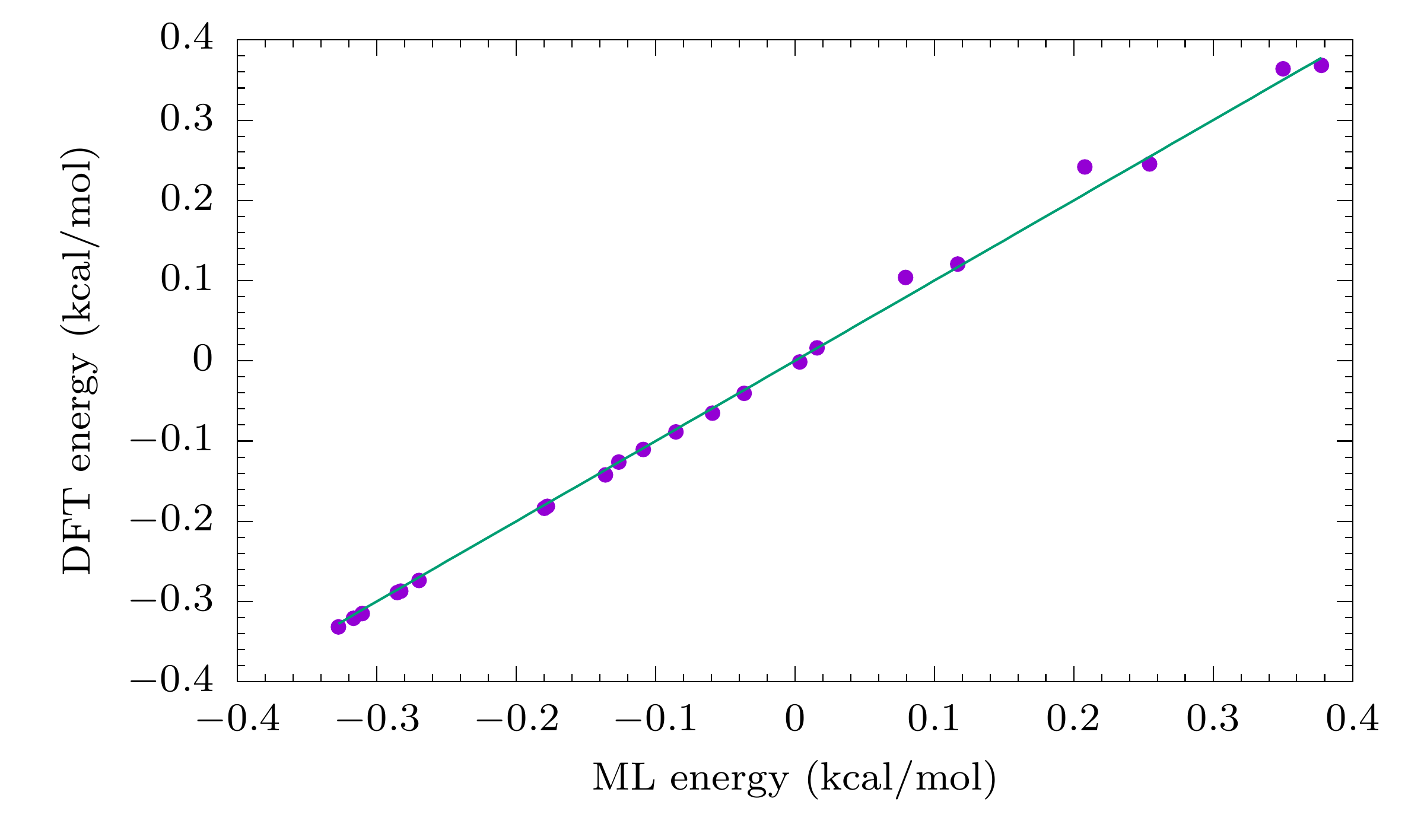}\\
    \includegraphics[scale=0.65]{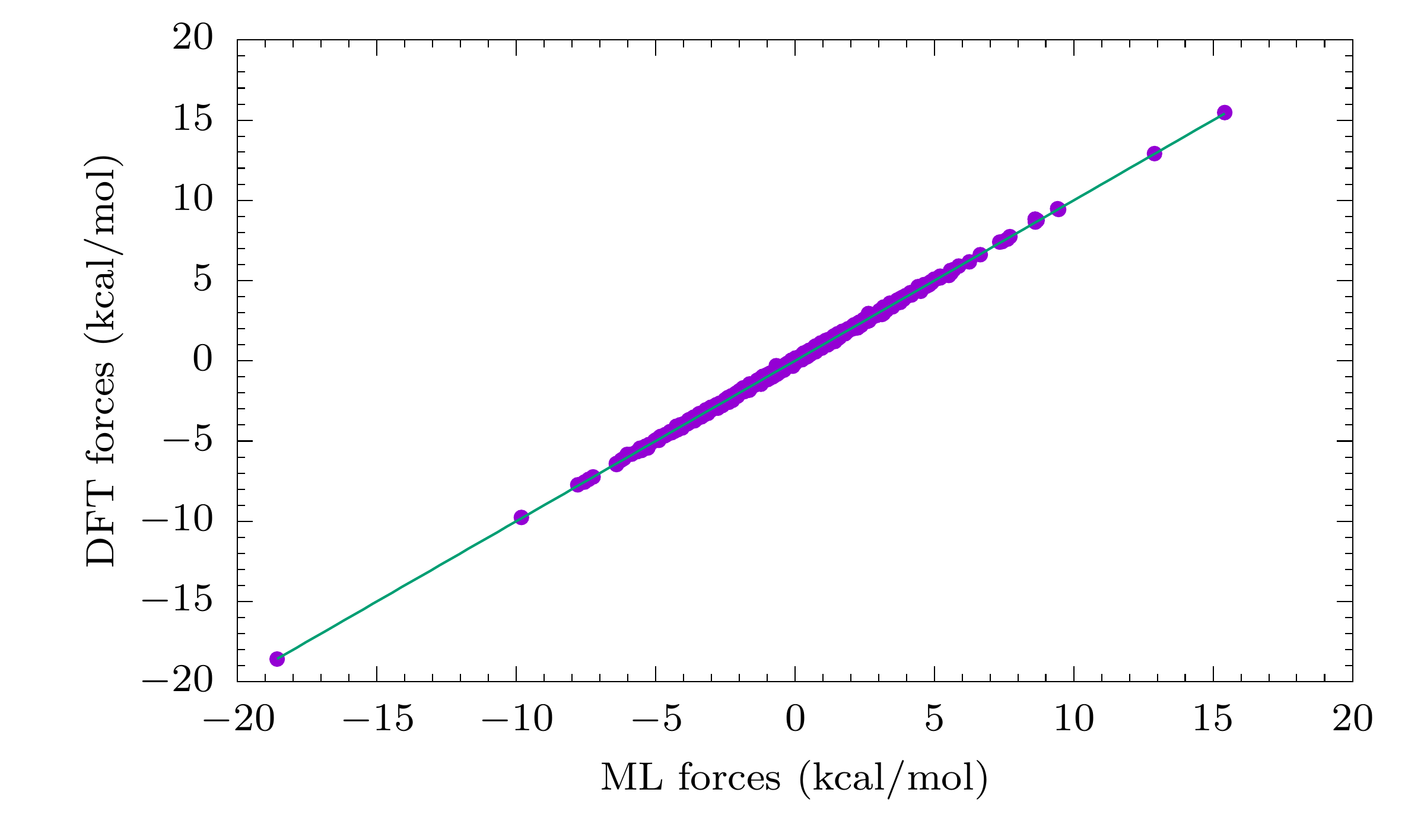}\\
    \includegraphics[scale=0.65]{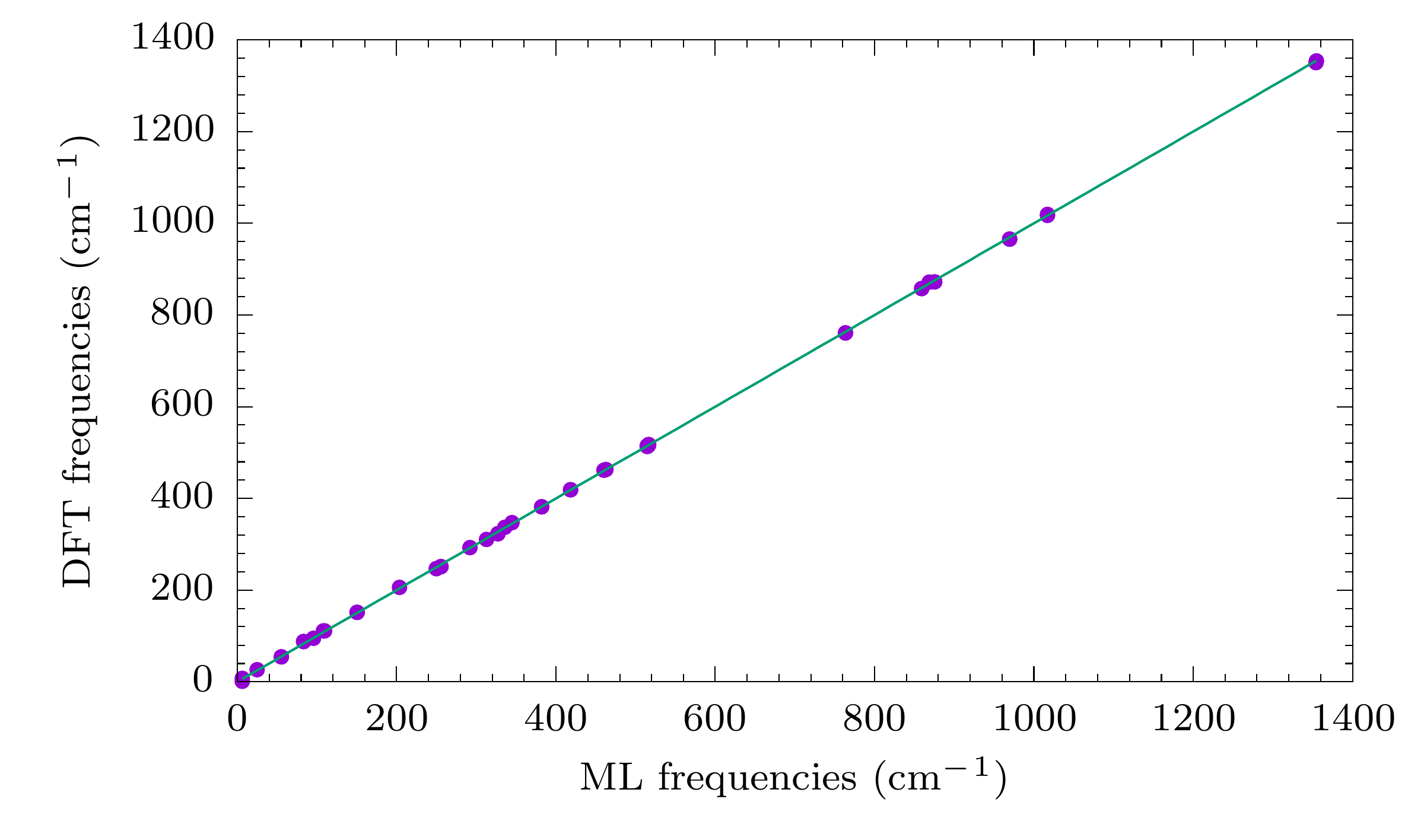}
    \includegraphics[scale=0.65]{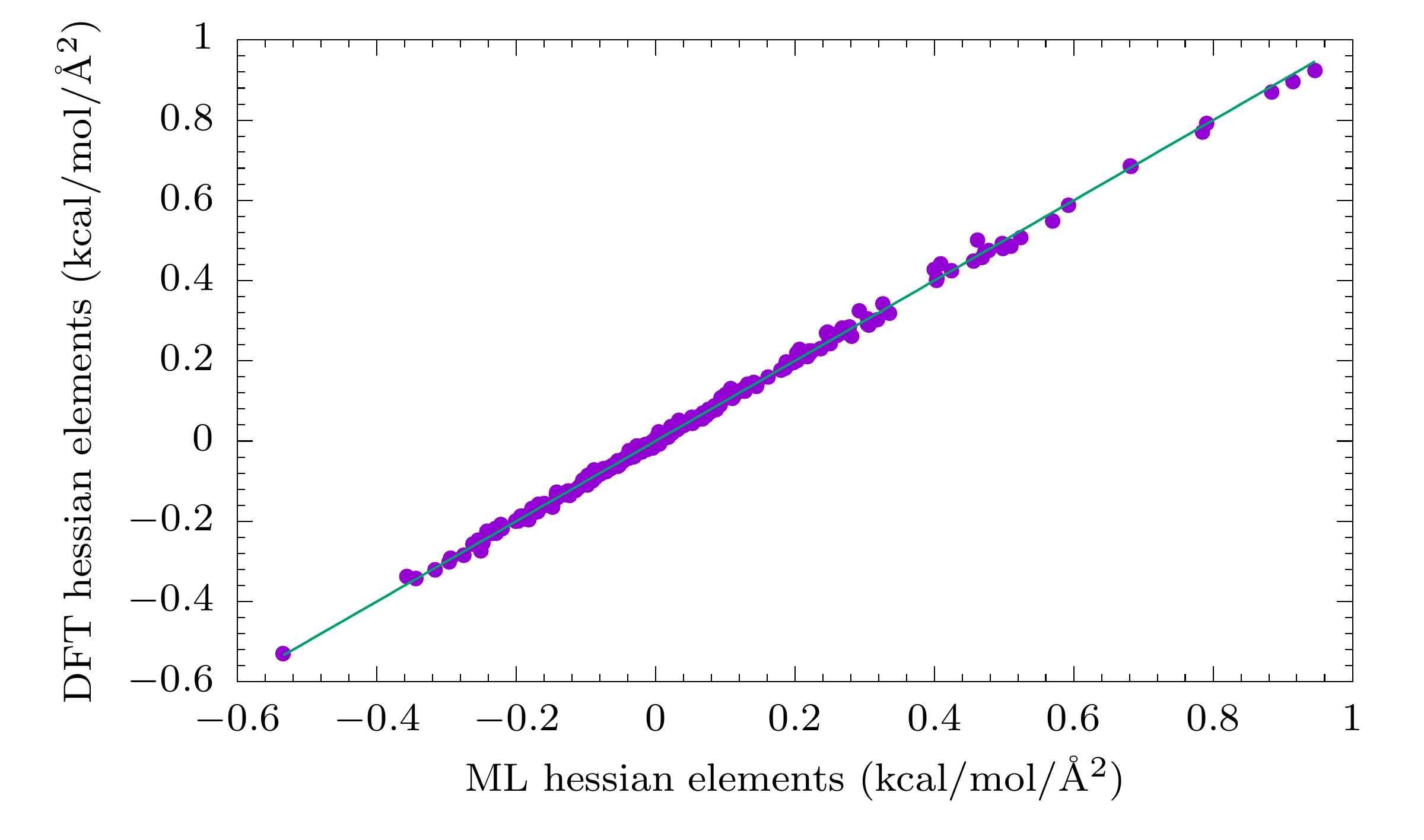}
    \caption{\textbf{Energy, forces, phonons and hessian.} $\delta =2.5$}
\end{figure}

\begin{figure}[H]
    \centering
    \includegraphics[scale=0.65]{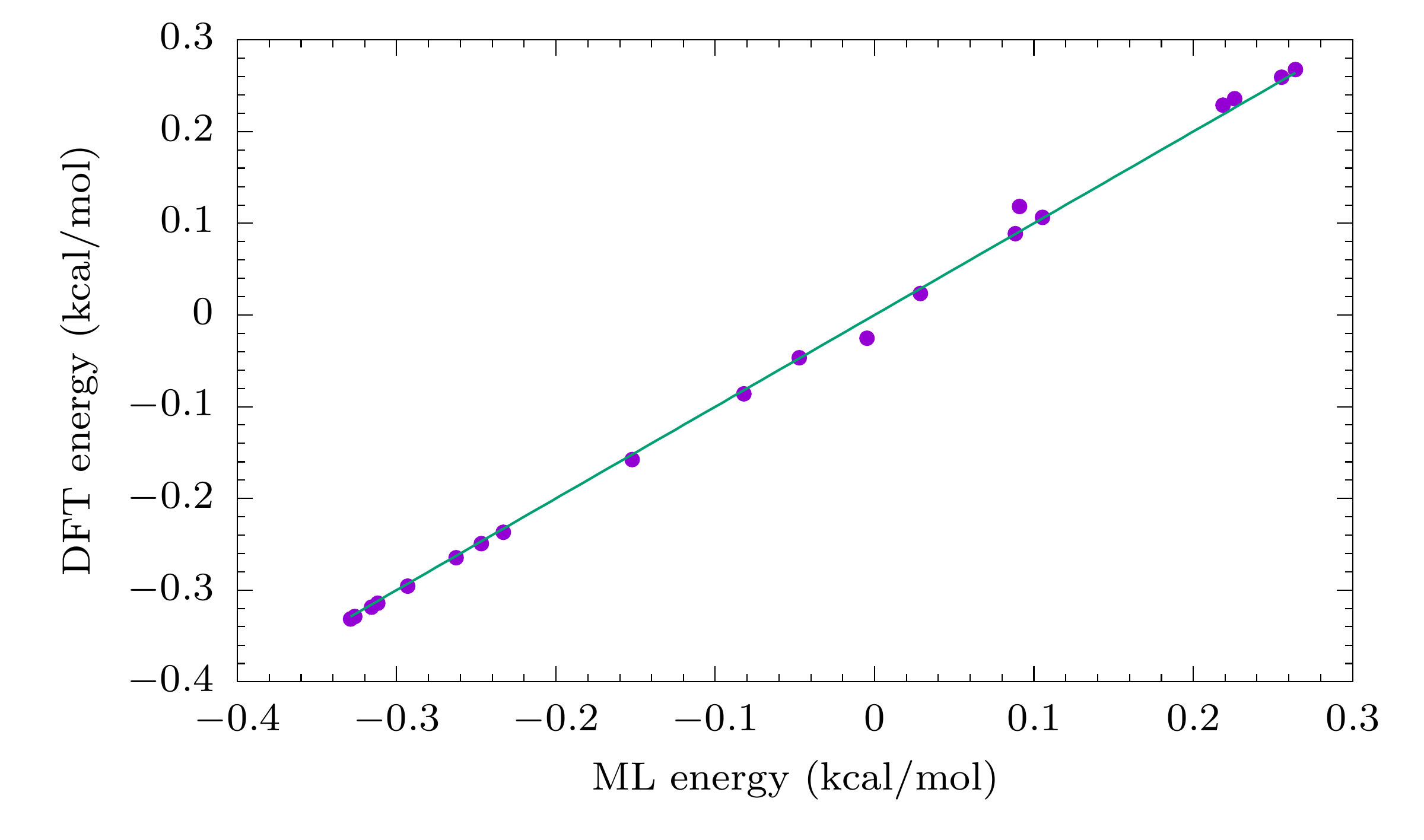}\\
    \includegraphics[scale=0.65]{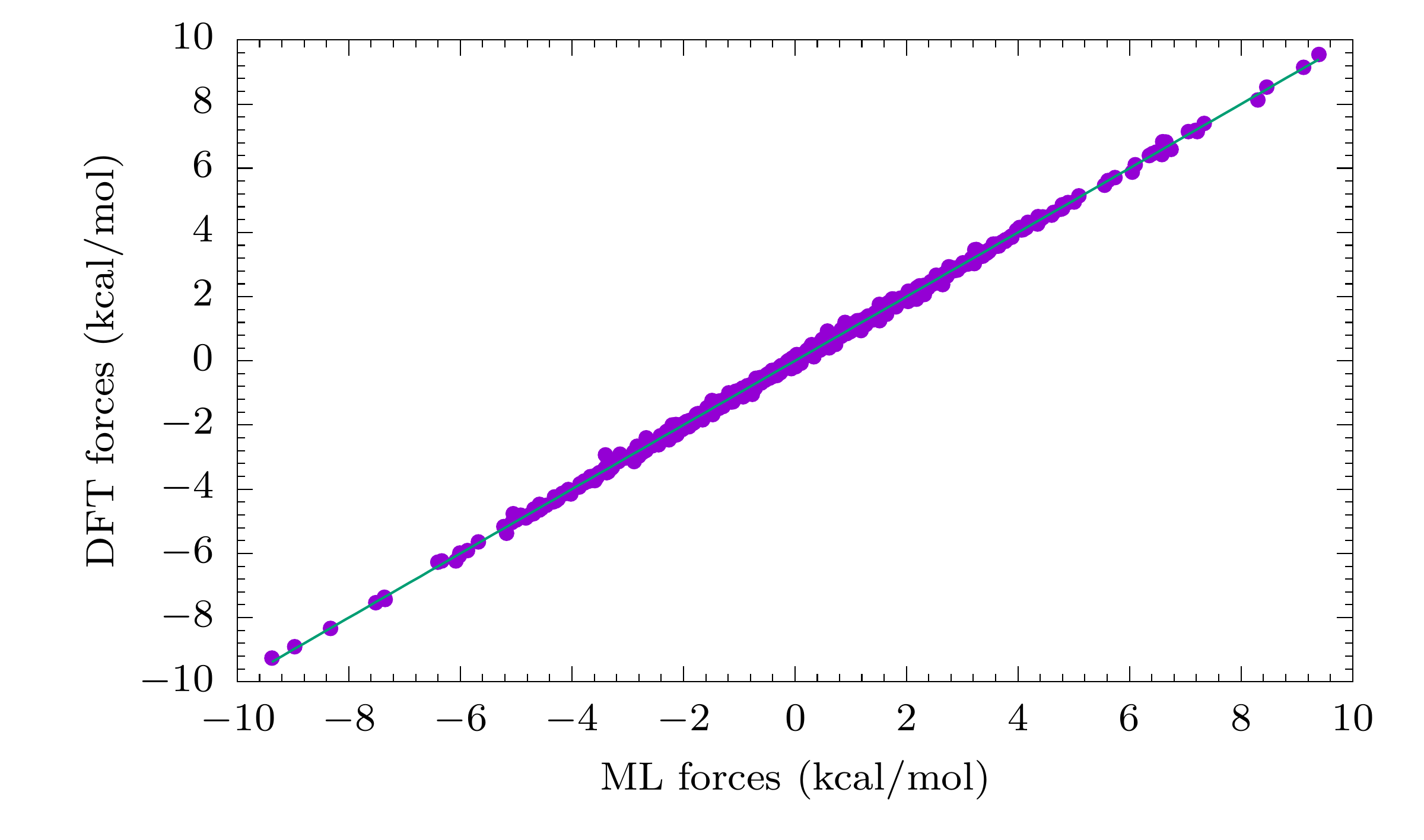}\\
    \includegraphics[scale=0.65]{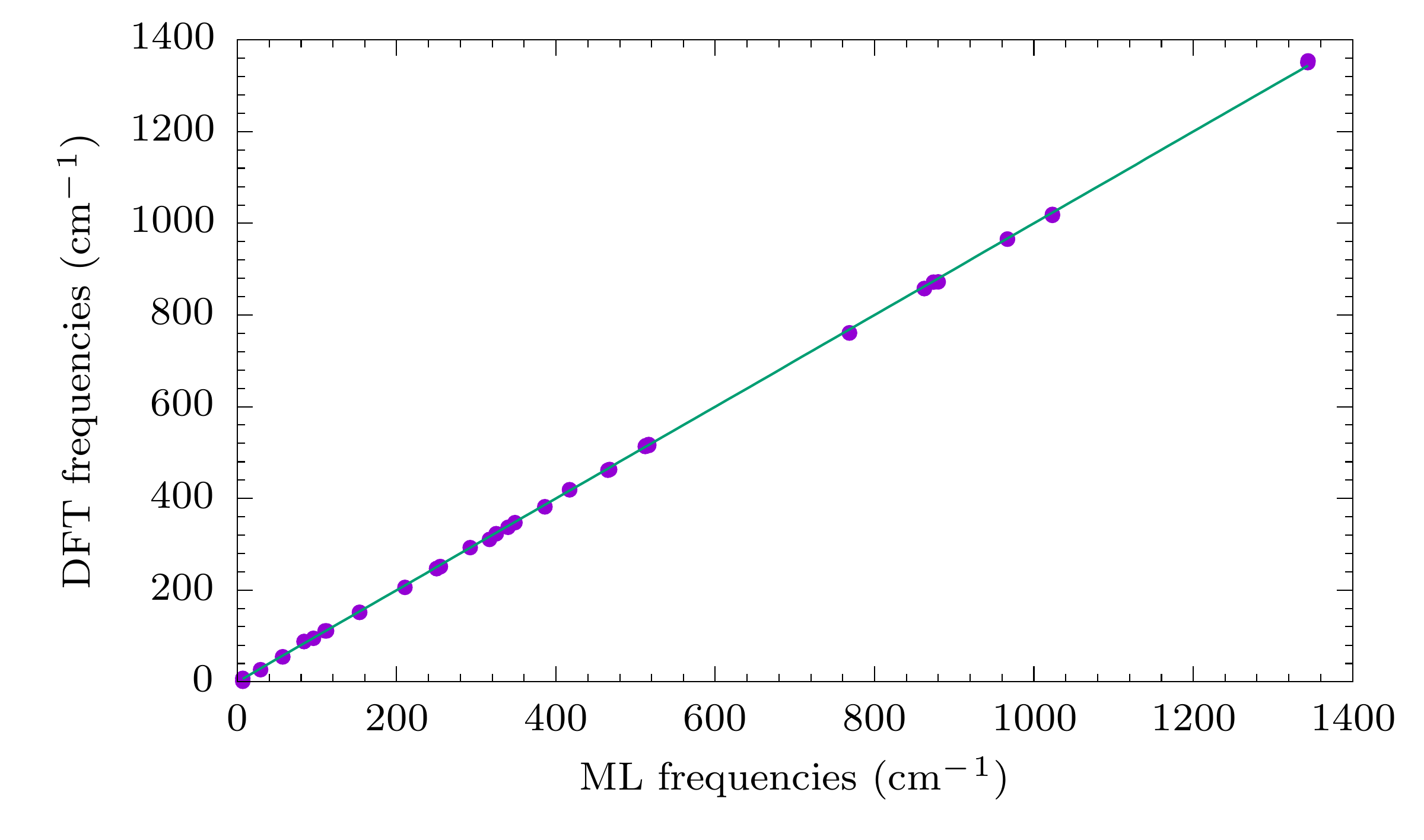}
    \includegraphics[scale=0.65]{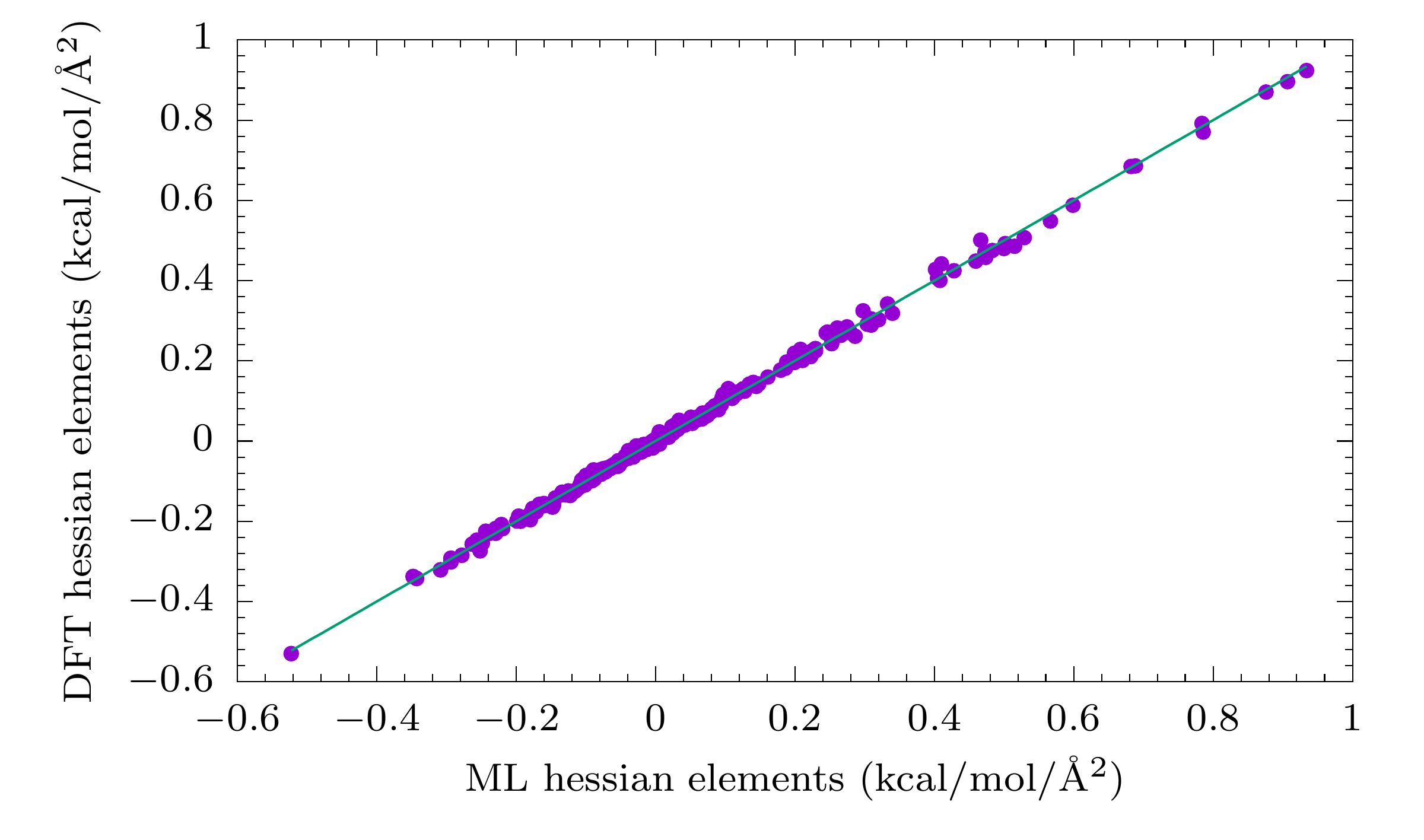}
    \caption{\textbf{Energy, forces, phonons and hessian.} $\delta =2.75$}
\end{figure}

\begin{figure}[H]
    \centering
    \includegraphics[scale=0.65]{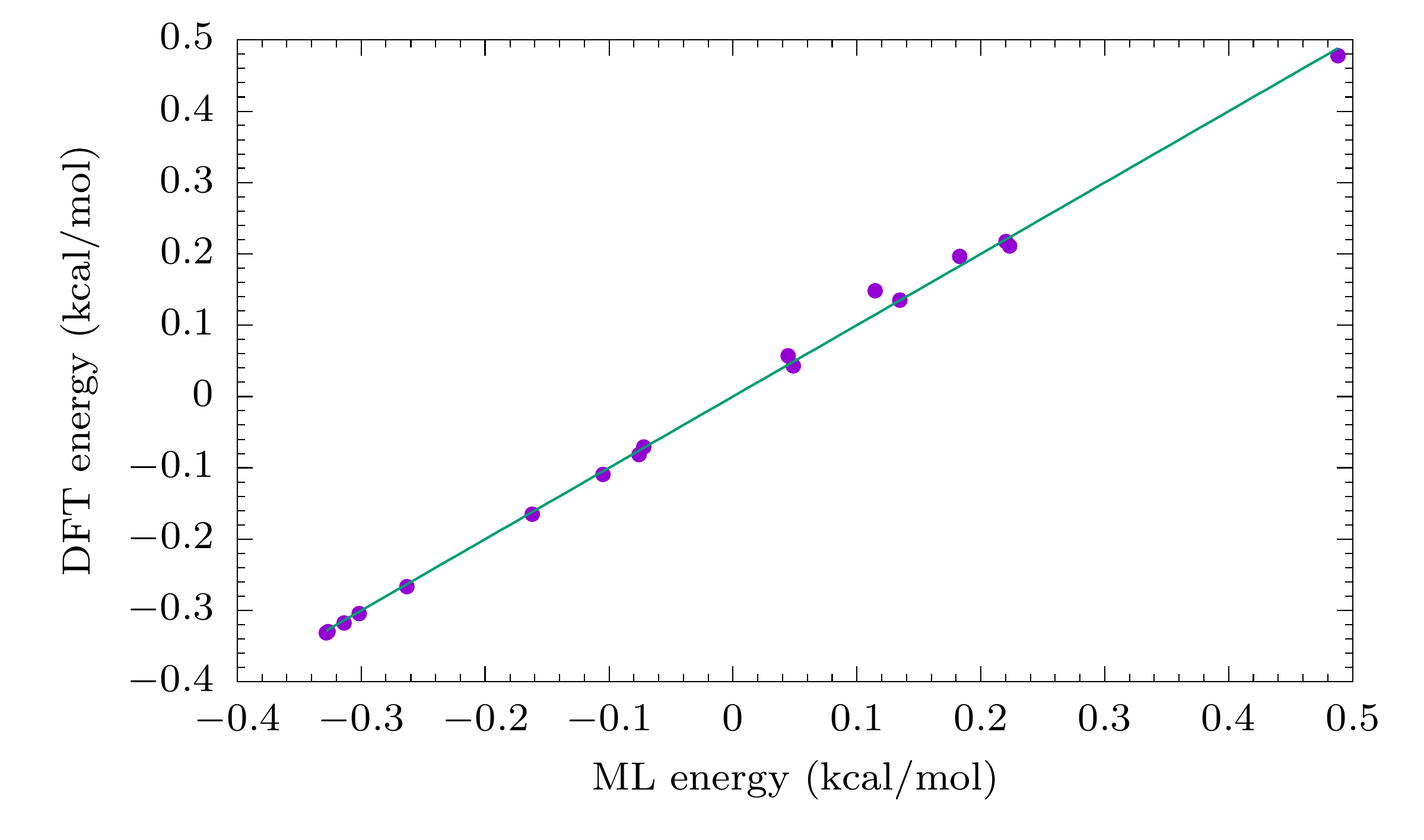}\\
    \includegraphics[scale=0.65]{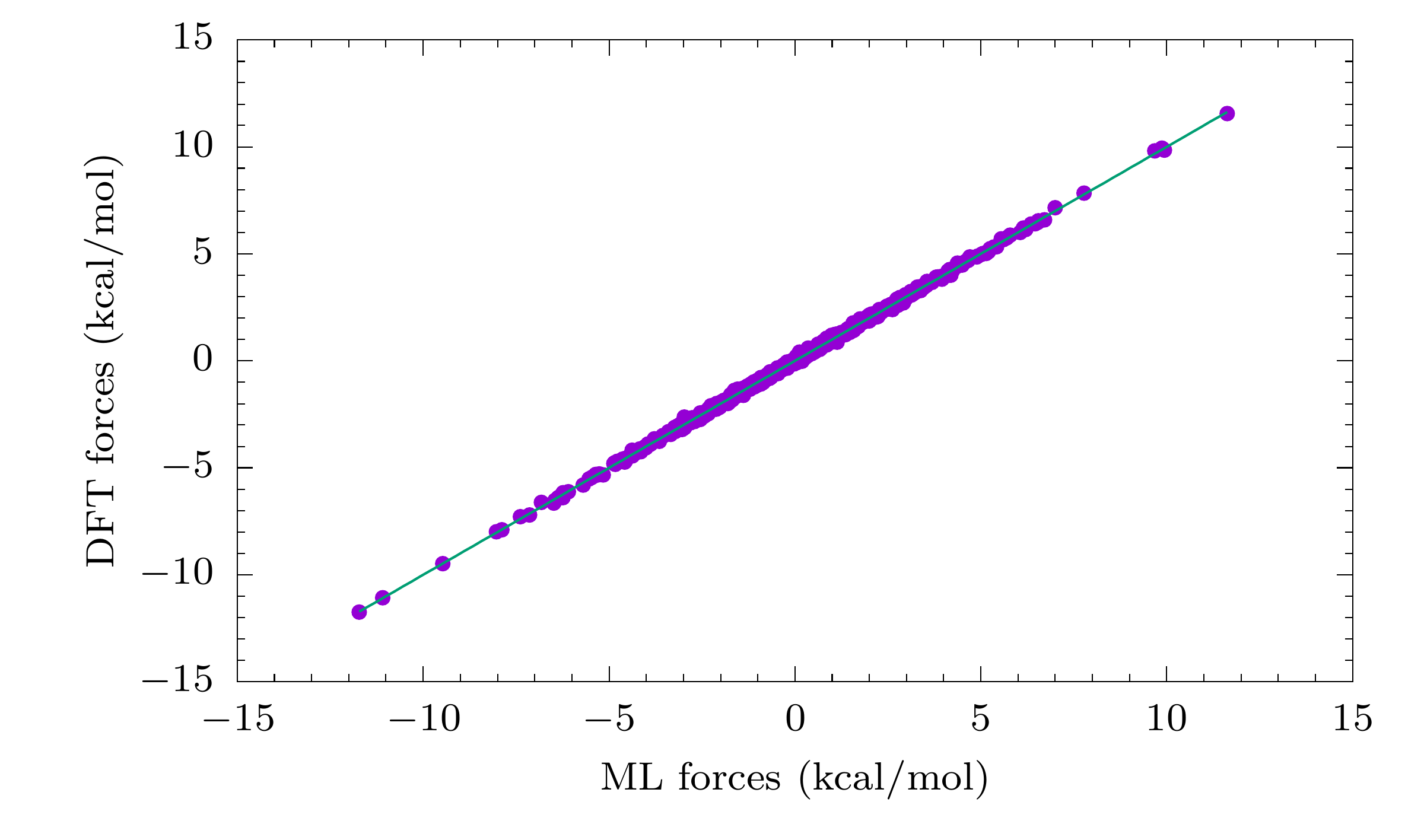}\\
    \includegraphics[scale=0.65]{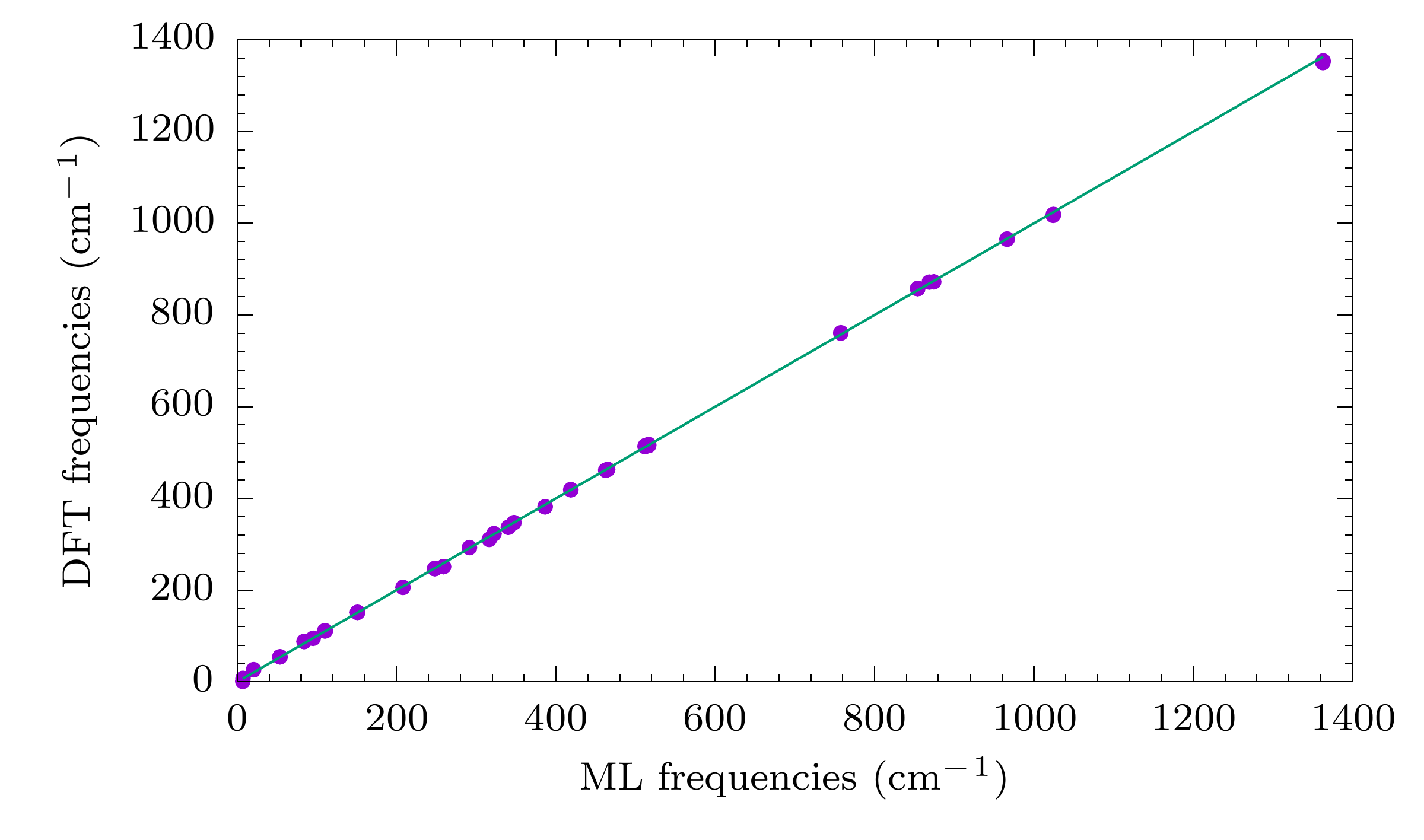}
    \includegraphics[scale=0.65]{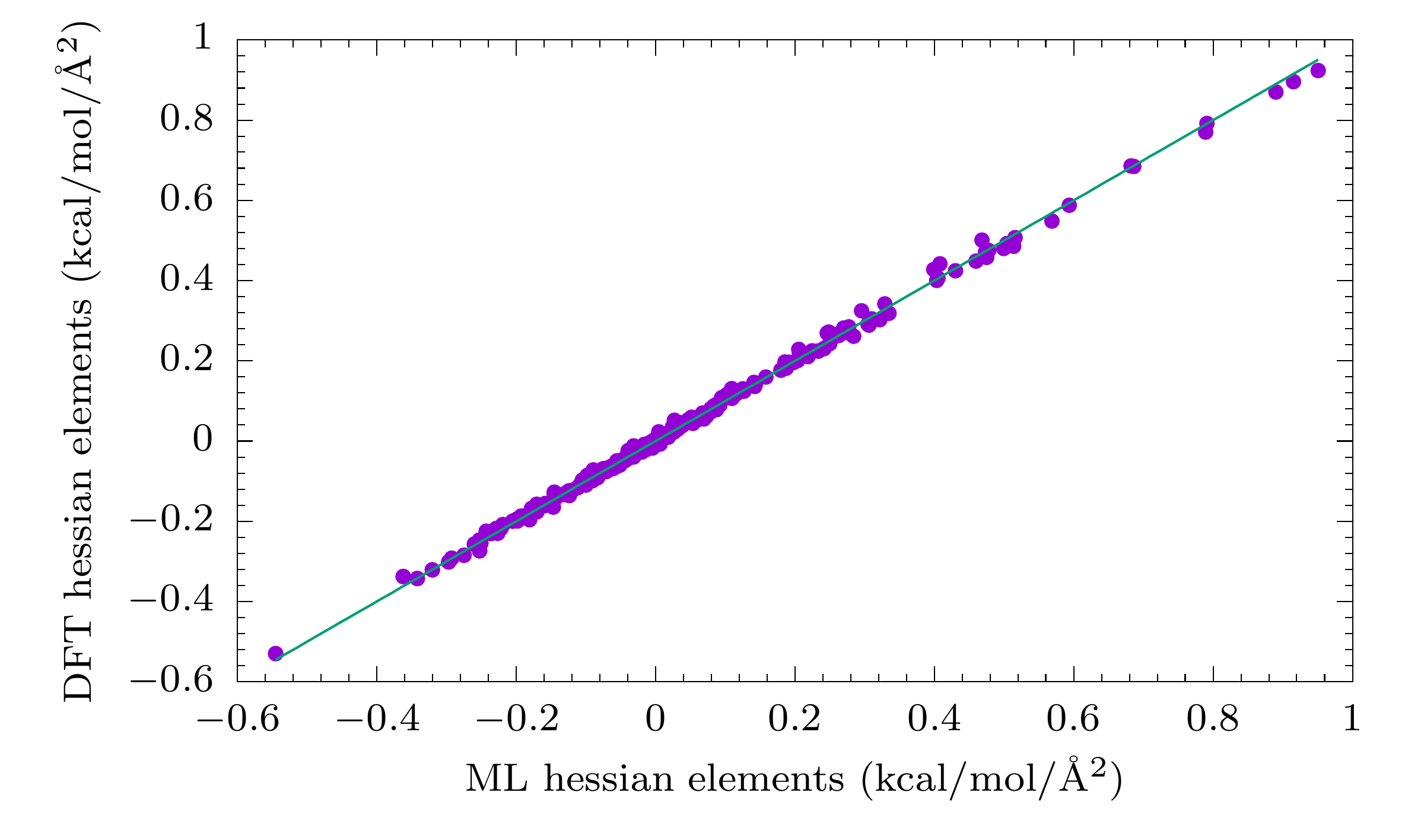}
    \caption{\textbf{Energy, forces, phonons and hessian.} $\delta =3$}
\end{figure}

\begin{figure}[H]
    \centering
    \includegraphics[scale=0.65]{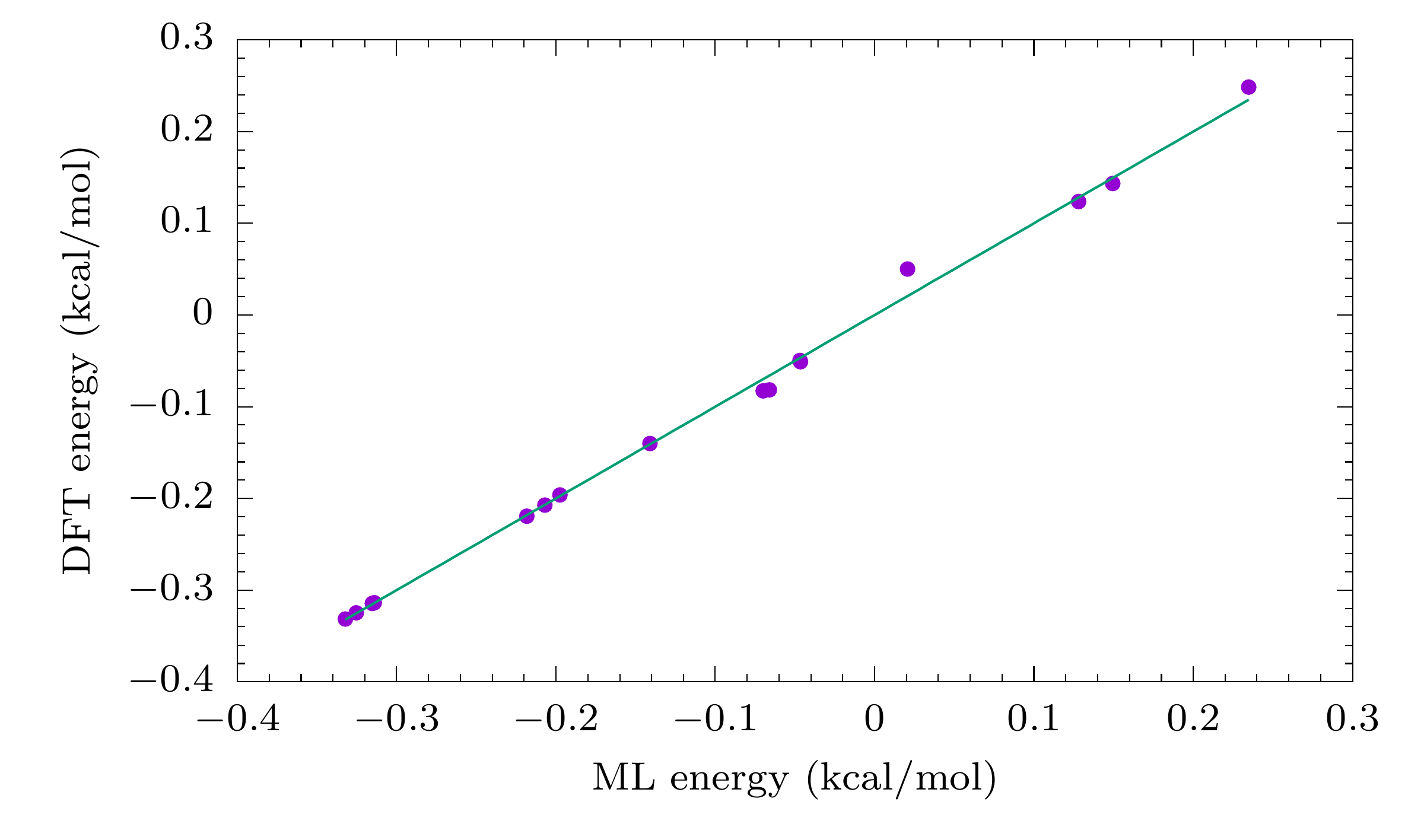}\\
    \includegraphics[scale=0.65]{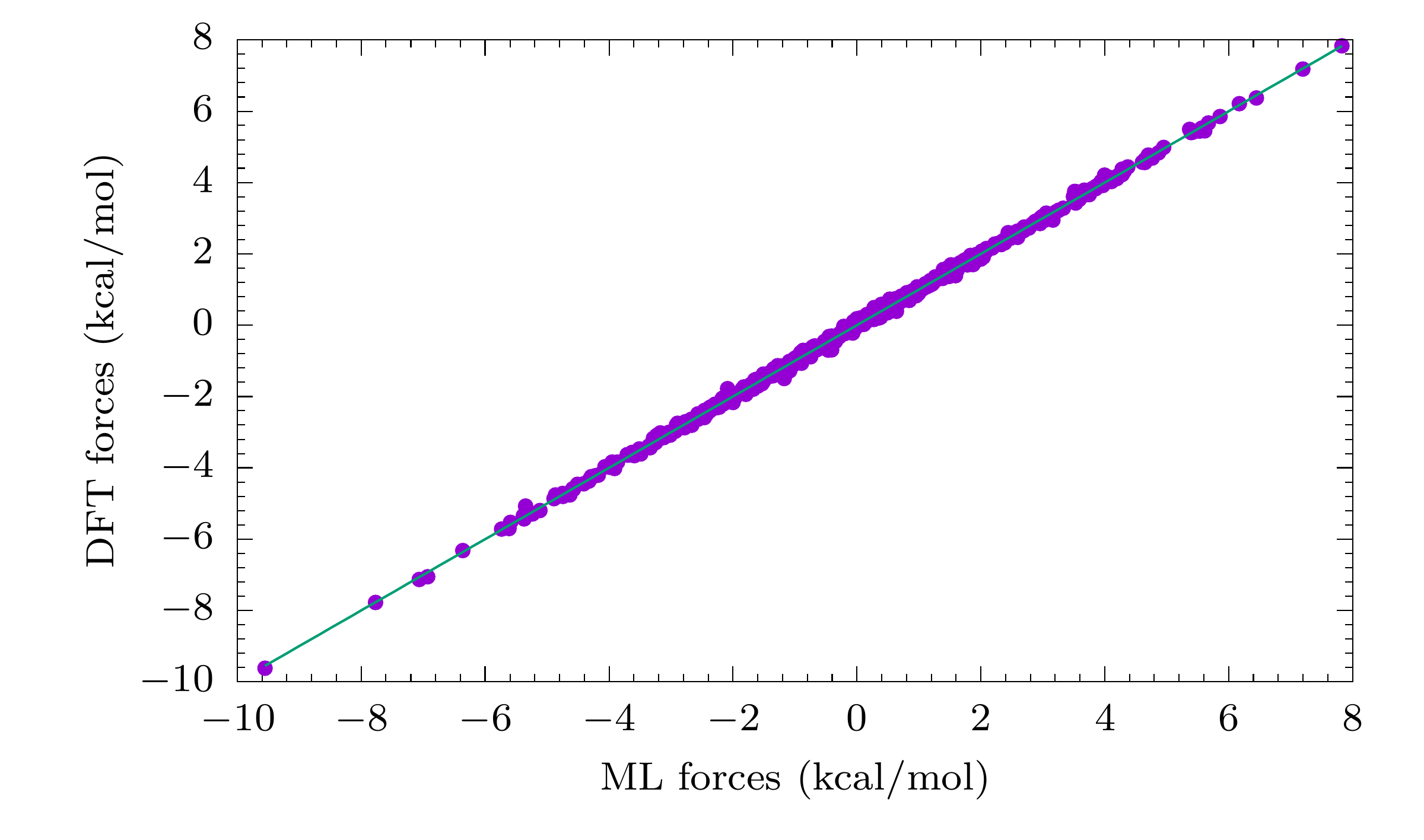}\\
    \includegraphics[scale=0.65]{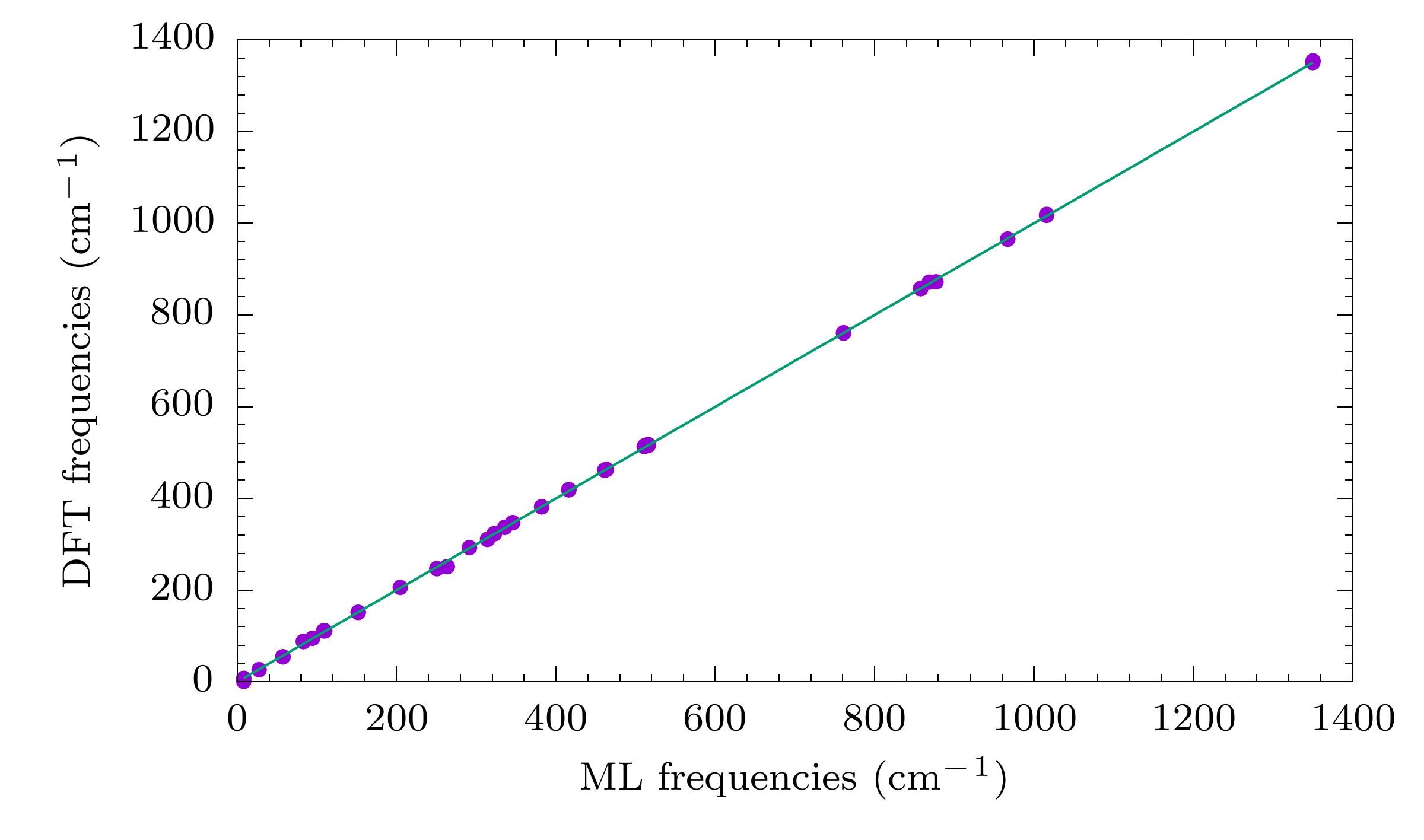}
    \includegraphics[scale=0.65]{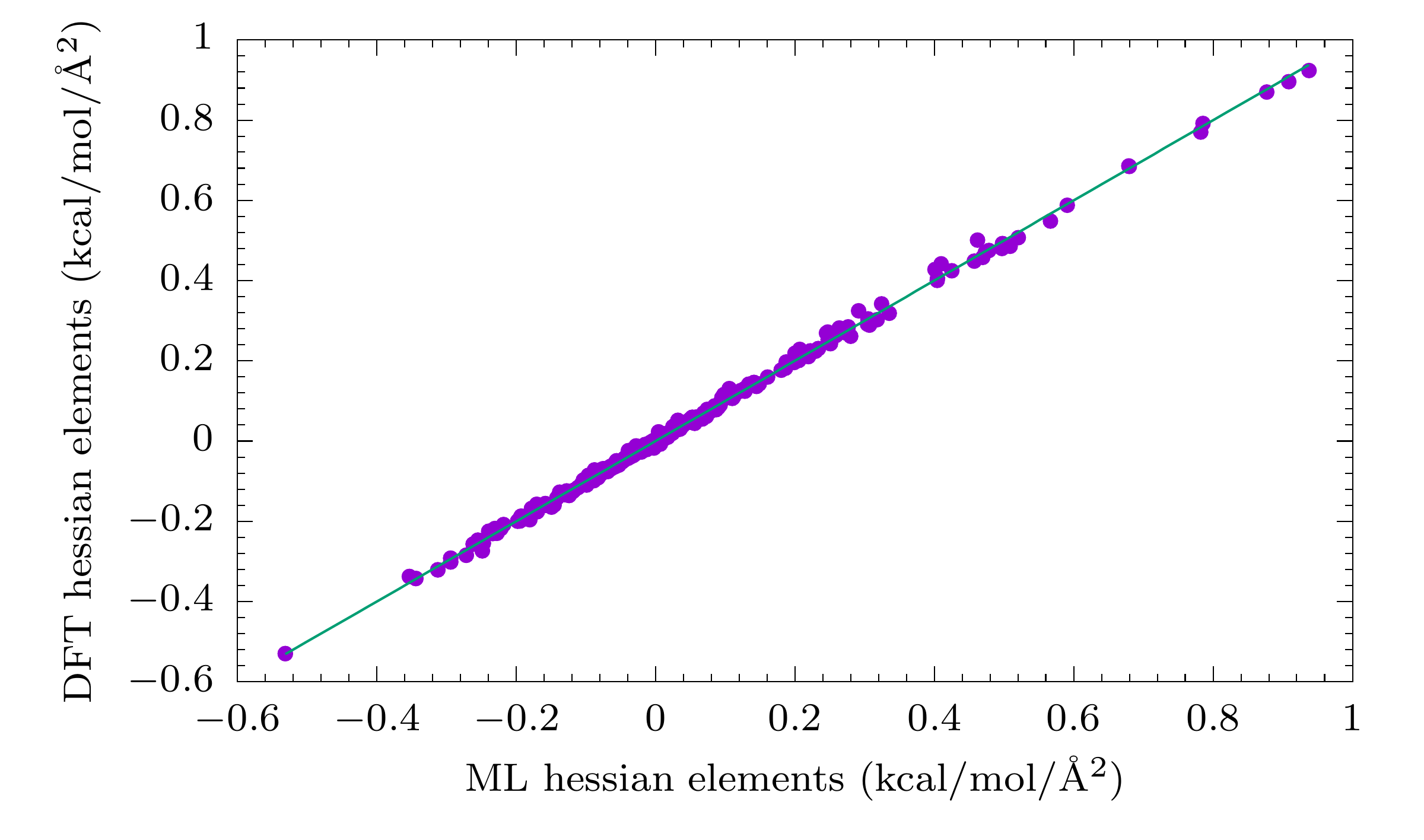}
    \caption{\textbf{Energy, forces, phonons and hessian.} $\delta =3.25$}
\end{figure}

\subsection{Compound 3}

\begin{figure}[H]
    \centering
    \includegraphics[scale=0.65]{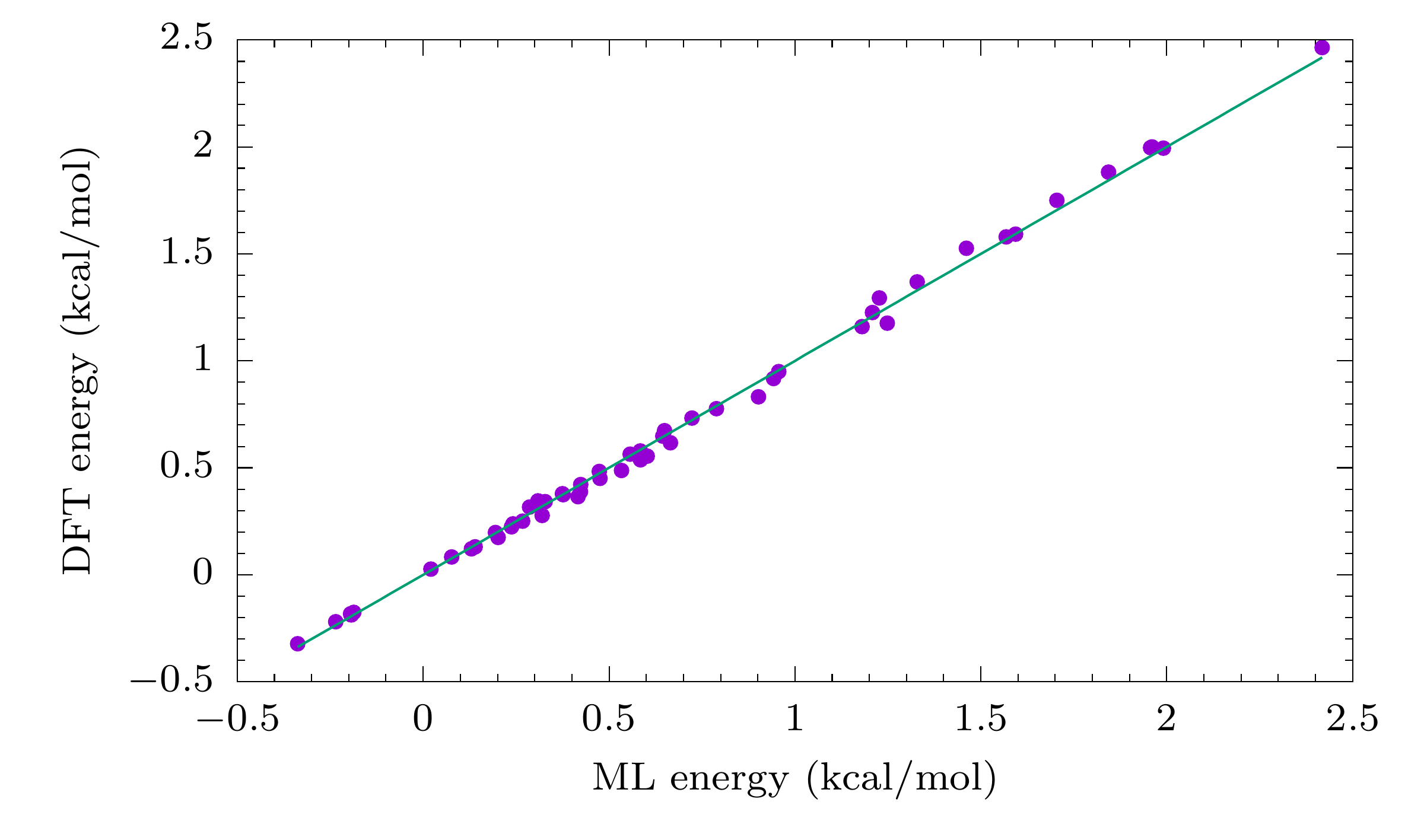}\\
    \includegraphics[scale=0.65]{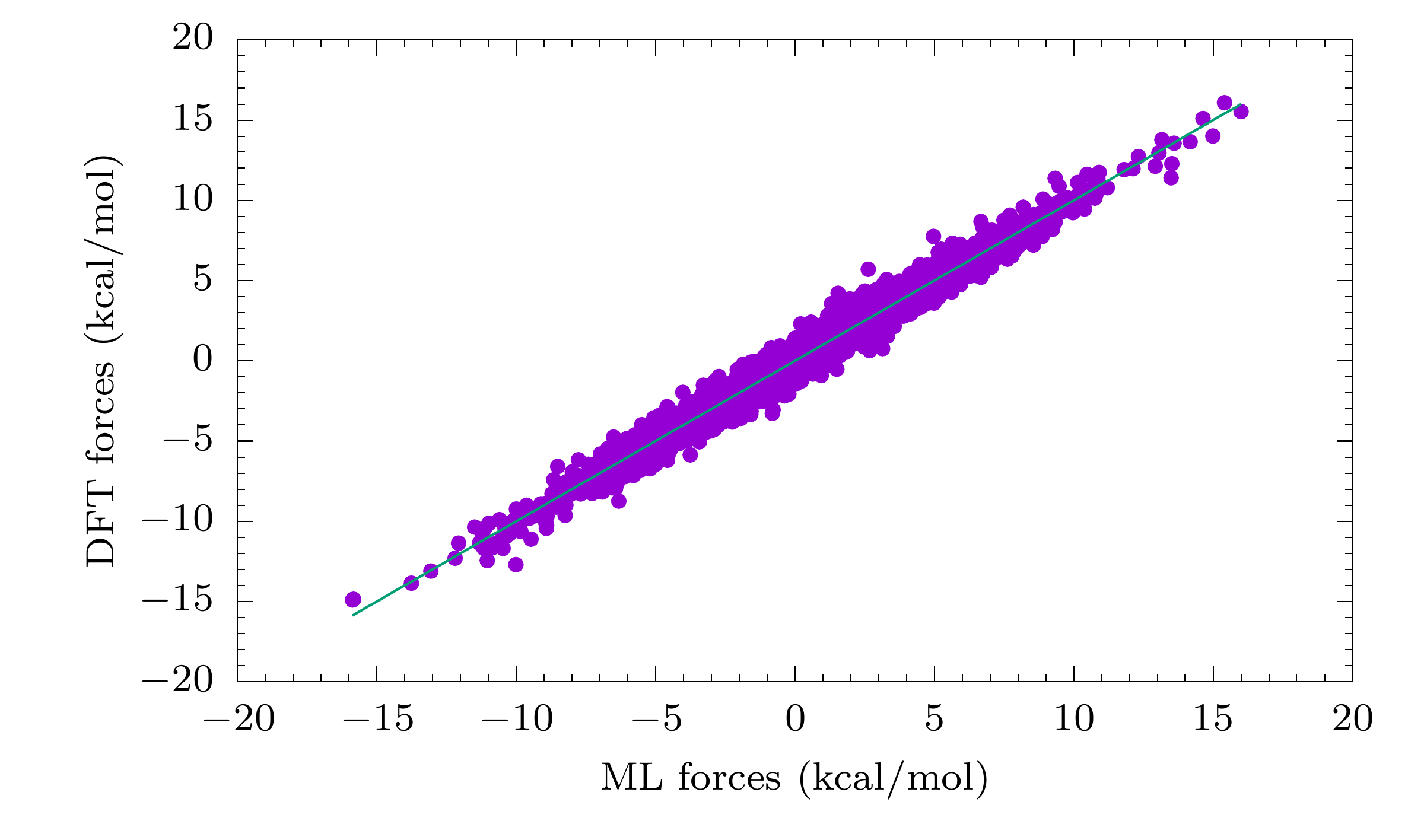}\\
    \includegraphics[scale=0.65]{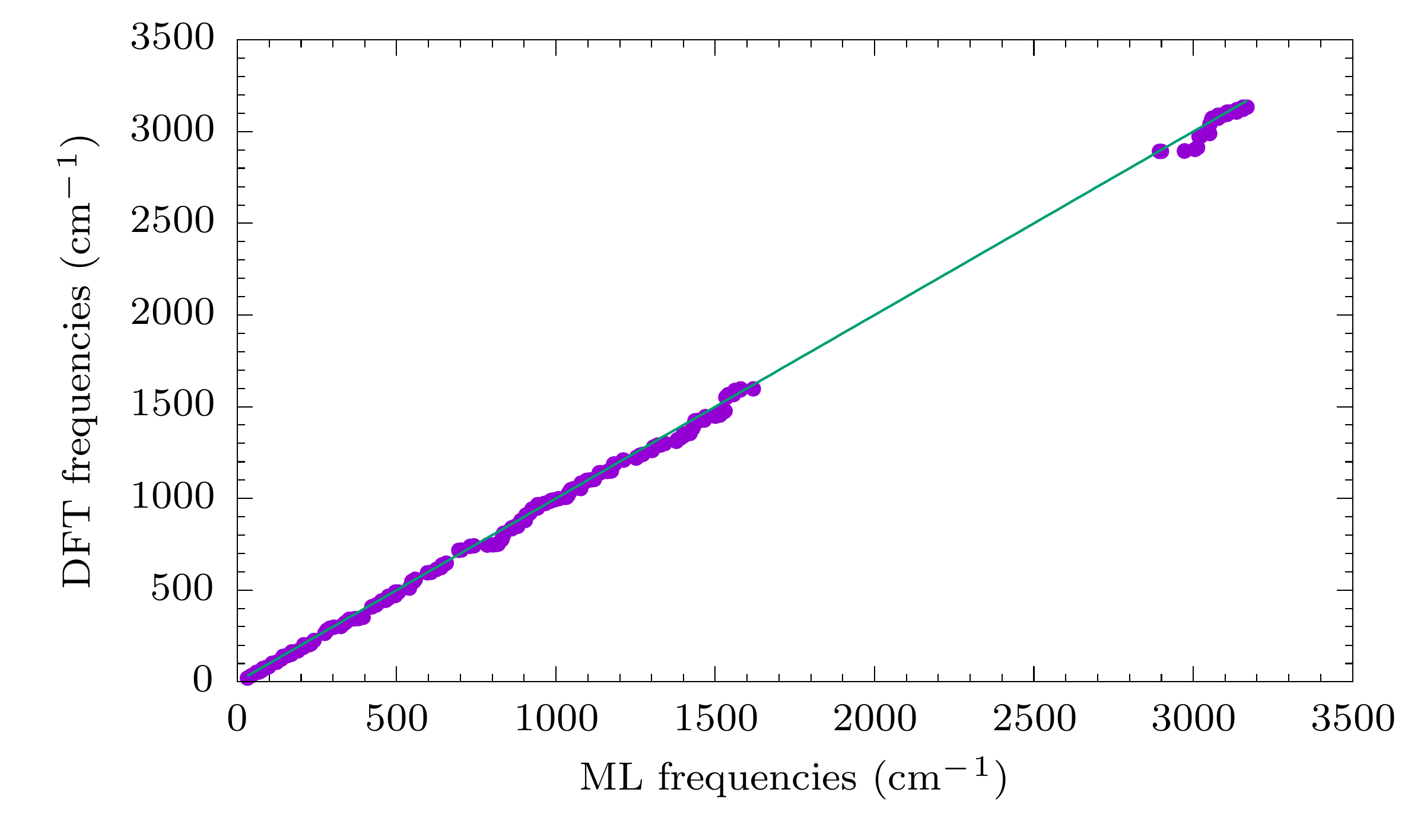}
    \includegraphics[scale=0.65]{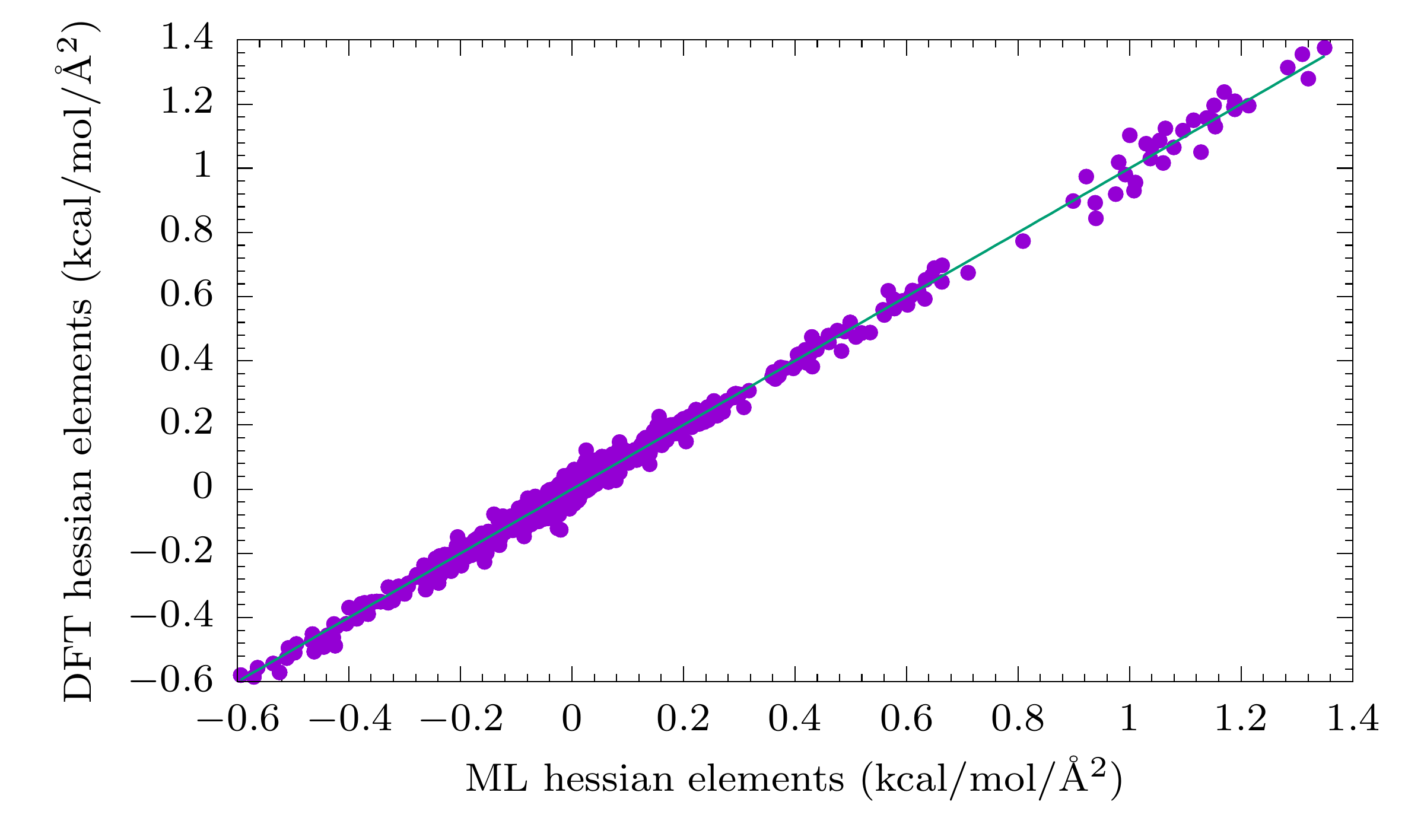}
    \caption{\textbf{Energy, forces, phonons and hessian.} $\delta =2.5$}
\end{figure}

\begin{figure}[H]
    \centering
    \includegraphics[scale=0.65]{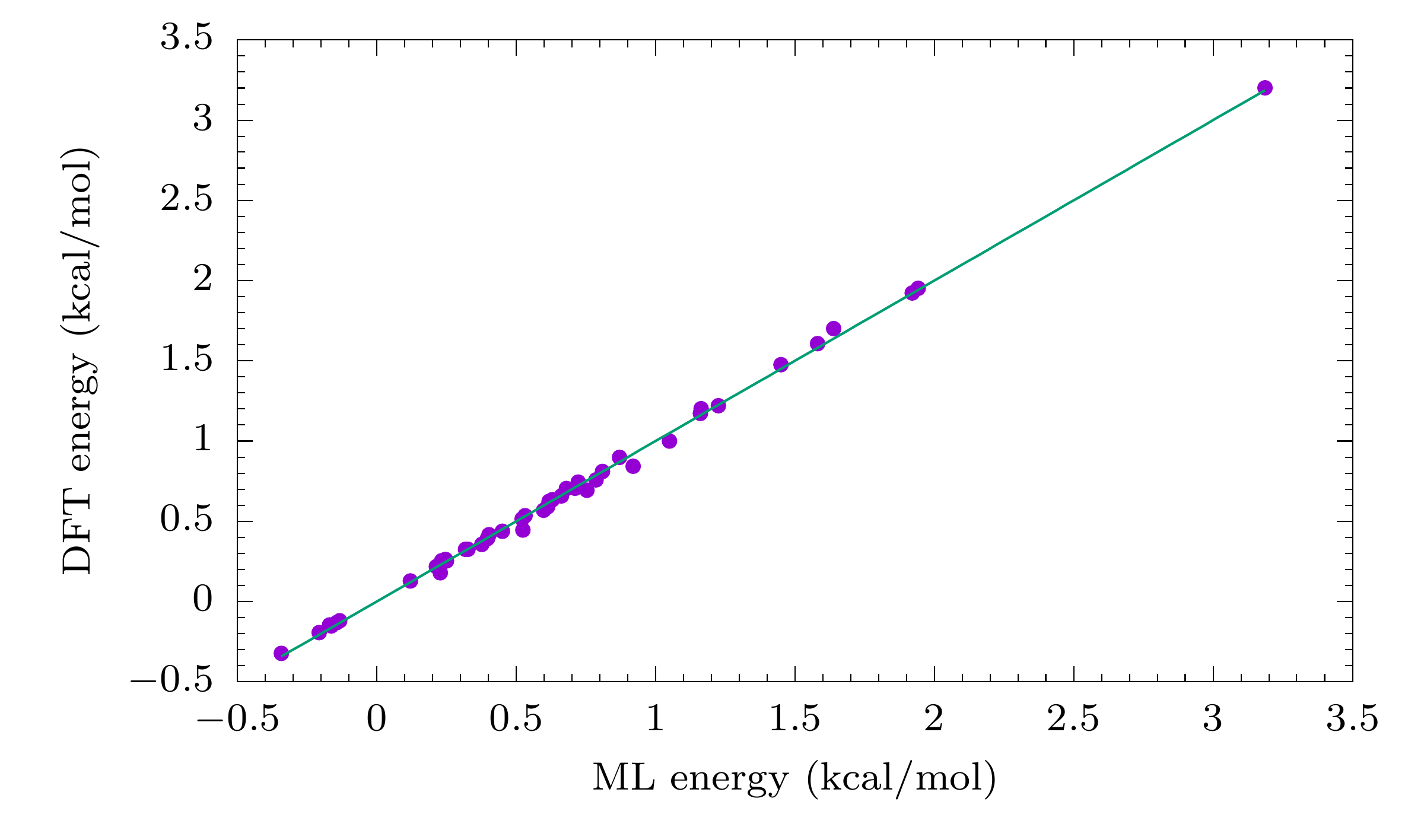}\\
    \includegraphics[scale=0.65]{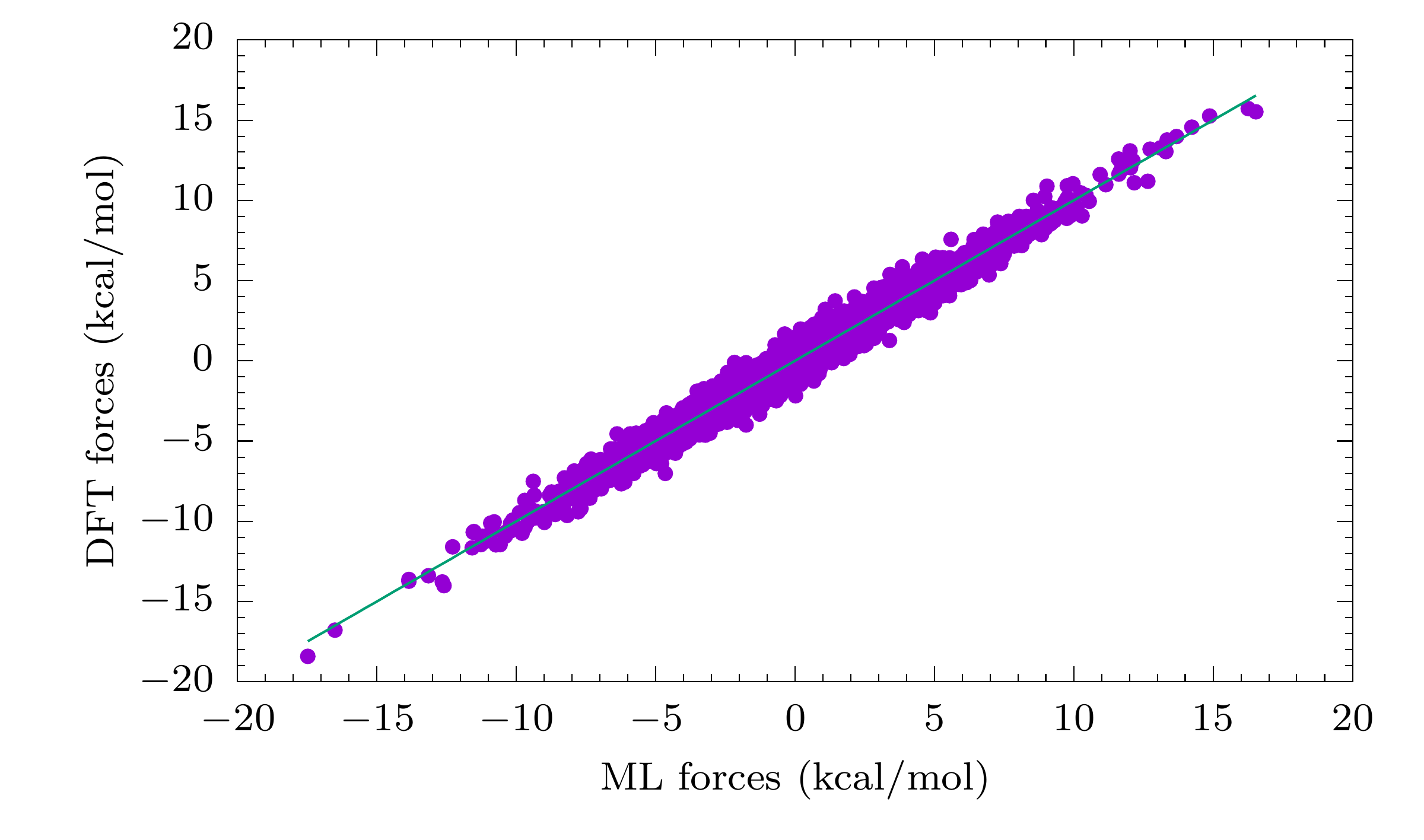}\\
    \includegraphics[scale=0.65]{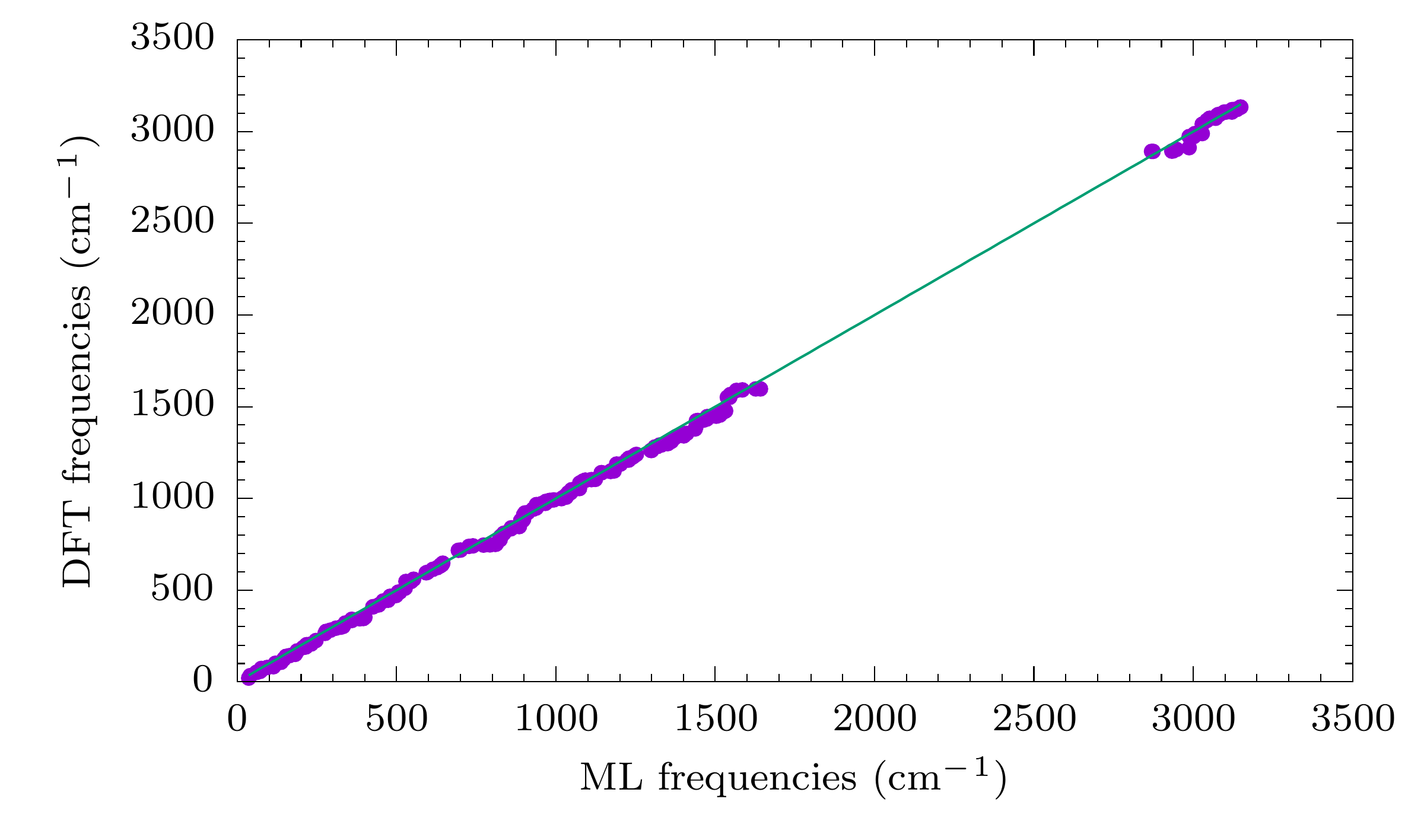}
    \includegraphics[scale=0.65]{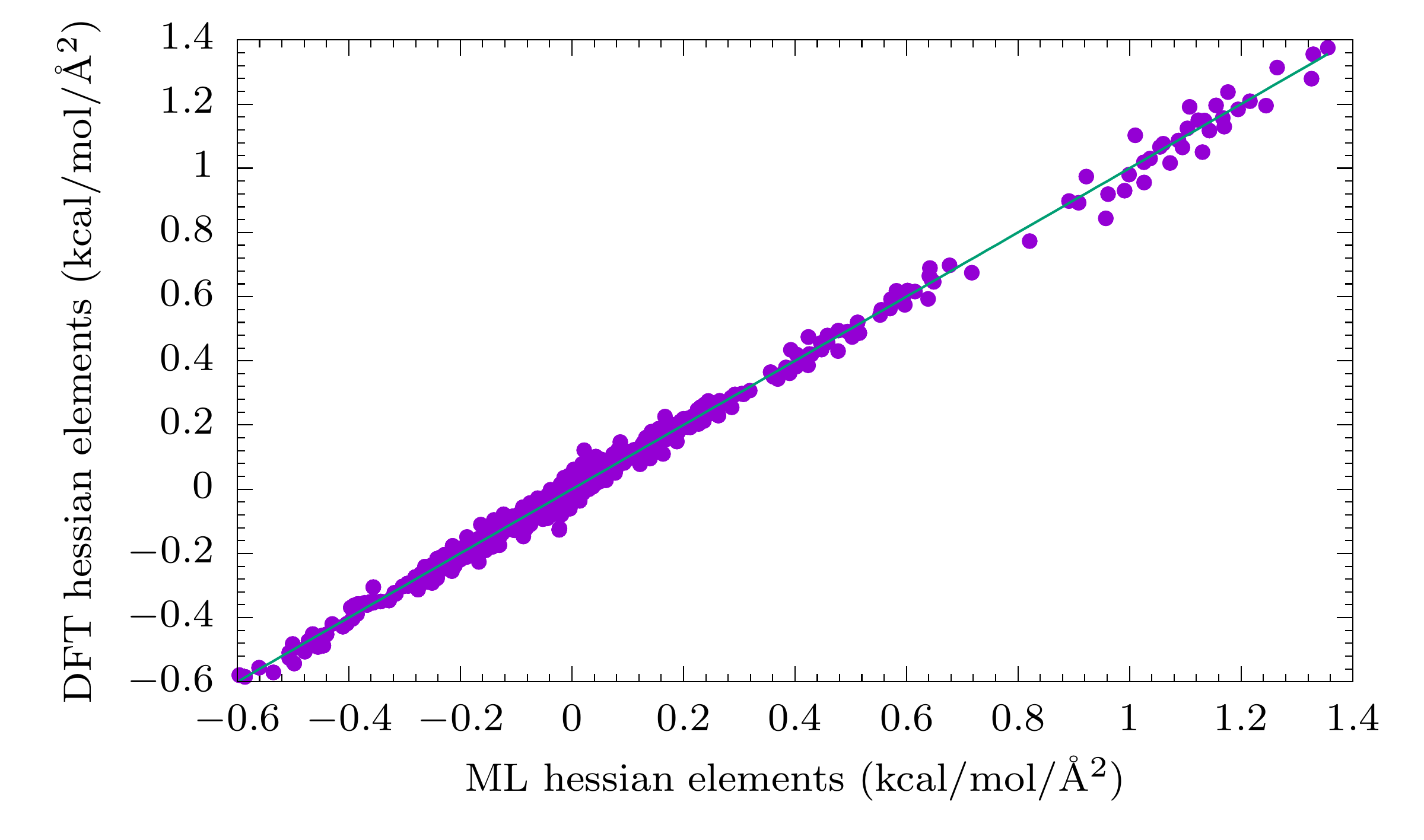}
    \caption{\textbf{Energy, forces, phonons and hessian.} $\delta =2.75$}
\end{figure}

\begin{figure}[H]
    \centering
    \includegraphics[scale=0.65]{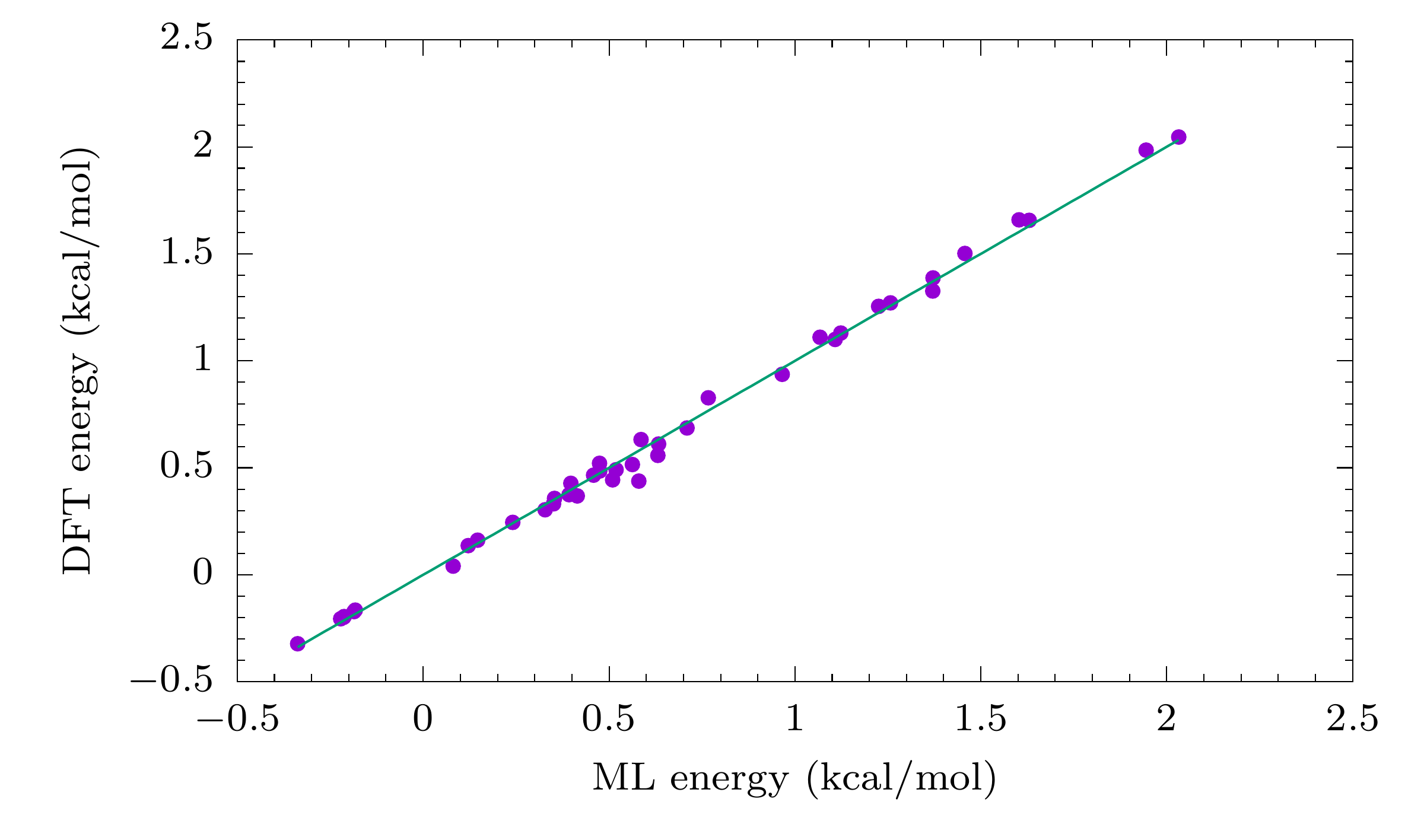}\\
    \includegraphics[scale=0.65]{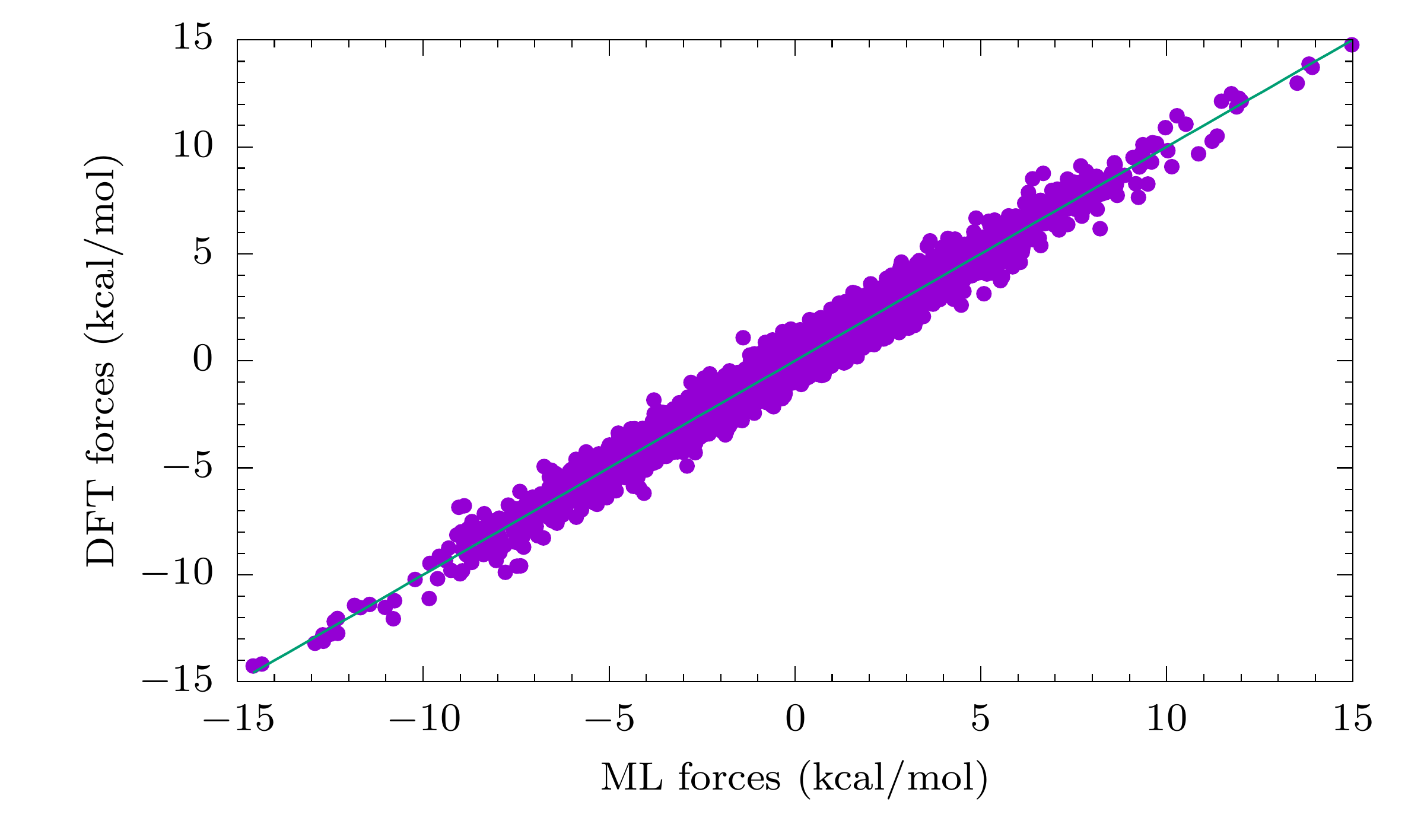}\\
    \includegraphics[scale=0.65]{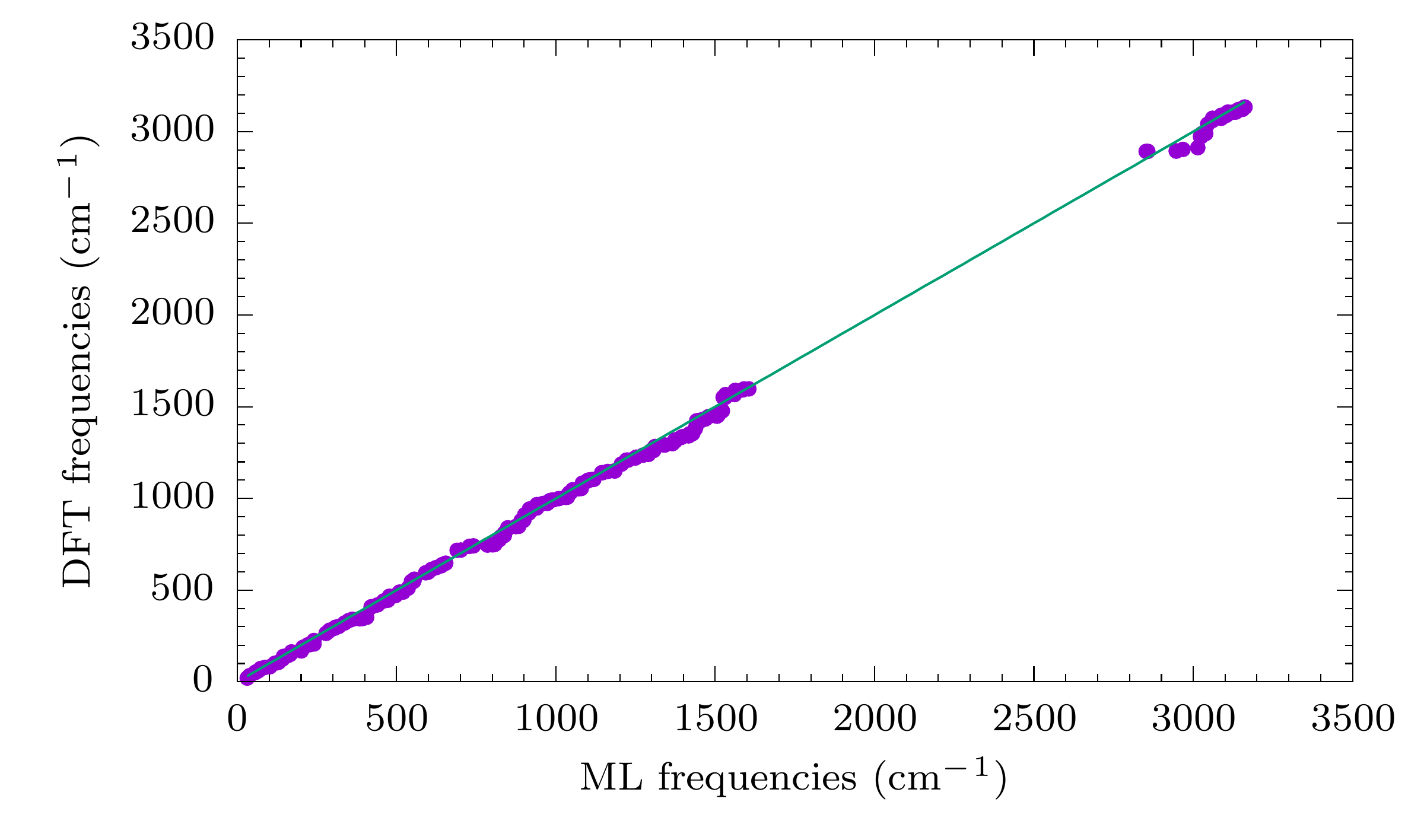}
    \includegraphics[scale=0.65]{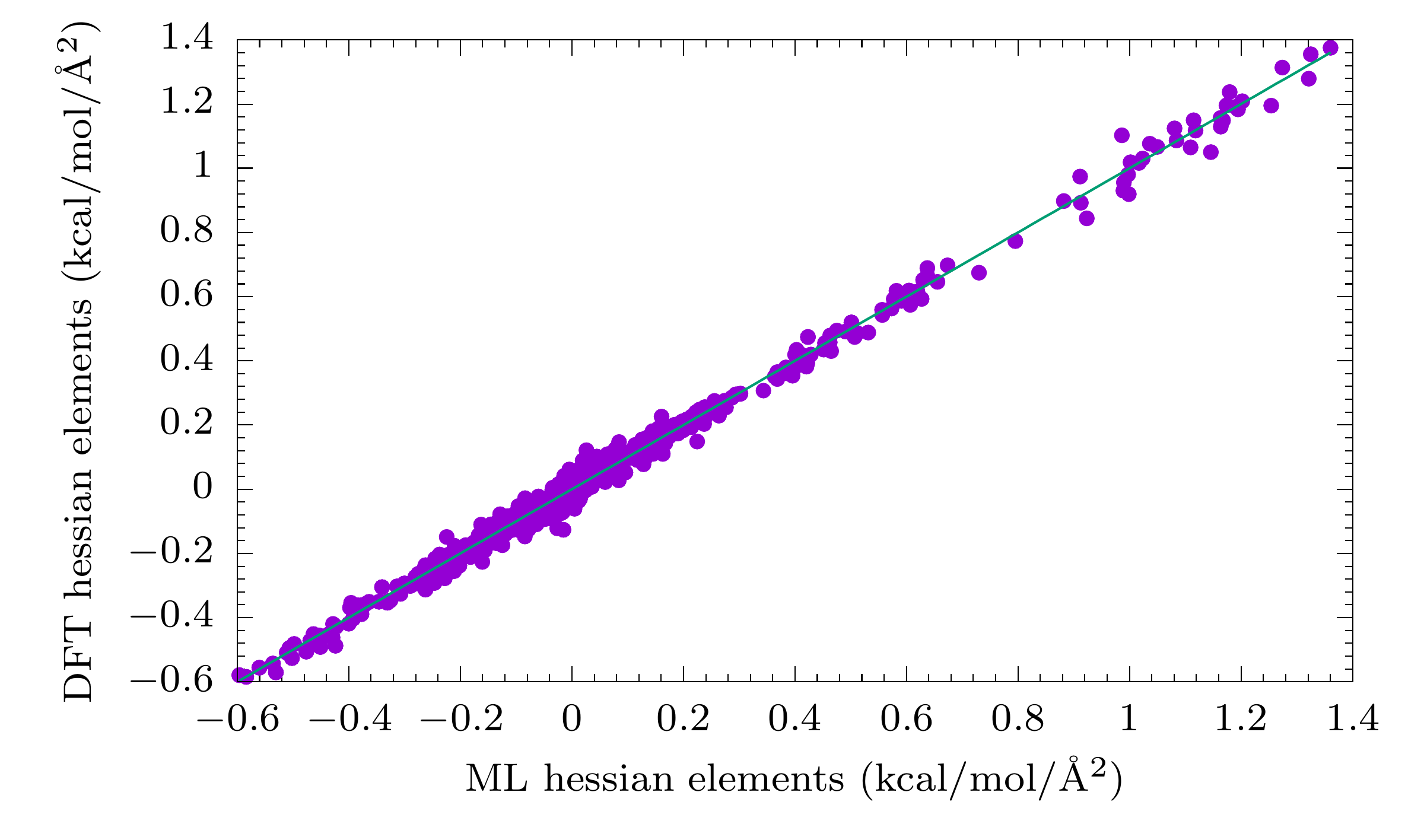}
    \caption{\textbf{Energy, forces, phonons and hessian.} $\delta =3$}
\end{figure}

\begin{figure}[H]
    \centering
    \includegraphics[scale=0.65]{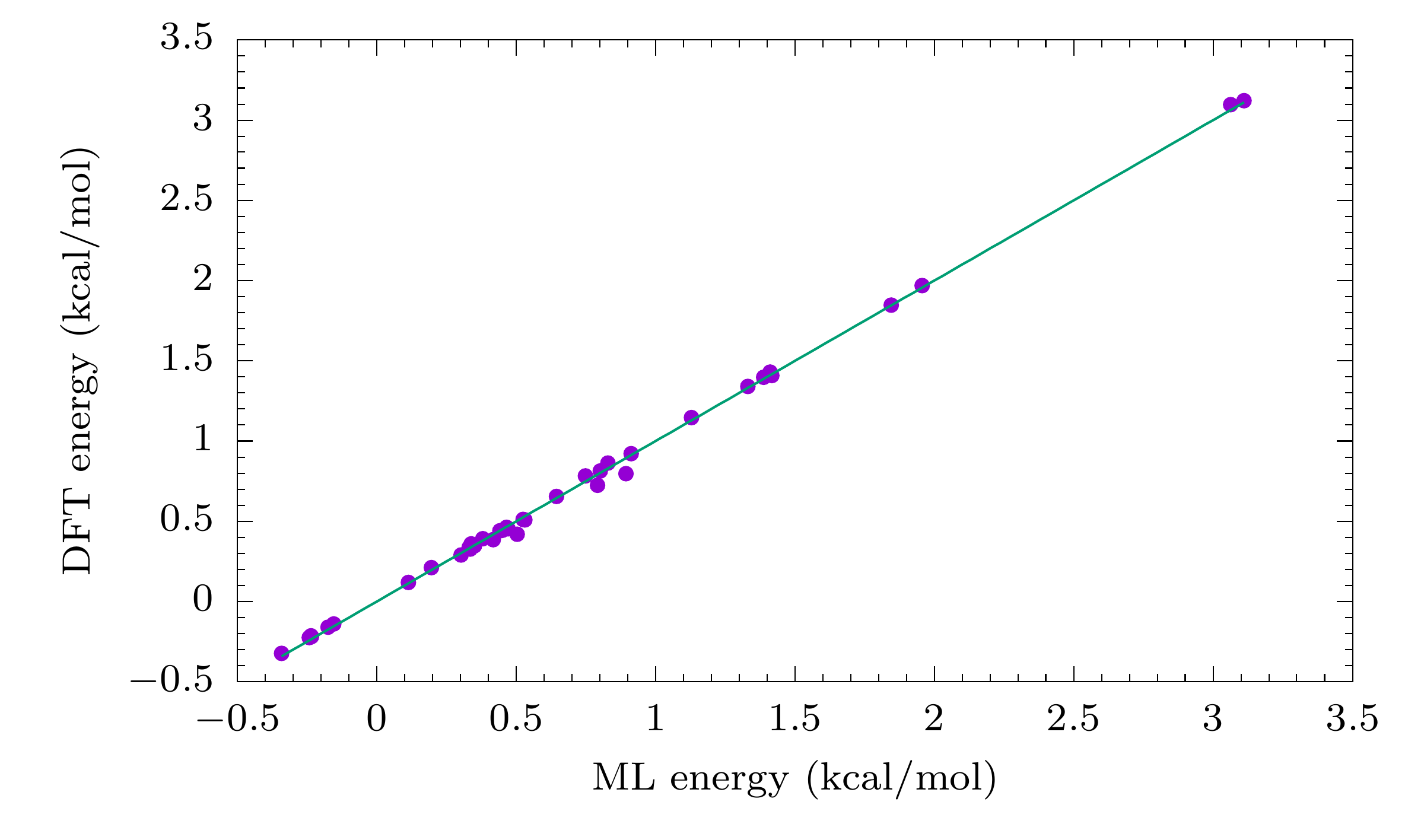}\\
    \includegraphics[scale=0.65]{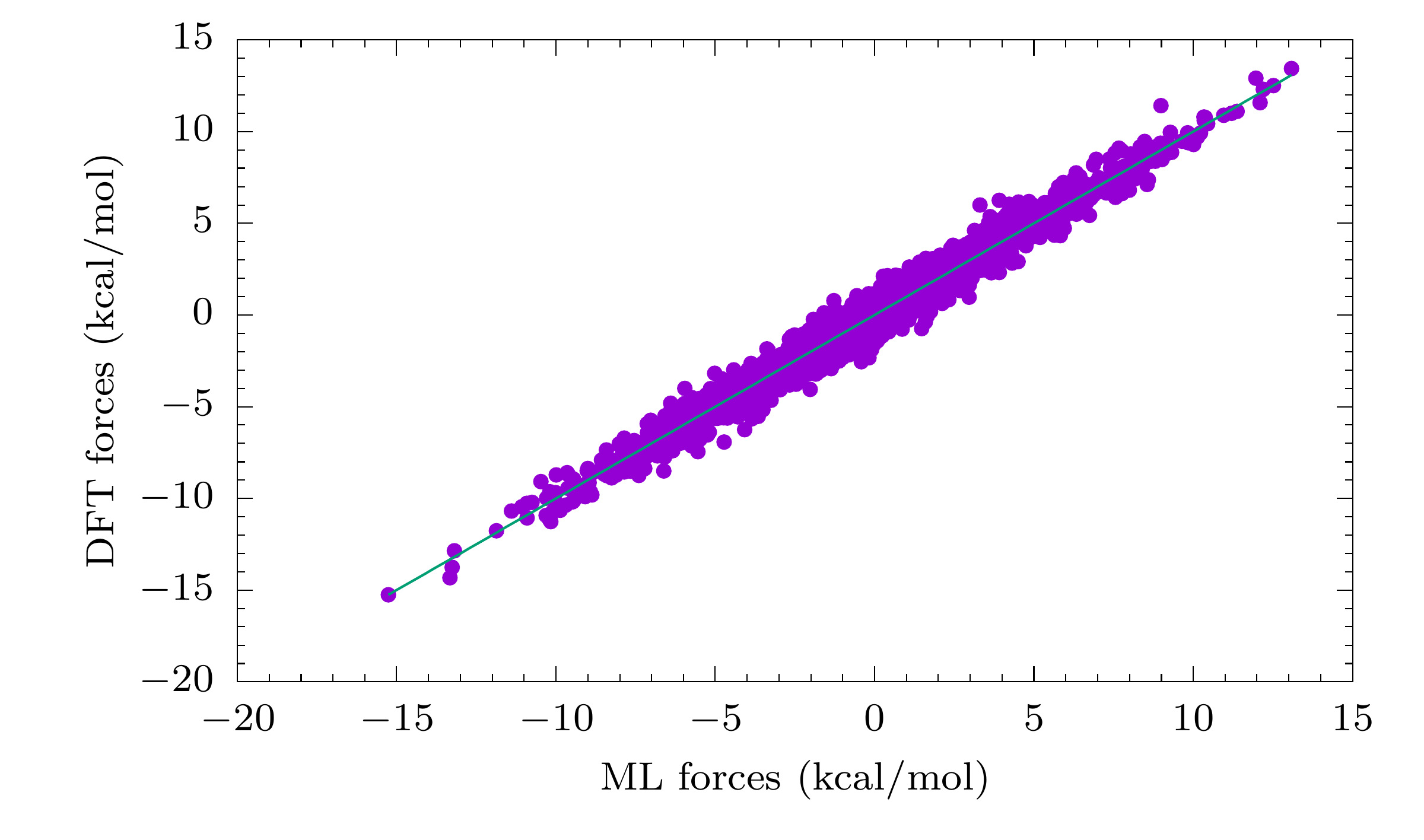}\\
    \includegraphics[scale=0.65]{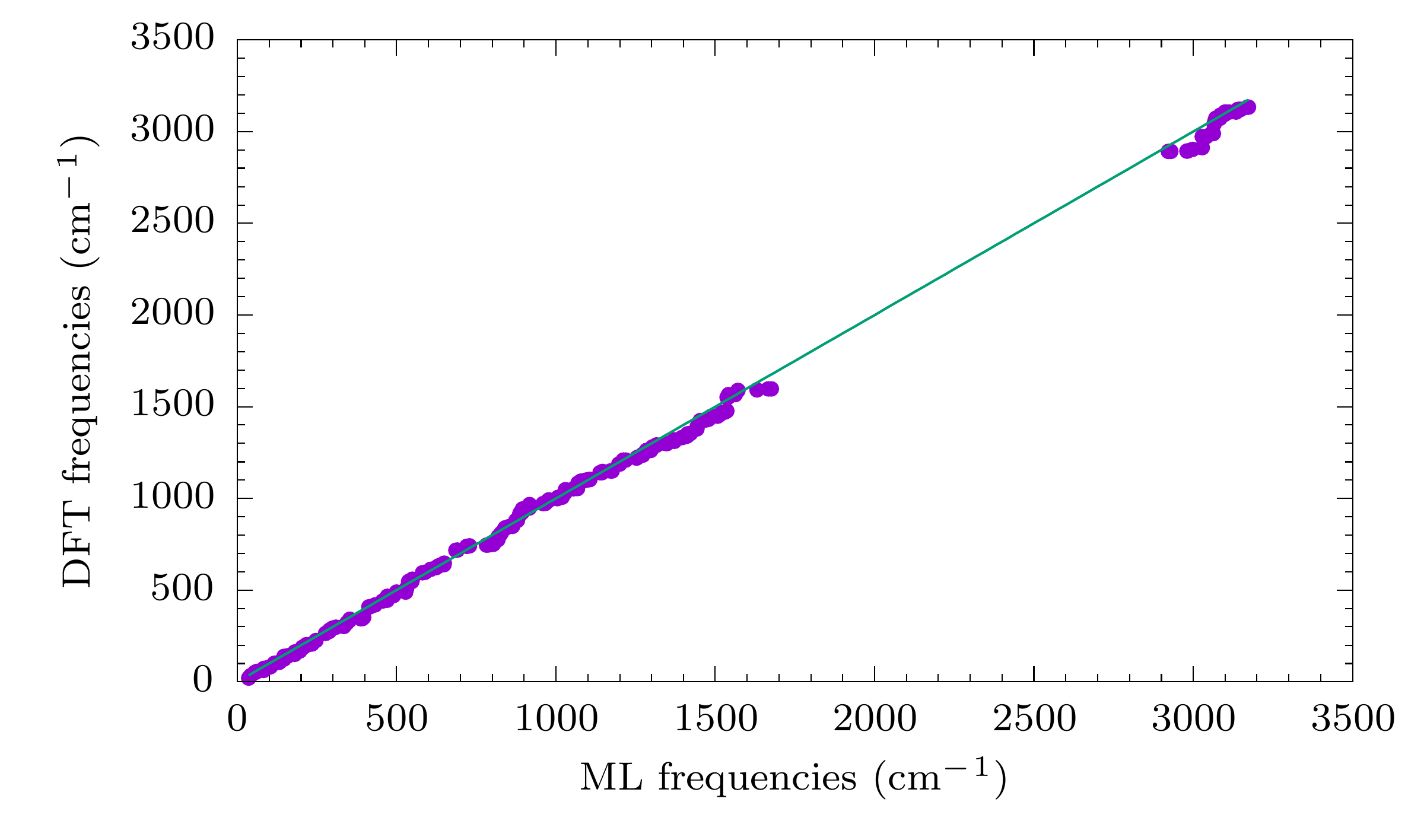}
    \includegraphics[scale=0.65]{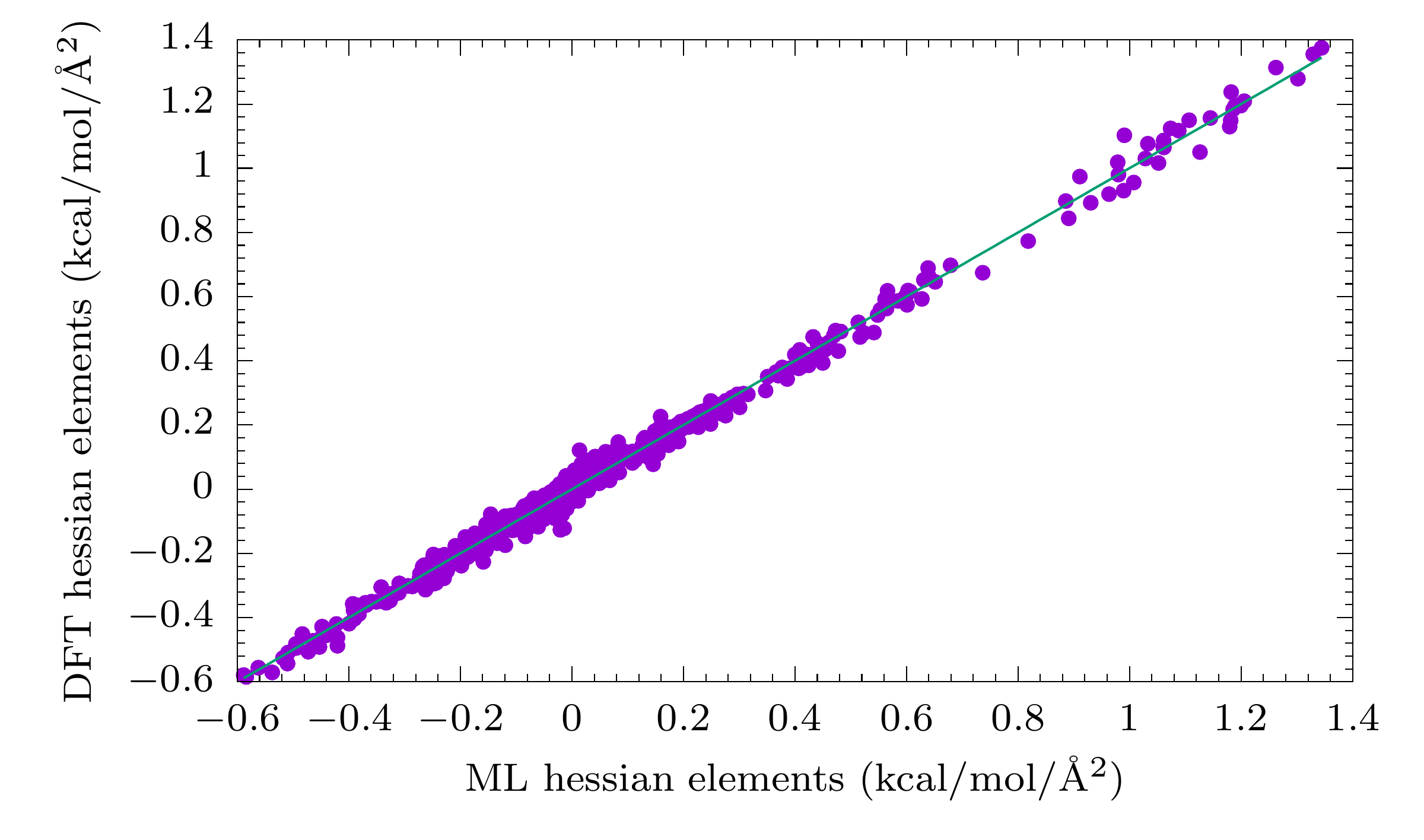}
    \caption{\textbf{Energy, forces, phonons and hessian.} $\delta =3.25$}
\end{figure}

\begin{table*} [t]
\caption{\textbf{Optimization results for the selected compounds} The RMSE for phonons and hessians are reported in cm$^{-1} $ and kcal/mol/\AA$^{2} $.
The training set size (TSS) selected by the active learning algorithm is also reported for different values of the threshold parameter $\delta$. The efficiency factor (E.F) is reported as a percentage.
%For comparison with the machine learning approach, in parentheses next to TSS for the smallest value of $\delta$, the number of \textit{ab initio} calculations required in a full \textit{ab initio} optimization.
}
\centering
\begin{tabular}{c c | c c c c}
\toprule
\textbf{Compound}  \hspace{0.1cm}  &  \hspace{0.1cm}  \textbf{$\delta$}  \hspace{0.1cm} &  \hspace{0.1cm}  \textbf{E.F.}  \hspace{0.1cm}  & \hspace{0.1cm} \textbf{TSS} \hspace{0.1cm} & \hspace{0.1cm} \textbf{RMSE Phonons } \hspace{0.1cm} & \hspace{0.1cm} \textbf{RMSE Hessian} \hspace{0.1cm}\\\hline

\midrule
\multirow{4}*{\textbf{1}}& 5 & 93\% & 9 & 30.8 & 0.013 \\
& 10 & 94\% & 8  & 57.07 & 0.032 \\
& 20 &97\% & 4 & 29.3 & 0.014 \\
& 30 & 98\%&3  & 42.17 & 0.019 \\ \hline
\midrule
\multirow{4}*{\textbf{2}}& 5 & 96\% & 14 & 23.54 & 0.007  \\
& 10 &97\% & 10   & 25.61 & 0.008 \\
& 20 &97\%  & 10  & 38 & 0.009 \\
& 30 &97\% & 9   & 38.3 & 0.009 \\ \hline
\midrule
\multirow{4}*{\textbf{3}}& 5 & 93\% & 26  & 52.12 & 0.011 \\
& 10 &95\%& 21  & 59.6 & 0.011\\
& 20 &95\% & 19   & 52.6 &0.012 \\
& 30 &96\% & 15   & 51.2 & 0.011 \\ \hline
%\multirow{4}*{\textbf{4}}& 5  & 21 (37) & 105.7 & 0.012  \\
%& 10  & 14  & 128.7 & 0.013 \\
%& 20  & 9  & 162.2 & 0.02  \\
%& 30  & 10  & 179.4 & 0.02 \\ \hline
\bottomrule\\
\label{SI:optimization_phonons_3_compounds}
\end{tabular}
\end{table*}

\begin{table*}[t]
\caption{\textbf{Phonon results after molecular dynamics at 50 K} The RMSE for phonons and hessians are reported in cm$^{-1} $ and kcal/mol/\AA$^{2} $. The efficiency factor (E.F) is reported as a percentage.
The training set size (TSS) selected by the active learning algorithm is also reported for different values of the threshold parameter $\delta$.
}
\centering
\begin{tabular}{c c |c c c c }
\toprule
\textbf{Compound}  \hspace{0.1cm}  &  \hspace{0.1cm}  \textbf{$\delta$}  \hspace{0.1cm}  & \hspace{0.1cm} \textbf{E.F.} \hspace{0.1cm} &  \hspace{0.1cm}  \textbf{TSS}  \hspace{0.1cm} & \hspace{0.1cm} \textbf{RMSE Phonons } \hspace{0.1cm} & \hspace{0.1cm} \textbf{RMSE Hessian} \hspace{0.1cm}\\\hline

\midrule
\multirow{4}*{\textbf{1}}& 2.5 & 80\% & 22 & 2.47 & 0.004 \\
& 2.75  & 82\%& 20 & 3.63 & 0.004 \\
& 3 & 84\% & 17 & 3.89 & 0.004 \\
& 3.25 & 85\%& 16 & 3.55 & 0.004 \\ \hline
\midrule
\multirow{4}*{\textbf{2}}& 2.5  & 81\%& 54& 14.8 & 0.005  \\
& 2.75  & 84\%& 44 & 21.8 & 0.005 \\
& 3  & 85\%& 43 & 21.15 & 0.005 \\
& 3.25  & 87\%& 36 & 19.5 & 0.005 \\ \hline
\midrule
\multirow{4}*{\textbf{3}}& 2.5  & 83\%& 53 & 23.29 &0.006 \\
& 2.75 & 85\%& 44 & 21.9 & 0.006\\
& 3  & 85\%& 41 & 24.1 &0.007 \\
& 3.25  & 87\%& 38 & 28.5 & 0.007\\ \hline
%\multirow{4}*{\textbf{4}}& 2.5  & 32 & 58.54 &  0.012 \\
%& 2.75  & 29  & 52.71 & 0.010 \\
%& 3  & 27  & 61.22 & 0.010  \\
%& 3.25  & 22  & 54.26 & 0.010 \\ \hline
\bottomrule\\
\label{SI:Phonons_after_MD_3_compounds}
\end{tabular}
\end{table*}

\clearpage

\section{Machine learning of spin-phonon coupling}

\begin{figure}[H]
    \centering
    \includegraphics[scale=0.7]{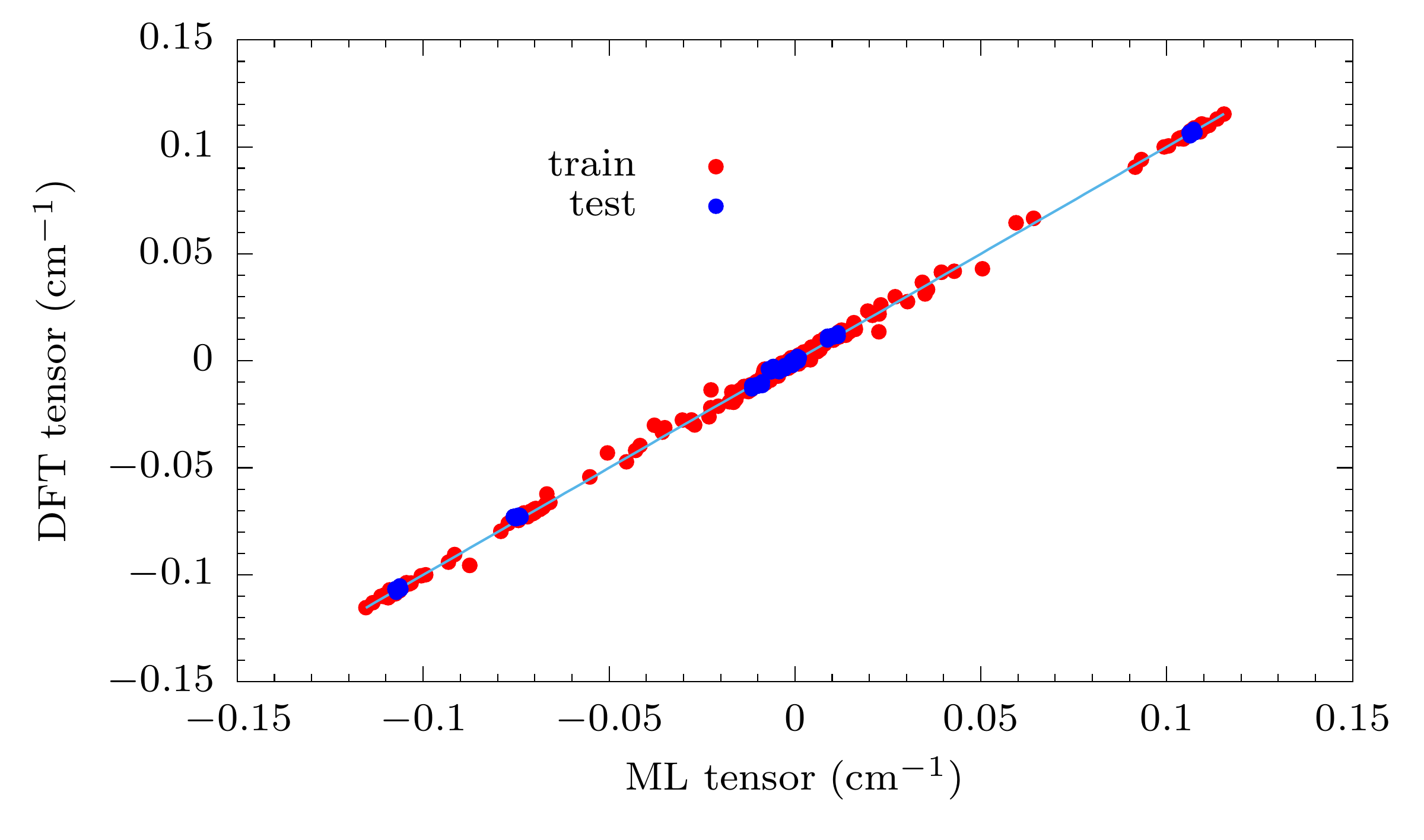}\\
    \includegraphics[scale=0.7]{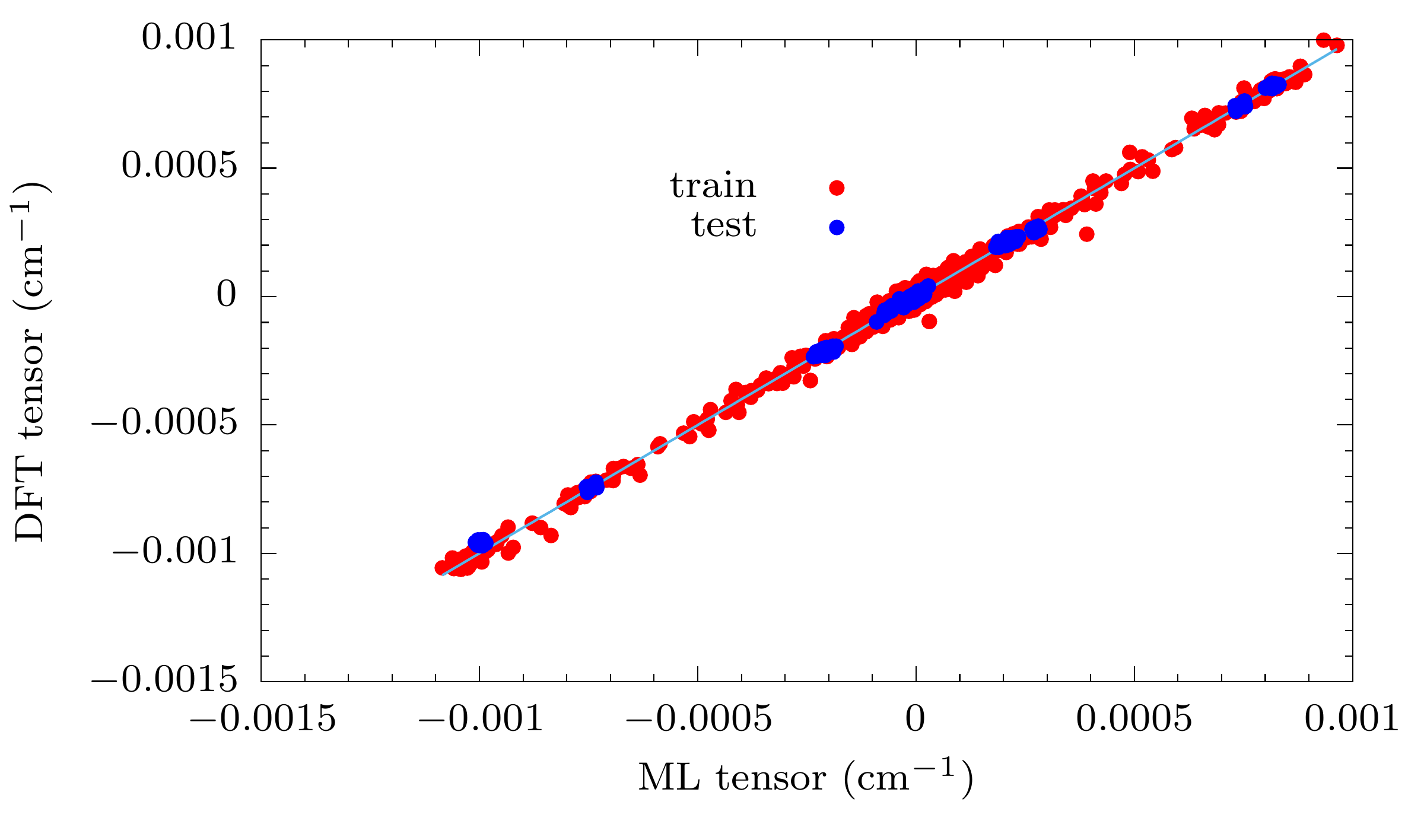}
    \caption{\textbf{Fit tensors.} Training and test error for the fit of the spherical tensor components in which the tesseral functions  are decomposed ($l=4,6$ in order) }
    \label{SI:fit_tensor_spin_relaxation}
\end{figure}

\begin{table} [t]
\caption{\textbf{RMSE on spin Hamiltonian tensor fit} The unit of measure is cm$^{-1}$. The results for \textbf{3} are reported for the different order of the tensor $l=2,4,6$, in order.  }
\centering
\begin{tabular}{c |c c c}
\toprule
\textbf{Comp}  \hspace{0.1cm}   & \hspace{0.1cm} \textbf{RMSE Train } \hspace{0.1cm} & \hspace{0.1cm} \textbf{RMSE Test} \hspace{0.1cm} \\\hline
\midrule
 \textbf{1} & 0.36 & 0.19 \\\hline
 \textbf{2} & 0.18  &  0.10  \\\hline
\multirow{3}*{\textbf{3}} & 3.5 $\cdot 10^{-2}$  &  2.0 $\cdot 10^{-2}$  \\
& 1.2 $\cdot 10^{-3}$ & 9.5 $\cdot 10^{-4}$ \\
& 2.0 $\cdot 10^{-5}$ & 1.0 $\cdot 10^{-5}$ \\ \hline
\midrule
\bottomrule 
\end{tabular}
\label{SI:tensor_training_test_prediction}
\end{table}

\begin{figure}
    \centering
    \includegraphics[scale=0.7]{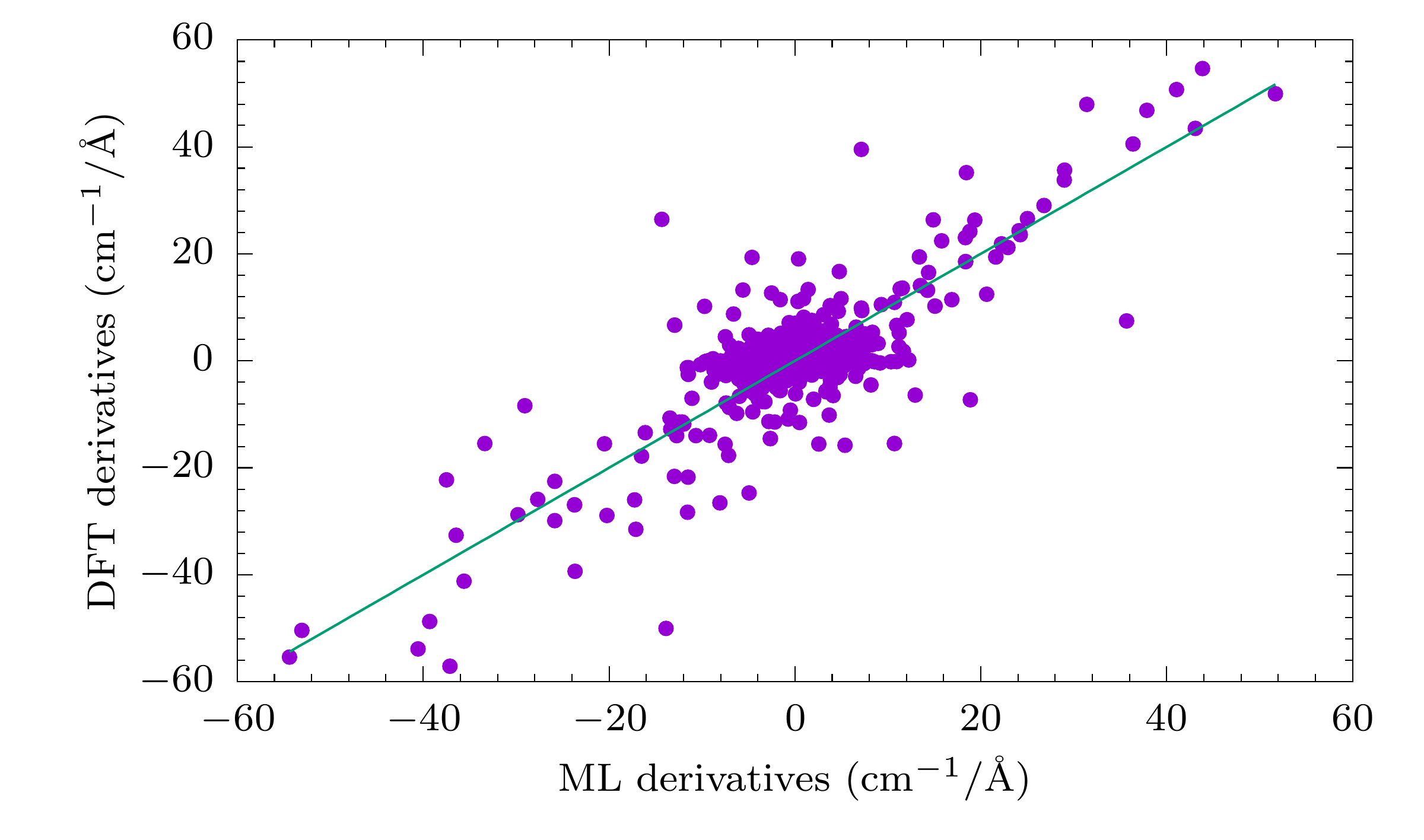}\\
    \includegraphics[scale=0.7]{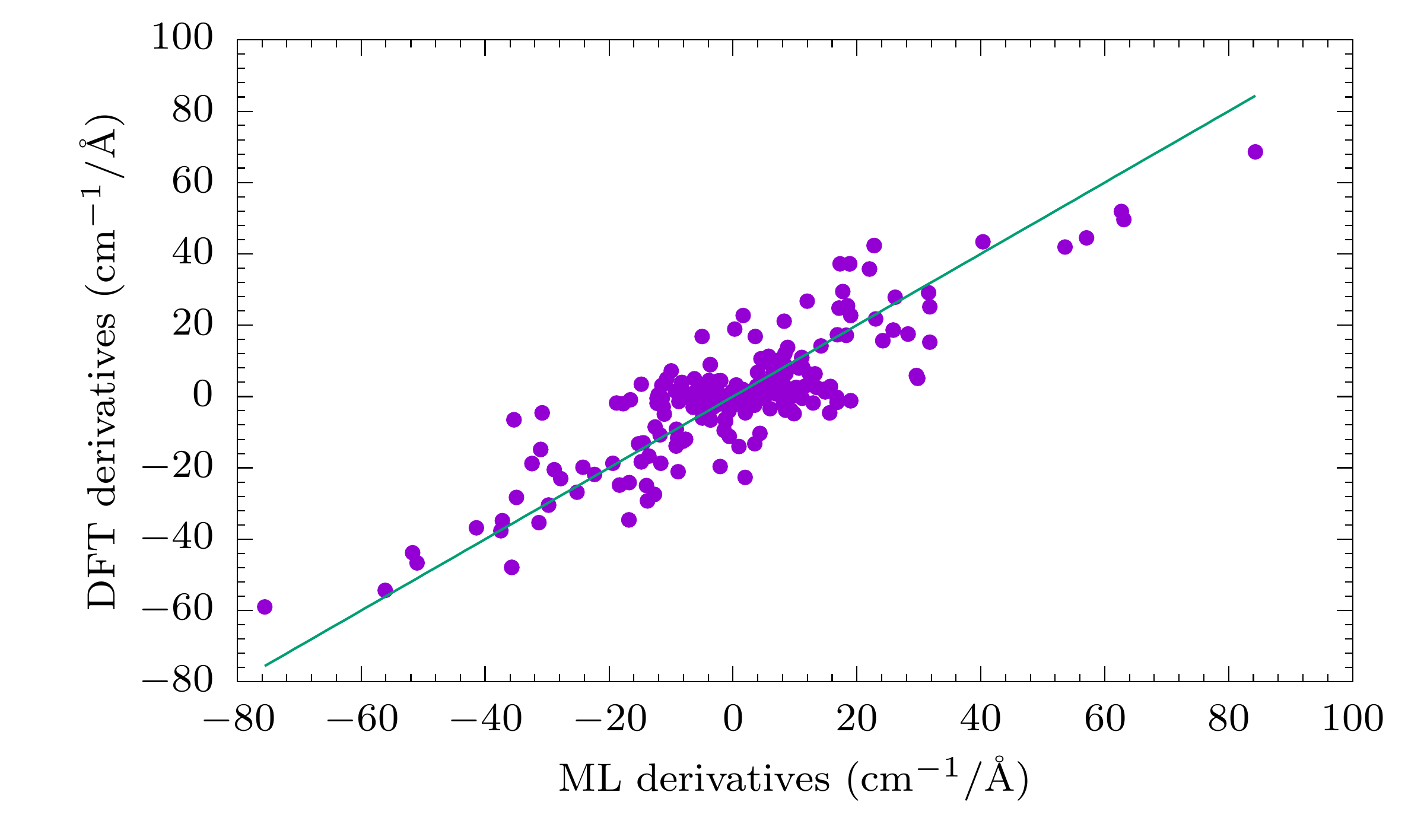}\\
    \includegraphics[scale=0.7]{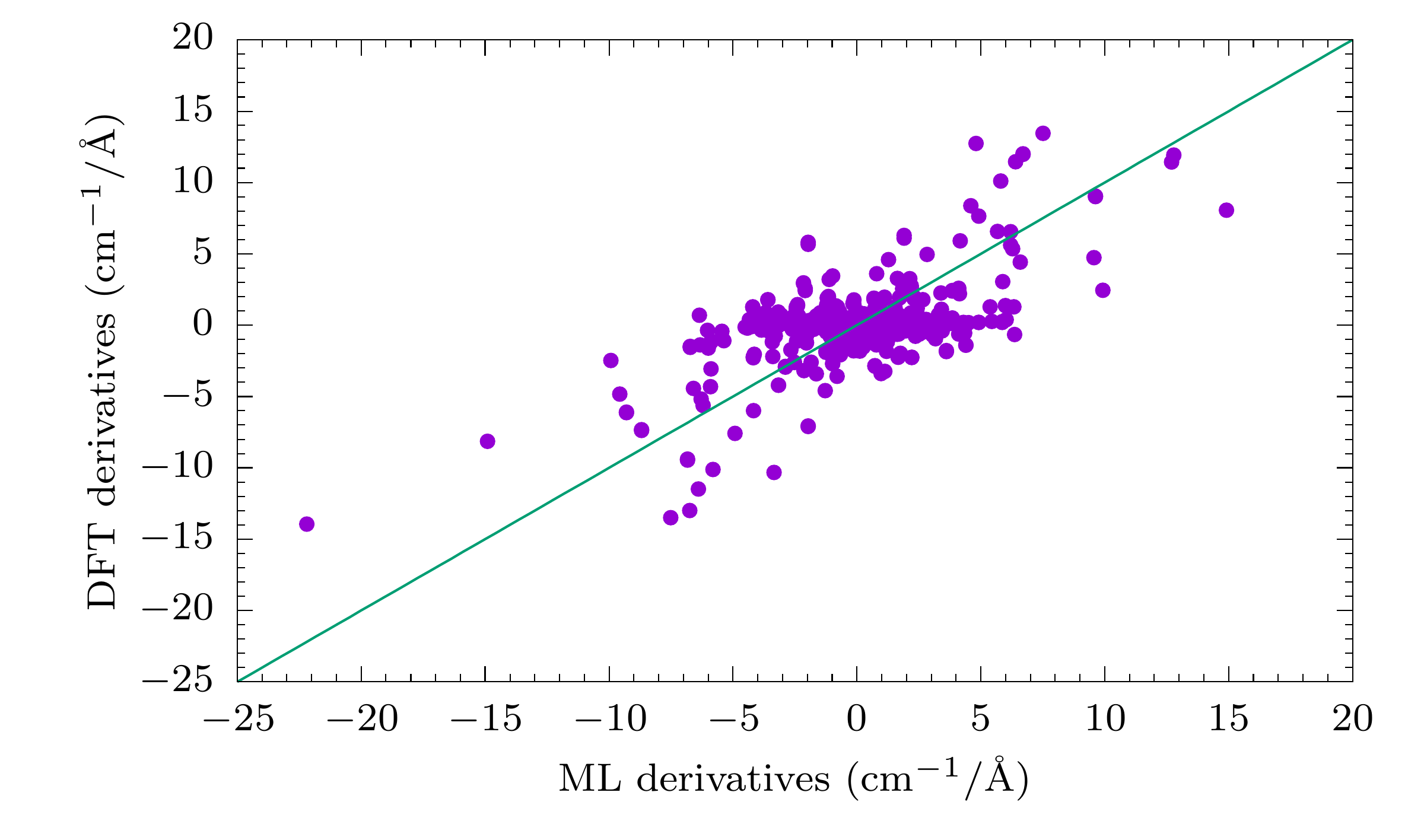}\\
    \includegraphics[scale=0.7]{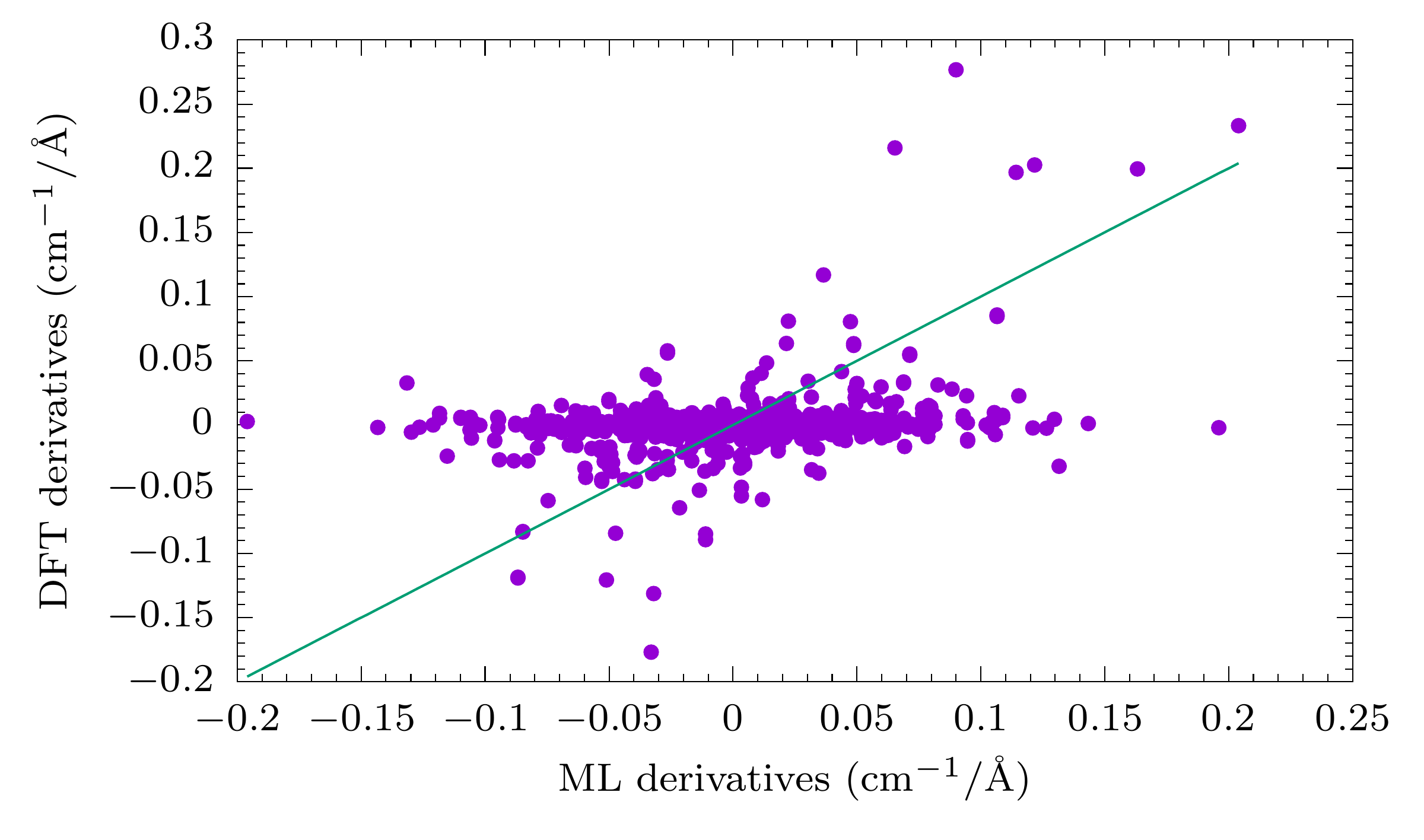}\\
    \includegraphics[scale=0.7]{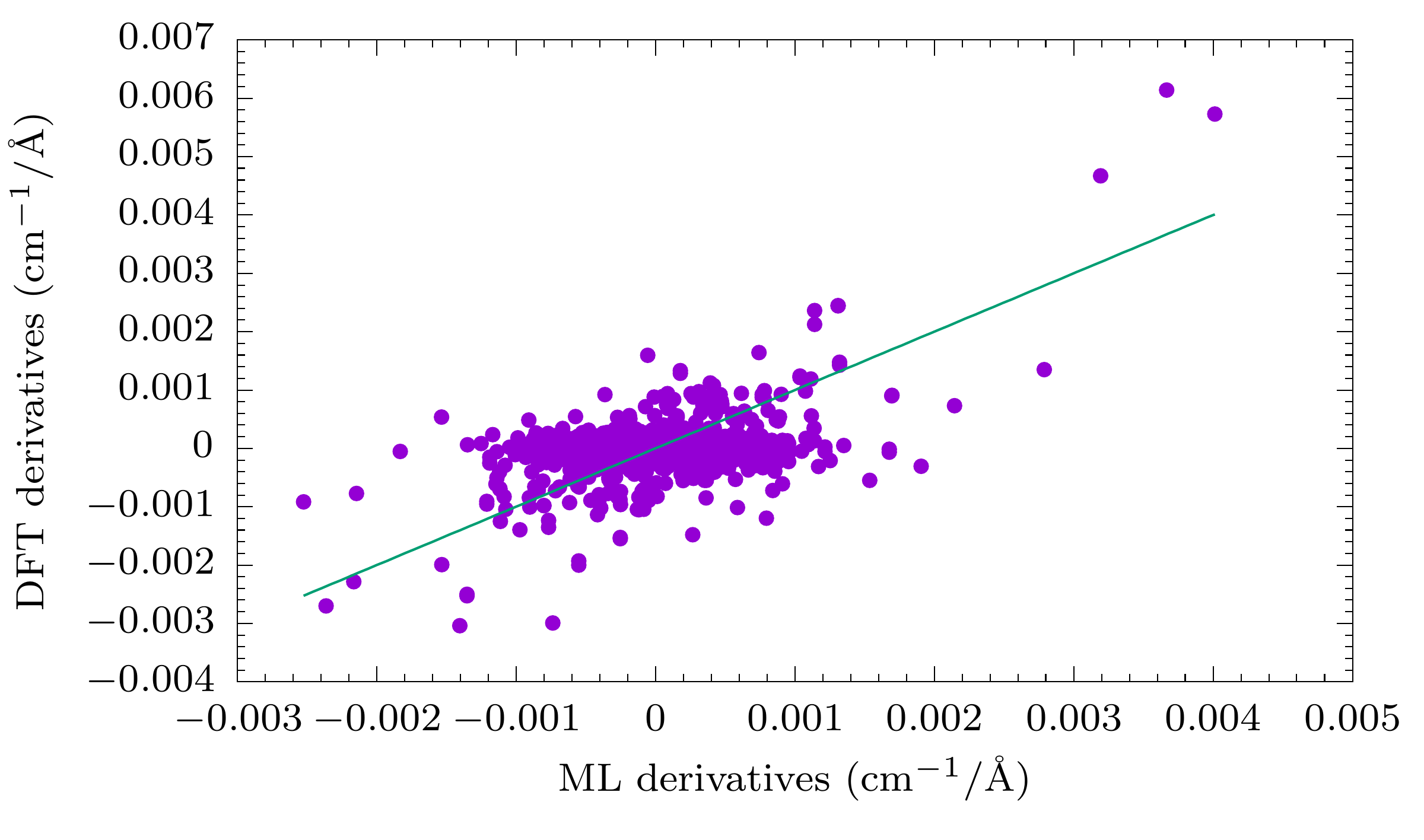}
    \caption{\textbf{Derivatives of the spin Hamiltonian $\partial \hat{H}_{\mathrm{0}} / \partial X_i$ with ML tensorial predictions }}
    \label{SI:cartesian_gradients_ML_predictions}
\end{figure}

\begin{figure}
    \centering
    \includegraphics[scale=0.7]{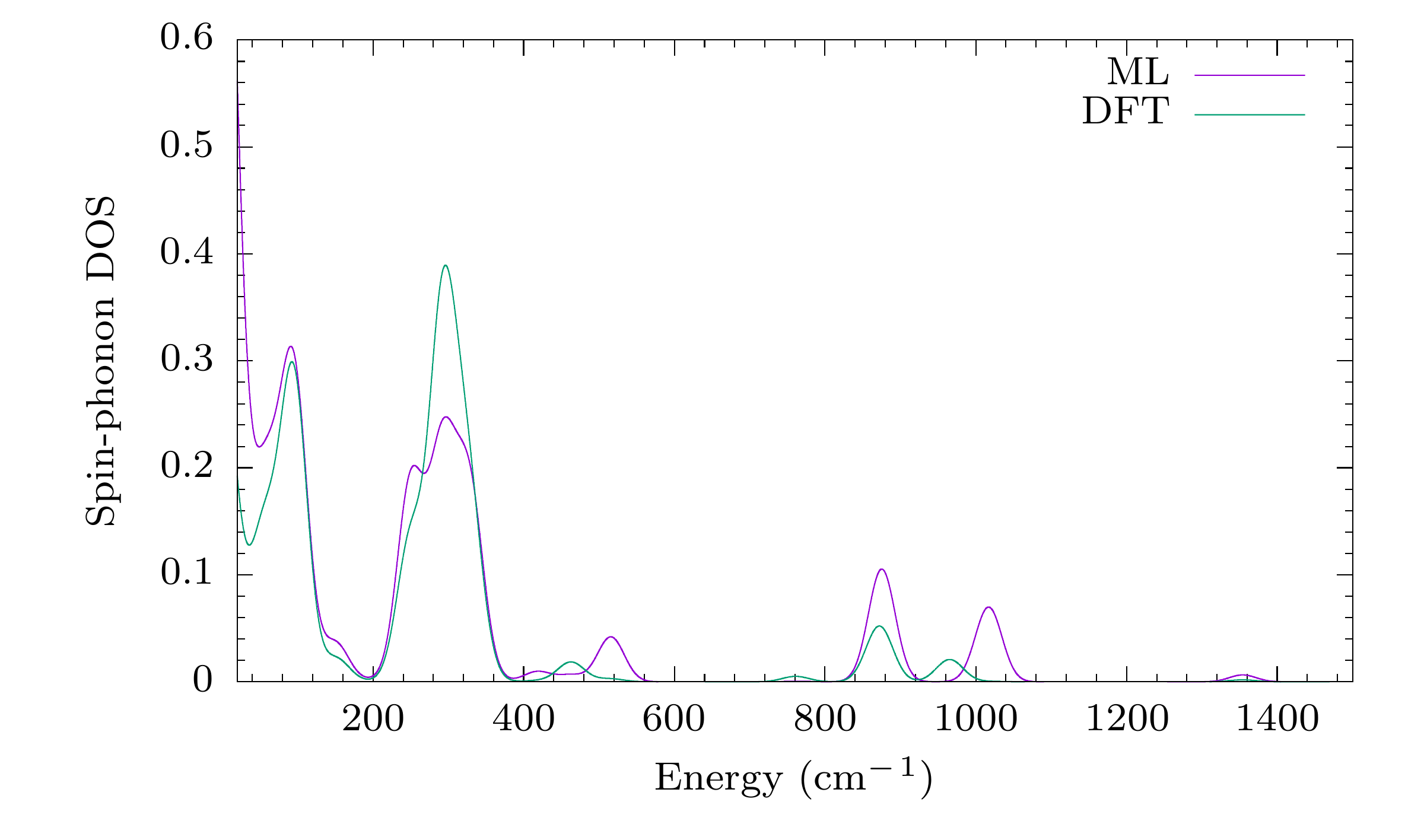}\\
    \includegraphics[scale=0.7]{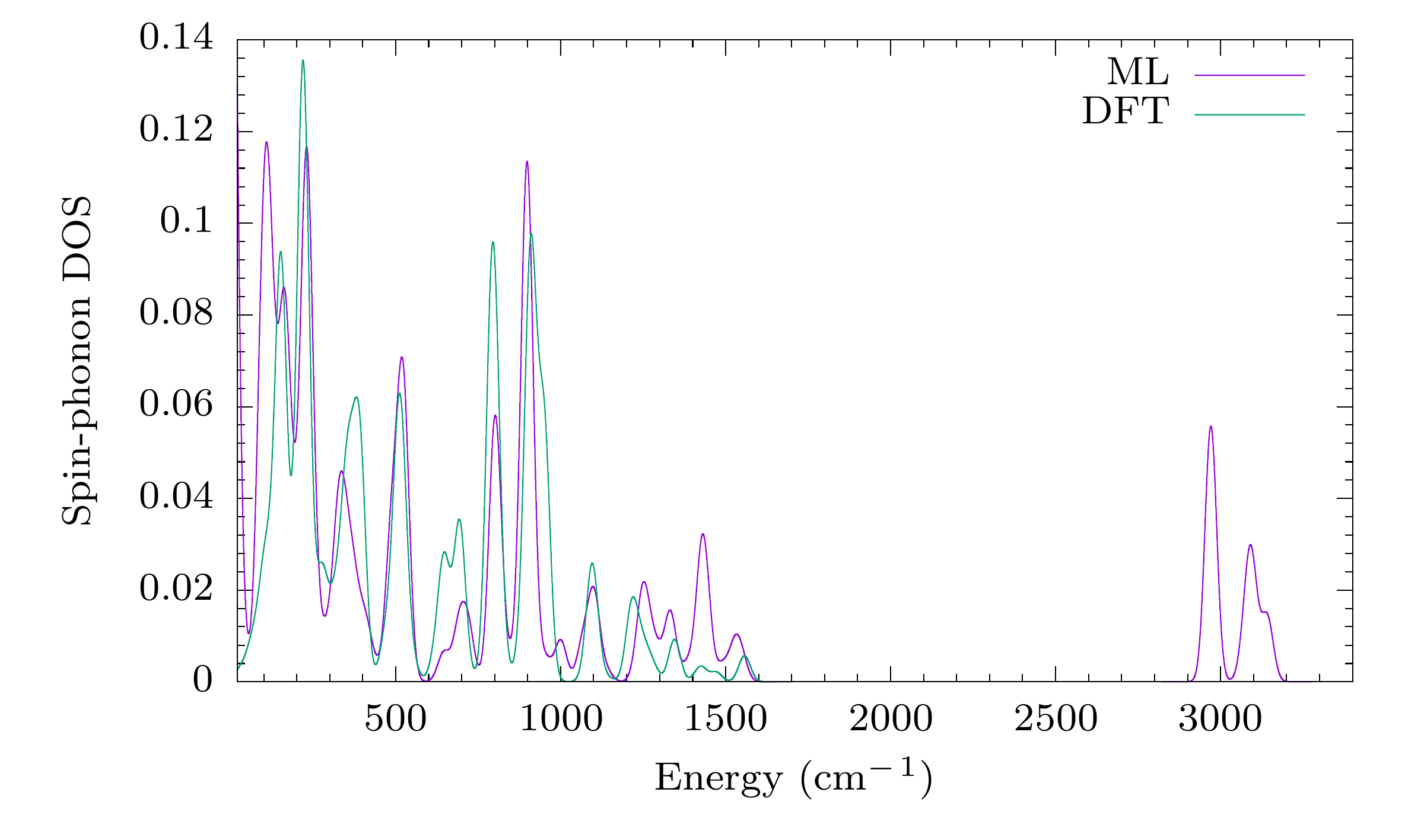}\\
    \includegraphics[scale=0.7]{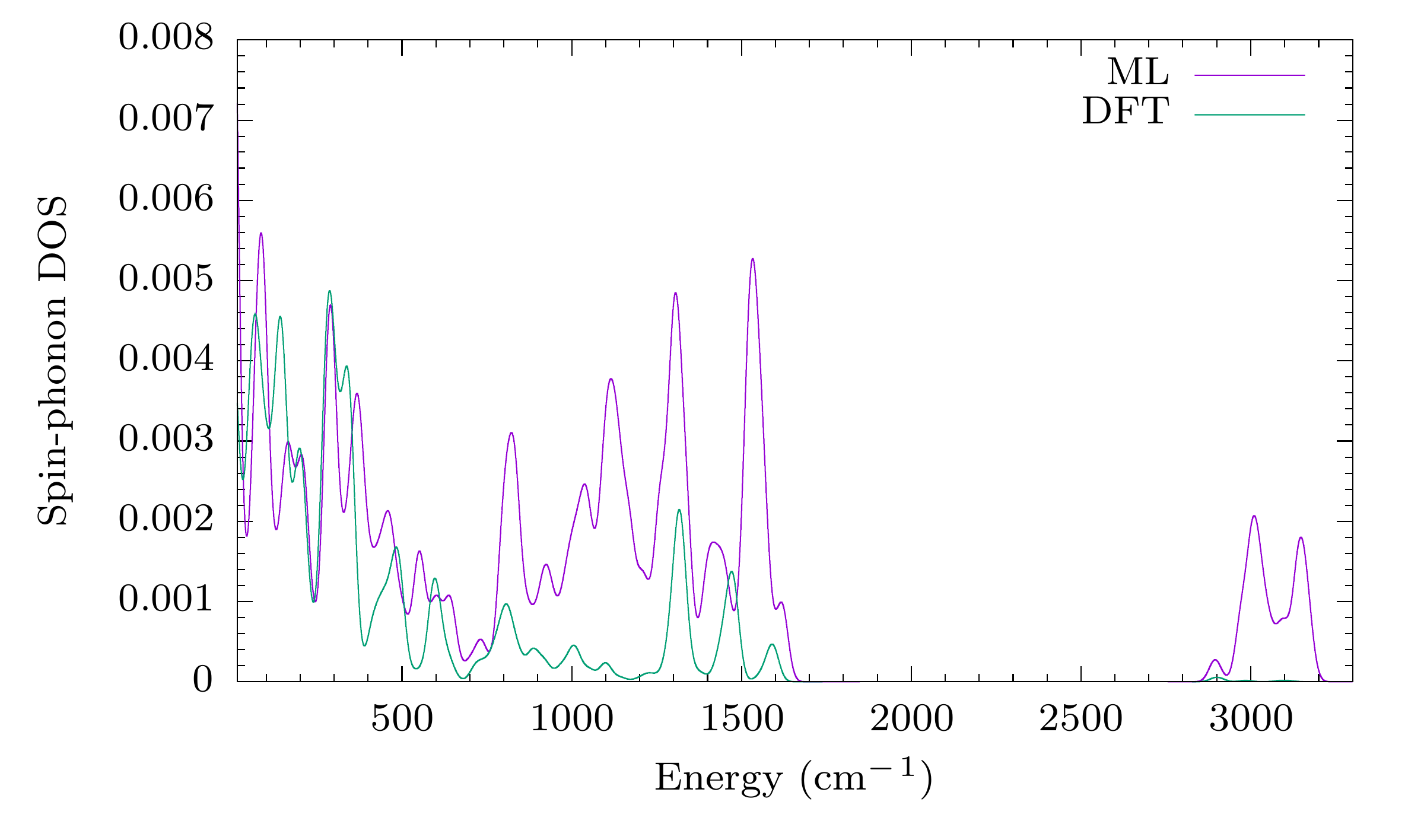}\\
    \includegraphics[scale=0.7]{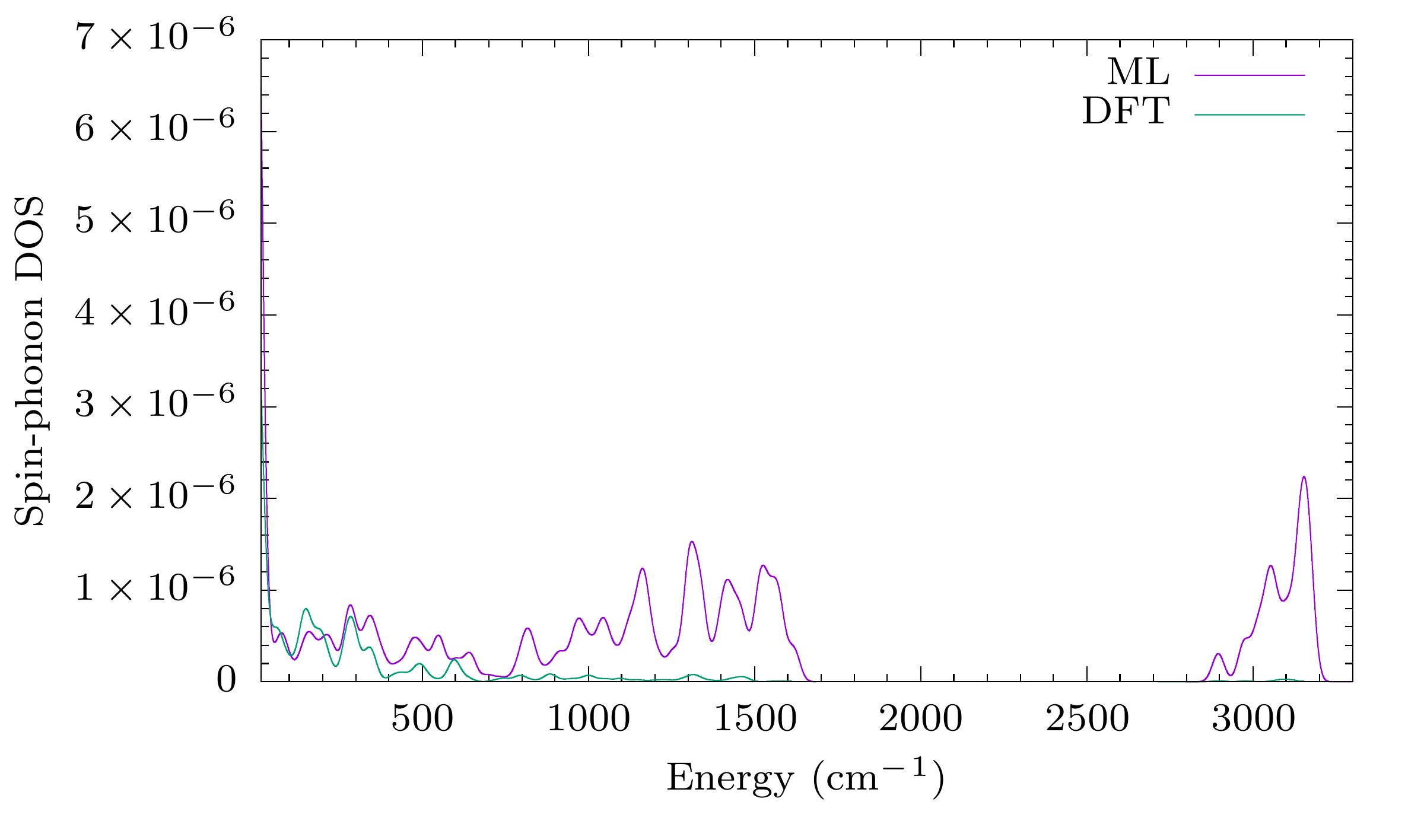}\\
    \includegraphics[scale=0.7]{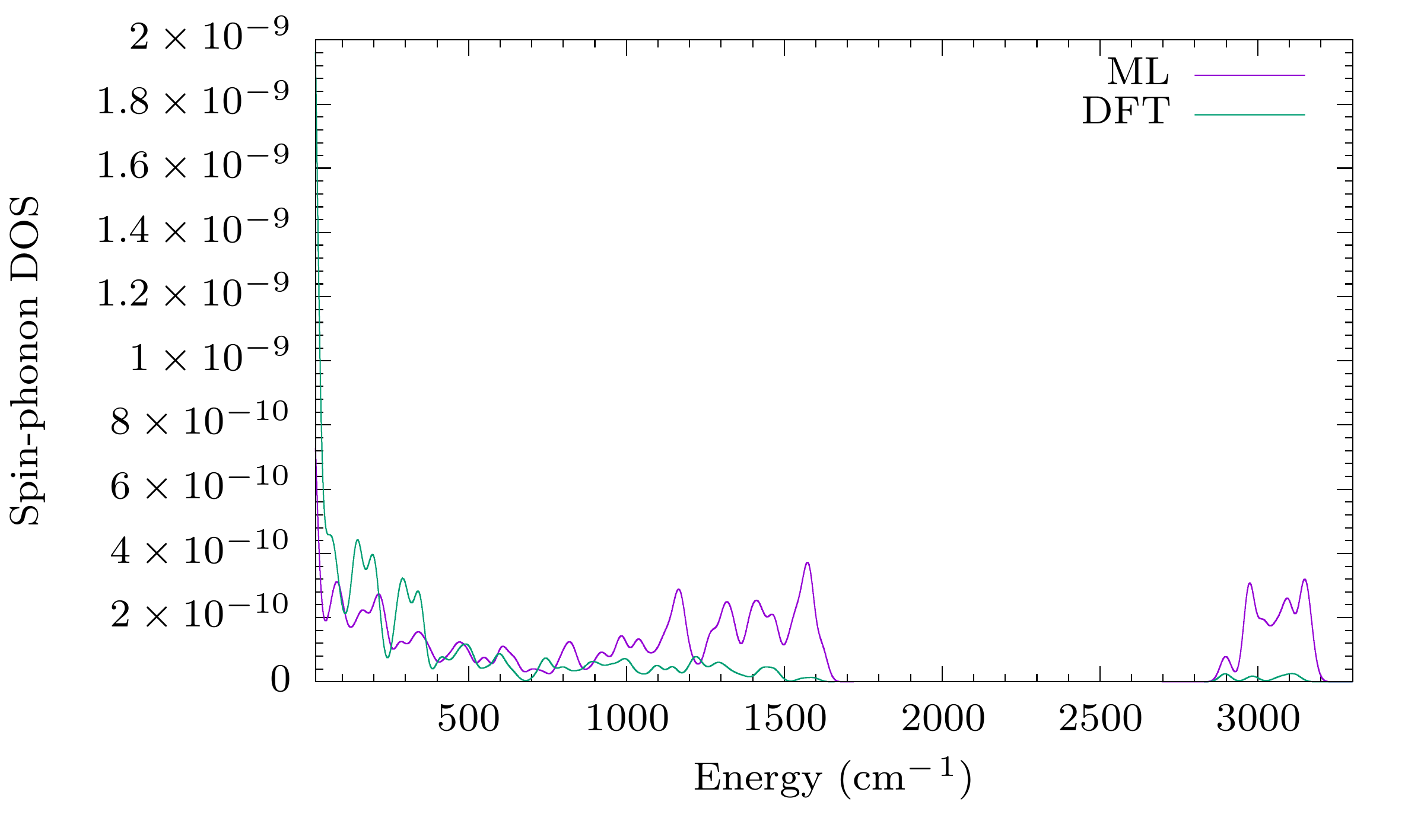}
    \caption{\textbf{Spin-phonon densities} Comparison between the spin-phonon densities calaculated with ML predicted Spin Hamiltonian tensors and phonons and with a full quantum mechanical approach.for \textbf{1},\textbf{2} and \textbf{3}, in order. For \textbf{3}, the graphs for the tesseral functions with $l=2,4,6$ are reported in order.}
    \label{SI:spin_phonon_DOS_ML_CAS}
\end{figure}

\clearpage

\section{Machine learning the full spin-potential energy surface}

\subsection{Compound 1}
\begin{figure}[H]
    \centering
    \includegraphics[scale=0.65]{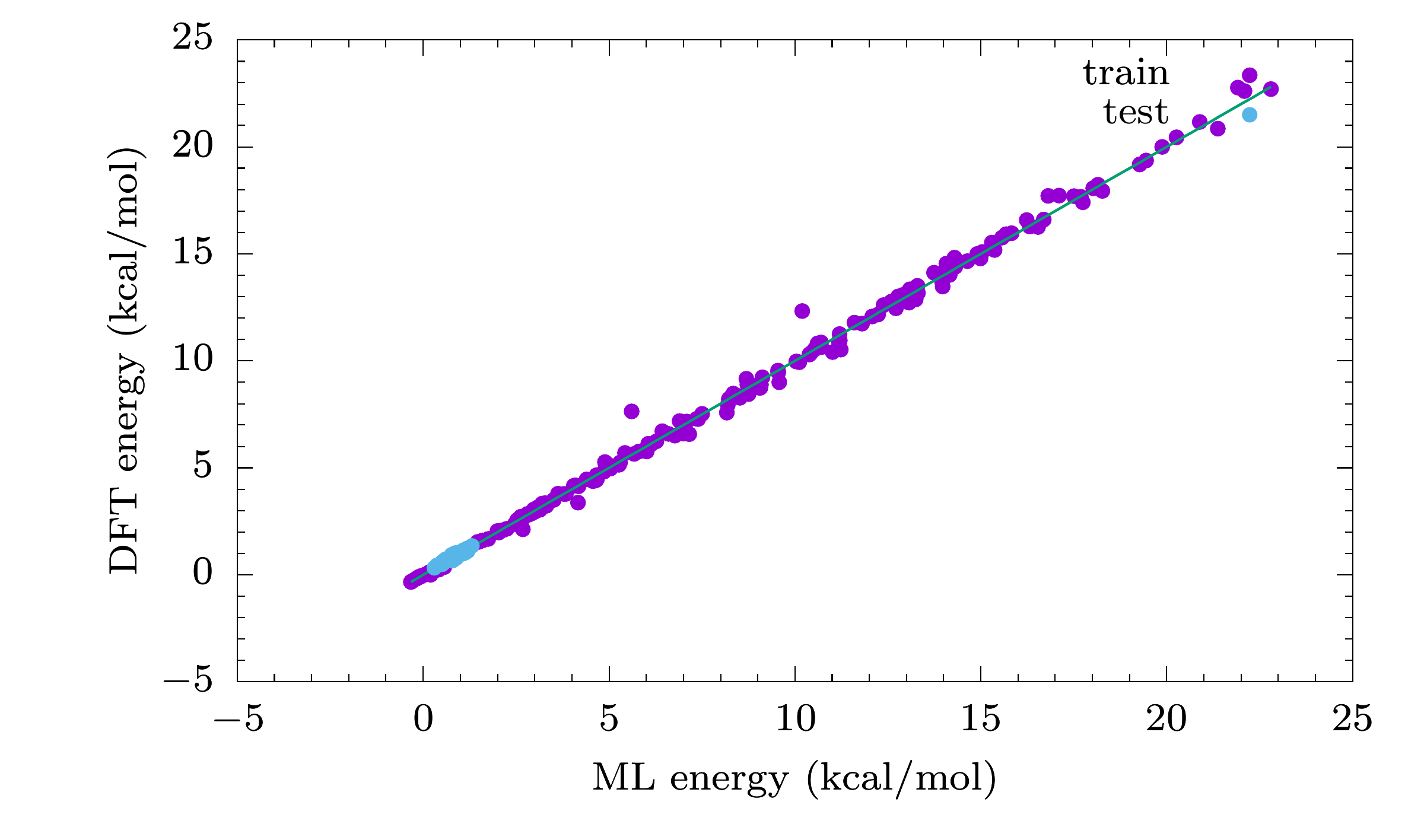}\\
    \includegraphics[scale=0.65]{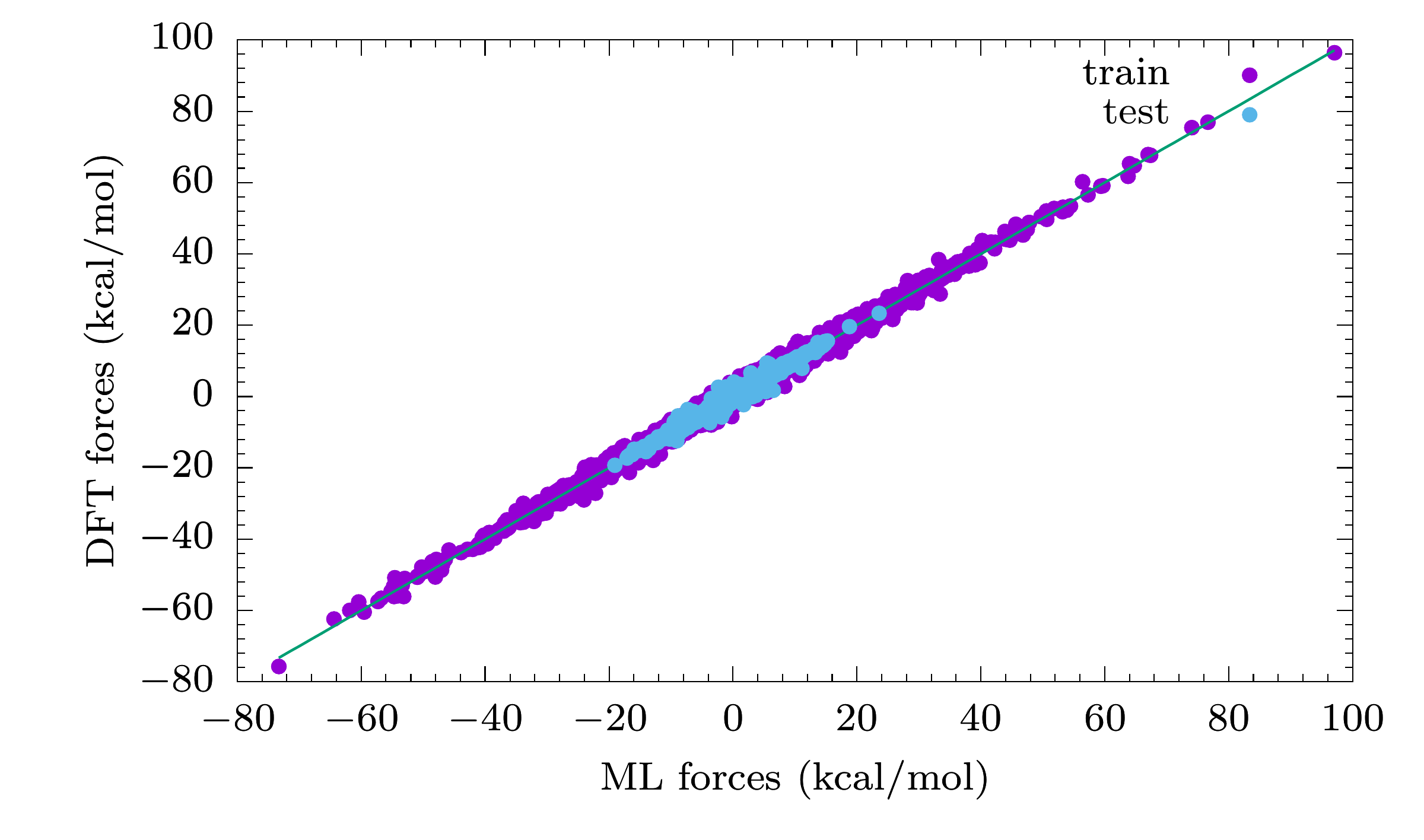}\\
    \includegraphics[scale=0.65]{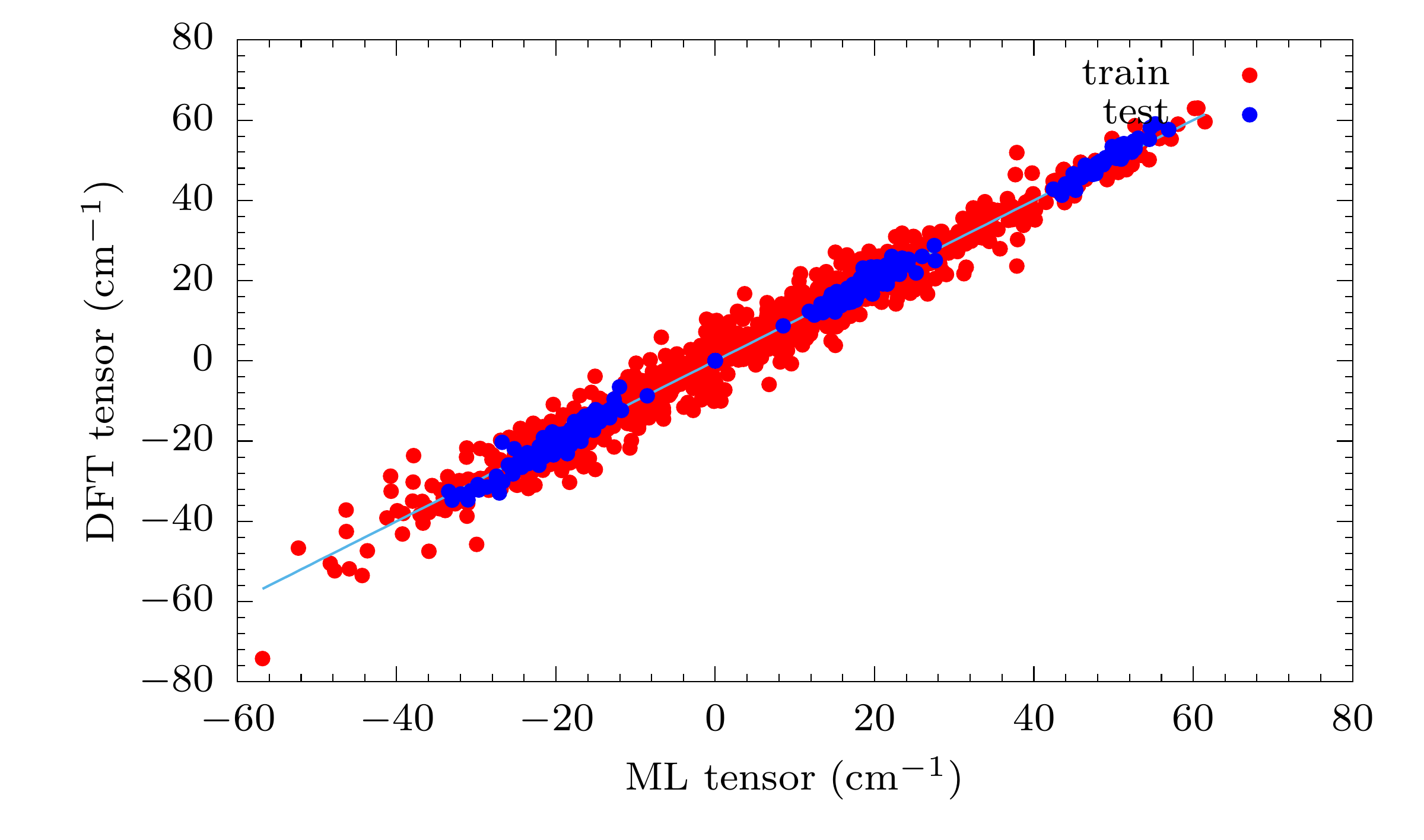}
    \caption{\textbf{Energy, forces and tensors} T=25 K}
\end{figure}

\begin{figure}[H]
    \centering
    \includegraphics[scale=0.65]{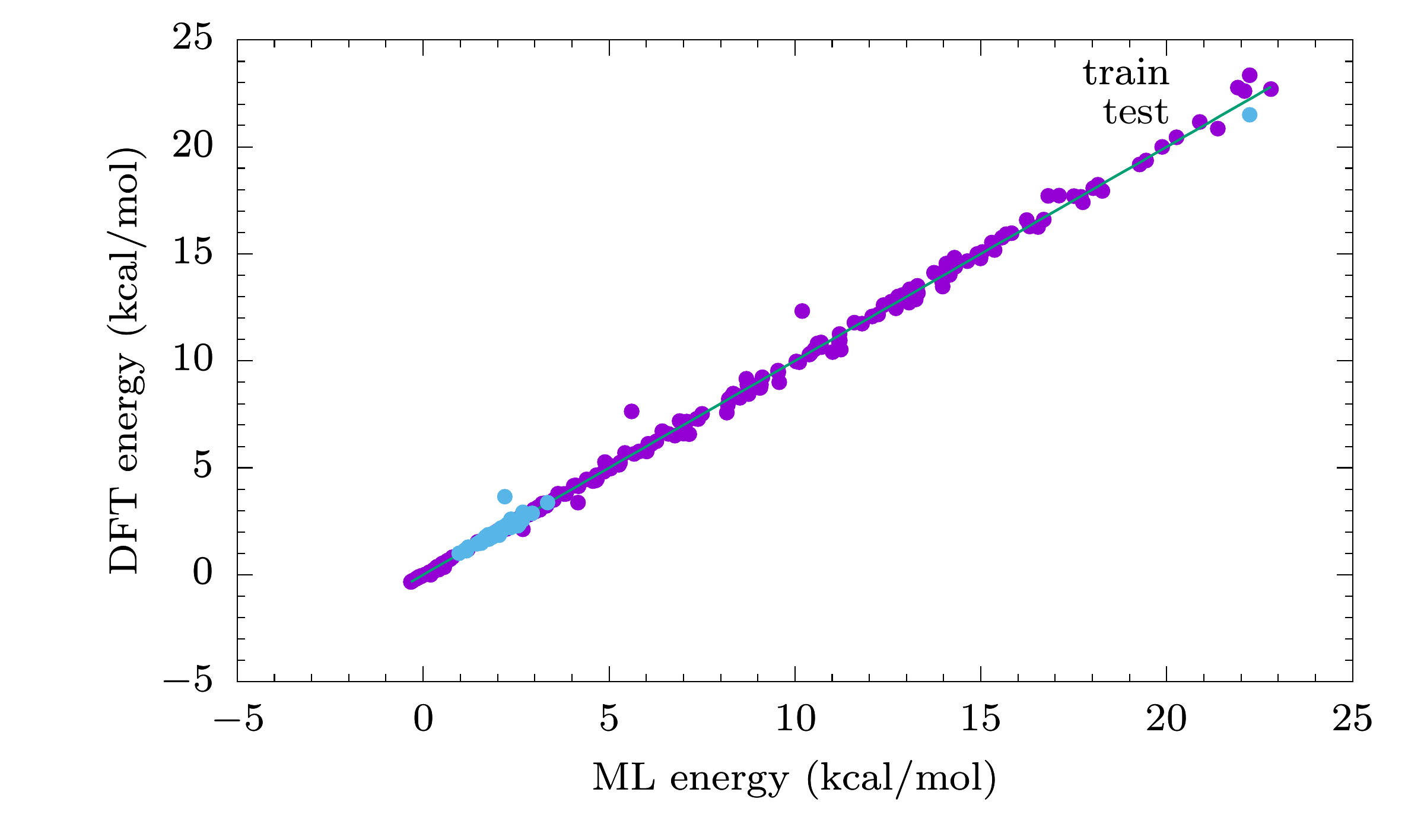}\\
    \includegraphics[scale=0.65]{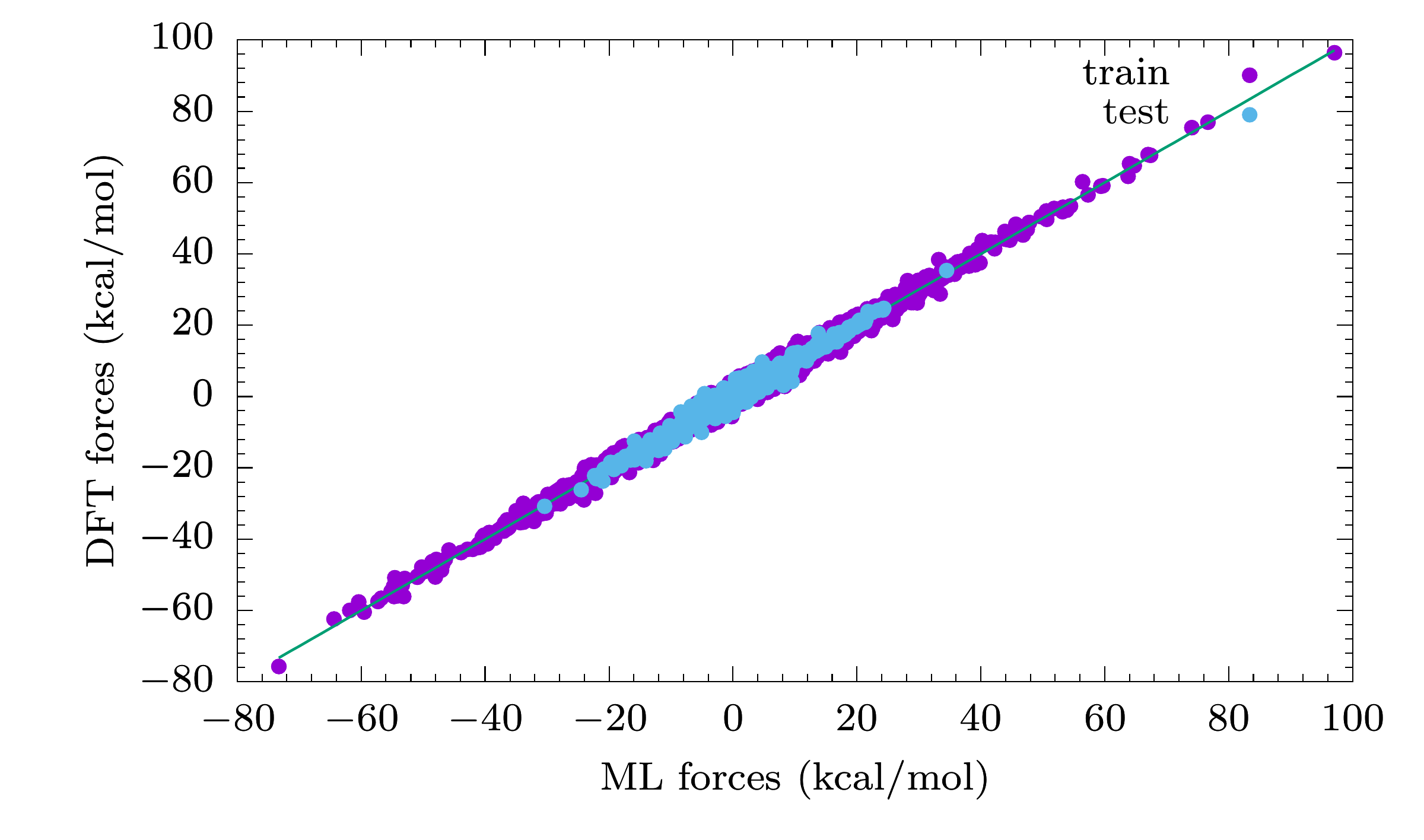}\\
    \includegraphics[scale=0.65]{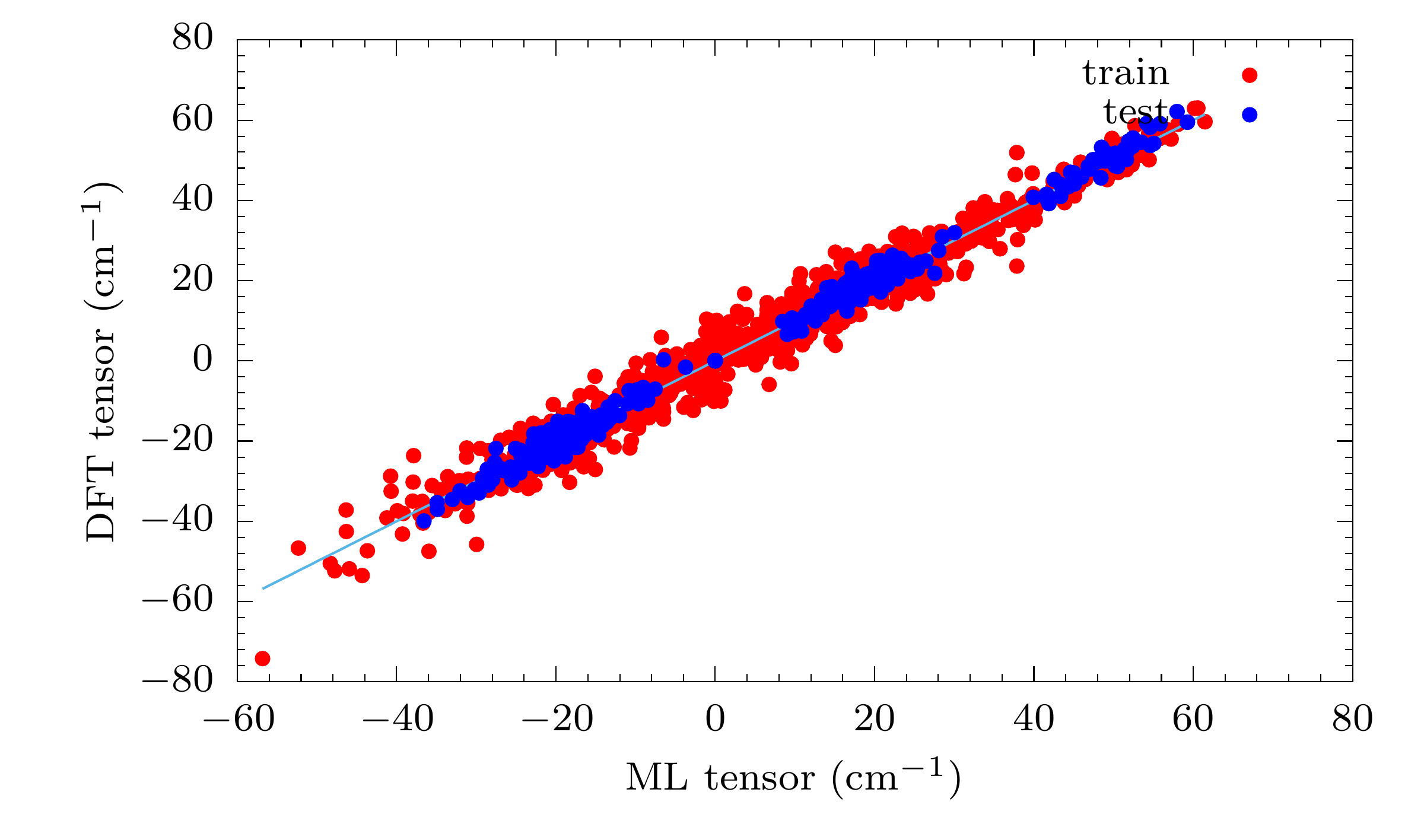}
    \caption{\textbf{Energy, forces and tensors} T=50 K}
\end{figure}

\begin{figure}[H]
    \centering
    \includegraphics[scale=0.65]{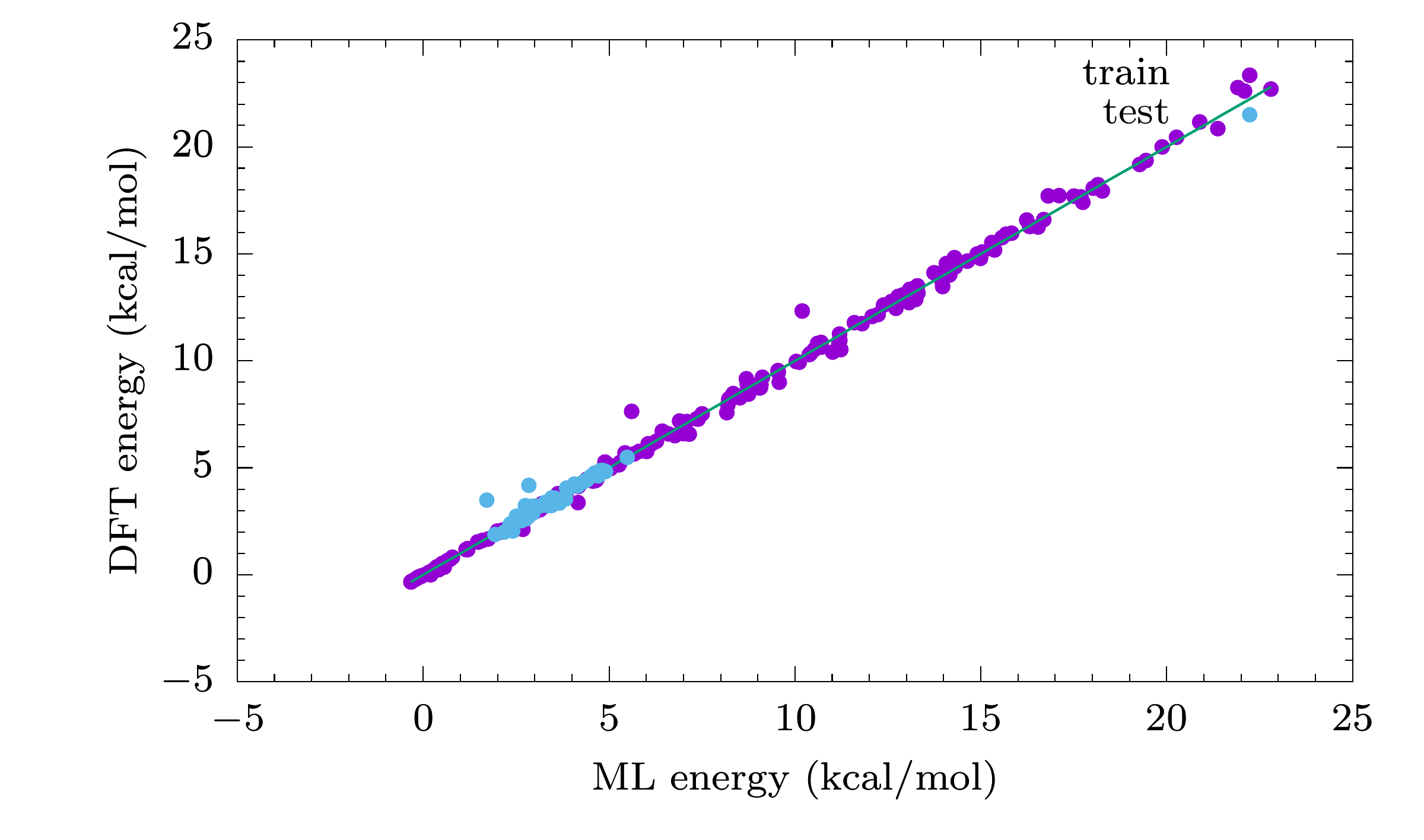}\\
    \includegraphics[scale=0.65]{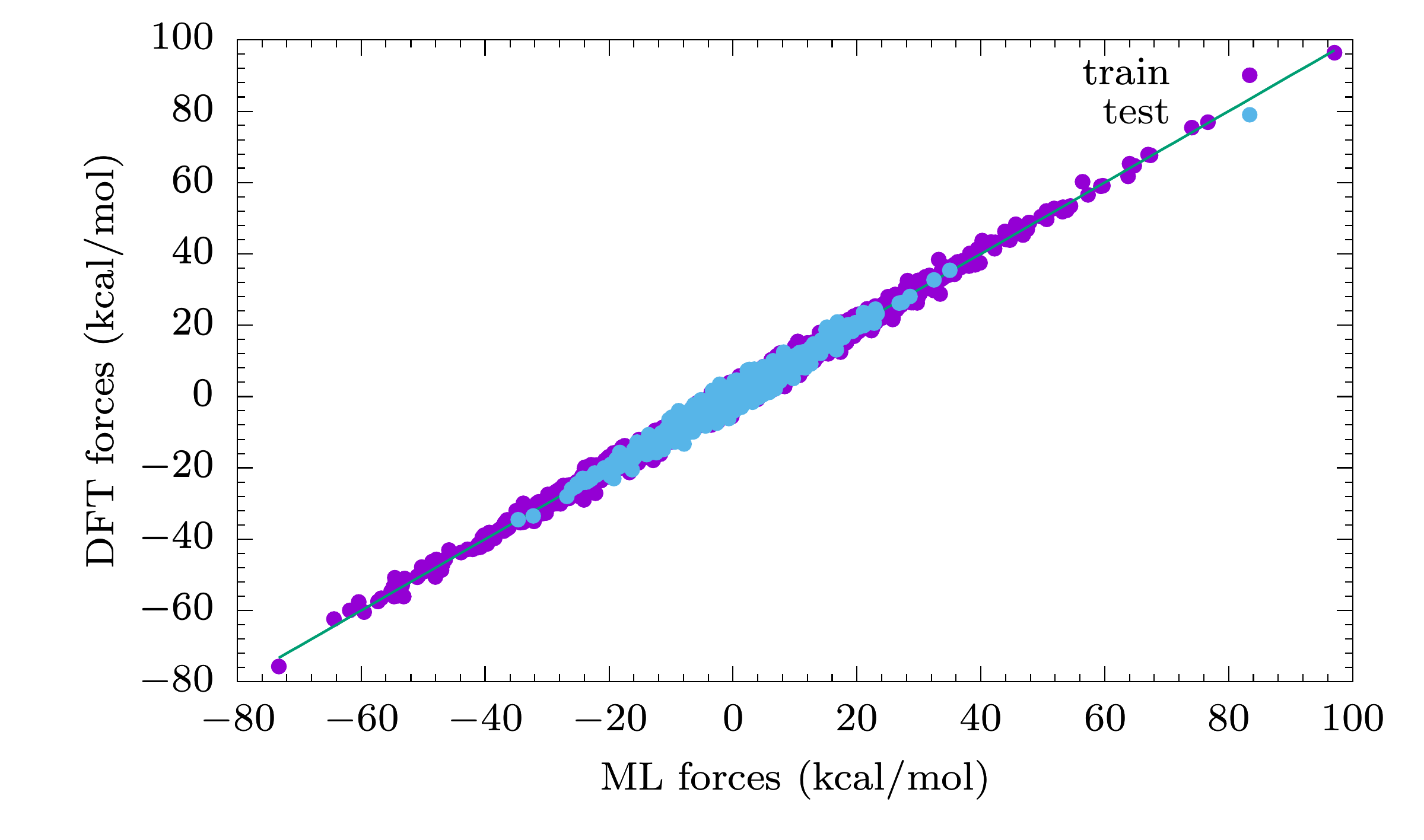}\\
    \includegraphics[scale=0.65]{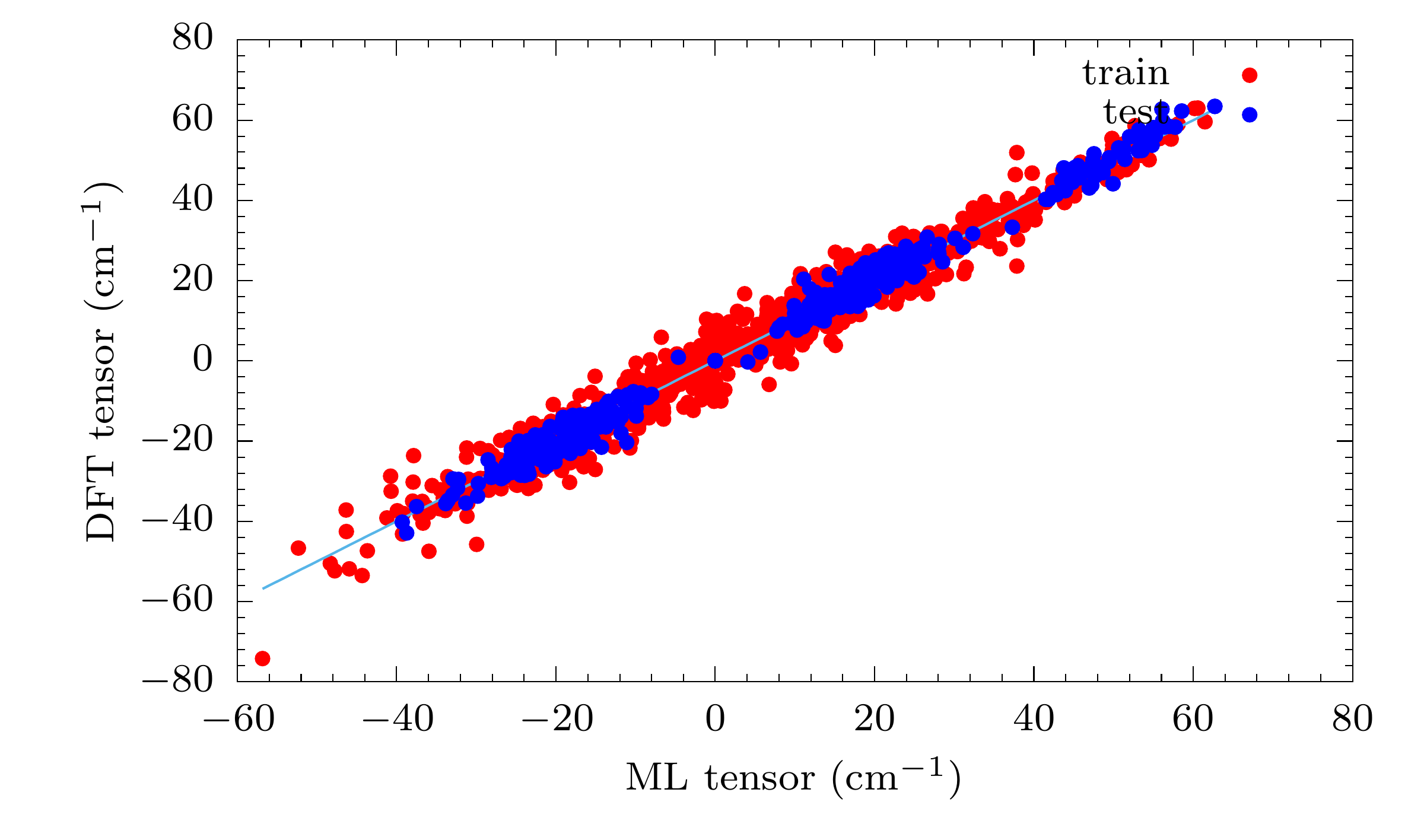}
    \caption{\textbf{Energy, forces and tensors} T=75 K}
\end{figure}

\begin{figure}[H]
    \centering
    \includegraphics[scale=0.65]{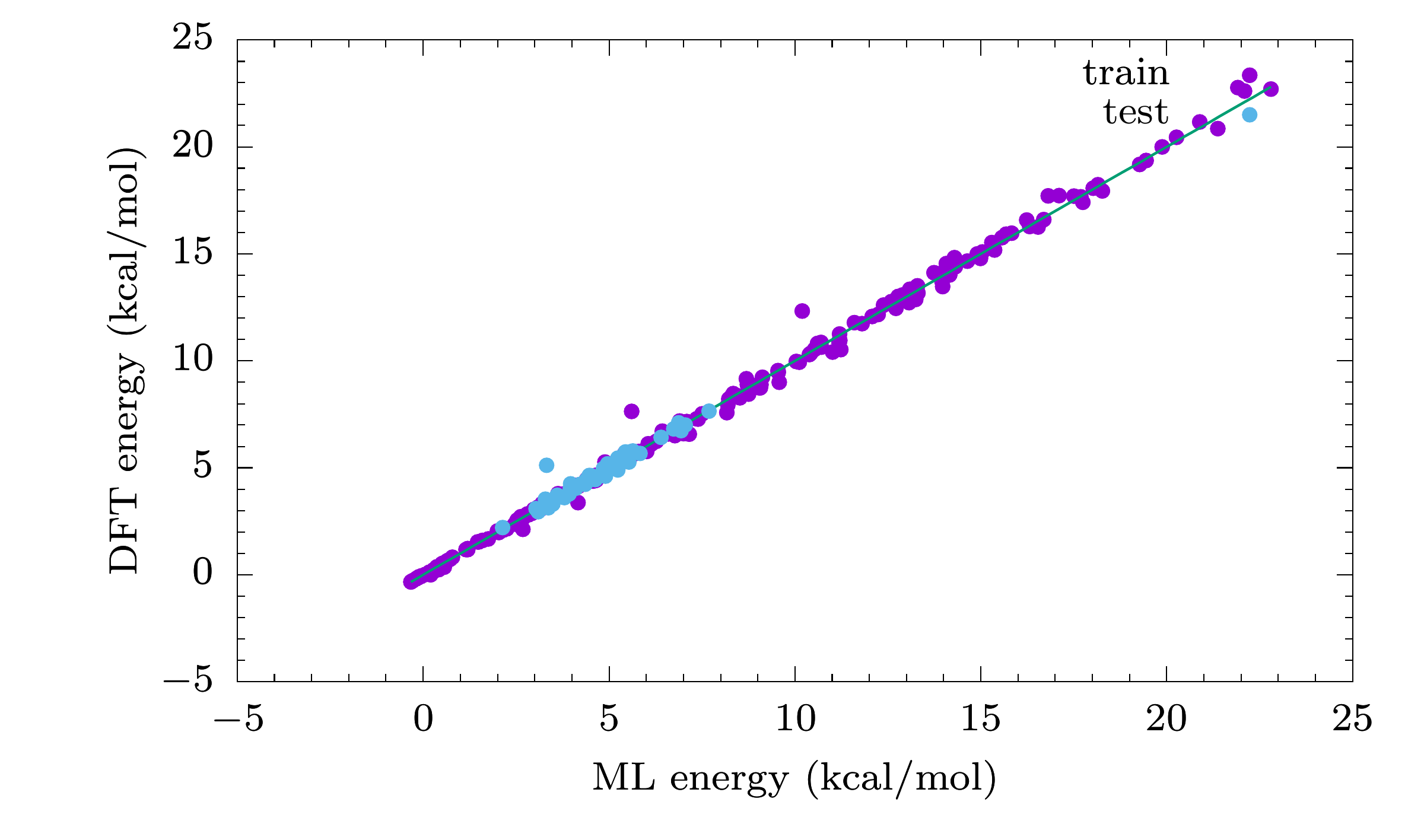}\\
    \includegraphics[scale=0.65]{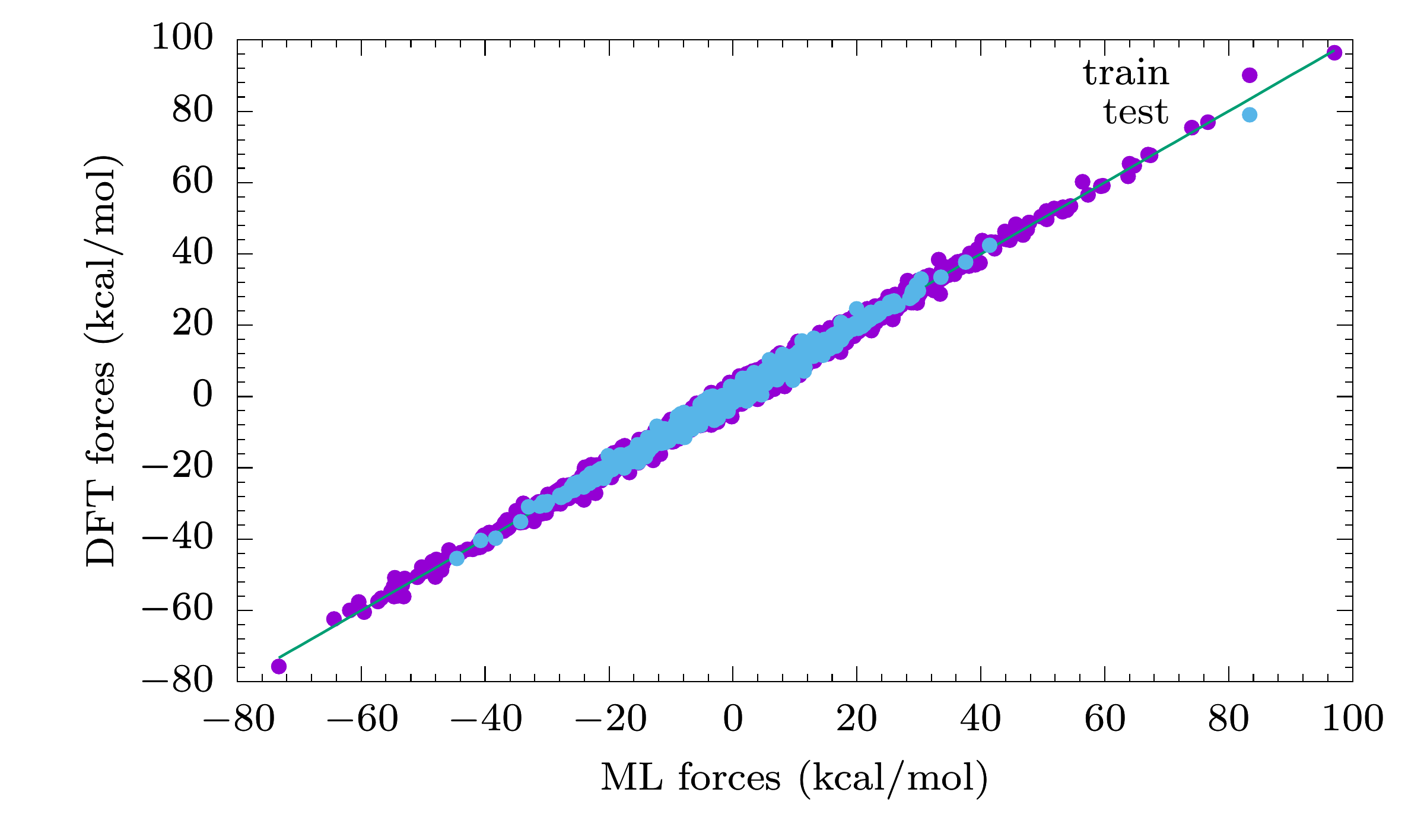}\\
    \includegraphics[scale=0.65]{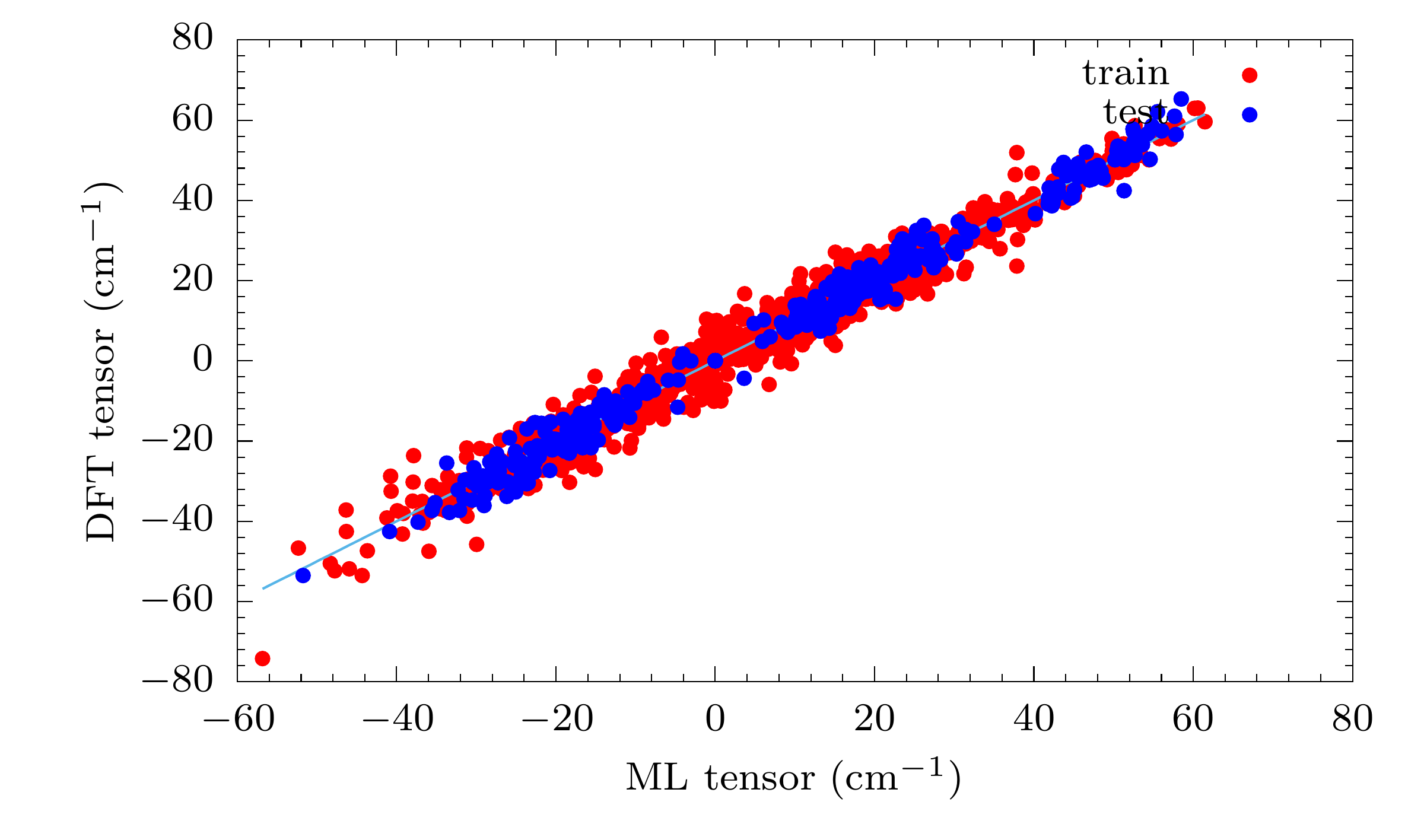}
    \caption{\textbf{Energy, forces and tensors} T=100 K}
\end{figure}

\subsection{Compound 2}

\begin{figure}[H]
    \centering
    \includegraphics[scale=0.65]{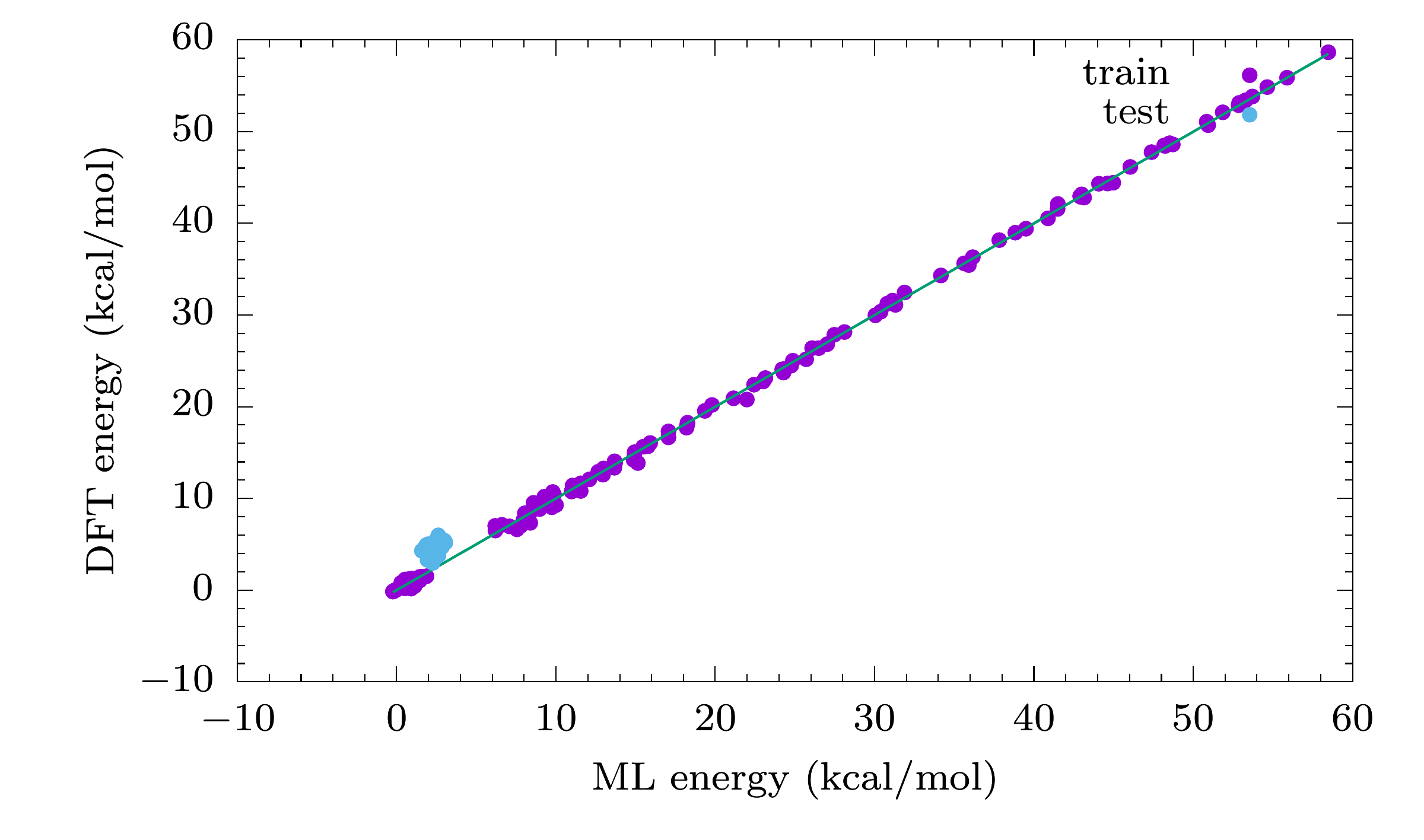}\\
    \includegraphics[scale=0.65]{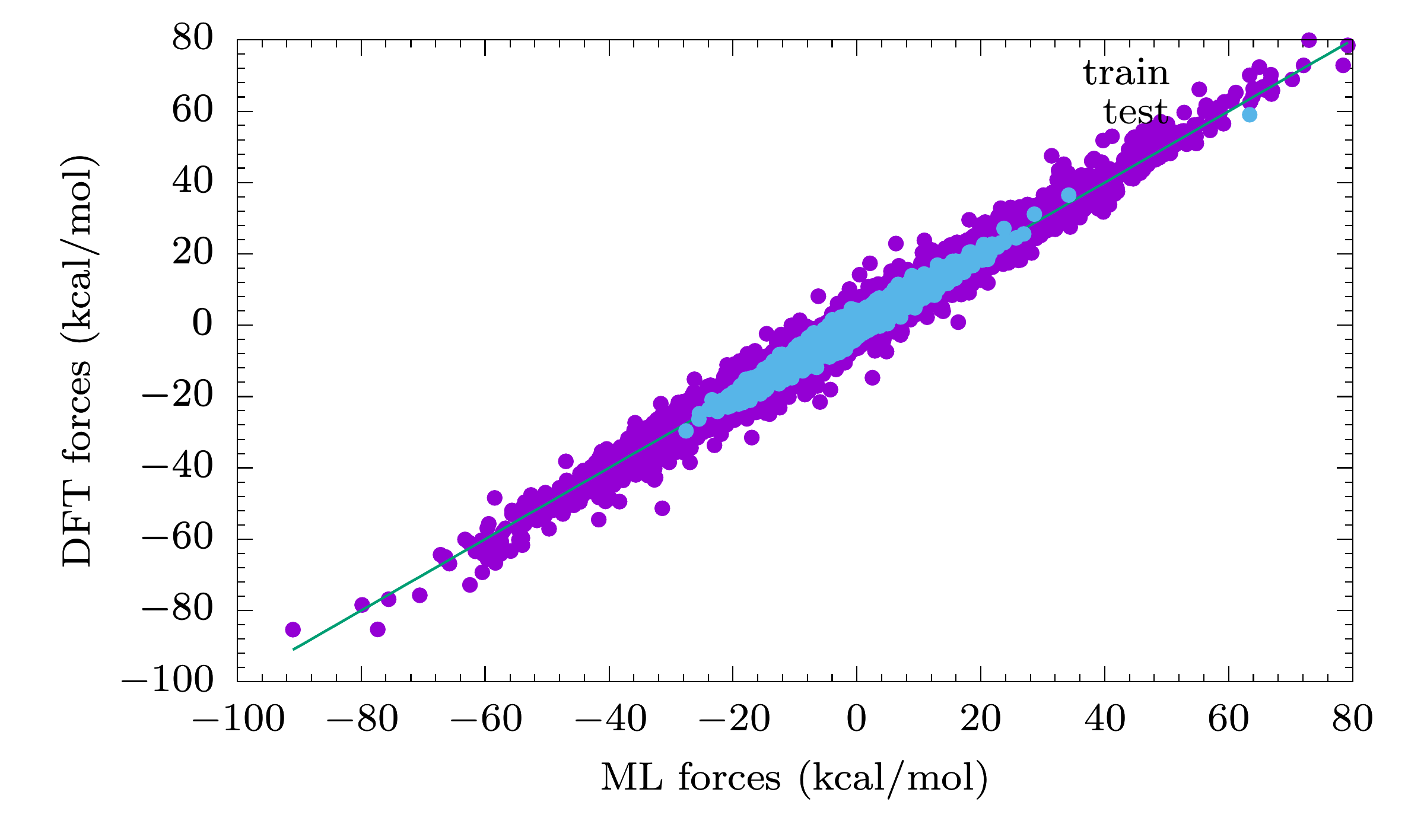}\\
    \includegraphics[scale=0.65]{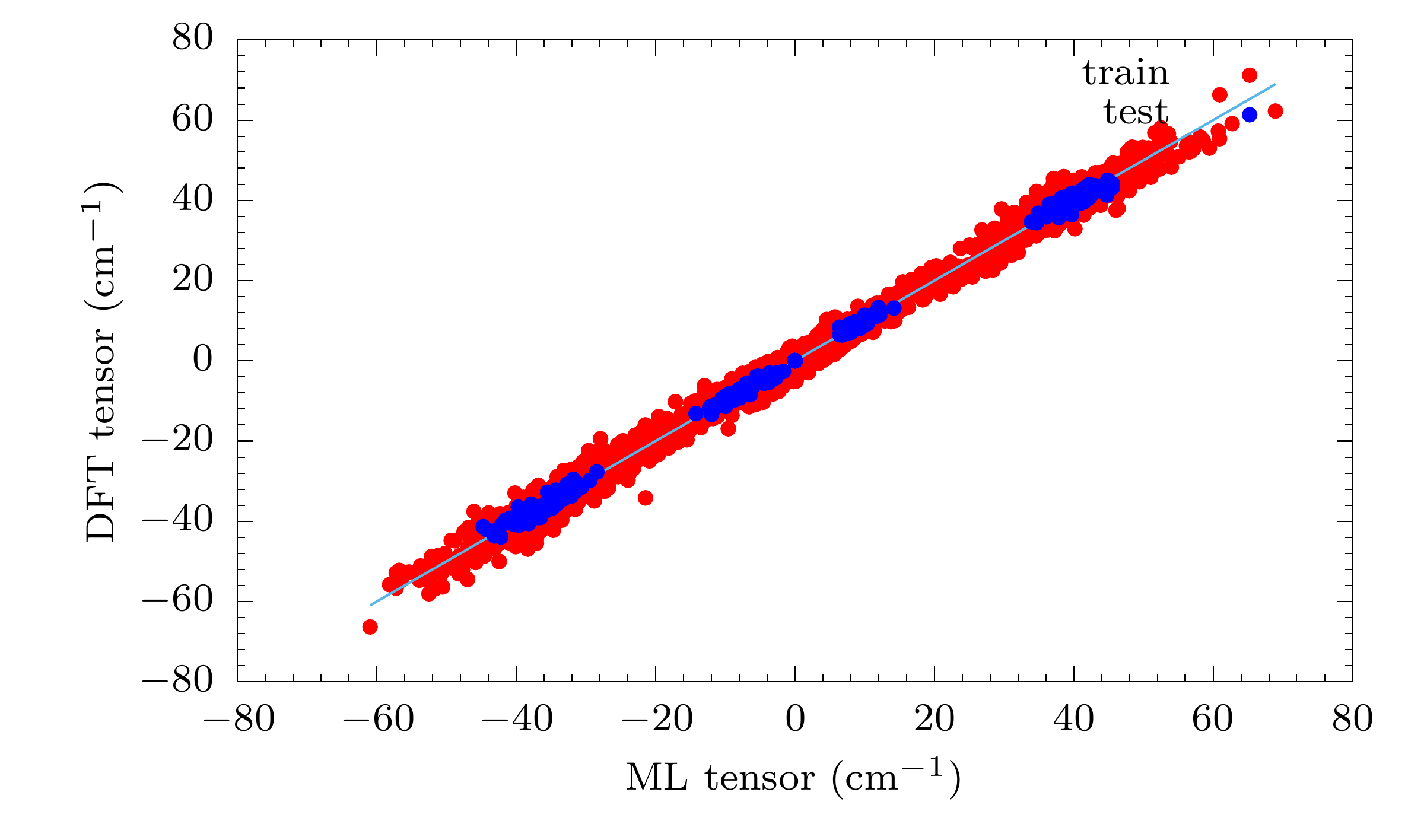}
    \caption{\textbf{Energy, forces and tensors} T=25 K}
\end{figure}

\begin{figure}[H]
    \centering
    \includegraphics[scale=0.65]{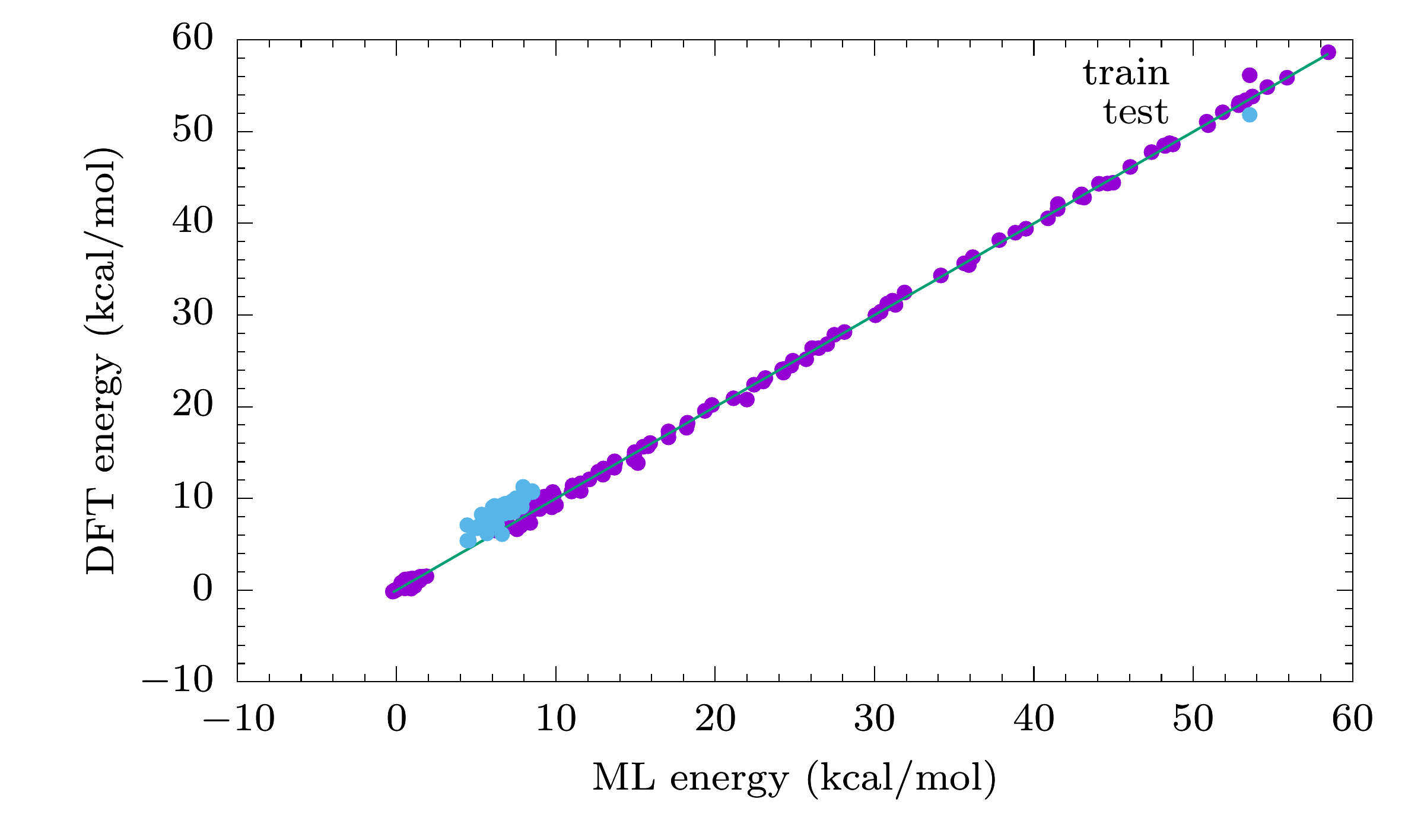}\\
    \includegraphics[scale=0.65]{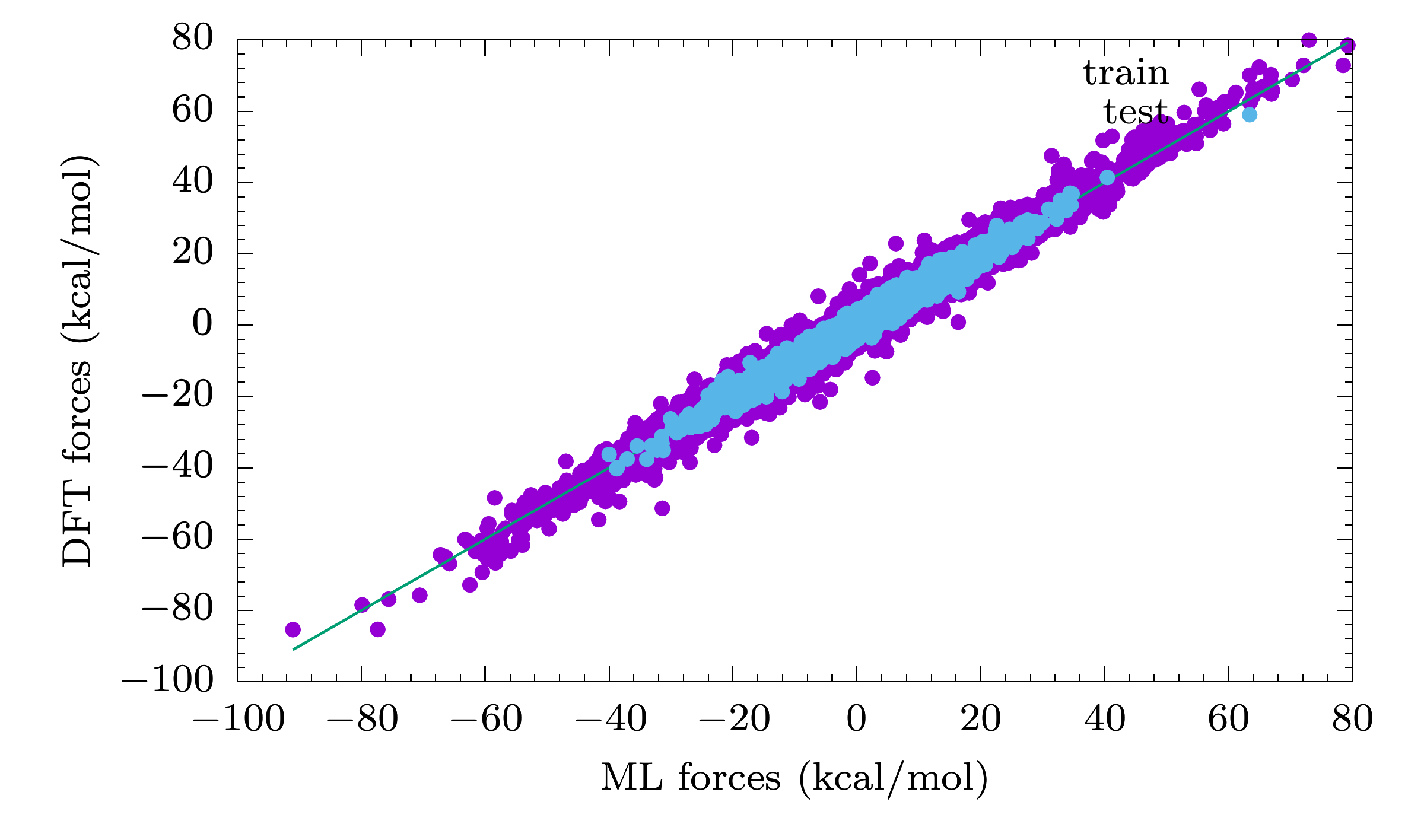}\\
    \includegraphics[scale=0.65]{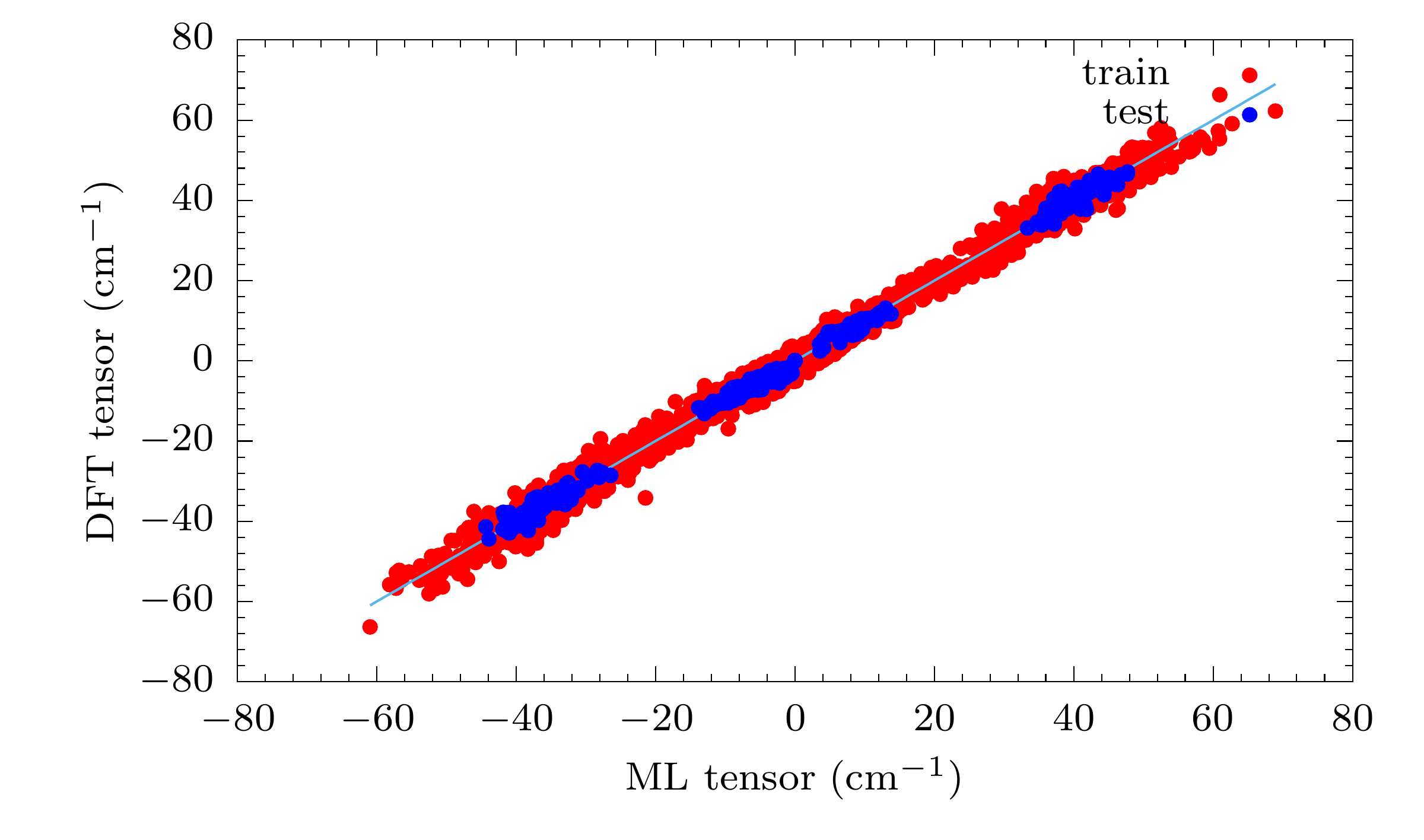}
    \caption{\textbf{Energy, forces and tensors} T=50 K}
\end{figure}

\begin{figure}[H]
    \centering
    \includegraphics[scale=0.65]{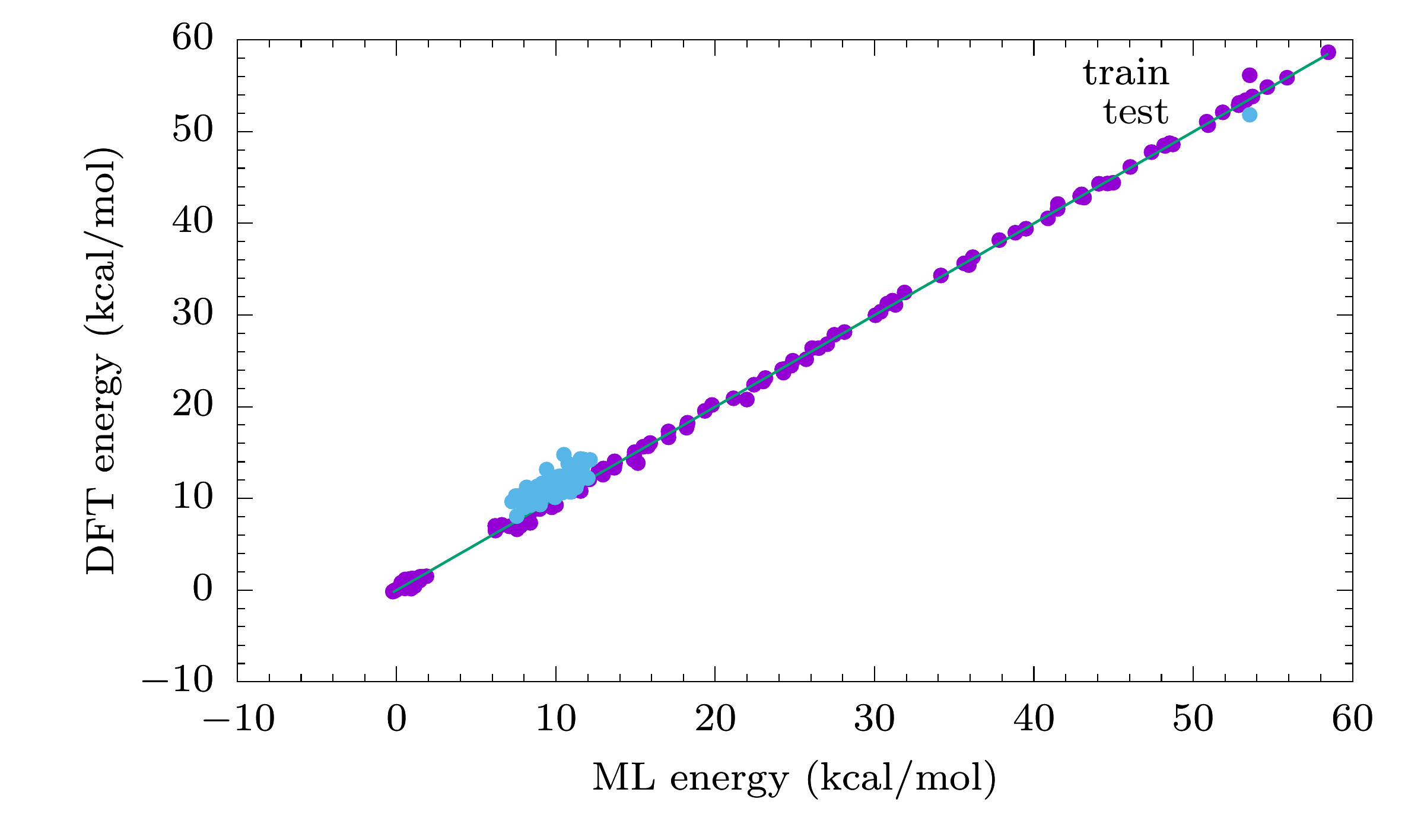}\\
    \includegraphics[scale=0.65]{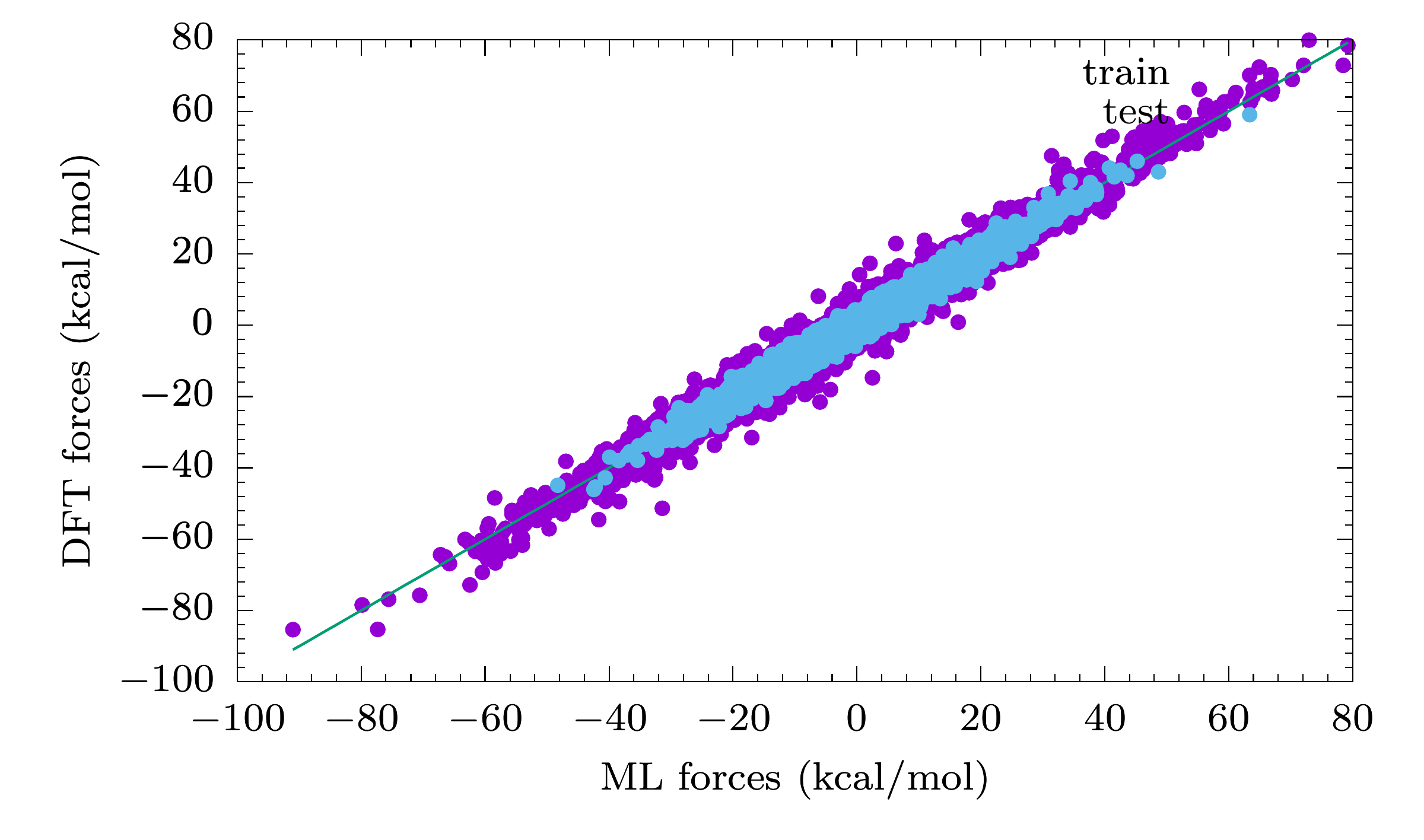}\\
    \includegraphics[scale=0.65]{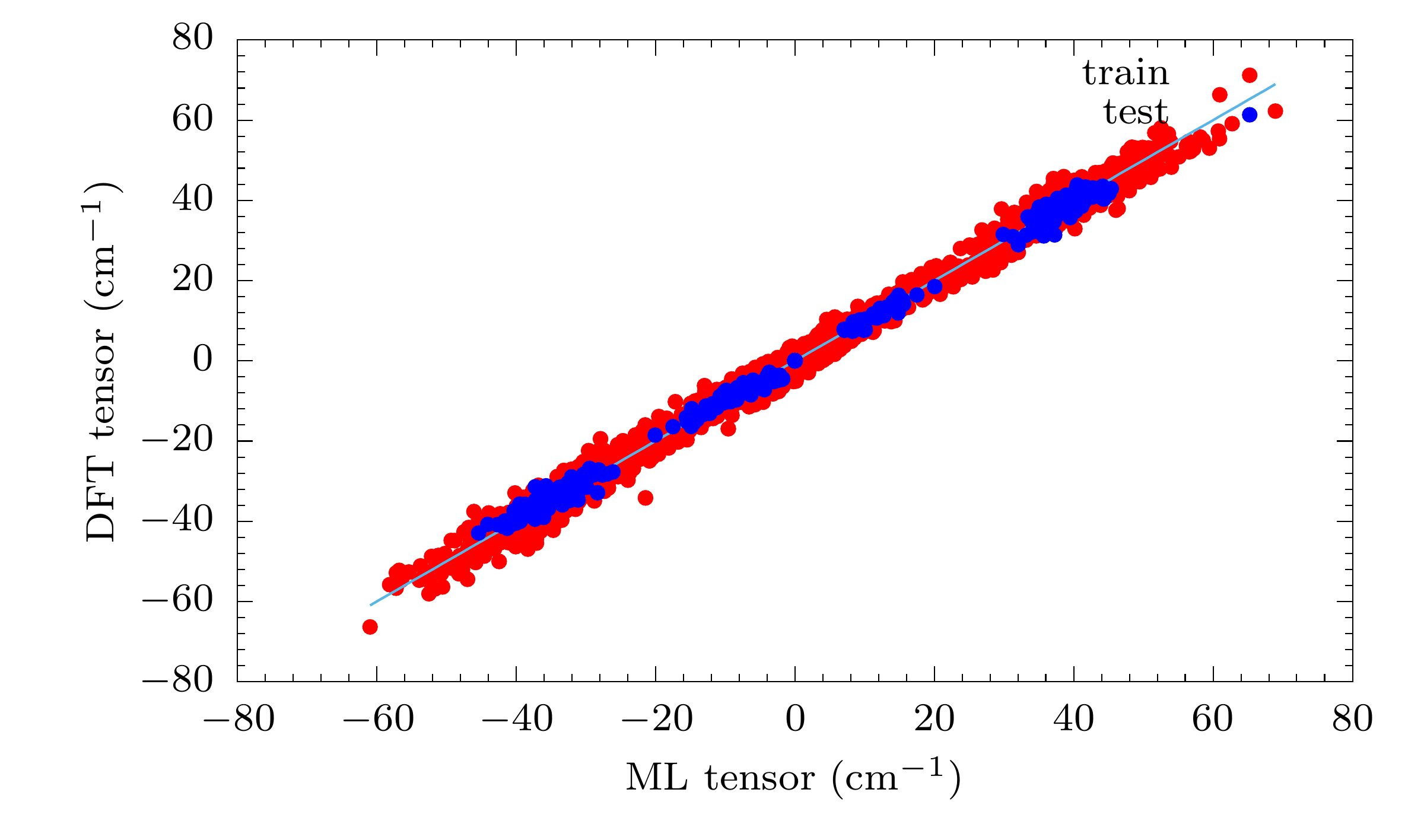}
    \caption{\textbf{Energy, forces and tensors} T=75 K}
\end{figure}

\begin{figure}[H]
    \centering
    \includegraphics[scale=0.65]{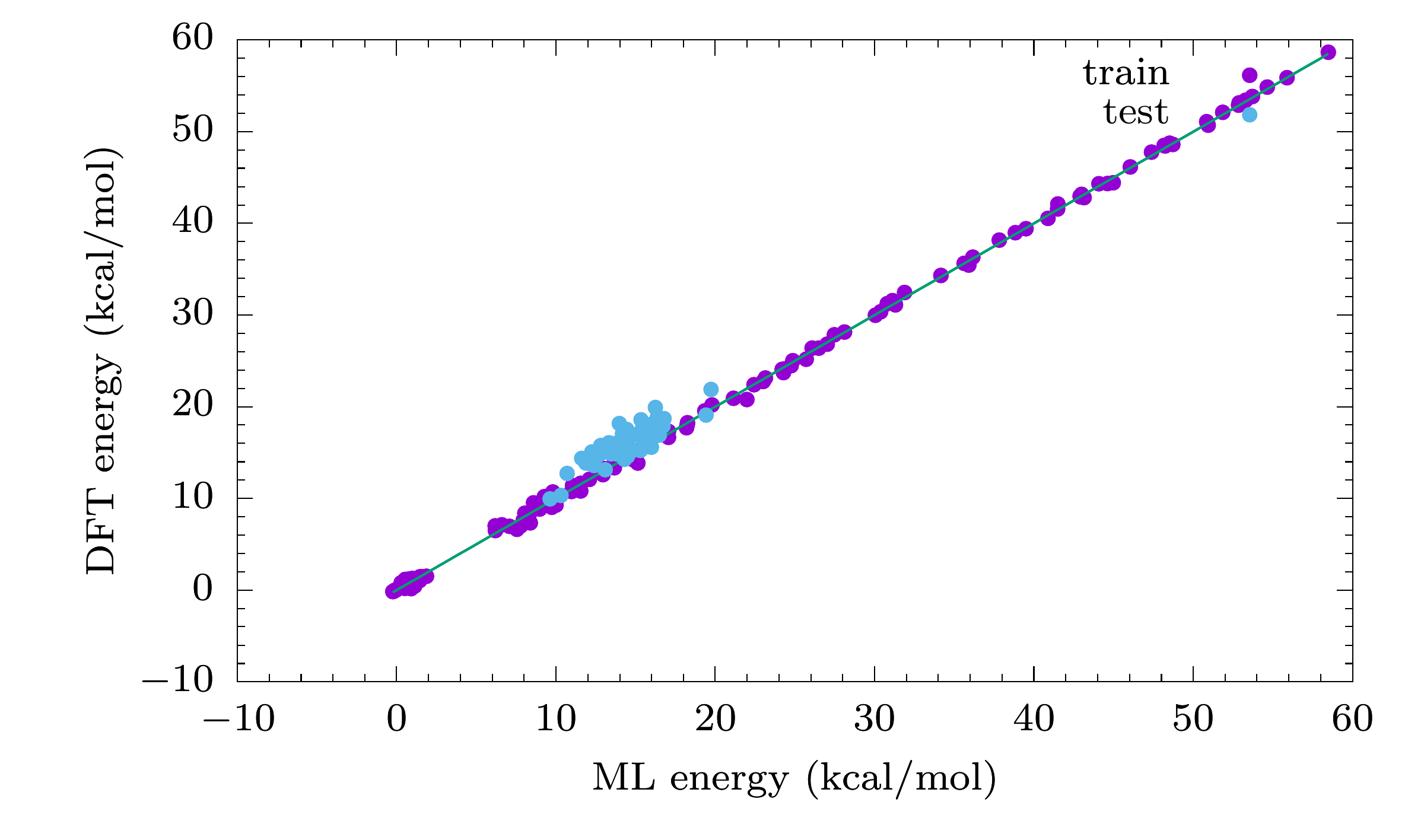}\\
    \includegraphics[scale=0.65]{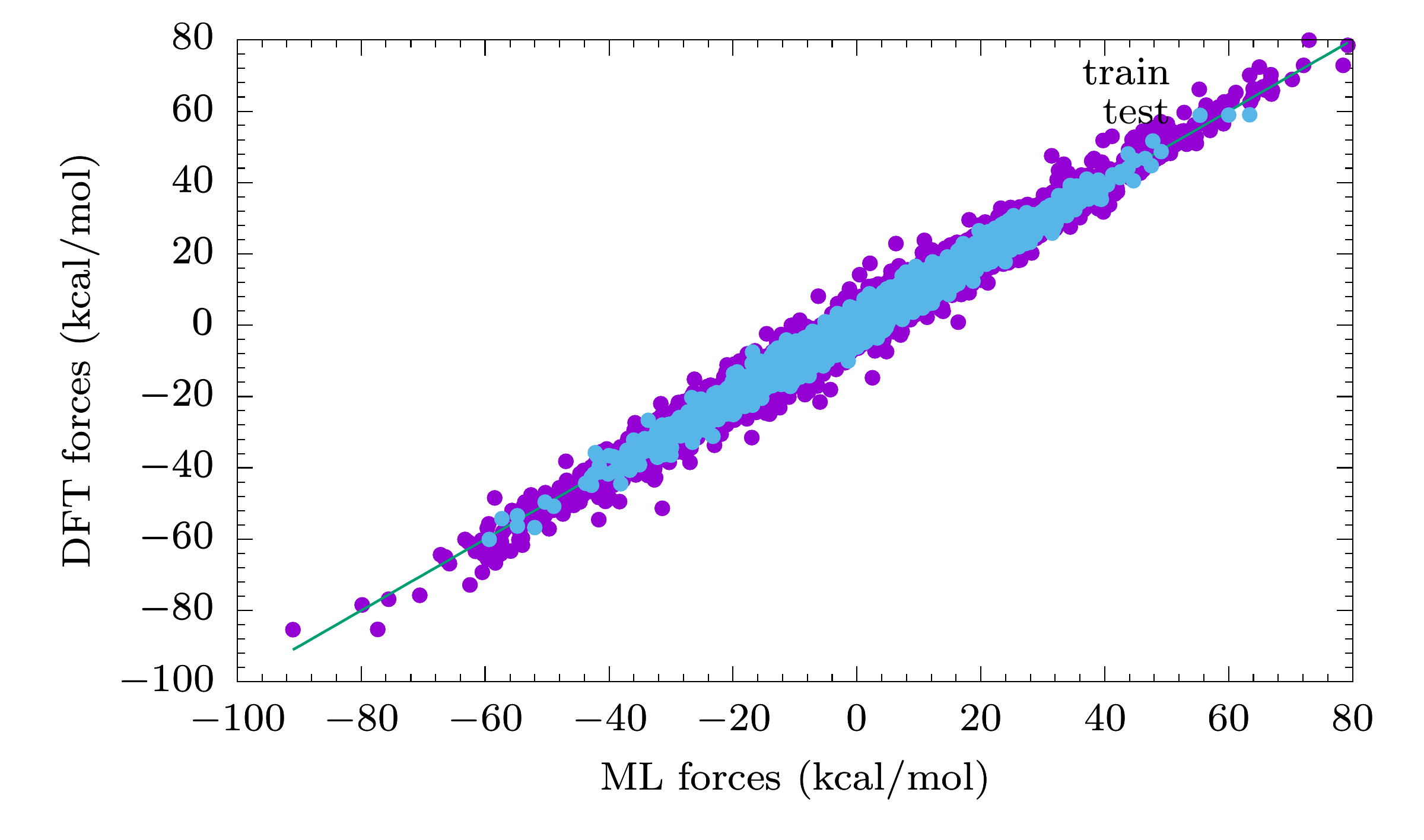}\\
    \includegraphics[scale=0.65]{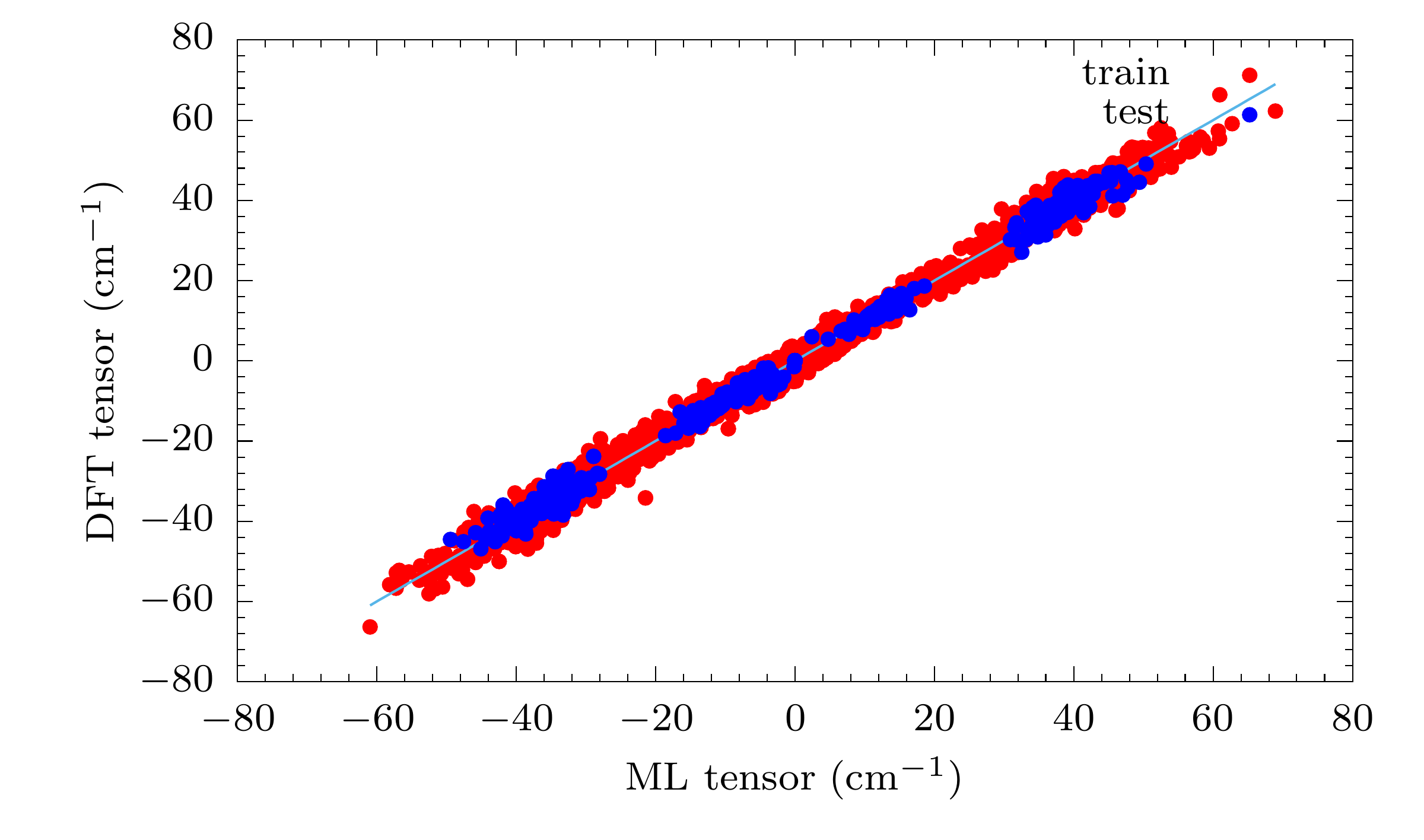}
    \caption{\textbf{Energy, forces and tensors} T=100 K}
\end{figure}

\subsection{Compound 3}

\begin{figure}[H]
    \centering
    \includegraphics[scale=0.65]{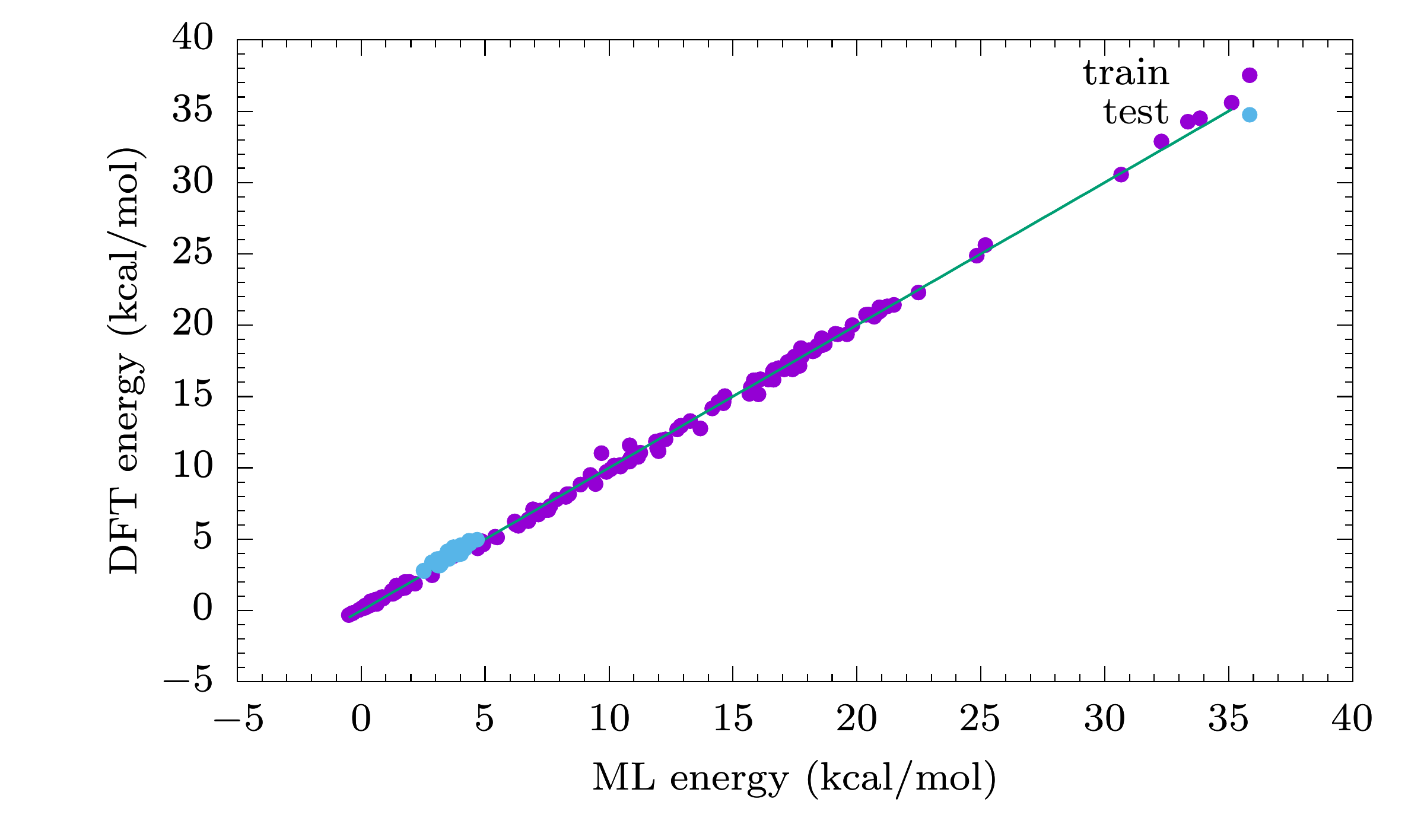}\\
    \includegraphics[scale=0.65]{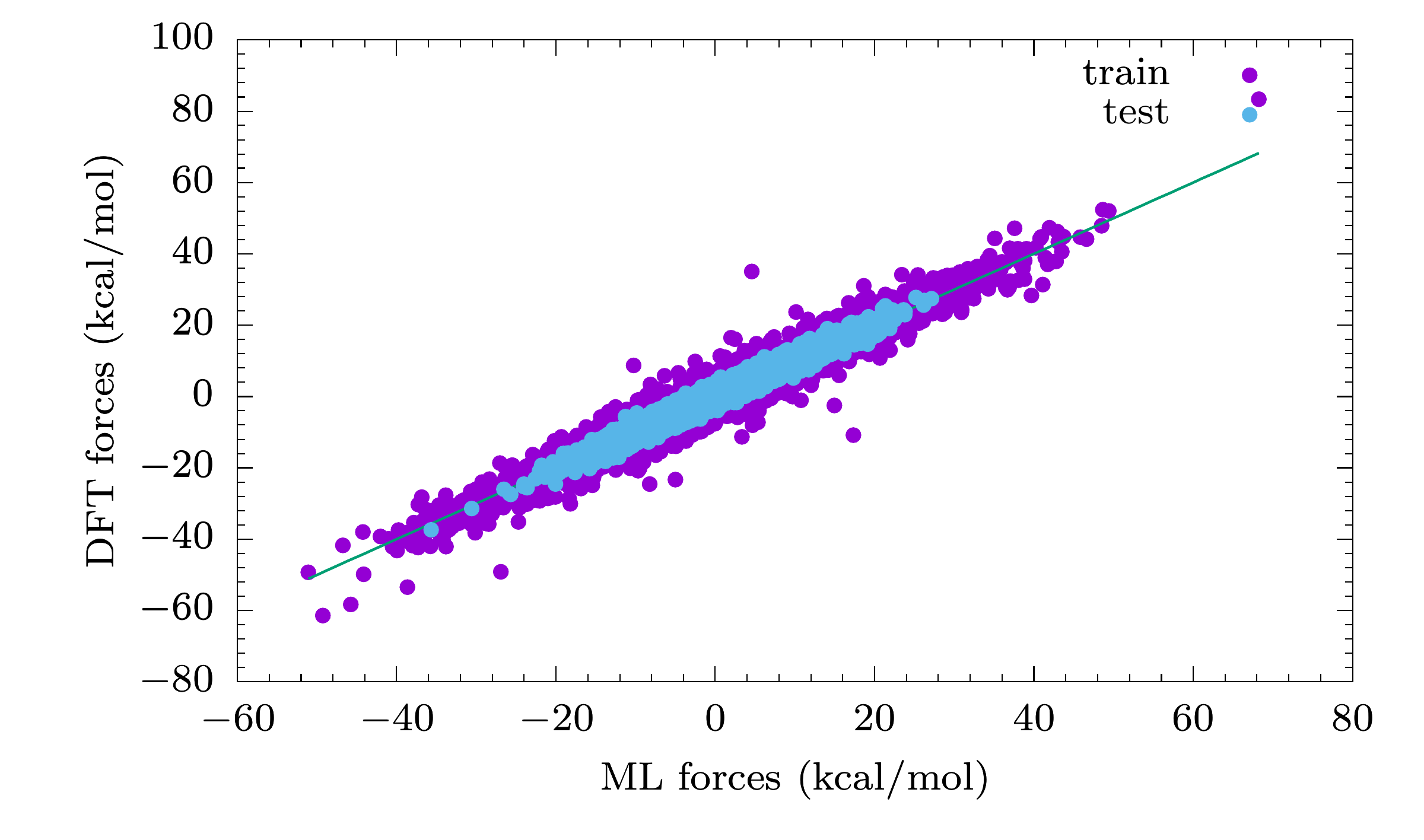}\\
    \includegraphics[scale=0.65]{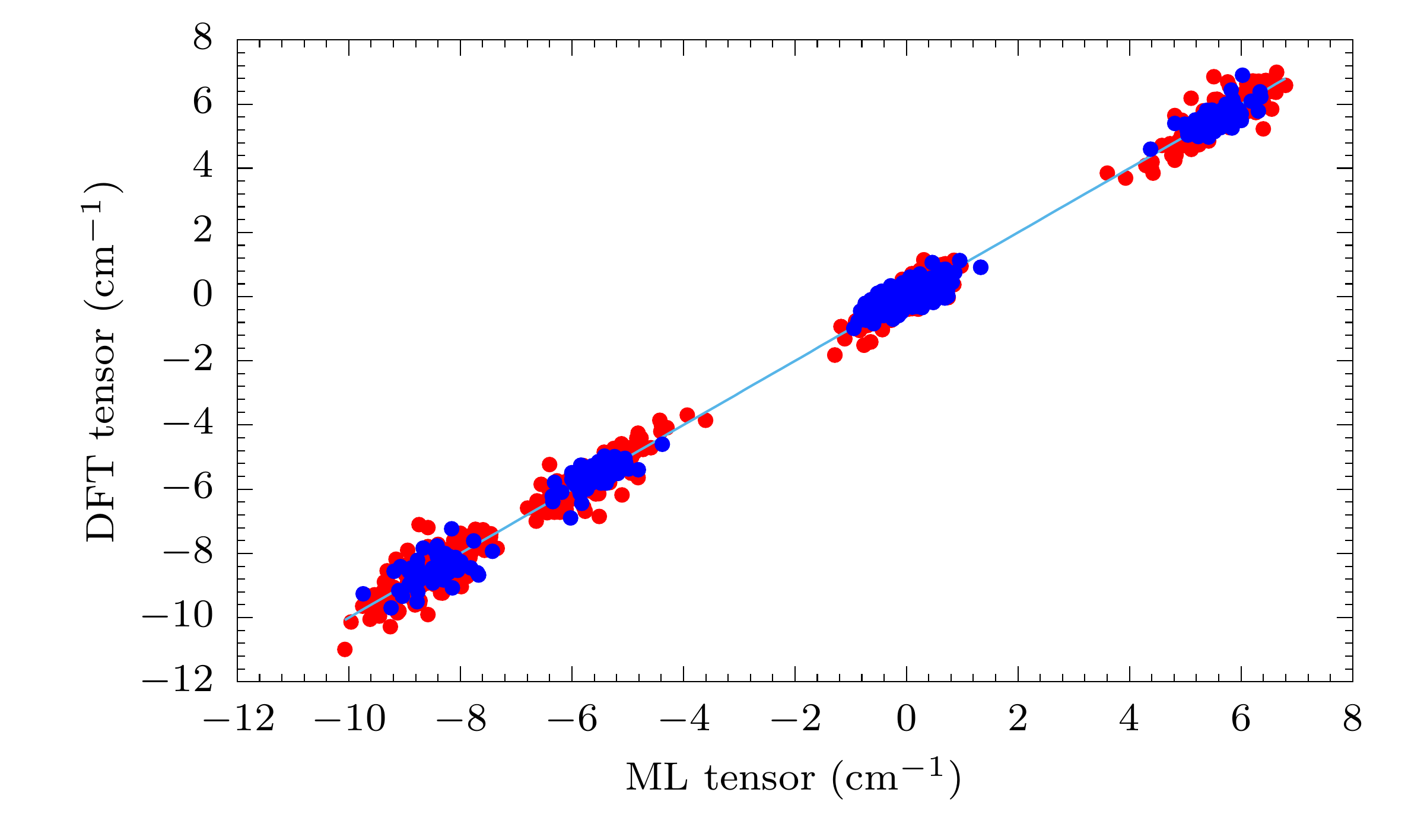}\\
    \includegraphics[scale=0.65]{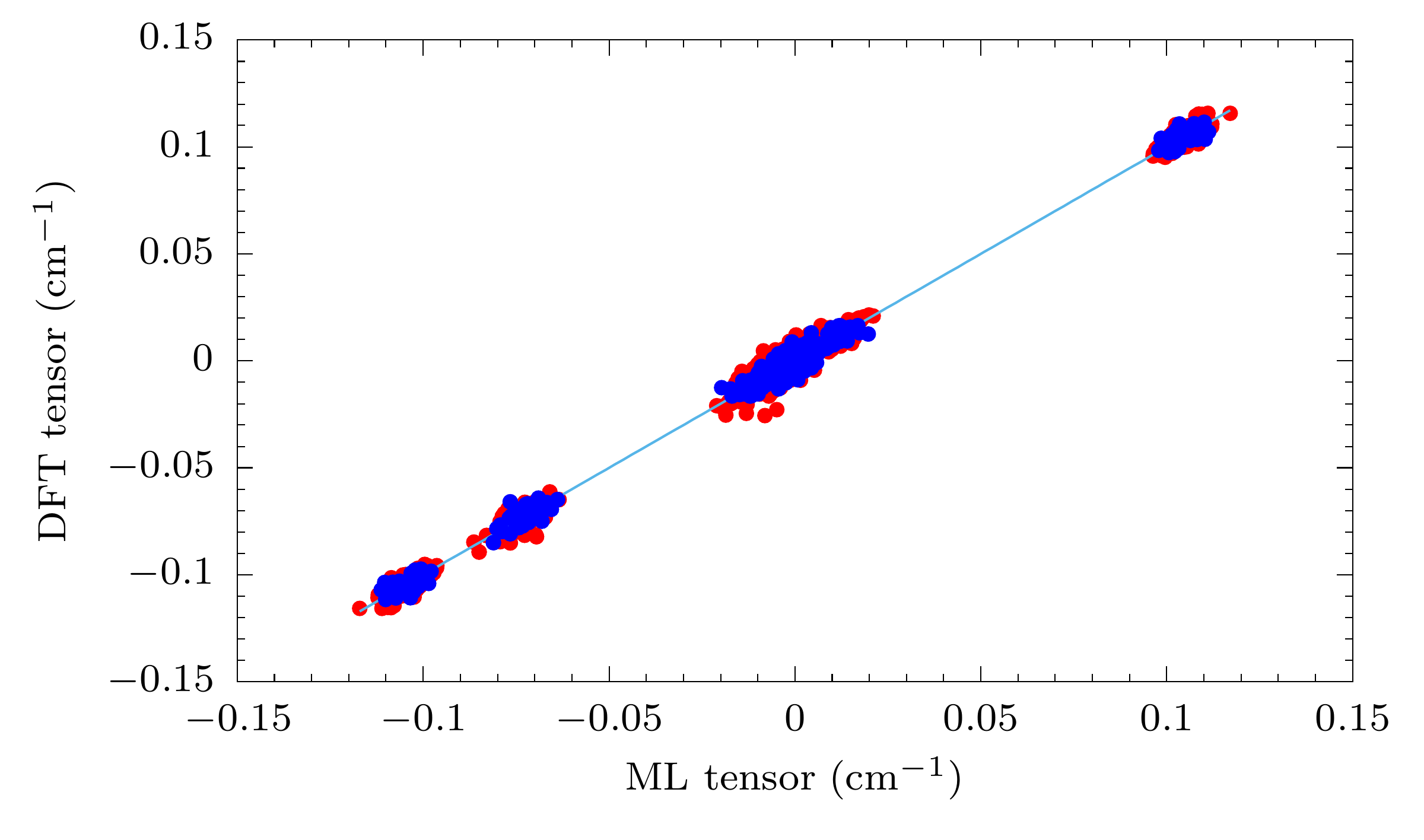}\\
    \includegraphics[scale=0.65]{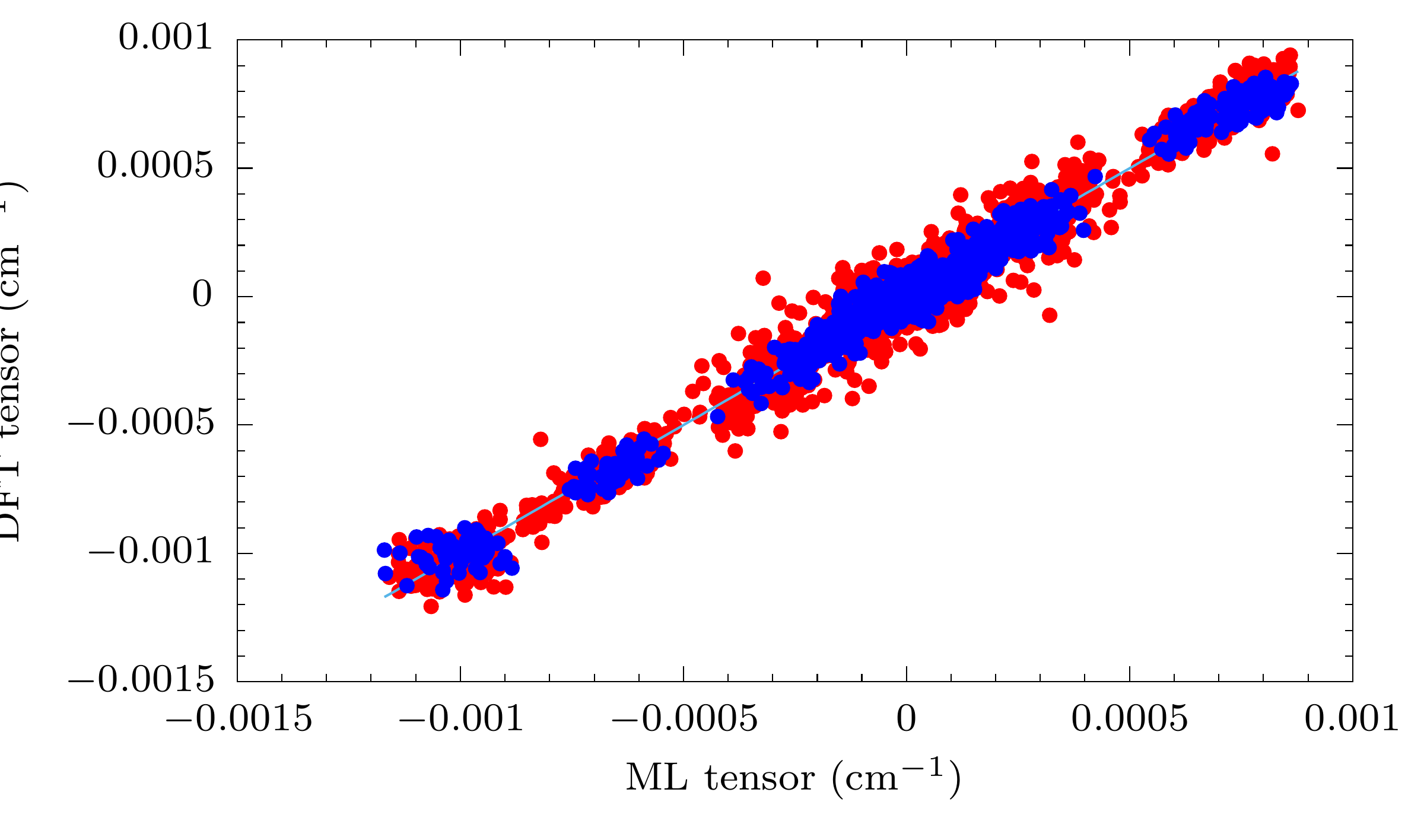}
    \caption{\textbf{Energy, forces and tensors} T=25 K}
\end{figure}

\begin{figure}[H]
    \centering
    \includegraphics[scale=0.65]{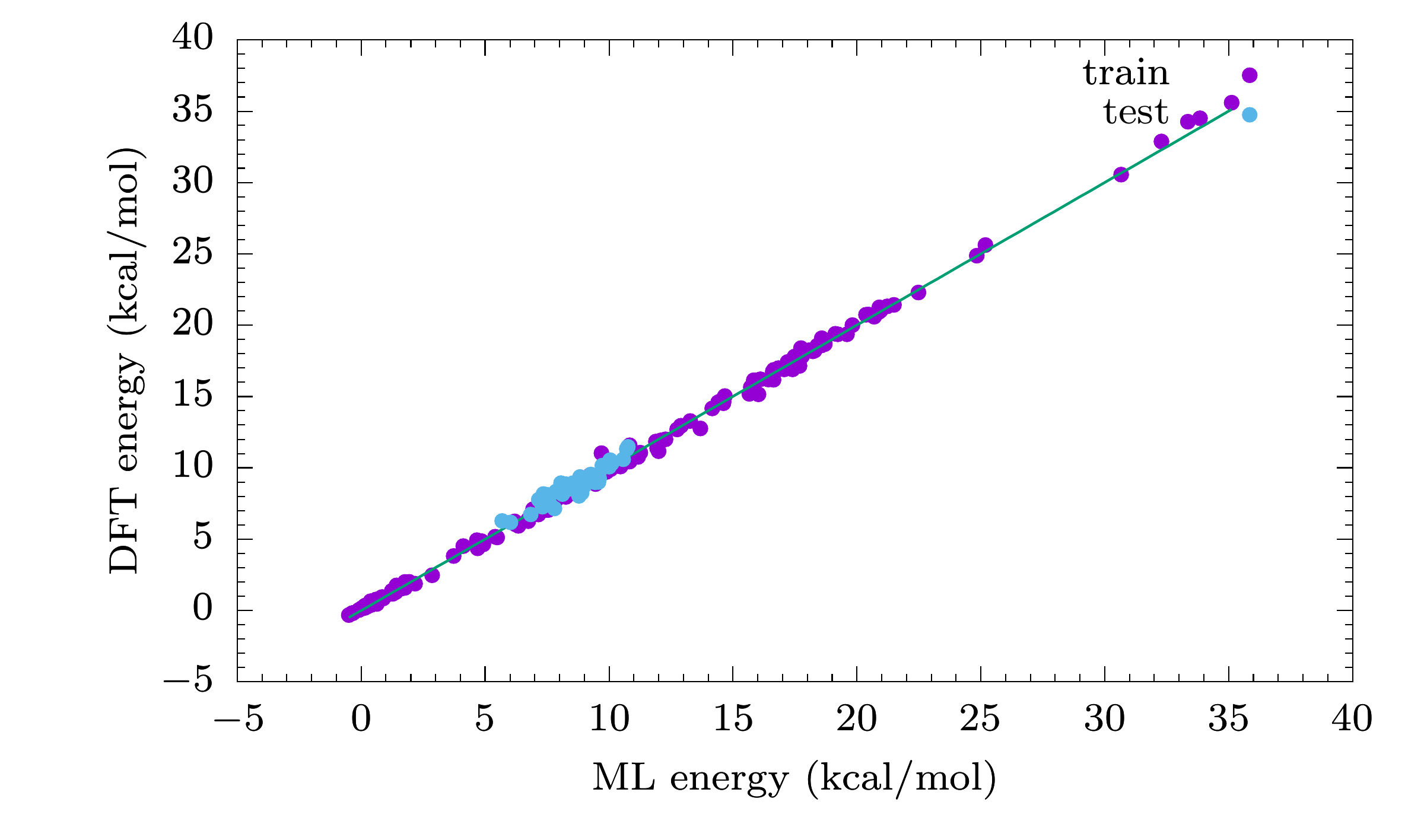}\\
    \includegraphics[scale=0.65]{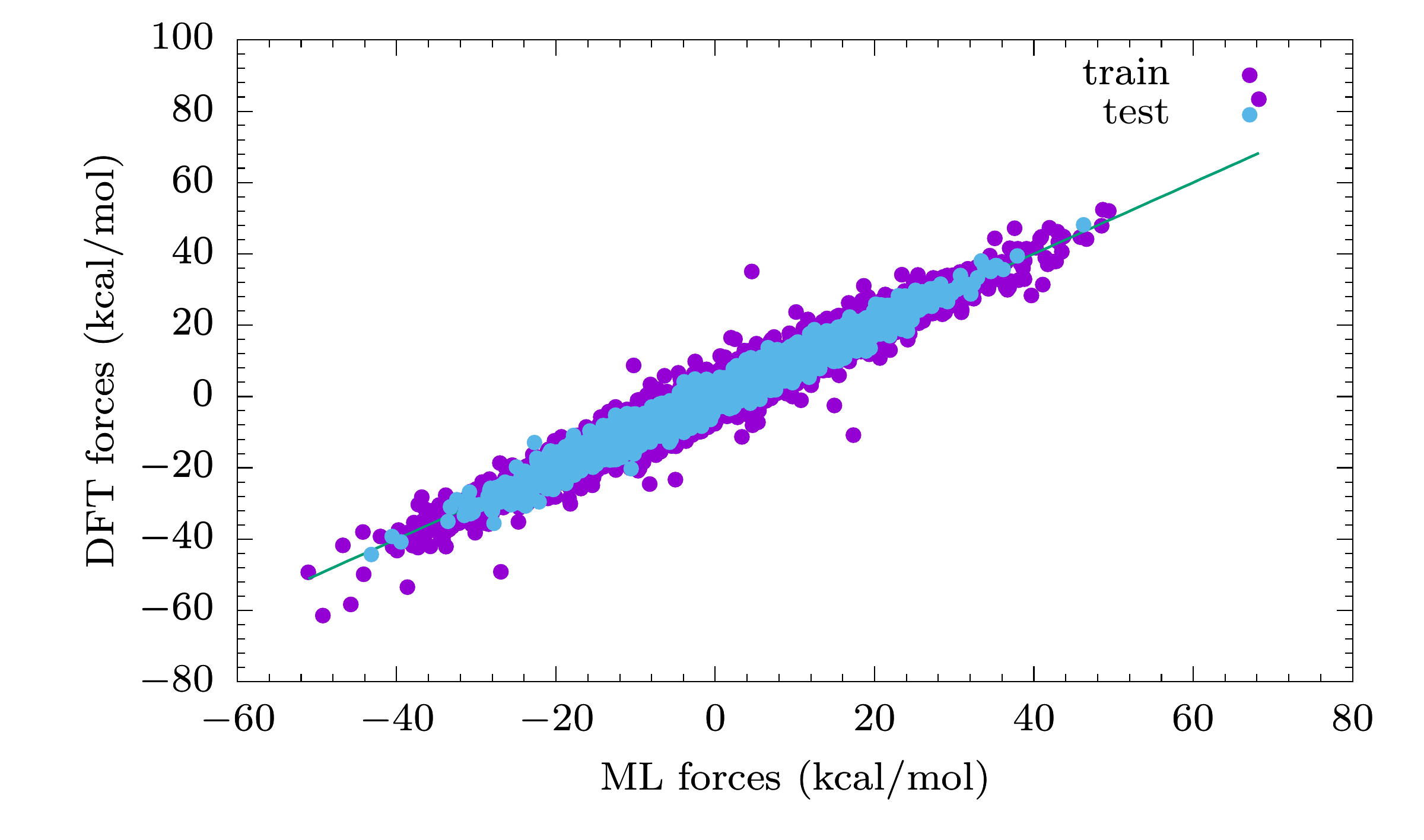}\\
    \includegraphics[scale=0.65]{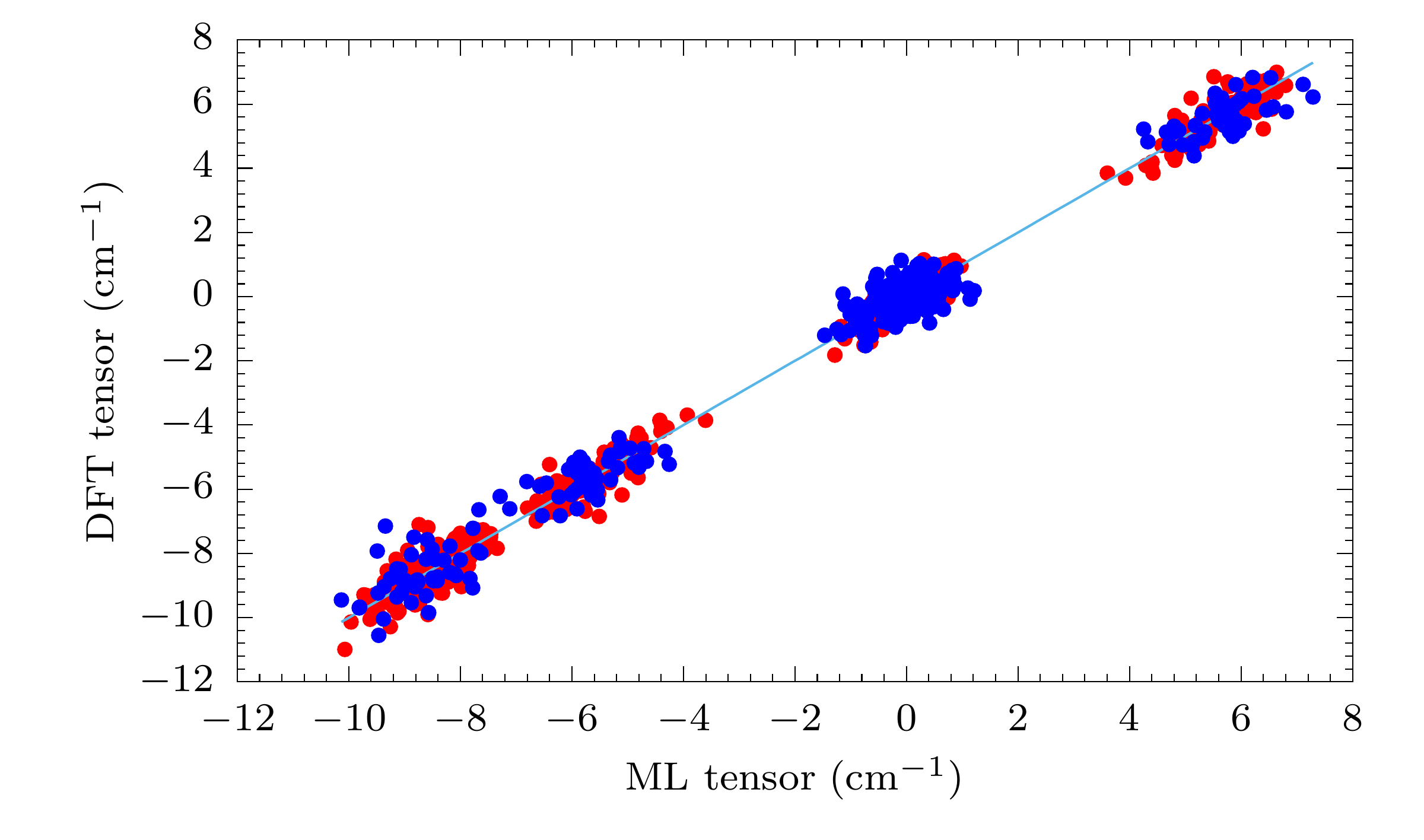}\\
    \includegraphics[scale=0.65]{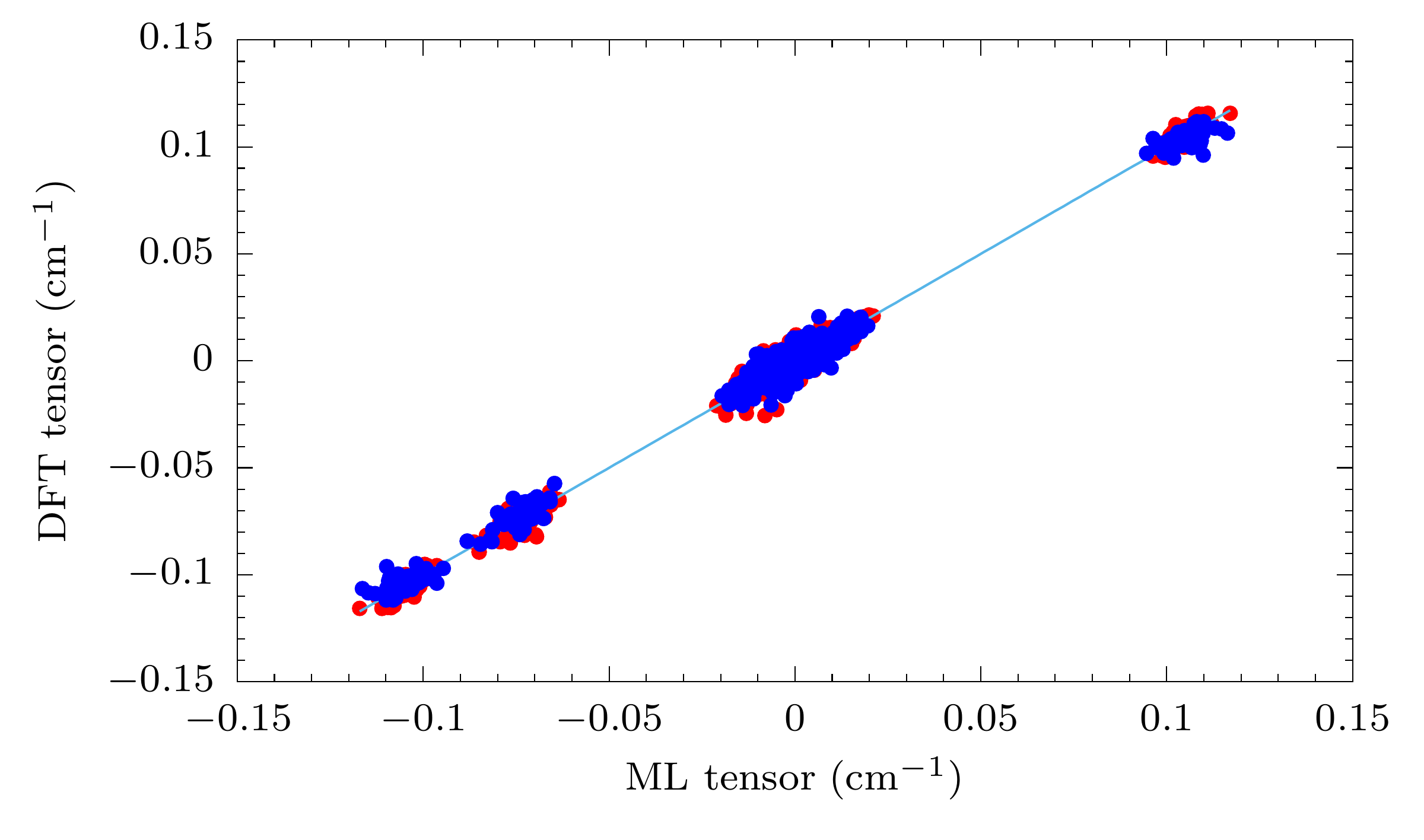}\\
    \includegraphics[scale=0.65]{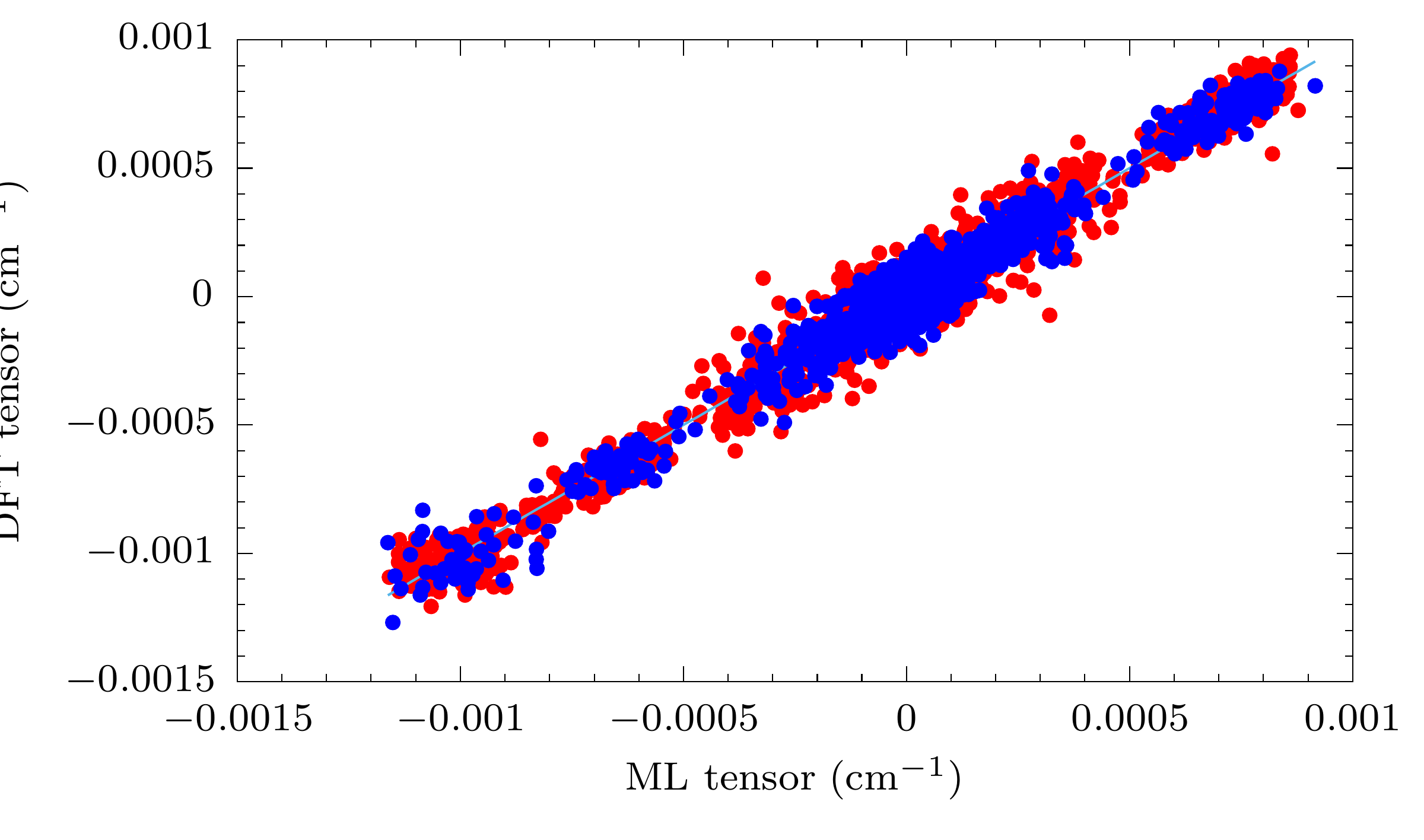}
    \caption{\textbf{Energy, forces and tensors} T=50 K}
\end{figure}

\begin{figure}[H]
    \centering
    \includegraphics[scale=0.65]{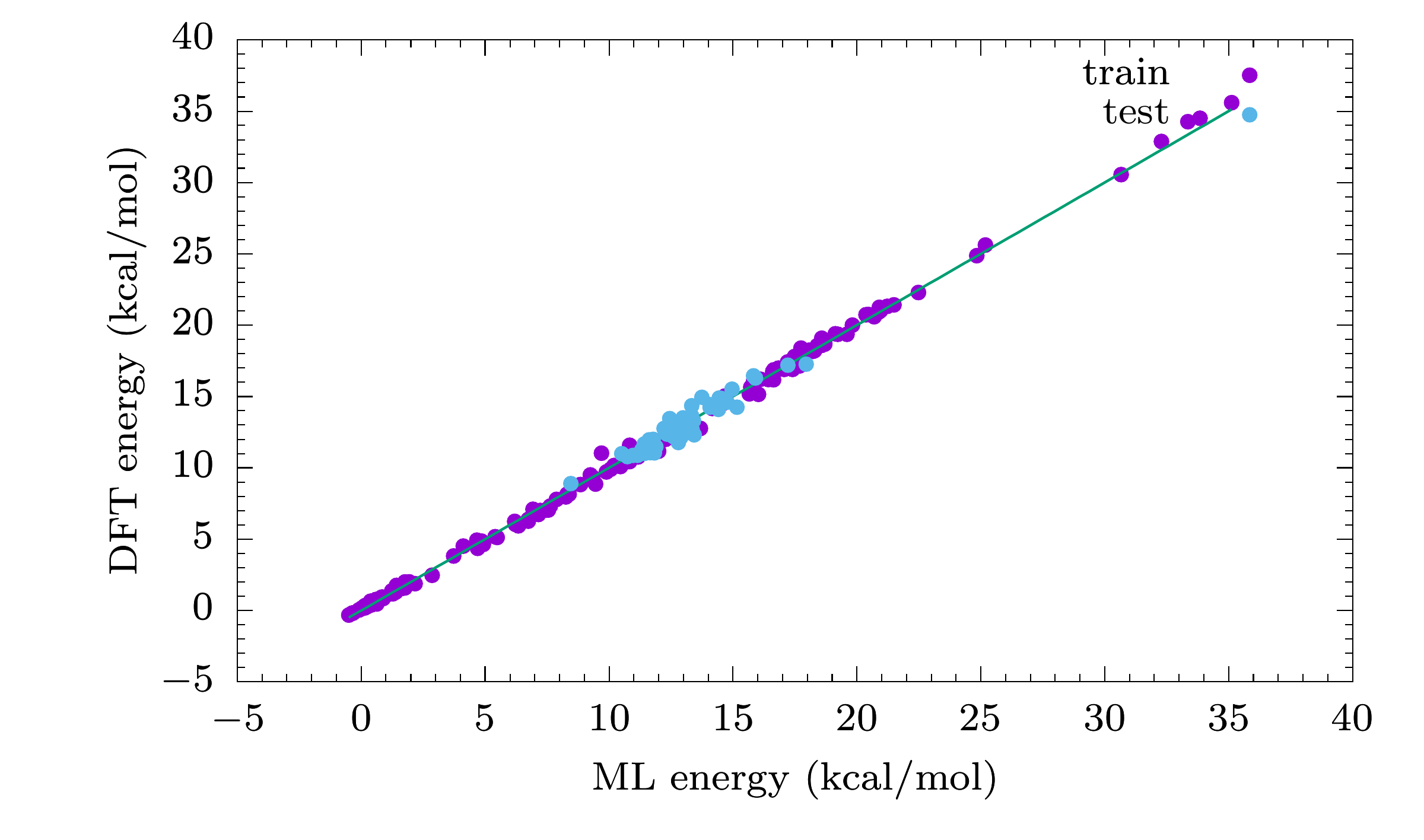}\\
    \includegraphics[scale=0.65]{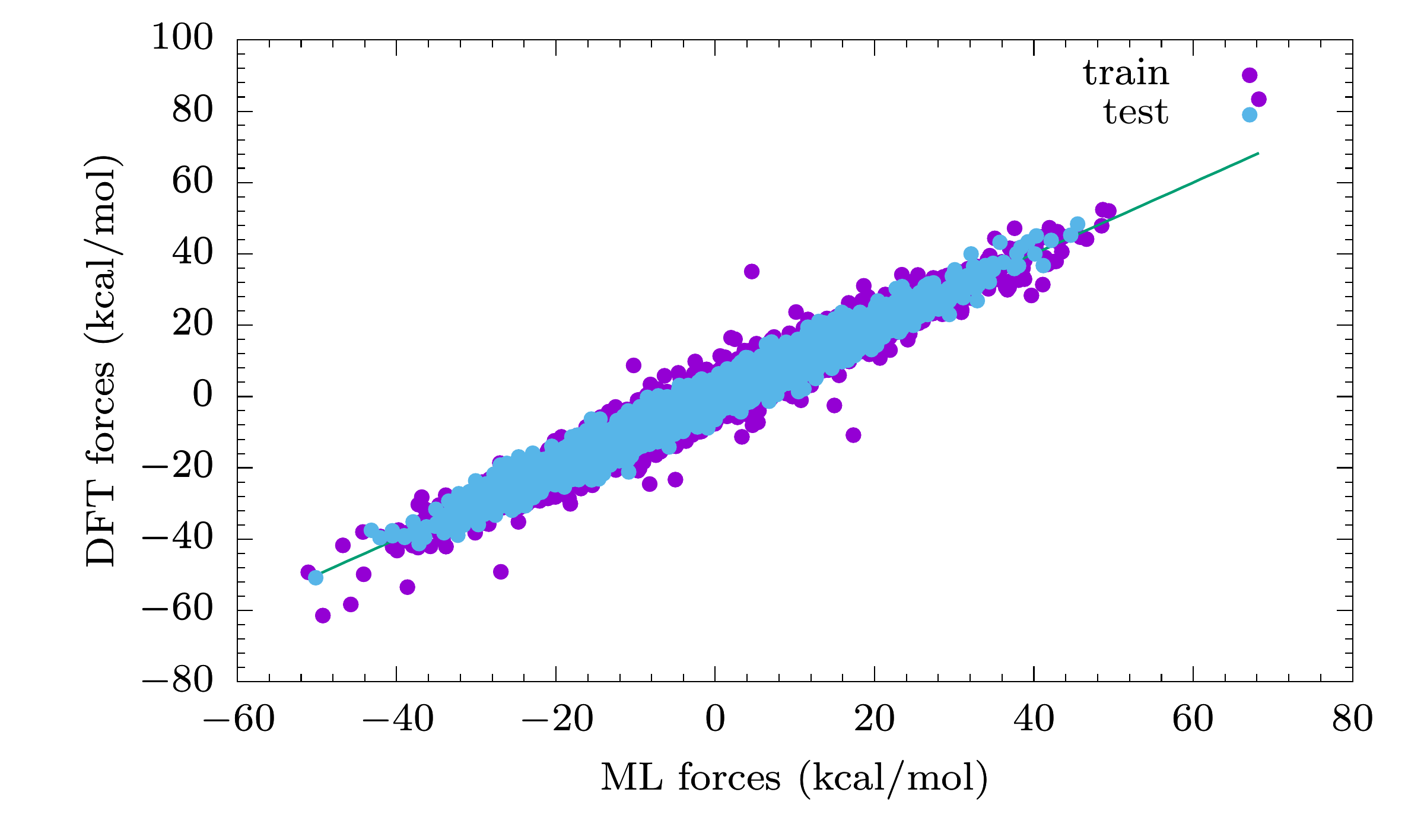}\\
    \includegraphics[scale=0.65]{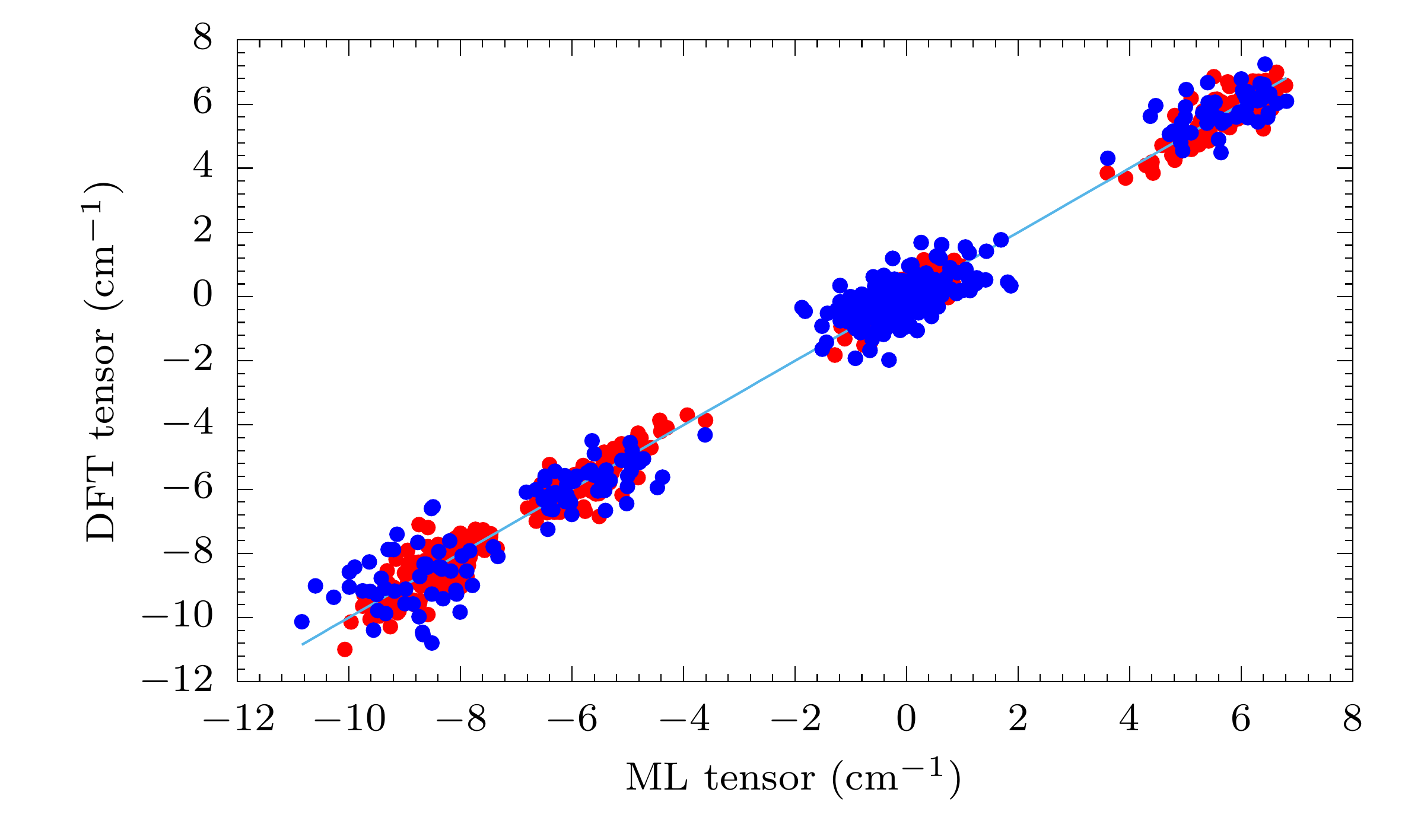}\\
    \includegraphics[scale=0.65]{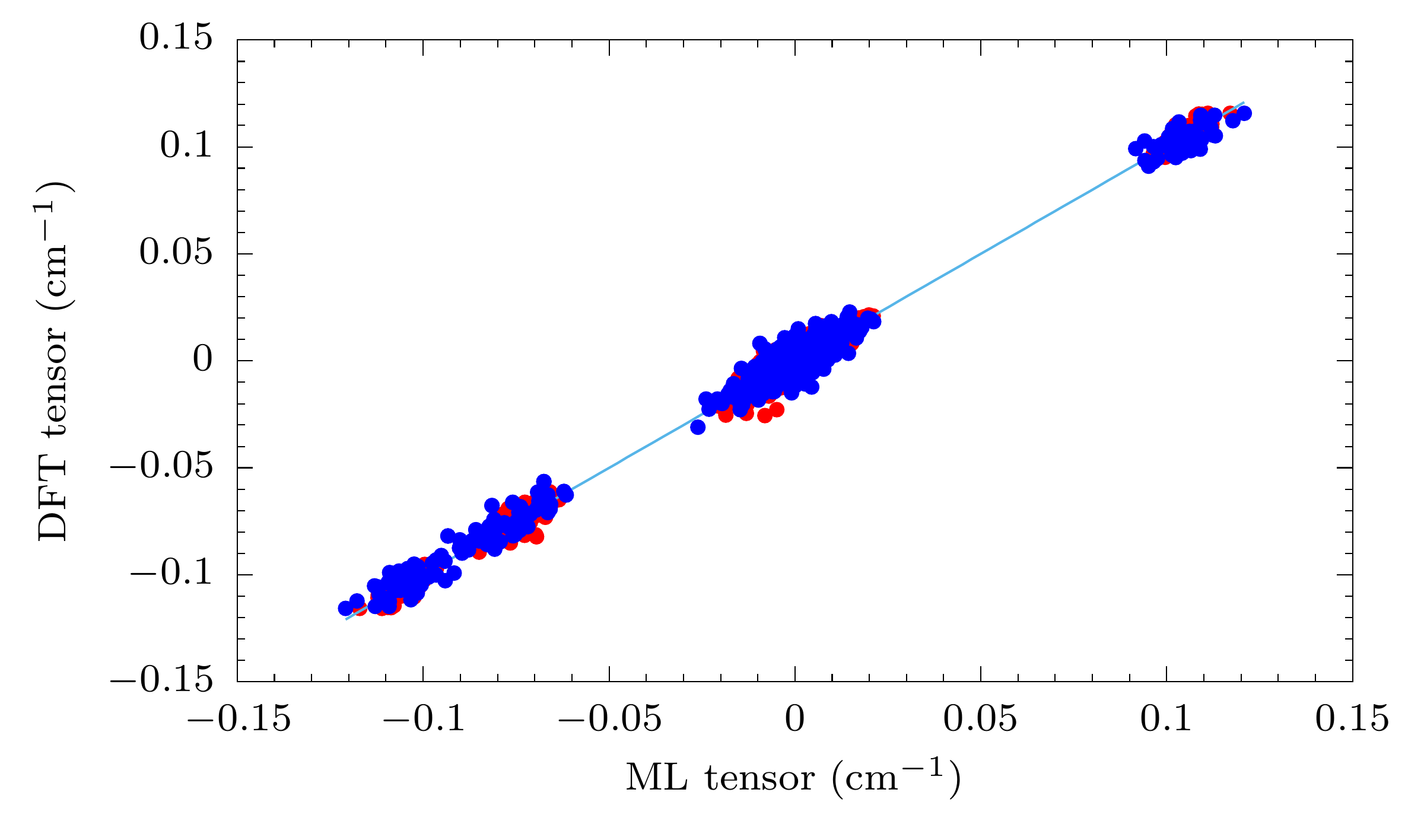}\\
    \includegraphics[scale=0.65]{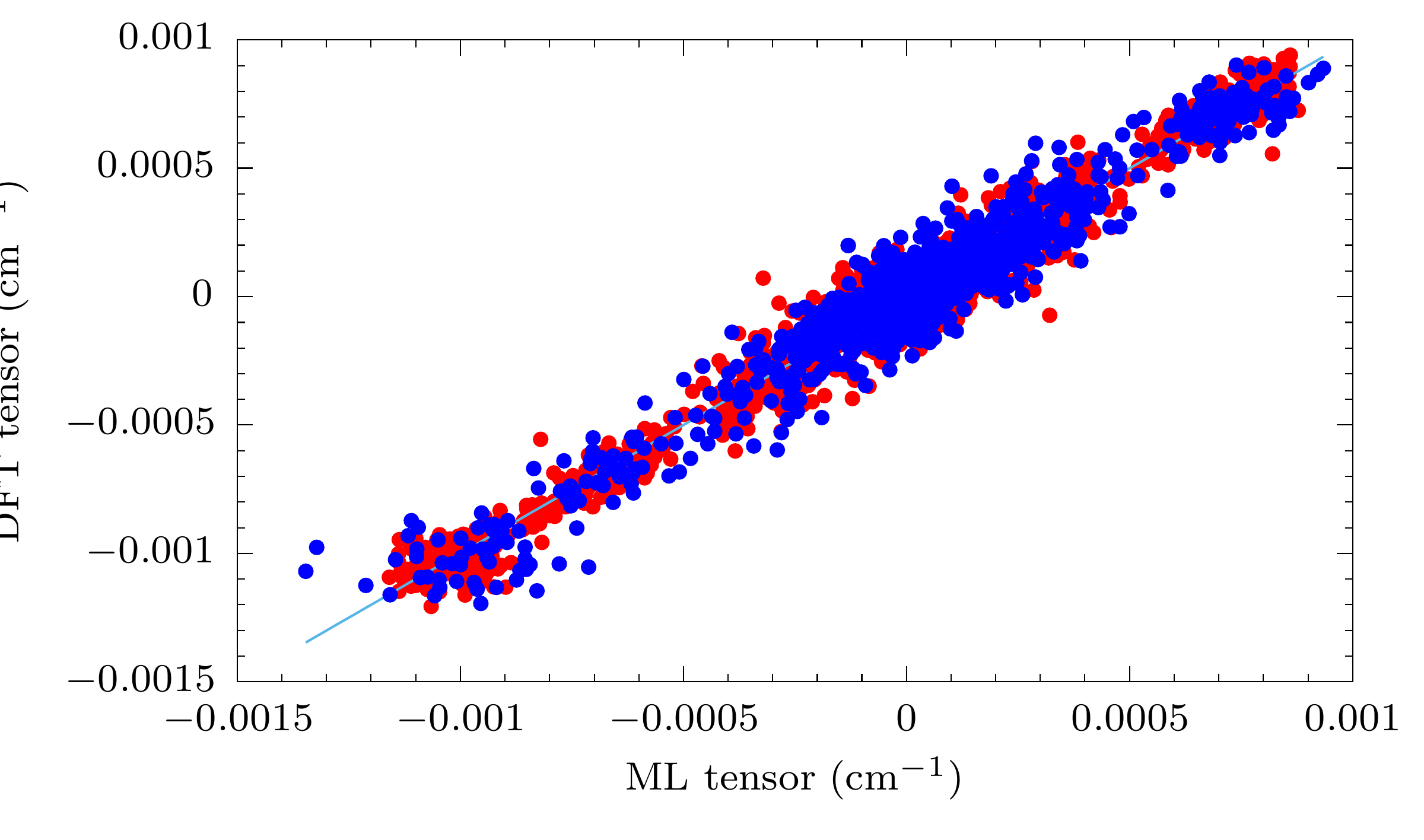}
    \caption{\textbf{Energy, forces and tensors} T=75 K}
\end{figure}

\begin{figure}[H]
    \centering
    \includegraphics[scale=0.65]{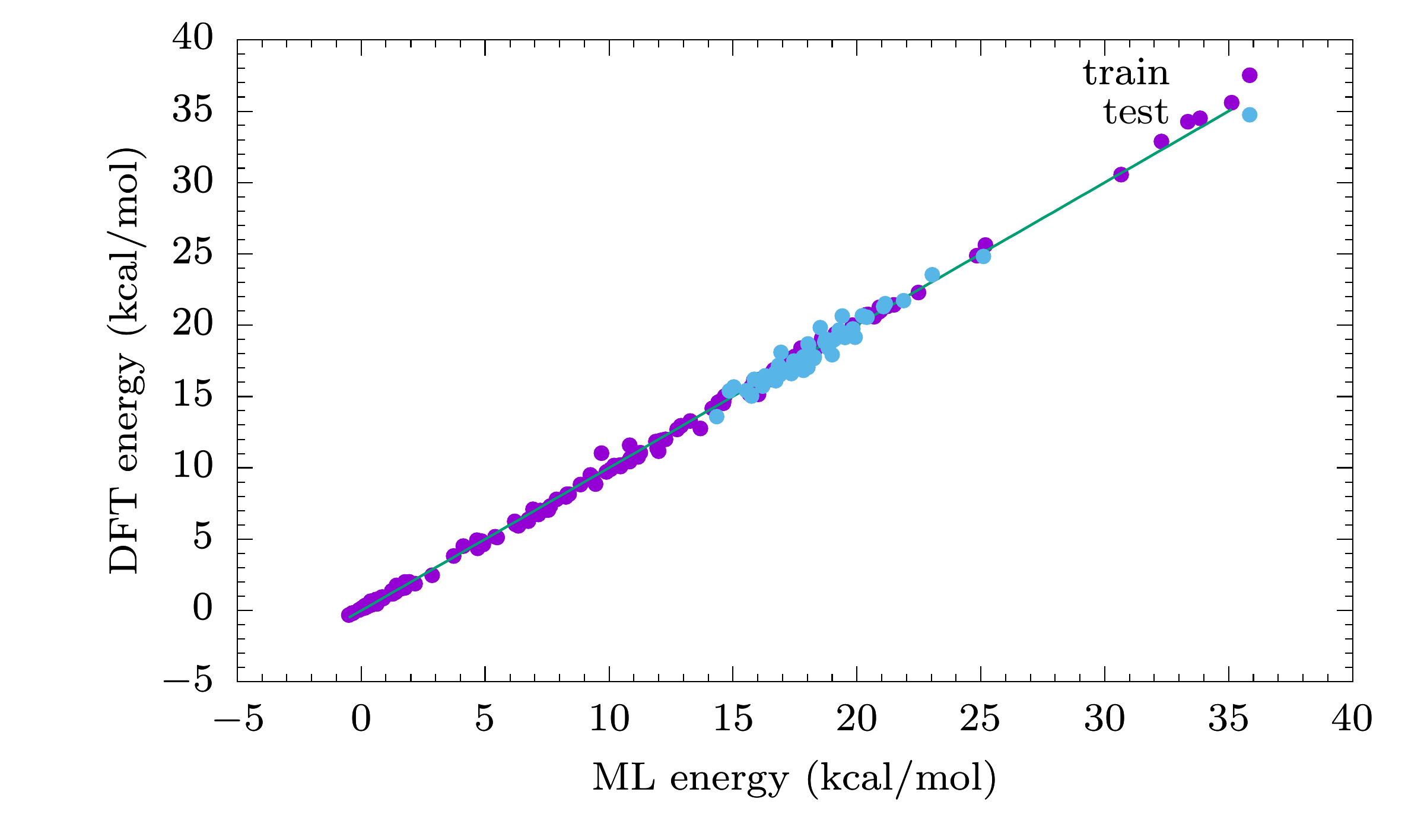}\\
    \includegraphics[scale=0.65]{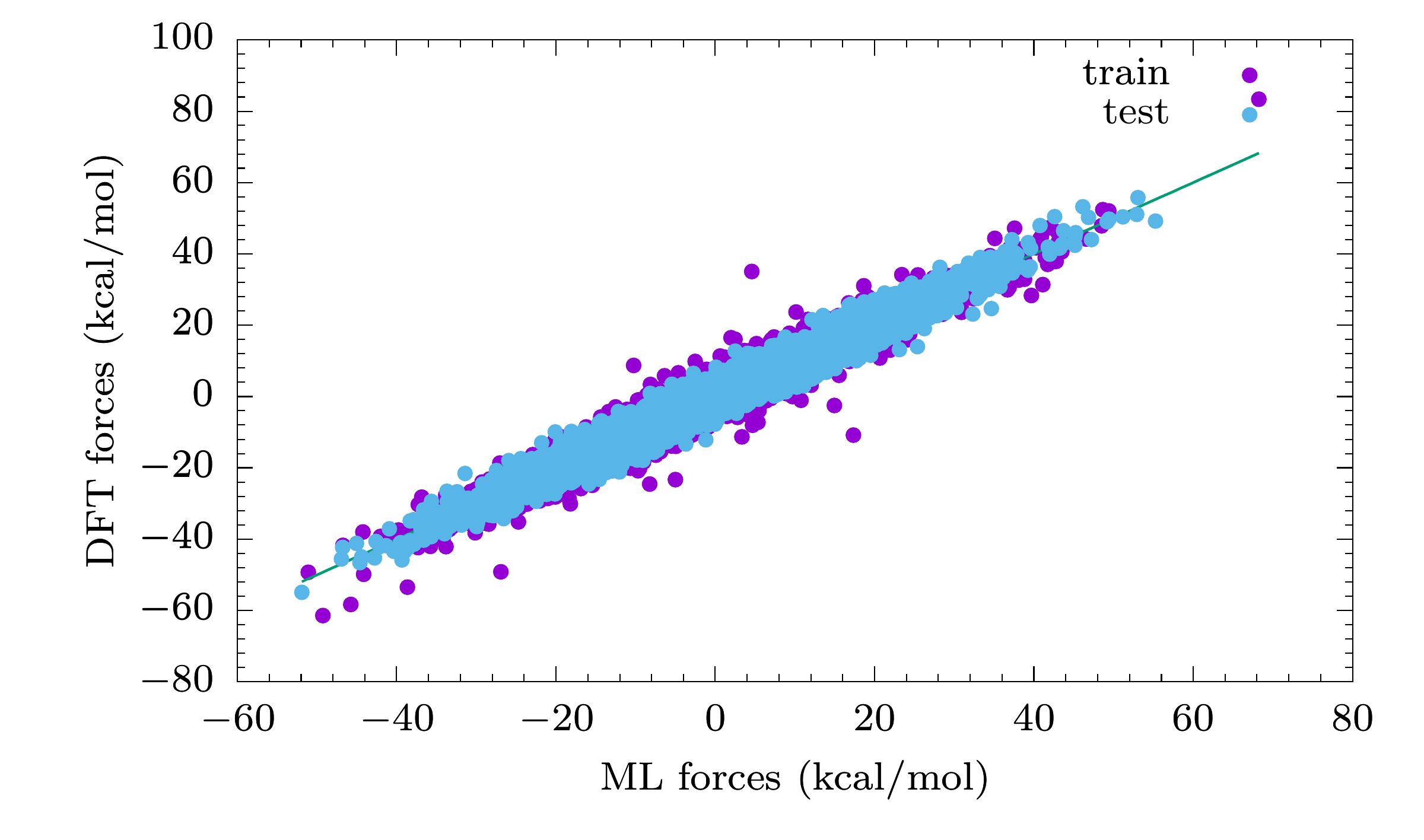}\\
    \includegraphics[scale=0.65]{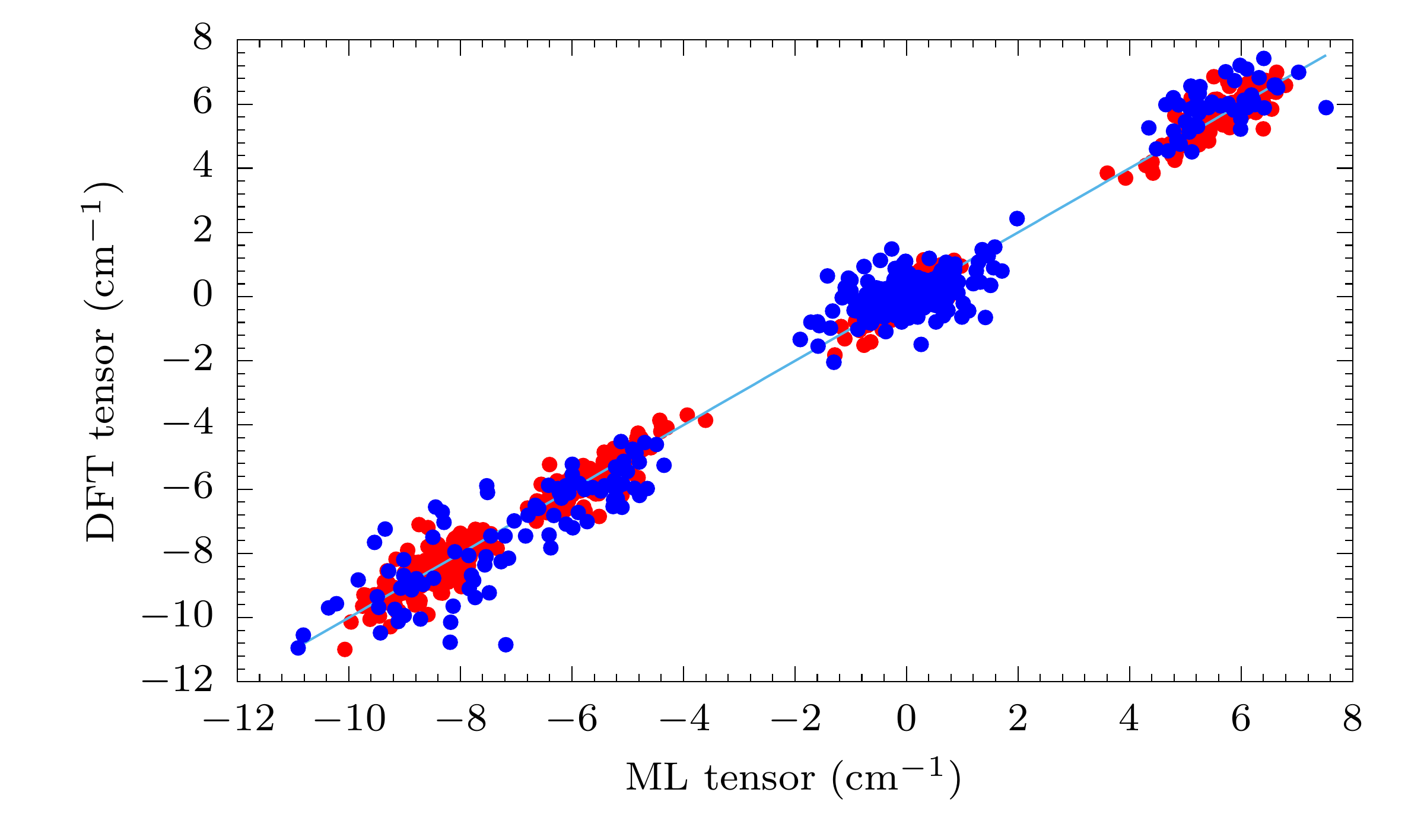}\\
    \includegraphics[scale=0.65]{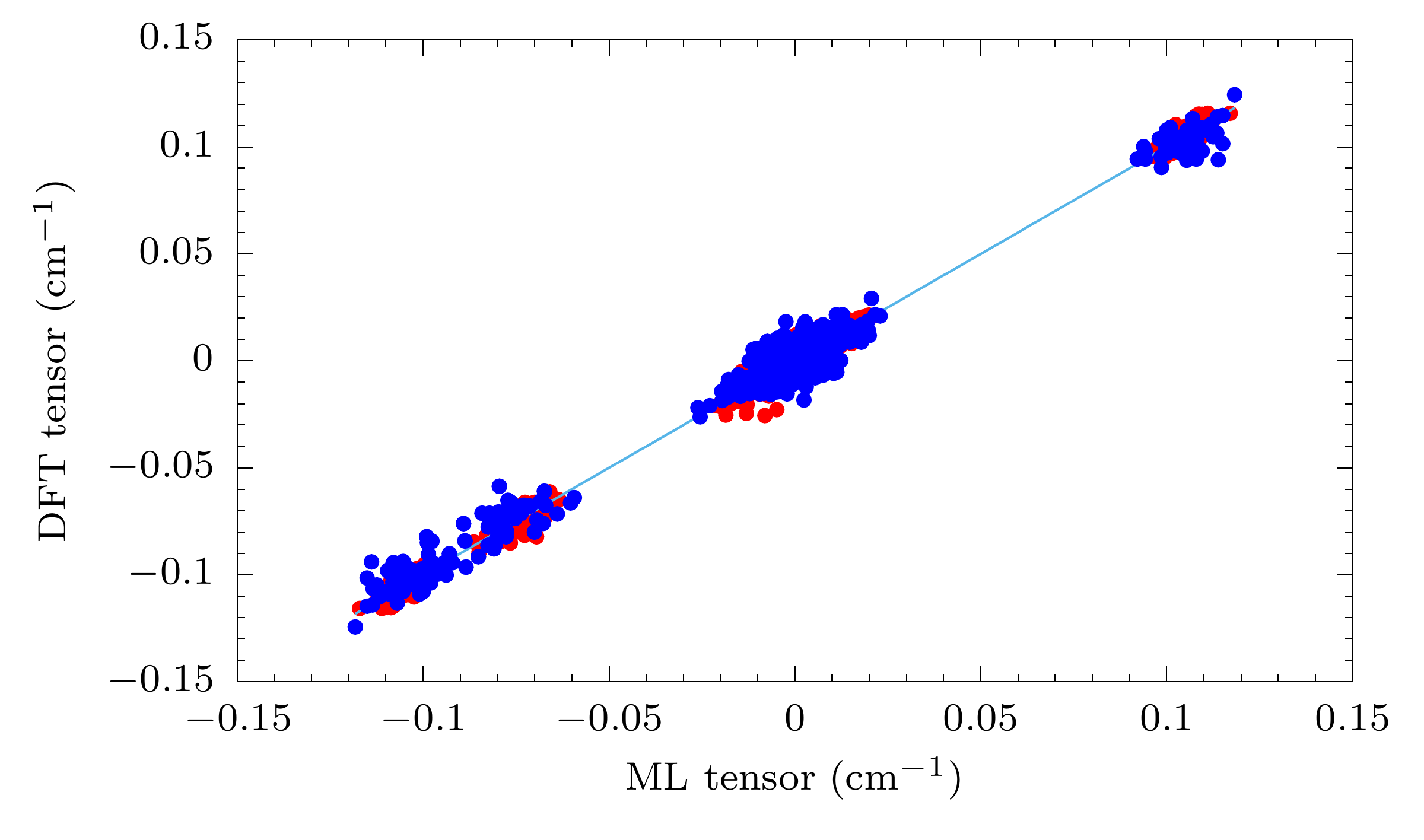}\\
    \includegraphics[scale=0.65]{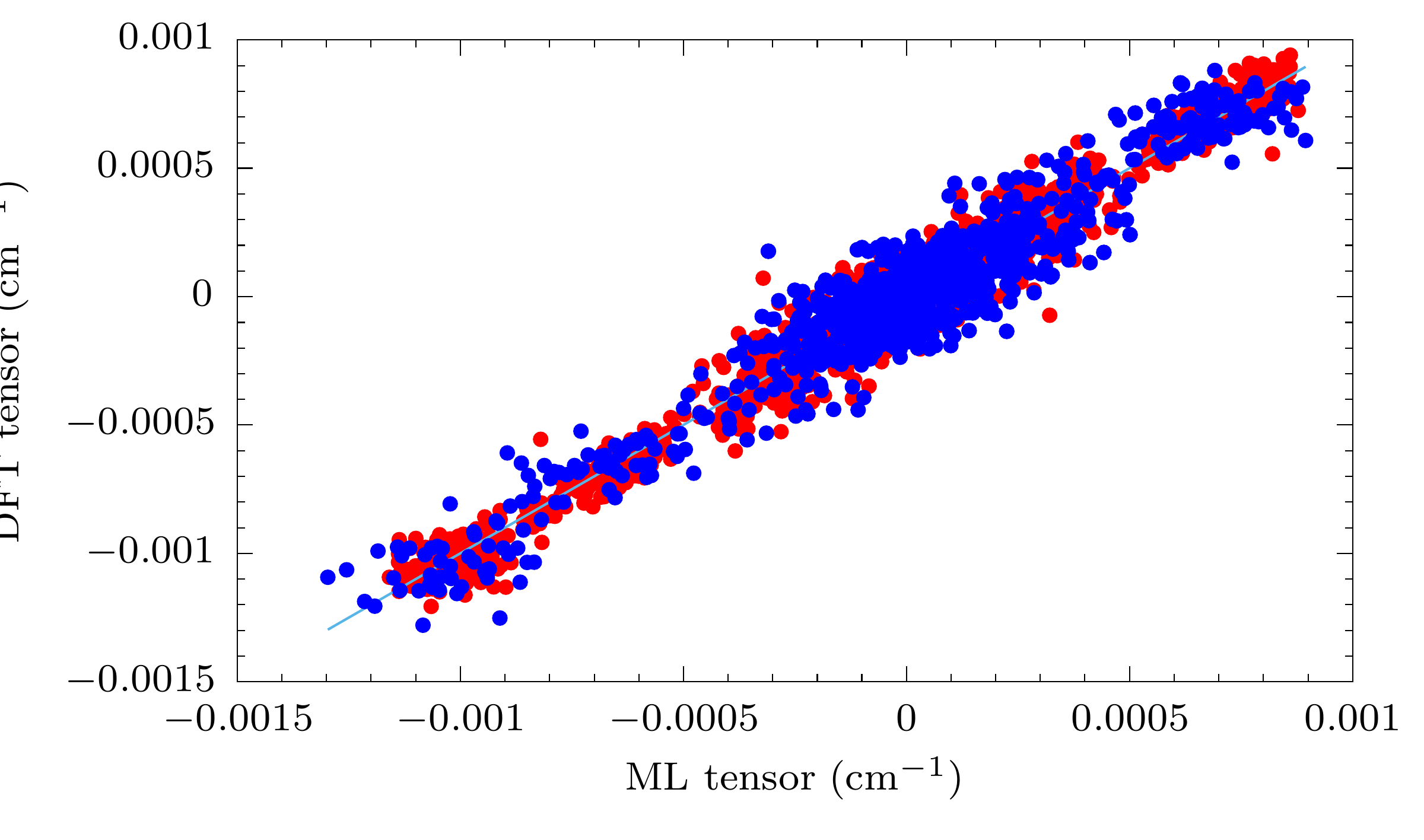}
    \caption{\textbf{Energy, forces and tensors} T=100 K}
\end{figure}

 \begin{figure}[H]
     \centering
     \includegraphics[scale=0.65]{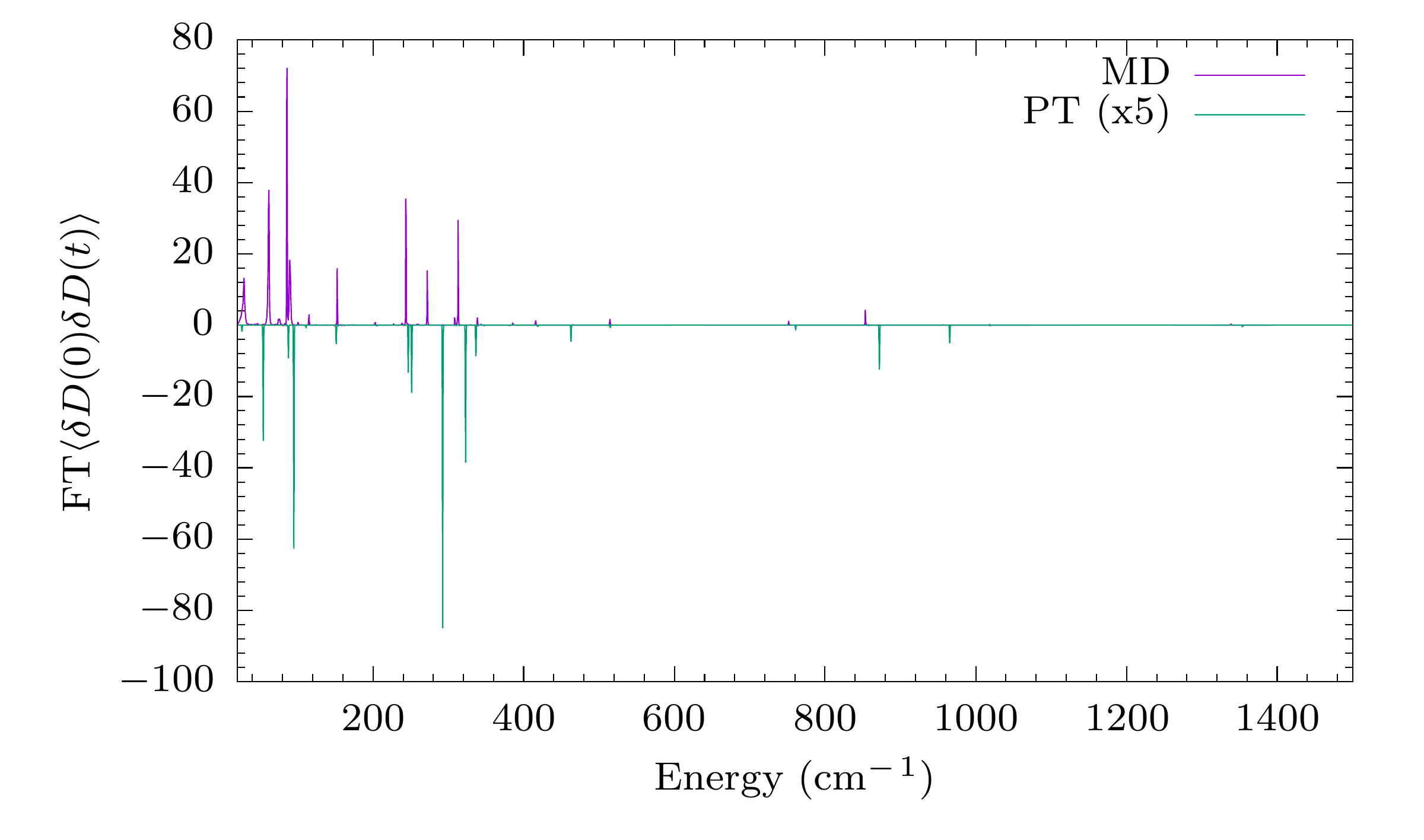}
     \includegraphics[scale=0.65]{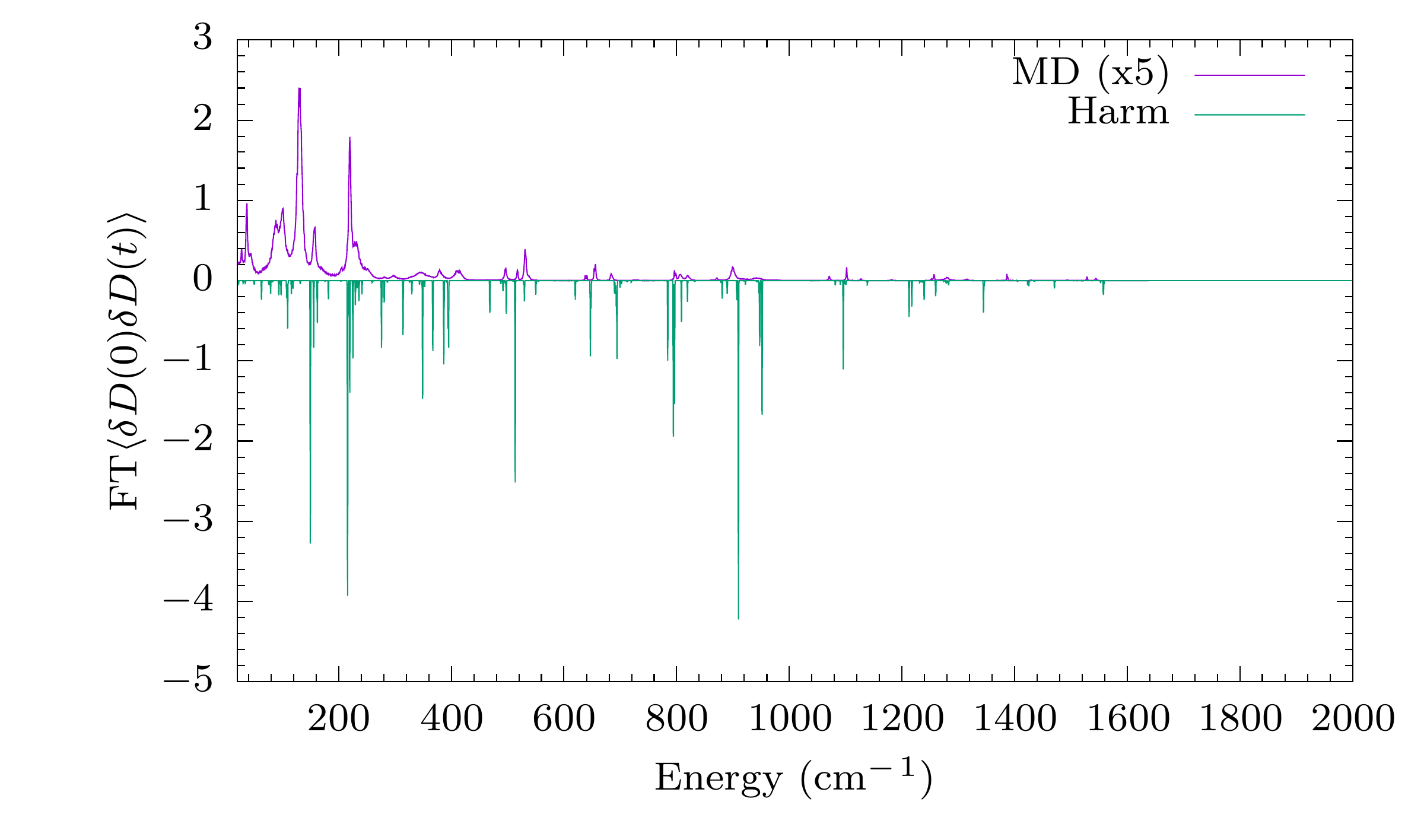}\\
     \includegraphics[scale=0.65]{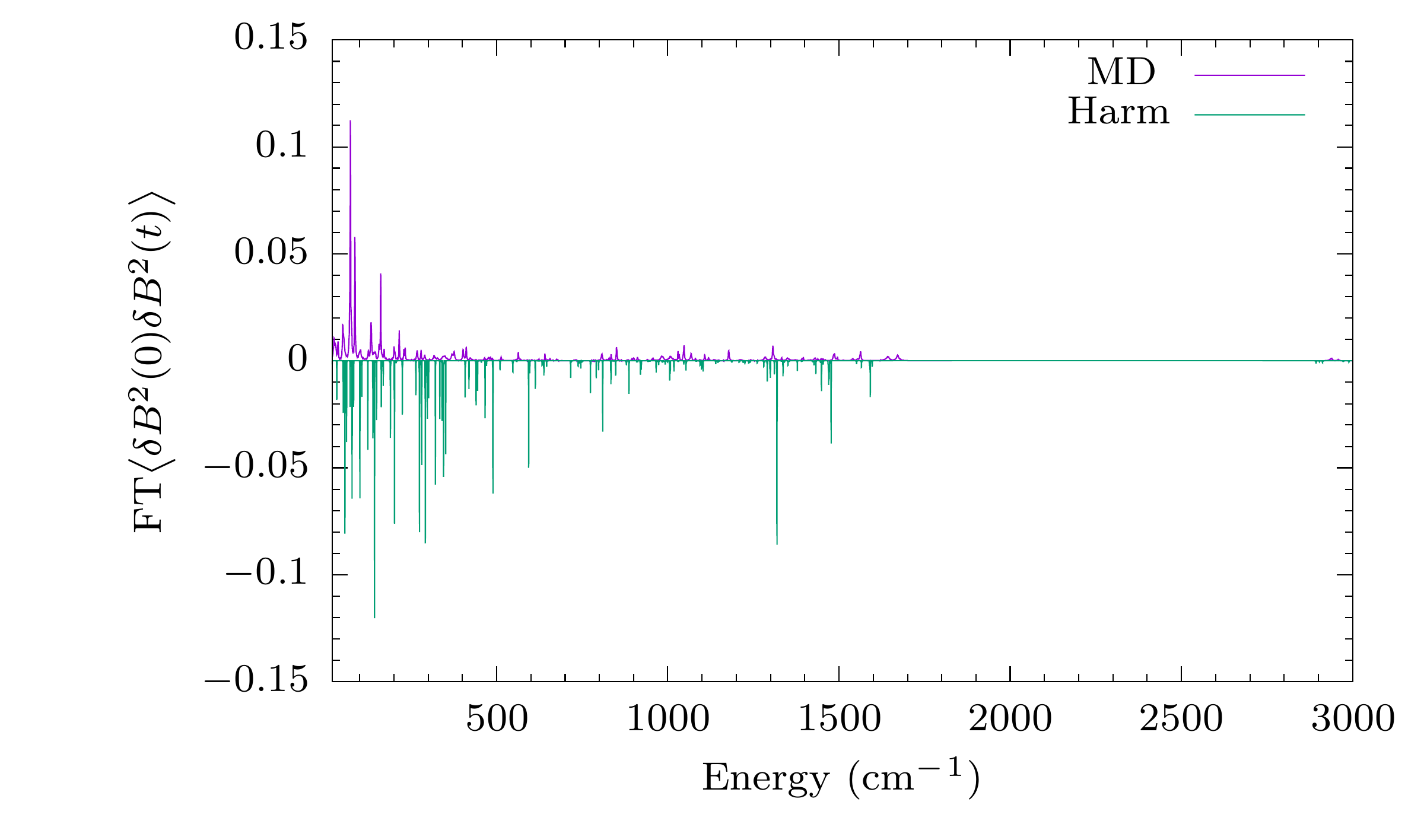}\\
     \includegraphics[scale=0.65]{Images_ESI/Corr_functions/Comp_1_25K/FT_final-1.png}\\
     \includegraphics[scale=0.65]{Images_ESI/Corr_functions/Comp_1_25K/FT_final-1.png}
     \caption{\textbf{Fourier transforms of the correlation functions at 25 K: comparison with the harmonic case}. Results for \textbf{1},\textbf{2} and \textbf{3} in order. For \textbf{3} the results for $l=2,4,6$ are reported, in order.}
 \end{figure}

  \begin{figure}[H]
     \centering
     \includegraphics[scale=0.65]{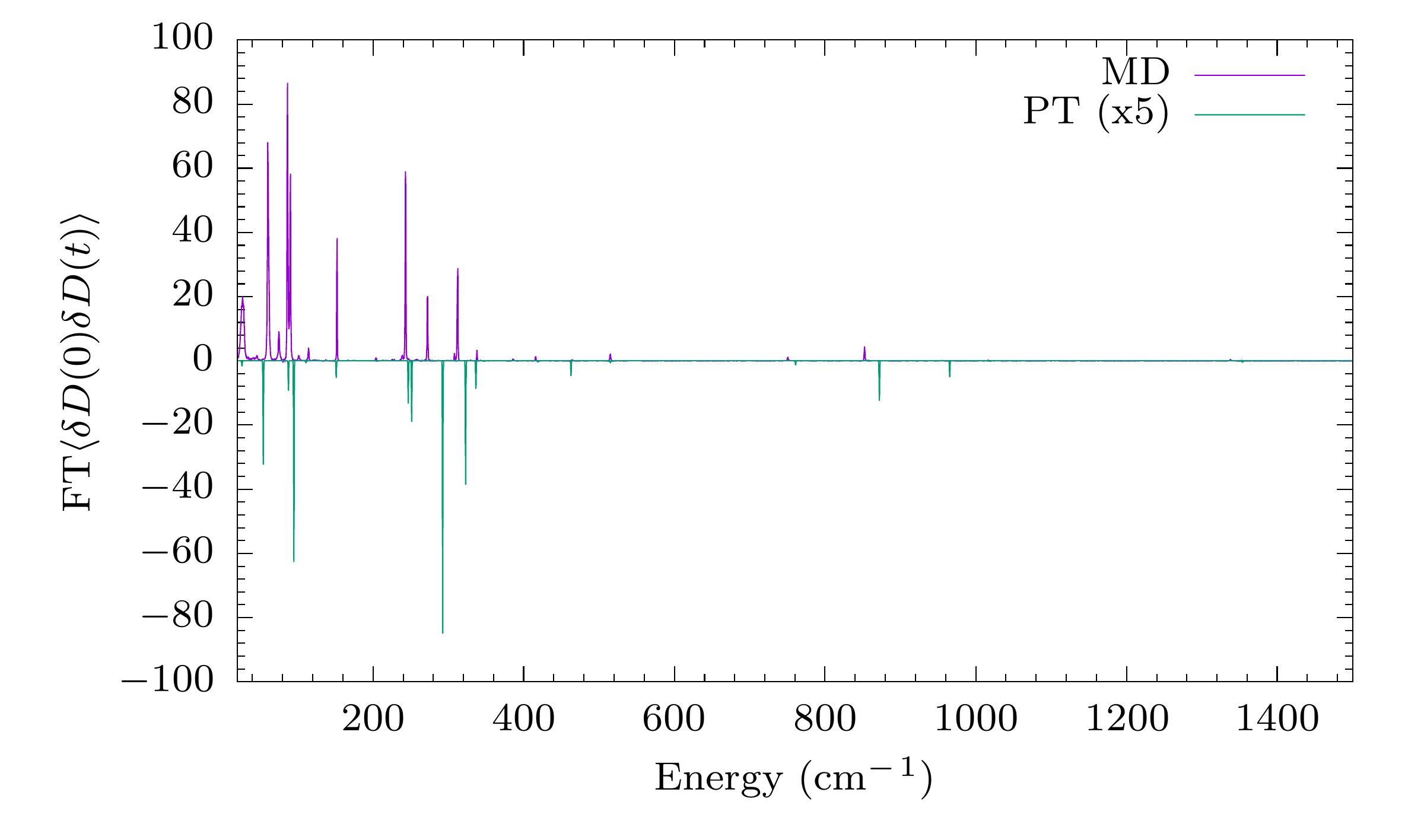}
     \includegraphics[scale=0.65]{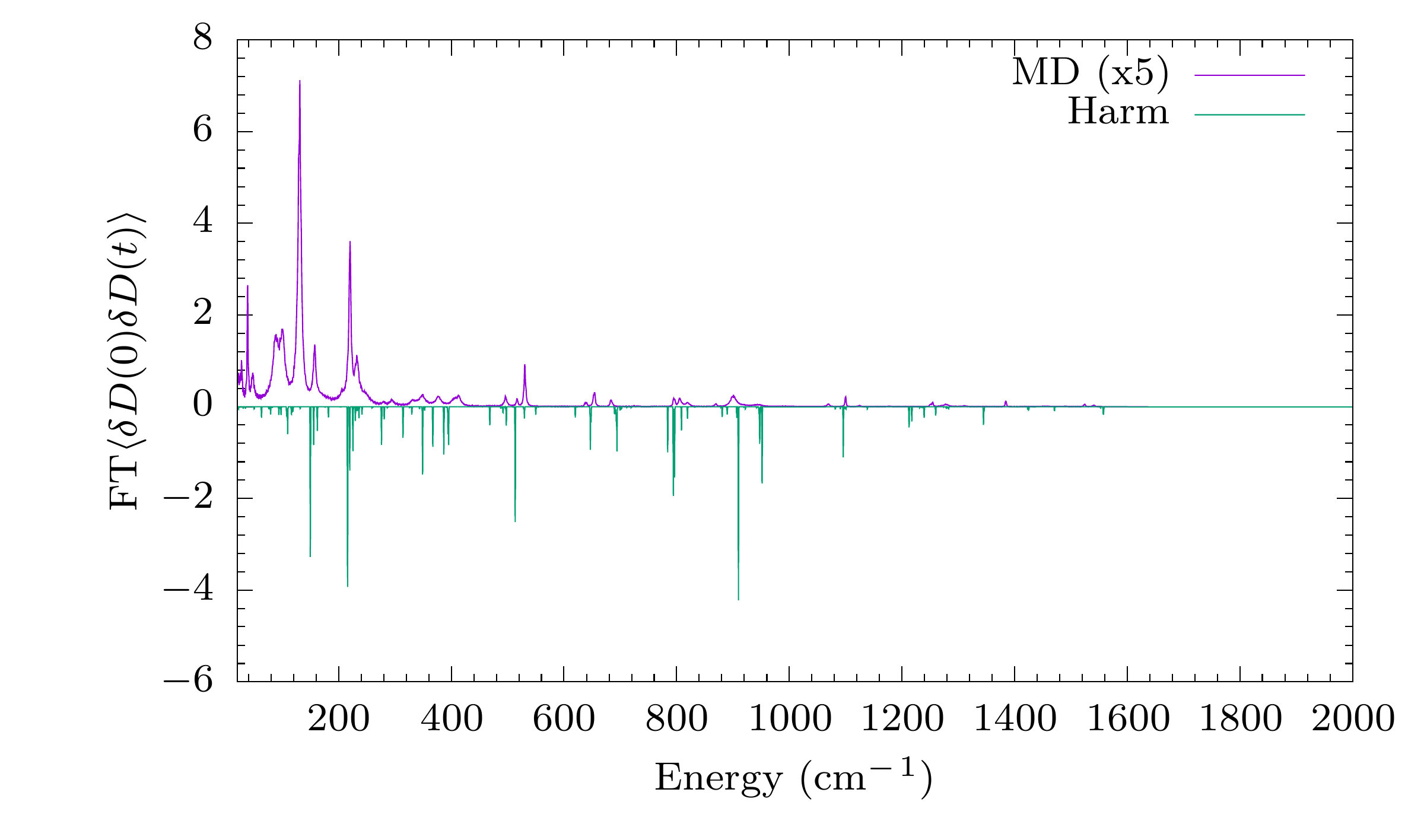}\\
     \includegraphics[scale=0.65]{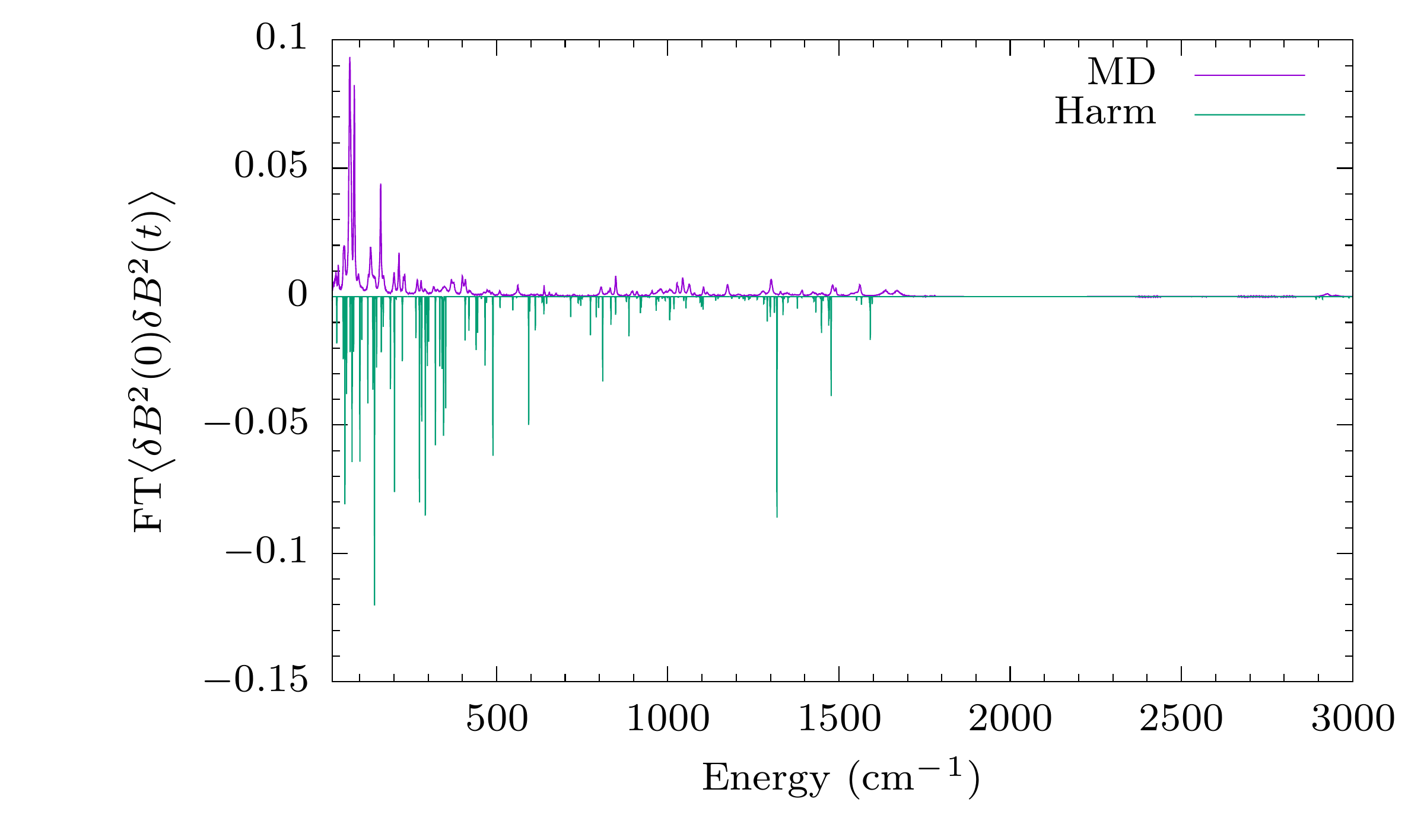}\\
     \includegraphics[scale=0.65]{Images_ESI/Corr_functions/Comp_1_50K/FT_final-1.png}\\
     \includegraphics[scale=0.65]{Images_ESI/Corr_functions/Comp_1_50K/FT_final-1.png}
     \caption{\textbf{Fourier transforms of the correlation functions at 50 K: comparison with the harmonic case}. Results for \textbf{1},\textbf{2} and \textbf{3} in order. For \textbf{3} the results for $l=2,4,6$ are reported, in order.}
 \end{figure}

  \begin{figure}[H]
     \centering
     \includegraphics[scale=0.65]{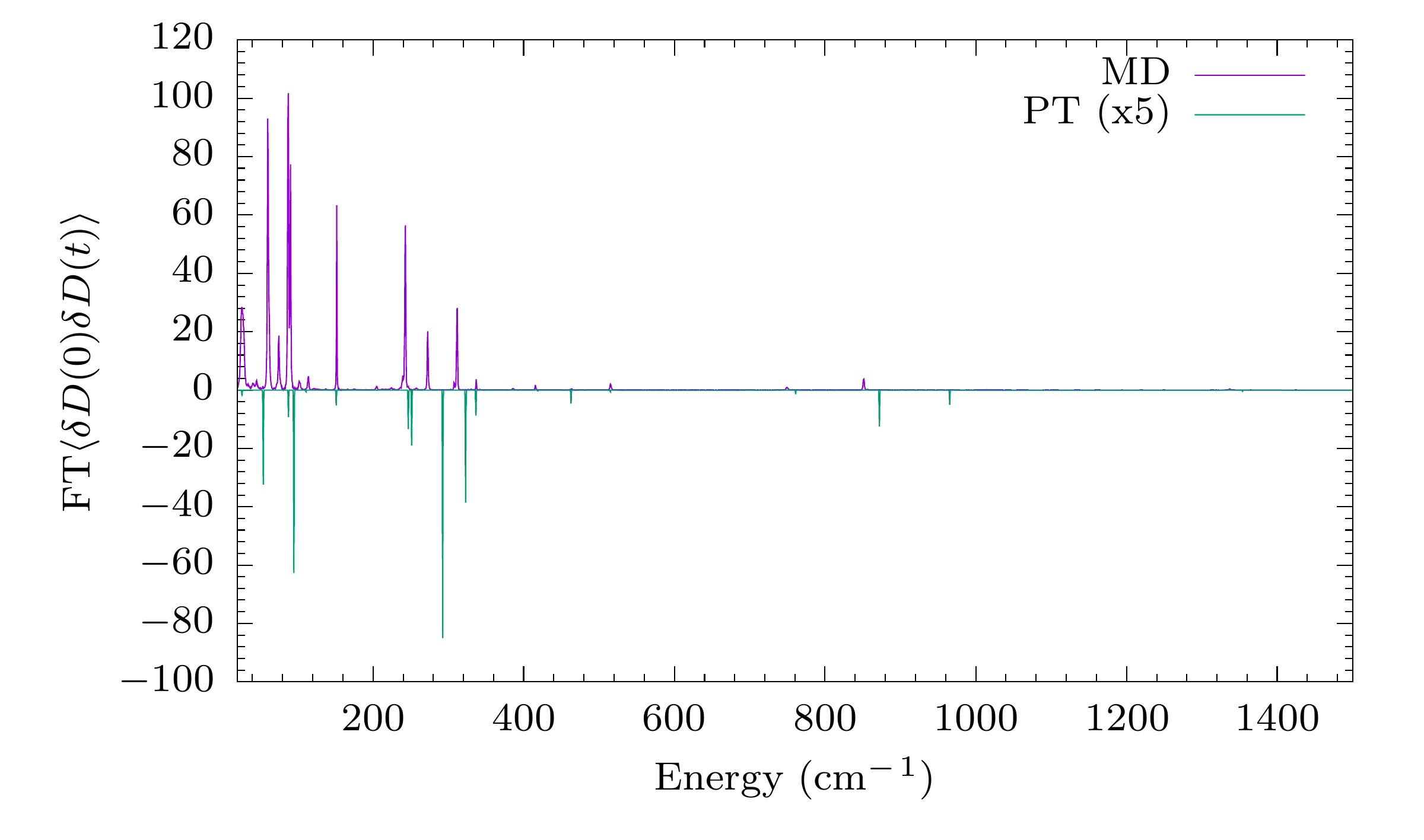}
     \includegraphics[scale=0.65]{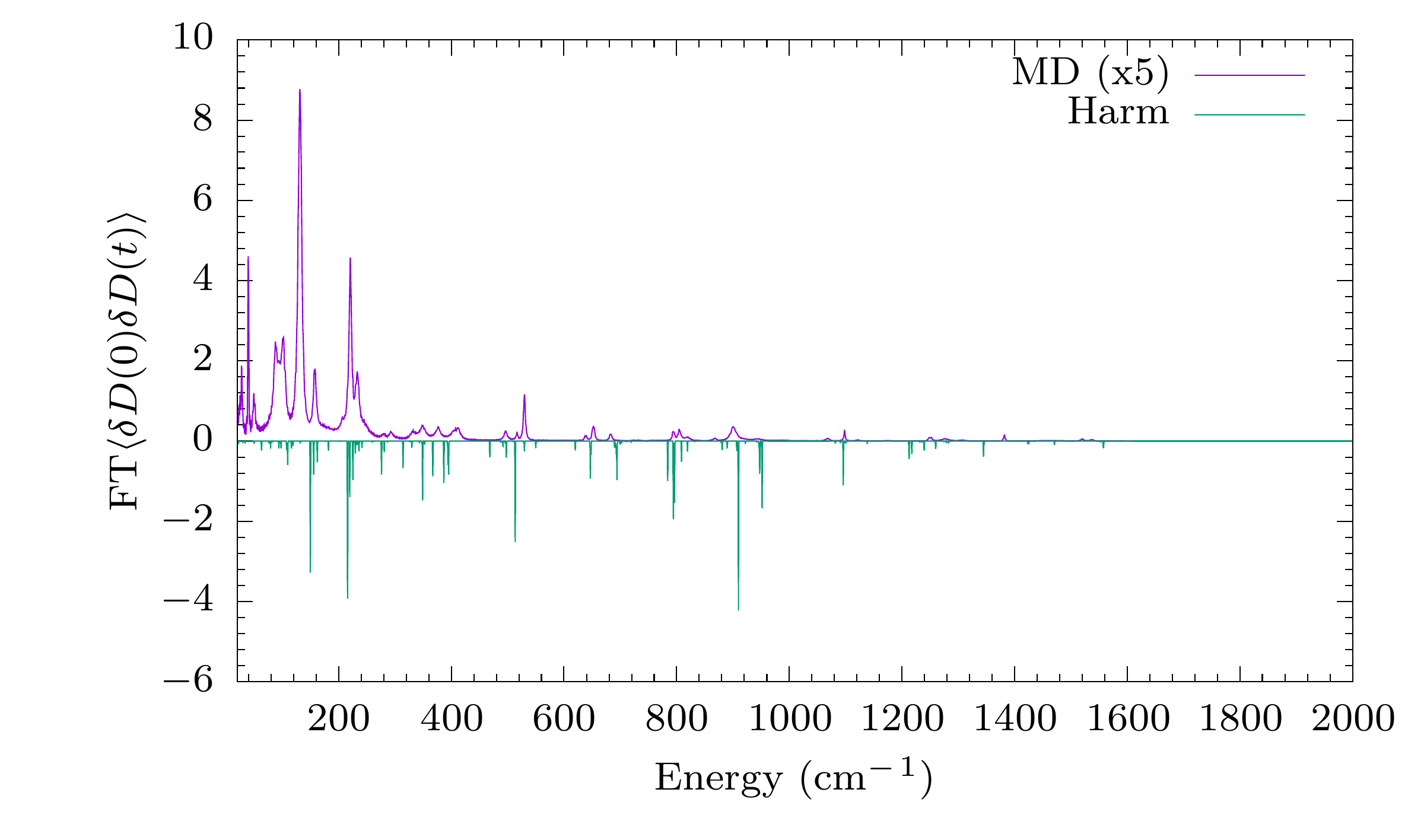}\\
     \includegraphics[scale=0.65]{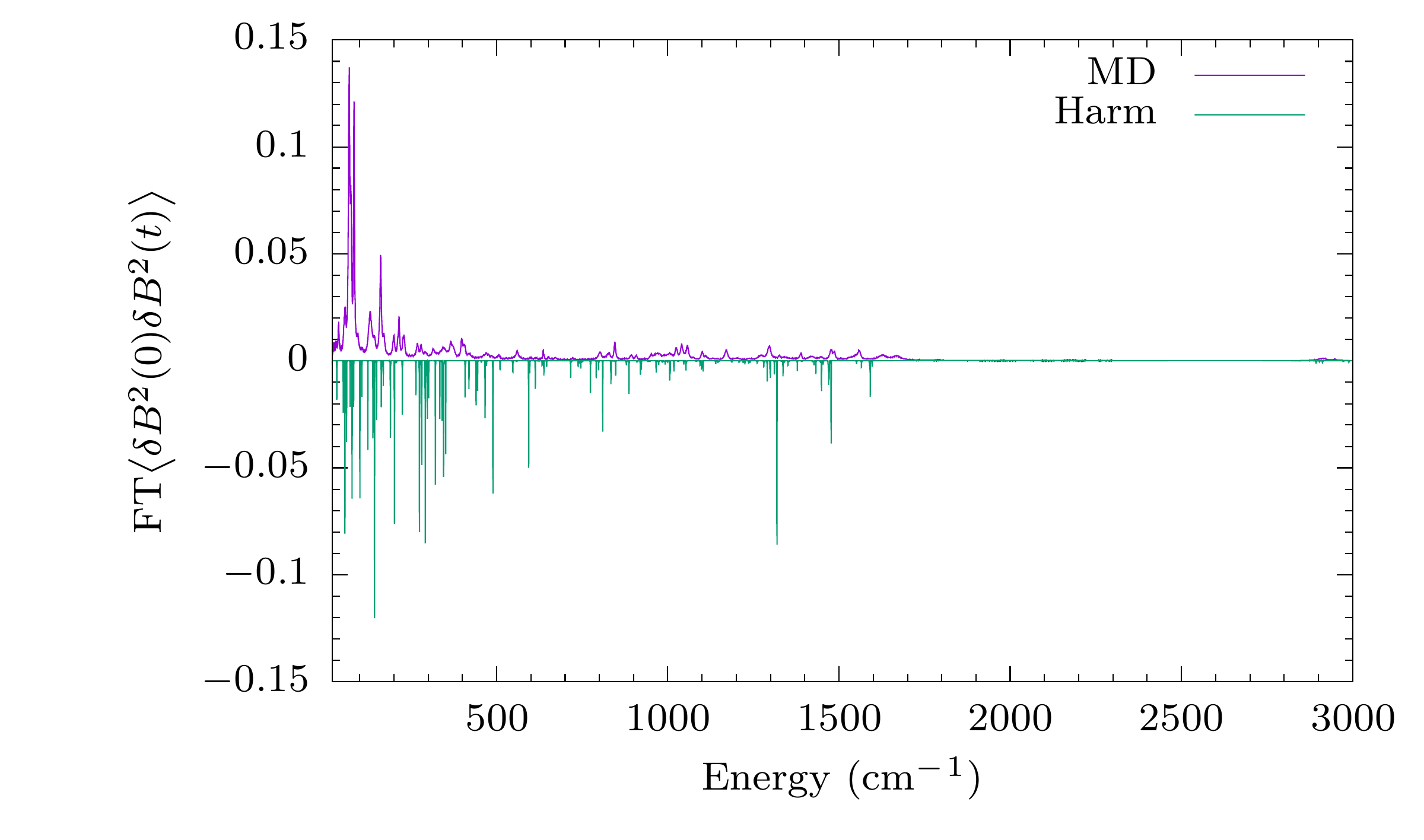}\\
     \includegraphics[scale=0.65]{Images_ESI/Corr_functions/Comp_1_75K/FT_final-1.png}\\
     \includegraphics[scale=0.65]{Images_ESI/Corr_functions/Comp_1_75K/FT_final-1.png}
     \caption{\textbf{Fourier transforms of the correlation functions at 75 K: comparison with the harmonic case}. Results for \textbf{1},\textbf{2} and \textbf{3} in order. For \textbf{3} the results for $l=2,4,6$ are reported, in order.}
 \end{figure}

  \begin{figure}[H]
     \centering
     \includegraphics[scale=0.65]{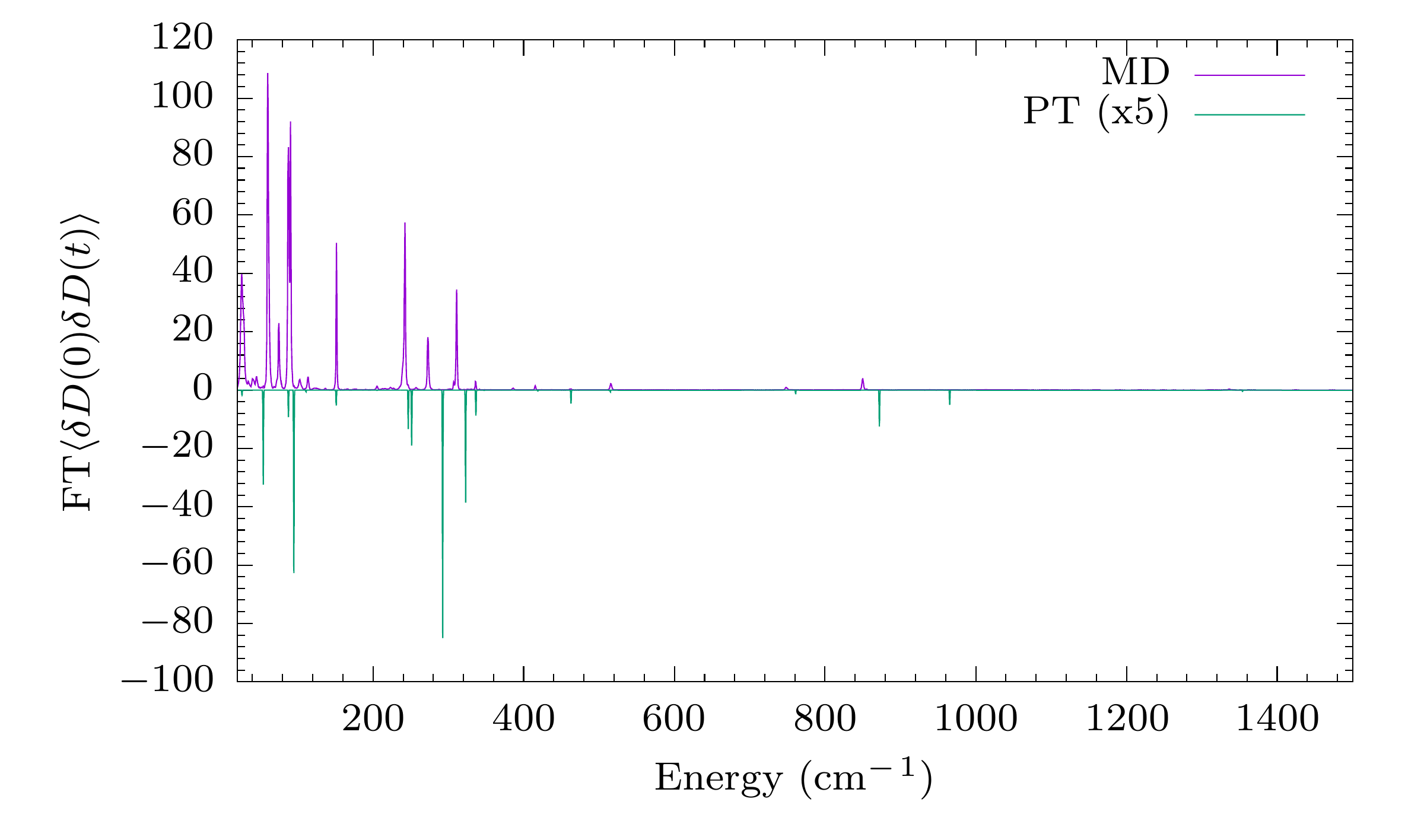}
     \includegraphics[scale=0.65]{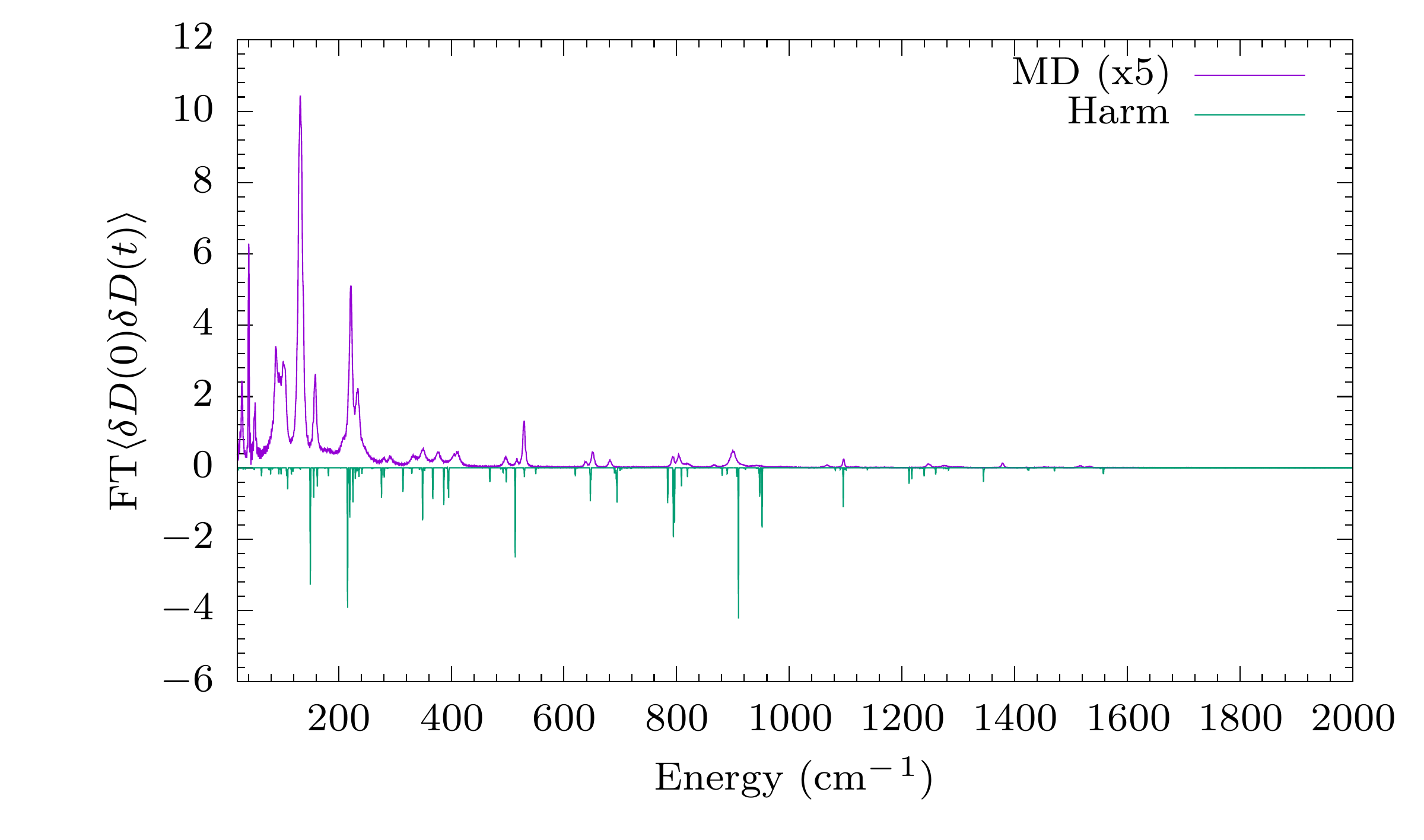}\\
     \includegraphics[scale=0.65]{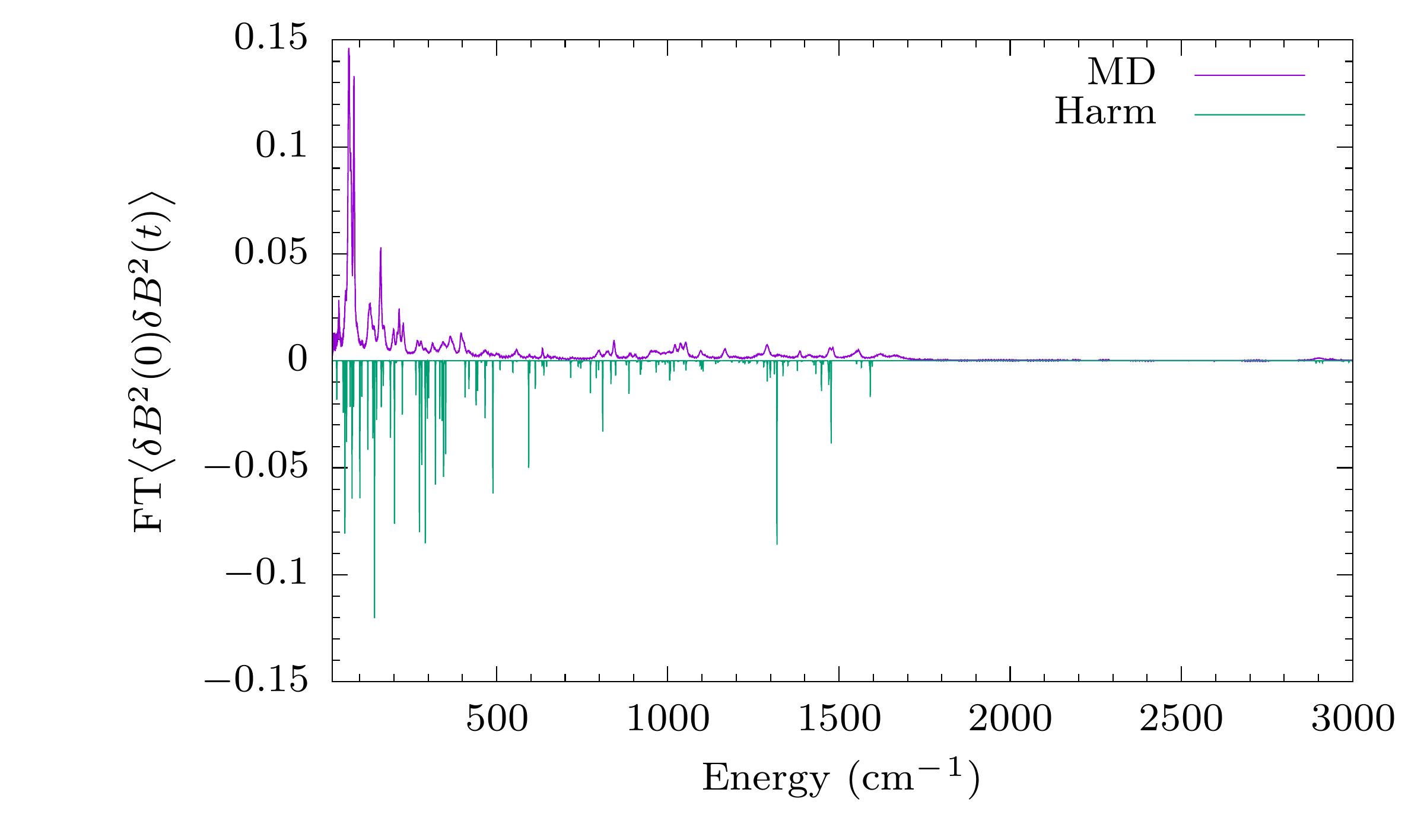}\\
     \includegraphics[scale=0.65]{Images_ESI/Corr_functions/Comp_1_100K/FT_final-1.png}\\
     \includegraphics[scale=0.65]{Images_ESI/Corr_functions/Comp_1_100K/FT_final-1.png}
     \caption{\textbf{Fourier transforms of the correlation functions at 100 K: comparison with the harmonic case}. Results for \textbf{1},\textbf{2} and \textbf{3} in order. For \textbf{3} the results for $l=2,4,6$ are reported, in order.}
 \end{figure}

\begin{table*} [t]
\caption{\textbf{RMSE on training set for energies/forces and tensorial properties for MD
trajectories at four different temperatures} The unit of measure is cm$^{-1}$. The results for \textbf{3} are reported for the different order of the tensor $l=2,4,6$, in order. The training set size (TSS) is reported for the AL on energy/forces and in parentheses for the spin Hamiltonian tensors.  }
\centering
\begin{tabular}{c |c c c c}
\toprule
\textbf{Comp}  \hspace{0.1cm} & \hspace{0.1cm} \textbf{TSS} \hspace{0.1cm}  & \hspace{0.1cm} \textbf{RMSE E } \hspace{0.1cm} & \hspace{0.1cm} \textbf{RMSE F} \hspace{0.1cm}  & \hspace{0.1cm} \textbf{RMSE D/B} \hspace{0.1cm} \\\hline
\midrule
 \textbf{1} & 196(144) & 0.30 & 0.87 & 3.62\\\hline
 \textbf{2} &  157 (338) & 0.38  & 1.68 & 2.23  \\\hline
\multirow{3}*{\textbf{3}} & \multirow{3}*{167 (161)}  & \multirow{3}*{0.30}  & \multirow{3}*{1.76}  &  0.26 \\
& &  &  & \num{2.5e-3} \\
& &  &  & \num{6.0e-5} \\ \hline
\midrule
\bottomrule 
\end{tabular}
\label{SI:training_set_RMSE_correlations}
\end{table*}

\begin{table*}[t]
\caption{\textbf{RMSE on test set for energies/forces and tensorial properties for MD trajectories at four different temperatures.} The RMSE for energies and forces is reported in kcal/mol/ and kcal/mol/\AA \, respectively; the RMSE for tensorial properties is reported in cm$^{-1} $. Temperatures of MD simulations are reported in K. 
%On the row corresponding to 25K, the training error is reported before the slash.
}
\centering
\begin{tabular}{c c| c c c c}
\toprule
\textbf{Compound}  \hspace{0.1cm}  &  \hspace{0.1cm}  \textbf{Temp} \hspace{0.1cm}  &  \hspace{0.1cm} \textbf{RMSE E} \hspace{0.1cm} & \hspace{0.1cm} \textbf{RMSE F} \hspace{0.1cm} & \hspace{0.1cm}  \textbf{RMSE D or B} \hspace{0.1cm}  \\\hline

\midrule
\multirow{4}*{\textbf{1}}& 25 & 0.07 & 0.58 & 1.49 \\
& 50& 0.23& 0.73 & 1.91\\
&75 &0.35 & 0.99 & 2.46 \\
& 100&  0.30 & 0.82 & 2.90 \\ \hline
\midrule
\multirow{4}*{\textbf{2}}&25 & 2.38& 1.17 & 0.97\\
&50 & 2.09 &1.33 &1.31 \\
& 75& 2.02 & 1.48&1.47\\
& 100& 1.96 &1.68 &1.90  \\ \hline
\midrule
\multirow{4}*{\textbf{3}}& 25  & 0.36 & 1.14 & 0.31 \,\num{2.7e-3} \,\num{5e-5} \\
& 50 &  0.44 & 1.60 & 0.50 \, \num{4.1e-3} \,\num{7.0e-5} \\
& 75 &  0.51& 1.93 & 0.62 \,\num{4.6e-3} \,\num{9.7e-5} \\
& 100 &  0.57 & 2.24 & 0.74 \,\num{5.9e-3} \,\num{1.2e-4}  \\ \hline
%\multirow{4}*{\textbf{4}}& 25 &\multirow{4}*{} & & &   \\
%& 50 & & & &  \\
%& 75 & & & &  \\
%& 100& & & &  \\ \hline
\bottomrule\\
\label{SI:test_set_RMSE_correlation}
\end{tabular}
\end{table*}

%\end{document}

\end{document}